\documentclass[epj]{svjour}
\usepackage{latexsym}
\usepackage{graphicx}
\usepackage{color}
\begin{document}
\title{Bond orientational ordering in liquids: Towards a unified description of water-like anomalies, 
liquid-liquid transition, glass transition, and crystallization}
\subtitle{Bond orientational ordering in liquids}
\author{Hajime Tanaka \thanks{e-mail: tanaka@iis.u-tokyo.ac.jp}
}                     
%
%
\institute{Institute of Industrial Science, University of Tokyo, 4-6-1 Komaba, Meguro-ku, Tokyo 153-8505, Japan}
\date{Received: June 22, 2012}
%
\abstract{
There are at least three fundamental states of matter, depending upon temperature and 
pressure: gas, liquid, and solid (crystal). These states are separated by first-order phase 
transitions between them. In both gas and liquid phases the complete translational and rotational 
symmetry exist, whereas in a solid phase both symmetries are broken. 
In intermediate phases between liquid and solid, which include liquid crystal and plastic crystal phases, 
only one of the two symmetries is preserved. Among the fundamental states of matter, the liquid state is most poorly understood. 
We argue that it is crucial for a better understanding of liquid to recognize that 
a liquid generally has a tendency to have local structural order and its presence 
is intrinsic and universal to any liquid. 
Such structural ordering is a consequence of many body correlations, more specifically, bond angle correlations, 
which we believe are crucial for the description of the liquid state. 
We show that this physical picture may naturally explain 
difficult unsolved problems associated with 
the liquid state, such as anomalies of water-type 
liquids (water, Si, Ge, ...), liquid-liquid transition, liquid-glass transition, crystallization and quasicrystal formation,  
in a unified manner.  
In other words, we need a new order parameter representing low local free-energy configuration, 
which is bond orientational order parameter in many cases, 
in addition to density order parameter for the physical description of these phenomena. 
Here we review our two-order-parameter model of liquid and consider 
how transient local structural ordering is linked to all of the above-mentioned phenomena. 
The relationship between these phenomena are also discussed. 
}
%

\maketitle

\sloppy

\section{Introduction}

We usually have an impression that 
liquid is in a completely random disordered state 
and has perfect translational and rotational symmetry. 
This may be true at a high-temperature or low-density limit near the gas-liquid transition. 
However, this picture is not necessarily correct. 
We have accumulated a number of evidence for local structural ordering 
in liquid. This is particularly well-known for liquids such as water and Si \cite{DebenedettiB} 
and metallic liquids \cite{Frank}. 
For example, a liquid often exhibits a shoulder around the main peak of the scattering function 
and/or a prepeak in a low $k$ side of the main peak, which are direct signatures of local structural ordering. 
Numerical simulations provide even more direct evidence for the presence of 
local (or mesoscopic) structural ordering and allow us to study the details of such a structure and its lifetime. 
Recently confocal microscopy also enables us to experimentally access the structure and dynamics 
of a colloidal suspension at a single-particle level \cite{Kegel,Weitz,schall2007,gasser,hunter2011physics}. 
Thus, we can even directly reveal such long-lived local structural ordering experimentally in colloidal gels and glasses \cite{PaddyNM}. 
Such a signature of local structural ordering is more pronounced at lower temperatures 
below the melting point $T_{\rm m}$ of an equilibrium crystal. 
However, it was also shown that such local structural order 
can exist even above $T_{\rm m}$, i.e., in an equilibrium liquid state. 
Thus, such ordering should not be regarded as something specific 
to a supercooled state or a glassy state. 
We also emphasize that this conclusion should be general and `not' restricted to some 
special families of liquids \cite{TanakaLLT}. 
For example, even hard spheres, which interact with the simplest interaction, have such a tendency of local structuring 
at a high density in order to lower the free energy, or to increase the entropy of a system `locally'  
\cite{TanakaNM,TanakaJSP,TanakaNara,MathieuNM,karayiannis2011,karayiannis2011s}: medium-range crystal-like bond orientational 
order and local icosahedral order, which compete with each other because of the 
mismatch in symmetry between them. 
Thus, we may say that local structural ordering, or  
the formation of locally favoured structures, is  
intrinsic and generic to a liquid state of any material 
\cite{TanakaLJPCM,TanakaGJPCM,TanakaLLT}. 
This forces us to change 
a simplified picture that liquid is in a perfectly disordered homogeneous state.

This view is not necessarily new and in particular a tendency of liquid to form local structural order 
by directional bondings such as covalent and hydrogen bondings  
has been recognized for a long time. For example, water has been known to form local 
tetrahedral order stabilized by hydrogen bonding \cite{Eisenberg,StillingerS,SoperR}. 
This was already recognized by R\"ontgen \cite{Rontgen}, which led him to the famous mixture model 
of water. Many molecular liquids are also expected to form some local structures 
stabilized by hydrogen bonding or by van der Waals interactions, although the details of such structures are difficult to 
figure out experimentally. 
This is due to (i) the difficulty to extract a non-periodic local structure from the spatial correlation of the scalar density field in the wave number ($k$) space, (ii) its vibrational distortion, and (iii) its short lifetime. 
Thus, it has not been widely recognized that the local structuring tendency is a universal key feature of liquids. 

Apparently, an atomic liquid, which has no obvious internal degrees of freedom, 
looks one of the simplest liquids. However, it has been known since a seminal work by Frank \cite{Frank} 
that for metallic liquids, in which atoms interact approximately by the 
Lennard-Jones potential, icosahedral order is favoured locally. 
This feature is further enhanced by the covalent nature of bonding between metals. 
Furthermore, some atomic liquids exhibit much more complex behaviours, because of anisotropic electronic 
interactions reflecting the symmetry of the electronic wave functions. 
Thus, some atomic liquids can hardly be regarded as 
simple Lennard-Jones liquids. 
For example, semimetals (Sb, Bi, Te, Ga, $\dots$) and some group-IV 
elements (Si, Ge, $\dots$) are famous for a number of unusual behaviours 
in this regard \cite{TanakaWPRB}. 
It is widely known that in a liquid state of these elements and chalcogenides, atoms tend to form 
local structures by covalent bondings. 
For example, the thermodynamic anomalies of liquid Te was analysed successfully in terms of a mixture 
model by Tsuchiya \cite{Tsuchiya}.  
These liquids including water are often called network-forming liquids. 
However, we prefer to regard such ordering 
as the formation of locally favoured structures rather than network formation 
(see, e.g., ref. \cite{ShintaniNP} and also below). 

\begin{figure}
\begin{center}
\includegraphics[width=5cm]{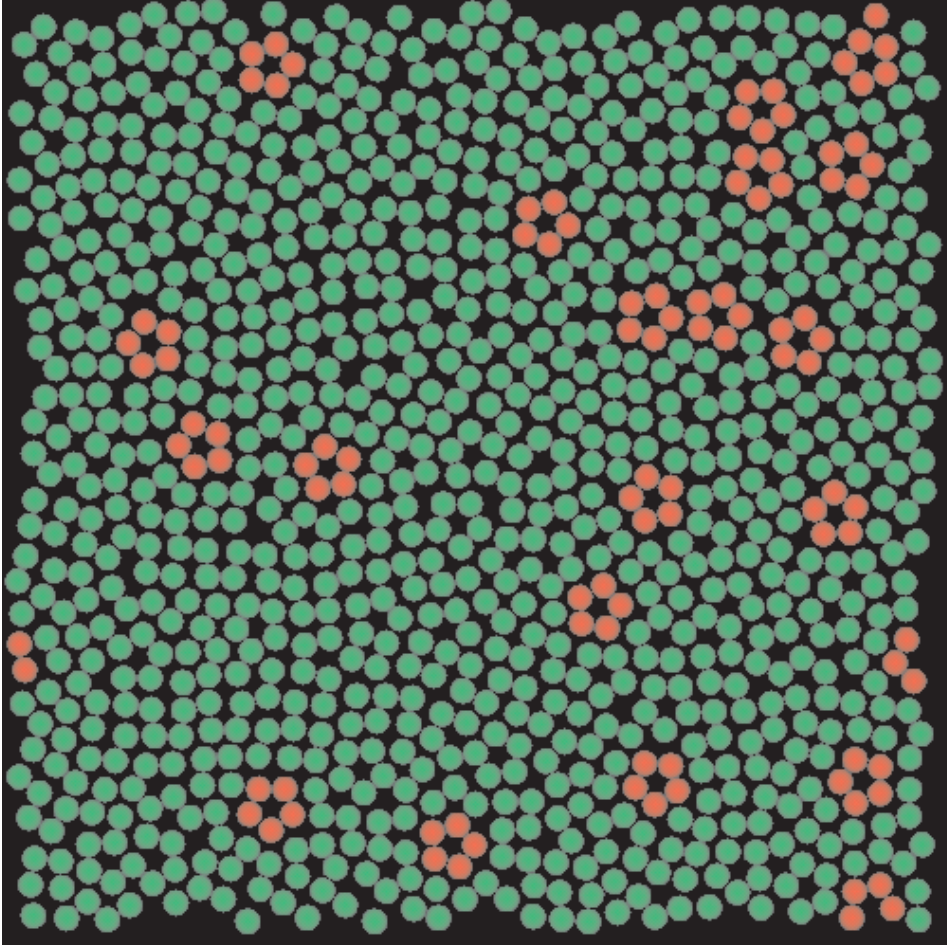}
\end{center}
\caption{(Colour on-line) Locally favoured structures (red pentagons) spontaneously formed 
in a sea of normal liquid structures. This is obtained by molecular dynamics 
simulations of spherical particles interacting with special anisotropic potential, 
which we call two-dimensional (2D) spin liquid \cite{ShintaniNP}. }
\label{fig:SRO}
\end{figure}

Despite much evidence for the presence of short-range ordering in liquid, 
it is not necessarily regarded as 
an intrinsic and universal feature of liquid until recently 
and the liquid state theory is basically described only by the scalar density field (in most cases by its 
pair correlation). 
The density functional theory and mode-coupling theory are successful theories along this line. 
Thus, it has been believed that the density order parameter can describe a gas-liquid transition, liquid-glass transition, 
and liquid-crystal transition which accompanies the break down of the translational symmetry.  
Thus, it was not so clear whether local structural ordering has important and fundamental roles in the behaviour of liquids or not. 
However, the above-mentioned examples clearly indicate that the state of a liquid cannot be described by 
two-body correlations of the scalar density field alone. 
Based on this recognition, some time ago we proposed that we need bond order parameter(s) 
representing local and mesoscopic structural ordering for the physical description of 
the liquid state, 
in addition to the scalar density order parameter \cite{TanakaLJPCM,TanakaGJPCM}. 
Our basic picture of the liquid state can be seen in fig. \ref{fig:SRO}, where 
pentagons (represented by red colour particles), which we call locally favoured structures, 
are created and annihilated in a sea of normal-liquid structures. 
In this 2D spin liquid, while further decreasing the temperature, crystal-like bond orientational order 
also develops and competes with locally favoured structures of five-fold symmetry (see below) \cite{ShintaniNP}. 
We proposed that bond orientational ordering also plays crucial roles under its coupling to 
density order parameter in various phenomena observed in a liquid state. 
Here we demonstrate that this two-order-parameter model may naturally describe 
water-like anomalies \cite{TanakaWEPL,TanakaWJCP,TanakaWPRB,TanakaWJPCM},  
liquid-liquid transition \cite{TanakaLJPCM,TanakaLLT,TKM,KuriSci,KuriButa}, 
glass transition \cite{TanakaGJCP1,TanakaGJCP2,ShintaniNP,STNM,TanakaGPRL,TanakaMJPCM,TanakaNM,TanakaJSP,TanakaNara}, 
crystal nucleation \cite{KTPNAS,Kawasaki3D,TanakaJSP,TanakaNara}, and quasicrystal formation \cite{TanakaMJPCM},  
within the same framework. 

In principle, we may also develop a theory by including appropriate many-body density correlations (three-body, four-body, $\cdots$) 
in addition to the two-body contribution, 
instead of introducing bond orientational order, but we think that the latter approach is physically and intuitively 
more appealing than the former approach.

As shown below, our model can explain these phenomena on an intuitive level as follows: 
\begin{itemize}
\item[(1)] Water-like thermodynamic anomaly 
of liquids is a result of the local ordering of bond order parameter 
\cite{TanakaWEPL,TanakaWJCP,TanakaWPRB}. 
\item[(2)] Liquid-liquid transition is a result of the gas-liquid-like cooperative 
ordering of bond order parameter (whereas a gas-liquid transition 
is that of density order parameter) \cite{TanakaLLT}.  
\item[(3)] Vitrification 
is a result of (i) competition between competing orderings, 
namely, between crystallization (long-range density and orientational ordering)  
and local bond ordering or (ii) random disorder effects on crystallization \cite{TanakaGJPCM,TanakaGPRL,TanakaMJPCM}.   
\item[(4)] Crystallization is initiated by the enhancement of the coherence of 
bond orientational ordering already developed in 
a supercooled liquid \cite{KTPNAS,Kawasaki3D,TanakaJSP,TanakaNara,russo2011,russoSM}, 
and `not' by density ordering. 
\end{itemize}
We also point out that a liquid having a strong tendency of short-range bond 
ordering may even achieve long-range bond ordering, if the local symmetry allows 
its growth to a (quasi-)long-range order. 
Such phenomena can be seen in (a) water and water-like tetrahedral 
liquids \cite{TanakaWPRB} and (b) metallic liquids \cite{TanakaMJPCM}. 
For case (a), a crystal having a larger specific volume than a liquid 
is formed (ice Ih in the case of water and the diamond-type crystal for Si), 
whereas for case (b) quasicrystal is formed. 
These ordered states can be viewed as results of long-range bond orientational ordering  
with and without periodicity, respectively. 

In the following, we mainly review our own works on the four topics (1)-(4) mentioned above, to draw a unified physical picture 
for these phenomena on the basis of the concept of spontaneous bond orientational ordering in a liquid. 
Thus the physical views on these topics may be highly biased. 
The more balanced views on these phenomena may be found in refs. \cite{MishimaR,AngellwaterR,DebenedettiR,DebenedettiB,SoperR}  on topic (1), refs. \cite{poole1997,DebenedettiB,MishimaR,mcmillan2007,Brazhkin} 
on topic (2), refs. \cite{AngellR,EdigerR,DebenedettiB,CavagnaR,BerthierR,GotzeB,berthier2011dynamical,binder2011glassy,das2011statistical} on topic (3), 
and refs. \cite{kelton2010,AuerR,SearR,GasserR,das2011statistical} on topic (4). 

The organization of this paper is as follows. 
In sec. 2 we describe our phenomenological two-order-parameter model of liquid. 
In sec. 3, we discuss thermodynamic and kinetic anomalies of water-type liquids. 
In sec. 4, we discuss liquid-liquid transition. In sec. 5, we discuss glass transition. 
In sec. \ref{sec:crystallization}, we discuss crystal nucleation assisted by medium-range crystal-like bond 
orientational order in a supercooled liquid. In sec. 7, we summarize our paper. 
The paper is organized so that each section can be read rather independently, but this results in some duplicated 
descriptions. 

\section{Phenomenological two-order-parameter model of liquid}

\subsection{Overview}
The standard liquid state theory has been developed on the basis of 
an ideal homogeneous liquid, and thus a random disordered structure has been assumed. 
This is the basis for the description of liquid by the two-body 
density correlator, or the radial distribution function $g(r)$. 
Recent studies indicate that this picture is not sufficient even for a hard-sphere 
liquid \cite{TanakaNM,TanakaJSP,TanakaNara}. The assumption of a homogeneous disordered structure of a liquid 
is always correct as the zero-th order approximation. 
However, we believe that a physical description beyond this is prerequisite for understanding 
unsolved fundamental problems in a liquid state, which include thermodynamic and kinetic 
anomalies of water-type liquids, liquid-liquid transition, 
liquid-glass transition, and crystal nucleation. 

As described above, 
a liquid is in a disordered state in the long range,  
but it can possess short-range and/or medium-range bond order. 
Such temporal bond orderings are induced to gain correlational entropy 
(e.g., for hard spheres) and/or by specific (often directional) energetic interactions 
between atoms or molecules that have the symmetry-selective 
nature. The latter may stem from interactions such as hydrogen and covalent bondings. 
The most well-known examples of local bond order are a tetrahedral structure 
for water, silicon, silica, and germania and an icosahedral structure 
for metallic glass formers. However, it should be noted that even hard-sphere liquids can possess 
local icosahedral order as well as medium-range crystal-like bond orientational order 
to gain the total entropy under a competition between correlational and 
configurational entropy \cite{TanakaJSP,TanakaNara,TanakaNM,Kawasaki3D}. 

\begin{figure}
\begin{center}
\includegraphics[width=7cm]{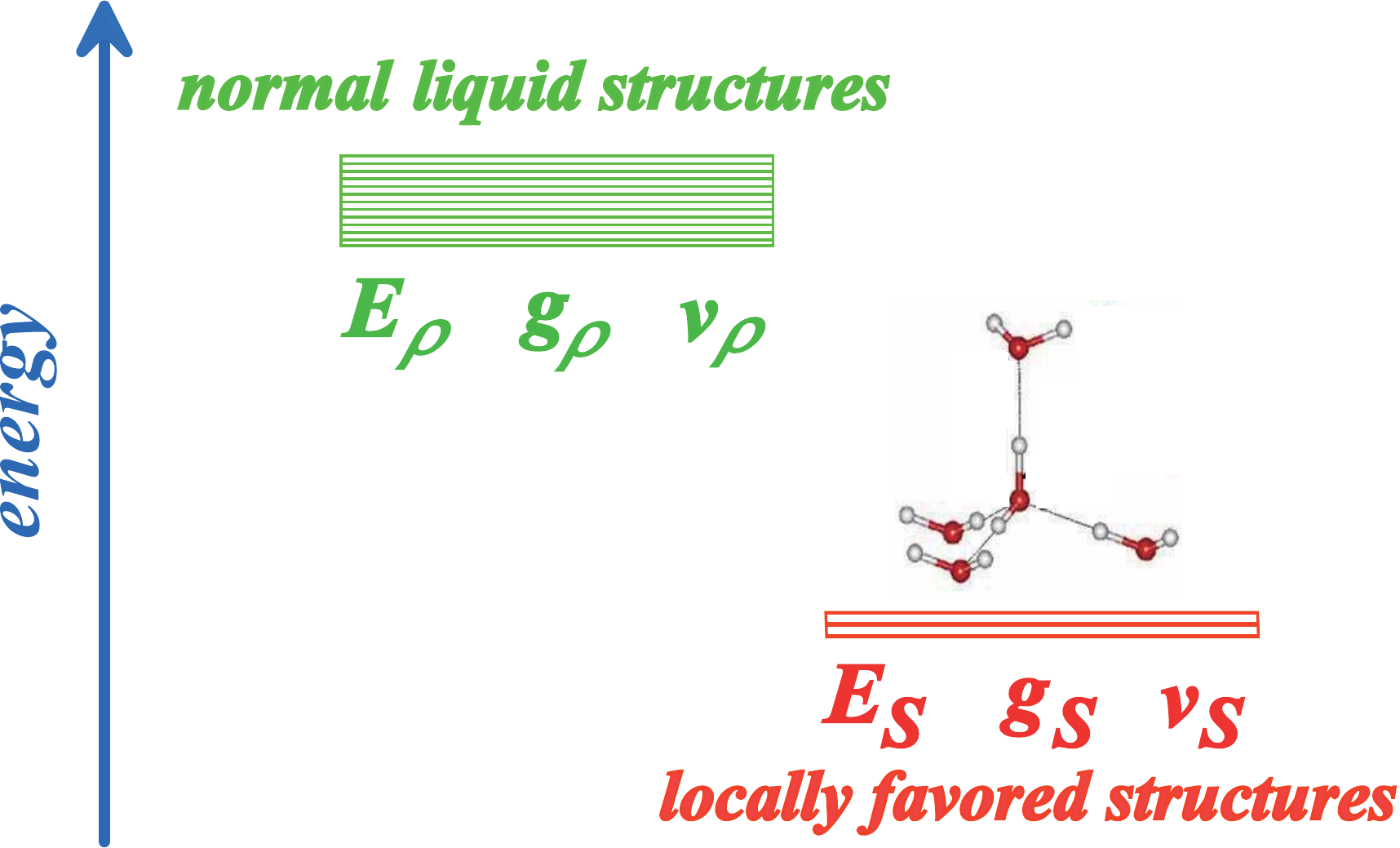}
\end{center}
  \caption{(Colour on-line) A two-state model for a liquid: One is normal-liquid structures 
  (energy $E_\rho$, degeneracy $g_\rho$, and specific volume $v_\rho$)
  and the other is locally favoured structures 
   (energy $E_S$, degeneracy $g_S$, and specific volume $v_S$). 
   For some liquids, there may be 
  more than two distinct energy states. }
  \label{twostate}
\end{figure}

As in the above case of hard spheres, there can be two types of bond orientational orderings, 
one of which is associated with local structural ordering and the other with medium-range crystal-like 
bond orientational order. On the basis of this physical picture, 
we express a liquid state by a simple two- (or multi-) state model 
with cooperativity of such bond orderings (see fig. \ref{twostate}). 
The first two-state model of liquid-liquid transition (LLT) was developed by Str\"assler 
and Kittel \cite{Kittel} and used by Rapoport \cite{Rapoport} to explain melting-curve maxima of 
atomic liquids, such as carbon, at high pressure. In these models, only short-range ordering was considered. 
Some time ago, we generalized this basic idea by introducing the bond order parameter(s) 
in addition to the density order parameter, and proposed the two- (or multi-) order-parameter model of liquid 
to explain not only LLT, but also water-like anomalies, liquid-glass transition and crystallization. 
Below we present a general framework of our model of liquid to describe these phenomena. 
We also show how these phenomena, which are apparently independent of each other, can be closely related to each other.

\subsection{Local bond ordering associated with 
the formation of locally favoured structures:Basis for water anomalies and liquid-liquid transition}
\label{sec:defineBOO}

First we focus on short-range bond orientational ordering. 
Our model \cite{TanakaWEPL,TanakaWJPCM,TanakaWJCP,TanakaWPRB,TanakaLLT,TanakaGJPCM} relies on 
a physical picture (see fig. \ref{fig:SRO}) 
that (i) there exist distinct locally favoured structures 
in a liquid and (ii) such structures are formed in a sea of normal liquid structures and 
its number density increases upon cooling since they 
are energetically (entropically for hard spheres) more favourable by $\Delta E$ than 
normal liquid structures: $\Delta E=E_\rho-E_S$ (see fig. \ref{twostate}), where 
$E_i$ is the energy of state $i$ ($i=\rho$ or $S$). 
Here normal liquid structures simply mean the background normal liquid structures.  
The specific volume and the entropy are larger 
and smaller for the former than for the latter, respectively, 
by $\Delta v=v_S-v_\rho$ and $\Delta \sigma=k_{\rm B} \ln (g_\rho/g_S)$. 
Here $v_i$ and $g_i$ are, respectively, the specific volume and the degree of the degeneracy of state $i$ ($i=\rho$ or $S$).
$\Delta v$ can be either positive or negative depending upon a system, 
whereas $\Delta \sigma$ is positive except for purely repulsive systems such as a hard-sphere liquid, where the gain of correlational entropy 
is the driving force of local structural ordering. 
We identify locally favoured structures as a minimum structural 
unit (symmetry element). 
It is tetrahedral order for water-type liquids, whereas  
icosahedron for metallic liquids \cite{Frank,tomida1995} and hard spheres \cite{MathieuNM,karayiannis2011,karayiannis2011s}. 
To express such short-range bond ordering in liquids, we introduce 
the so-called bond-orientational order parameter $Q_{lm}$ \cite{Steinhardt,NelsonB,sadoc1999}. 

Bond orientational order can be expressed by the distribution 
of bonds jointing a particle located at $\vec{r}$ 
to its nearest neighbours \cite{NelsonB}. Expanding the density 
$\rho(\vec{r}, \omega)$ of points pierced by these bonds 
on a small sphere inscribed about $\vec{r}$, we have 
\cite{NelsonB}
\begin{eqnarray}
\rho(\vec{r},\Omega)=\sum_{l=0}^{\infty}\sum_{m=-l}^{m=l} 
q_{lm}(\vec{r}) Y_{lm}(\Omega),
\end{eqnarray}
where the $Y_{lm}(\Omega)$ are spherical harmonics. 

We take the normalized average of $q_{lm}$ 
over a small volume located at $\vec{r}$, 
which we express by $\bar{q}_{lm}(\vec{r})$. Then, 
its rotationally invariant combination  can be defined as 
\begin{equation}
q_l(\vec{r})=[\frac{4 \pi}{2l+1} \sum_{m=-l}^{l}
|\bar{q}_{lm}(\vec{r})|^2]^{1/2}. 
\end{equation}
We can use the fraction of atoms (or particles) having $q_l(\vec{r})$ 
higher than a certain threshold value as the local bond order parameter 
$S$ (note that $S$ is ``not'' entropy and instead $\sigma$ represents entropy throughout this paper). 
If the two-state picture is correct, there should be a clear threshold value 
separating the two states.  
Note that $l=6$ for icosahedron \cite{Steinhardt}, whereas 
$l=3$ for tetrahedron \cite{Wang_Si}. 
For tetrahedrality, we can also define a more specific order parameter \cite{Chau,Errington}:
\begin{eqnarray}
q_{\rm tetra}=1-\frac{3}{8} \sum_{j=1}^{3} \sum_{k=j+1}^{4} \left(\cos \Psi_{jk}+\frac{1}{3}\right)^2. \nonumber 
\end{eqnarray}
In the case of water, $\Psi_{jk}$ is the angle formed by the lines joining the oxygen atom of a given water molecule and
those of its nearest neighbours $j$ and $k$. 

Here we also define other quantities characterizing bond orientational order.   
\begin{eqnarray}
		w_l =& \sum_{m_1+m_2+m_3=0} 
			\left( \begin{array}{ccc}
				\ell & \ell & \ell \\
				m_1 & m_2 & m_3 
			\end{array} \right)
			q_{l m_1} q_{l m_2} q_{l m_3}, \label{eq:wl}
\end{eqnarray}
Here the term in brackets in the above third-order rotational invariant is the Wigner 3-j symbol. 
Following \cite{lechner} we coarse-grain the tensorial bond orientaional order parameter over the neighbours:
\begin{equation}
	Q_{lm}(i) = \frac{1}{N_i+1}\left( q_{lm}(i) +  \sum_{j=0}^{N_i} q_{lm}(j)\right), 
	\label{eq:Qlm}
\end{equation}
and define coarse-grained invariants $Q_l$ and $W_l$ in the same way as the above. Structures with and without spatial extendability are then much easier to tell apart \cite{lechner}. We note that for non-extendable local structures like icosahedra, their $Q_l$ and $W_l$ are buried into the liquid distribution. 
In the following, we also use $Q_l$ to express $q_l$ unless we explicitly state $Q_l \equiv Q_l$. 

As can be seen above, both the scalar density field $\rho$ and the tensorial bond orientational 
order $\mbox{\boldmath$Q$}$ stem from the angle-dependent density field $\rho(\mbox{\boldmath$r$}, \Omega)$. 
With this orientational order parameter, the phenomenological liquid-state free energy functional associated with 
locally favoured structures 
is given by \cite{TanakaWEPL,TanakaWJPCM,TanakaWJCP,TanakaWPRB,TanakaLLT,TanakaGJPCM}
\begin{eqnarray}
f(S)&=&\int d \vec{r}\ [ -\Delta G 
S(\vec{r}) +J S(\vec{r}) 
(1-S(\vec{r})) \nonumber \\
&+&k_{\rm B} T (S(\vec{r})\ln S(\vec{r})+(1-S(\vec{r}))\ln (1-S(\vec{r}))) ],   \label{eq:fS}
\end{eqnarray}
where $\Delta G=\Delta E-T \Delta \sigma-\Delta v P$. $\Delta G$ is the free energy change 
associated with the formation of a locally favoured structure. $J$ represents 
the cooperativity, $k_{\rm B}$ is the Boltzmann constant, $T$ is the temperature, and 
$P$ is the pressure.

Next we consider density ordering, which describes crystallization 
\cite{TanakaGJPCM,TanakaGJCP1,TanakaGJCP2,TanakaJSP,TanakaNara}. 
Since we are interested only in equilibrium and supercooled liquid states, 
we do not consider a gas-liquid transition, which is also described by 
density ordering.

\subsection{Crystallization as cooperative translational and bond orientational ordering: Basis for glass transition and crystal nucleation} \label{sec:densitybond}

\subsubsection{Classical density functional theory of freezing}
\label{sec:DFT}

The free energy functional, denoted $F\{ \rho \}$, is
expanded functionally about a density, $\rho=\rho_l$, corresponding
to a liquid state lying on the liquidus line of the solid-liquid
coexistence phase diagram of a pure material. The expansion is performed in powers of $\delta \rho=\rho-\rho_l$. 
Then the free energy density can be written as \cite{ramakrishnan1979first,singh1991density} 
\begin{eqnarray}
\frac{F_x \{\rho \}}{k_{\rm B}T}= 
\int d\mbox{\boldmath$r$}\ [\rho(\mbox{\boldmath$r$})\ln (\frac{\rho(\mbox{\boldmath$r$})}{\rho_l})-\delta \rho(\mbox{\boldmath$r$})] \nonumber \\
+\sum_{n=2}^{\infty} \frac{1}{n \!} \prod_{i=1}^{n} \int d\mbox{\boldmath$r$}_i \delta \rho(\mbox{\boldmath$r$}_i)
C_n(\mbox{\boldmath$r$}_1,\mbox{\boldmath$r$}_2,\cdots), \label{eq:F_xrho}
\end{eqnarray}
where $F_x \{\rho \}$ is the free energy corresponding to the density $\rho(\mbox{\boldmath$r$})$ minus that at the constant density $\rho_l$, 
and the $C_n$ functions are $n$-point direct correlation functions of an isotropic fluid. Formally the correlation functions are 
defined by
\begin{eqnarray}
C_n(\mbox{\boldmath$r$}_1,\mbox{\boldmath$r$}_2,\cdots)=\frac{\delta^n \Phi}{\prod_{i=1}^{n} \delta \rho(\mbox{\boldmath$r$}_i)}, \label{eq:C_n}
\end{eqnarray}
where $\Phi$ represents the total potential energy of interactions between the particles in the system. 

The free energy of a liquid made of $m$ components can be simplified up to the second order as 
\begin{eqnarray}
\frac{F_x \{\rho \}}{k_{\rm B}T}= 
\sum_{i=1}^{m}\int d\mbox{\boldmath$r$}_i\ [\rho(\mbox{\boldmath$r$}_i)\ln \frac{\rho(\mbox{\boldmath$r$}_i)}{\rho_l}-\delta \rho(\mbox{\boldmath$r$}_i)] \nonumber \\
+\sum_{i,j} \int d\mbox{\boldmath$r$}_i d\mbox{\boldmath$r$}_j \delta \rho(\mbox{\boldmath$r$}_i)
C^{ij} \delta \rho(\mbox{\boldmath$r$}_i), \label{eq:F_x2}
\end{eqnarray}
where the sums are over the elements in a mixture, $\delta \rho_i=\rho_i-\rho^i_l$ and $\rho^i_l$
is the value of the number density of component
$i$ on the liquid-side of the liquid-solid coexistence line. The
function $C^{ij}$ is the two-point direct correlation function of 
between components $i$ and $j$ in an isotropic fluid. As in the
case of a single-component system, it can be assumed that 
$C^{ij}=C^{ij}(|\mbox{\boldmath$r$}_i-\mbox{\boldmath$r$}_j|)$. The next term in the expansion of eq. (\ref{eq:F_x2}) contains the
three-point correlation, the next after that, the four point, etc.
It should be noted that  these higher order correlations may be crucial for some systems. 

The free energy functional of a single-component system is considered in the limit that the series given in
eq. (\ref{eq:F_xrho}) can be truncated at $C_2=c$:
\begin{eqnarray}
F_x^2\{\rho \}=k_{\rm B}T 
\int d\mbox{\boldmath$r$}\ \rho(\mbox{\boldmath$r$})[\ln \frac{\rho(\mbox{\boldmath$r$})}{\rho_l}-1] \nonumber \\
+\int \int d\mbox{\boldmath$r$} d\mbox{\boldmath$r$}' \delta \rho(\mbox{\boldmath$r$})
c(\mbox{\boldmath$r$}-\mbox{\boldmath$r$}') \delta \rho(\mbox{\boldmath$r$}'). \label{eq:Fx}
\end{eqnarray}

To understand the basic features of this free energy 
functional it is useful to expand $F_x^2 \{\rho \}$ in terms of $\langle \delta \rho (\mbox{\boldmath$r$}) \rangle= \langle \rho (\mbox{\boldmath$r$}) \rangle -\rho_l$, 
where $\langle \rho (\mbox{\boldmath$r$}) \rangle$ is the locally averaged density. Below, we consider the symmetry selection upon freezing 
on the basis of this simple free energy.

\subsubsection{Alexander-McTague theory of liquid-solid transition}

To see the essence of density functional theories on solidification, we review the seminal argument 
by Alexander and McTague \cite{Alexander}, which puts a focus on the instability of 
density fluctuations around $k=k_0$ (see also \cite{Lubensky,sachdev1985order}). 
The reasonable approximation to the structure factor $\mathcal{S}(k)$ is to 
consider only the main maximum peak 
around $k=k_0$ and ignore minor peaks by assuming the following form: 
\begin{eqnarray}
\mathcal{S}(k)=\frac{k_{\rm B}T}{\tau+\kappa(k^2-k_0^2)}, \label{eq:sk}
\end{eqnarray}
where $\tau=a(T-T_m^\ast)$. 
We note that the ignorance of minor peaks throws away information on local structural ordering, 
which actually plays a key role as will be shown later. 
The temperature $T_m^\ast$ is the mean-field limit of stability of the 
liquid phase. 
Note that $\mathcal{S}(k)$ is the Fourier transformation of the two-point density correlator, 
\begin{eqnarray}
\mathcal{S}(\vec{r}, \vec{r}')=\langle \delta\rho(\vec{r}) \delta\rho(\vec{r}')\rangle. \nonumber
\end{eqnarray}
Since $\chi(\vec{r}, \vec{r}')=
k_{\rm B}T \mathcal{S}(\vec{r}, \vec{r}')$ is the functional 
derivative of the free energy with respect to the mean-field density fields 
$\langle \delta\rho(\vec{r})\rangle$ and 
$\langle \delta\rho(\vec{r}')\rangle$, a phenomenological free energy 
which gives eq. (\ref{eq:sk}) in the mean-field level is 
\begin{eqnarray}
F_\rho=\int d \vec{r} d \vec{r}' 
\langle \delta\rho(\vec{r})\rangle
\chi^{-1}(\vec{r}, \vec{r}')
\langle \delta\rho(\vec{r}')\rangle 
\nonumber \\
-w \int d \vec{r}\langle \delta\rho(\vec{r})\rangle^3
+u \int d \vec{r}\langle \delta\rho(\vec{r})\rangle^4, 
\label{eq:density}
\end{eqnarray}
where 
\begin{eqnarray}
c(\vec{r}, \vec{r}') \cong \chi^{-1}(\vec{r}, \vec{r}') =
[\tau+\kappa(\nabla^2+k_0^2)^2] \delta (\vec{r}
-\vec{r}'). \nonumber
\end{eqnarray}
Note that in general $c(\vec{r}, \vec{r}')=(a_0+a_1 \nabla^2+a_2 \nabla^4+ \cdots) \delta (\vec{r}-\vec{r}')$, 
where the gradients are with respect to $\vec{r}'$. 

Using the Fourier decomposition of eq. (\ref{eq:density}), 
we obtain
\begin{eqnarray}
f_\rho&=&\frac{F_\rho}{V}=\sum_{\vec{G}} \frac{1}{2}
\tau_{\vec{G}}|\rho_{\vec{G}}|^2 
\nonumber \\
&-&w(k_0) \sum_{\vec{G}_1, \vec{G}_2, 
\vec{G}_3}} \rho_{\vec{G}_1}
\rho_{\vec{G}_2} \rho_{\vec{G}_3}
\delta_{{\vec{G}_1+\vec{G}_2+\vec{G}_3,0}
\nonumber \\
&+&u(k_0) \sum_{\vec{G}_1, \vec{G}_2, 
\vec{G}_3, \vec{G}_4} \rho_{\vec{G}_1}
\rho_{\vec{G}_2} \rho_{\vec{G}_3} 
\rho_{\vec{G}_4}
\delta_{\vec{G}_1+\vec{G}_2+\vec{G}_3
+\vec{G}_4,0}, \nonumber
\end{eqnarray} 
where $\tau_{\vec{G}}=\tau+\kappa(G^2-k_0^2)^2$ and $V$ is the system volume. 
 
The existence of the third order term leads to the first-order transition, although the transition already has 
a fluctuation-induced first-order character. 
Here we note that the description of crystallization in the wavenumber space is motivated by a physical picture 
that crystallization is due to `translational ordering'. 

Alexander and McTague \cite{Alexander} 
assumed that the isotropic component dominates the quartic term. Under this assumption, the free energy 
$f_\rho$ is minimized by a set of $\rho_{\vec{G}_i}$ 
which maximizes the amplitude of the third order term. 
Then they considered ${\pm {\vec{G}}_i}$ parallel to the 
edges of a triangle, an octahedron (tetrahedron), and an icosahedron. 
We can add to this list a tetrahedral bi-pyramid and an idealized 
pentagonal bi-pyramid (see also \cite{Lubensky,sachdev1985order,Jaric}). 
These corresponds to 2D hexagonal lattice (2Dhex), 
bcc, icosahedral edge model (ieqc), 
3D hexagonal lattice (3Dhex), and idealized closed packing of tetrahedra 
(ideal), 
respectively. 
The result is $f_\rho^{bcc}<f_\rho^{3Dhex}<f_\rho^{ideal}
<f_\rho^{2Dhex}<f_\rho^{ieqc}$. 
This leads to the conclusion that bcc is most favoured whereas 
icosahedral edge model quasicrystalline ordering is least favoured. 

This conclusion is a direct consequence of the fact that the lowest order term 
of symmetry selective nature  
in the free energy [eq. (\ref{eq:density})] plays a major role in the 
earlier stage of growth of density fluctuations. 
However, we argue below that there may be
an additional important selection rule 
from the constraint of dense packing or directional bonding, under which crystallization 
usually takes place (except for a system interacting with 
rather long-range repulsions). 

For example, we need an additional mechanism stabilizing icosahedral order 
to explain the formation of icosahedral crystals. 
Introduction of bond orientational ordering is one natural resolution 
\cite{Steinhardt,Jaric}. Below we argue that bond orientational order 
may play a crucial role not only in quasicrystal formation, 
but also in crystallization and glass transition in general.

\subsubsection{Multiple types of bond orientational ordering in liquid}

In the above, we consider only translational ordering. 
Partly because translational ordering automatically accompanies orientational ordering, 
the importance of the latter has been overlooked in theories of solidification 
for a long time despite the recognition of its importance in 80s \cite{NelsonB,sadoc1999,Hess,Mitus,Mitus2,Mitus3,Haymet,Steinhardt,Jaric}.  
On the other hand, orientational order has often been used in simulations to detect crystal order (see, e.g., ref. \cite{tenWolde}). 
The liquid-solid transition accompanies the breakdown of both translational and rotational symmetry.  
Here we argue that orientaional order is crucial for our understanding of the state of liquid 
and the phenomena related to the liquid state such as glass transition and crystal nucleation. 

We stress that translational ordering is global in the sense 
it is described by periodic modulation with wave vectors  
$\vec{G}$, whereas bond orientational ordering can be defined locally 
around a particle. 
So their origins are essentially different. 
We emphasize that crystal nucleation is initiated from 
a small nucleus where translational order that is expressed as the Fourier components 
in the wave number space is not well developed yet. 
Obviously, directional bonding such as covalent and hydrogen bonding selects a special local symmetry, 
which can also be represented by bond orientational order. 
Bond orientational ordering also originates from the geometrical constraint 
from dense packing of disks or spheres which interact with hard-core 
interactions: the excluded volume effects.    
For 2D hard disk systems, for example, the most probable number of nearest neighbour 
particles is 6 and thus the relevant bond orientational order is 
represented by hexatic order parameter. 
This example tells us that bond orientational order 
should always play an important role in a densely packed state, 
in which crystallization usually takes place.

We argue that the relevant order parameters for describing glass transition 
is $\mbox{\boldmath$Q$}_{\rm CRY}$ 
whose symmetry is consistent with the equilibrium crystal, and bond orientational order $\mbox{\boldmath$Q$}_{\rm LFS}$ 
which has a symmetry of locally favoured structures. 
$\mbox{\boldmath$Q$}_{\rm CRY}$ and $\mbox{\boldmath$Q$}_{\rm LFS}$ are driven by parts of  
interparticle interactions compatible to the equilibrium crystal 
and those incompatible to it, respectively. 
In general, there can be more than two bond order parameters relevant to crystallization 
and vitrification, but for simplicity we stick to the simplest case. 

Bond orientational order parameters phenomenologically represent many-body particle correlations 
in a natural manner. Many-body interactions beyond pair correlations are key to the description 
of liquid (at low temperatures), where both structure and dynamics are ``strongly correlated''.  
Furthermore, as will be shown in sec. \ref{sec:crystallization}, 
it is bond orientational order and not translational order that triggers crystal nucleation. 
Thus, frustration or random disorder effects against crystallization, which lead to vitrification, 
should act primarily on crystal-like bond orientational order. 
In other words, the disturbance on crystal-like bond orientational order is enough to make 
a barrier for crystal nucleation high and avoid crystallization upon cooling.   

\begin{figure}
\begin{center}
\includegraphics[width=6cm]{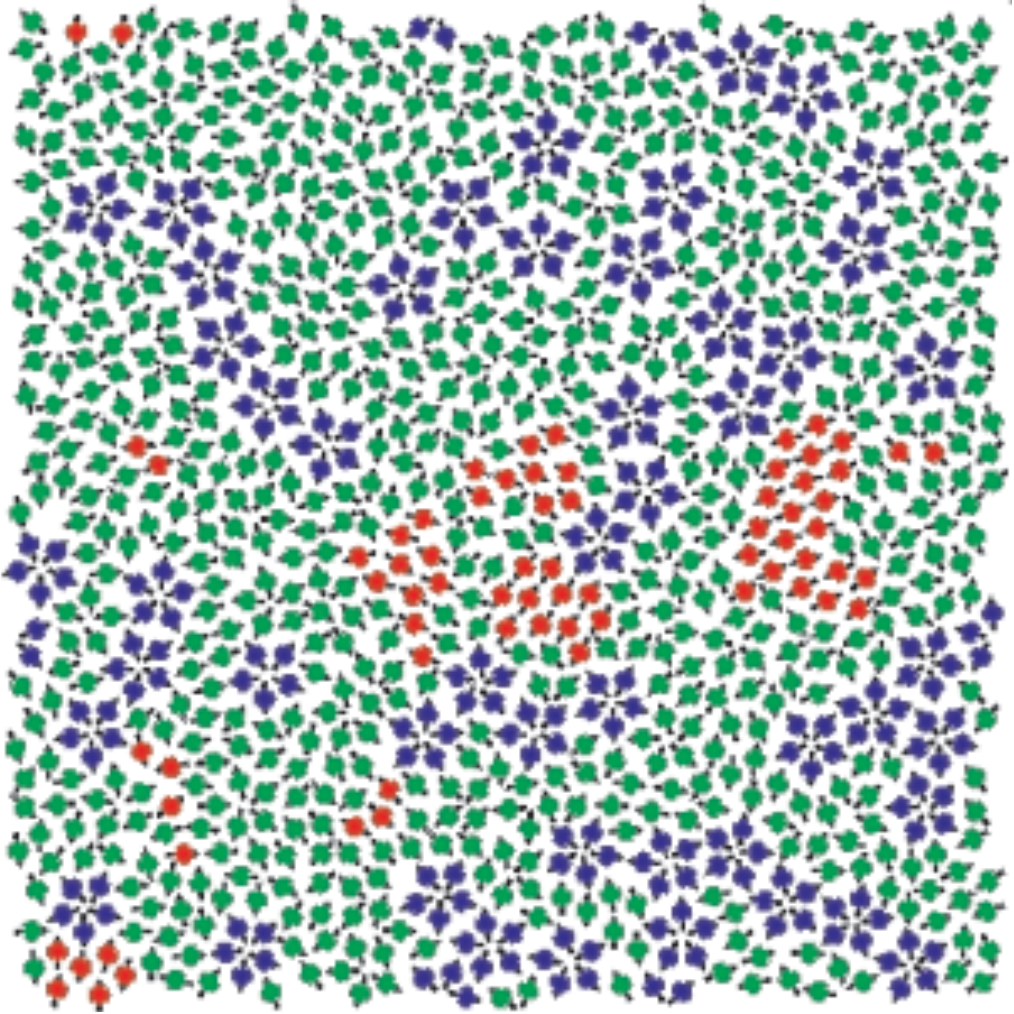}
\end{center}
\caption{(Colour on-line) A snapshot of 2D spin liquid in a supercooled state. 
Red particles have crystal-like bond orientational ordering 
(more specifically, antiferromagnetic order), which 
plays a crucial role in glass transition and crystallization, whereas 
blue particles are locally favoured structures with pentagonal symmetry, 
which plays a primary role in water-type anomalies and liquid-liquid transition. 
The latter also plays an important role in vitrification if it competes 
with crystallization, which is linked to the above bond orientational ordering (appeared red). 
Spin on a particle is also shown by an arrow in this figure. 
}
\label{fig:KTTP}
\end{figure}

Here we mention the difference in the nature of local structural ordering 
between (i) water-like anomalies and liquid-liquid transition and (ii)
liquid-glass transition and crystallization, which will be discussed in more detail 
later.  
This can be clearly seen in a snapshot of our molecular dynamics 
simulation (see fig. \ref{fig:KTTP}). 
In water-like anomalies and liquid-liquid transition, the 
formation of long-lived locally favoured structures, which can be expressed 
by $\mbox{\boldmath$Q$}_{\rm LFS}$, plays crucial roles. 
Here we note that for water-like liquids at ambient pressure $\mbox{\boldmath$Q$}_{\rm LFS} \cong \mbox{\boldmath$Q$}_{\rm CRY}$. 
The order parameter can be defined as the number density of 
locally favoured structures, which expresses the change in the `quantity'. 
More precisely, it is the rotationally invariant scalar parameter made from 
$\mbox{\boldmath$Q$}_{\rm LFS}$. 
In our view,  water-like anomalies is a consequence of the 
change in this scalar order parameter itself and its direct coupling to 
the thermodynamic quantities and transport coefficient, whereas liquid-liquid transition is a consequence of gas-liquid-like 
transition of the order parameter due to its cooperativity. 
In glass transition and crystallization, on the other hand, medium-range structural ordering, 
which is expressed by the tensorial order parameter $\mbox{\boldmath$Q$}_{\rm CRY}$, plays important roles under frustration with $\mbox{\boldmath$Q$}_{\rm LFS}$ or under random disorder effects. 
In this case, the order parameter 
is the degree of bond orientational order, which expresses the change 
in the `quality' rather than the `quantity'. 
The development of bond orientational order is not only responsible for 
slow dynamics associated with glass transition, but also plays 
an important role in helping crystal nucleation under its coupling to 
the density field. 

\subsubsection{Possible roles of bond orientational ordering in crystallization, quasicrystal formation, 
and glass transition}

The importance of bond orientational 
ordering was pointed out for crystal ordering 
\cite{Hess,Mitus,Mitus2,Mitus3} as well as for orientational 
ordering \cite{Haymet,Steinhardt,Jaric}.  
Nelson and Toner \cite{Toner} also considered a possible existence of residual 
bond angle order analogous to that found in a two-dimensional hexatic phase. 
They pointed out that a bulk phase with bond orientational order 
(``cubic'' liquid crystal) might be observable in supercooled liquids. 
For 3D crystallization, however, there now seems to be a consensus that 
bond orientational ordering is merely a consequence of translational ordering 
and does not play a crucial role in crystallization. 
Thus crystallization in 3D has usually been described by the density 
functional theory in which the only relevant order parameter is `scalar density'. 

To illustrate the importance of bond orientational ordering, here we consider the problem of 
quasicrystal formation as an example. 
Since icosahedral ordering is least favoured according to the above-mentioned Alexander-McTague argument, 
the question here is what physical mechanism makes icosahedral ordering 
more favourable than other crystalline structures. 
There were a few approaches to the theoretical description of 
quasicrystal formation from the standpoint of density ordering. 
Bak \cite{Bak}, Mermin and Troian \cite{Mermin}, 
and Kalguin et al. \cite{Kalugin} considered this problem on the basis of 
the Landau theory of solidification formulated by Alexander 
and McTague \cite{Alexander}. 
In order to bypass the original conclusion that a bcc structure should be 
generally favoured (see above), 
they either include higher-order terms or an additional 
component to the density. 

Unlike these approaches, Jar\'ic \cite{Jaric} 
proposed to introduce a bond orientational order parameter which tends to 
stabilize the quasicrystalline phase. He provided an interesting view of 
quasicrystal formation as an interplay between orientational order 
and translational order parameters (see also refs. \cite{Toner,Steinhardt}). 

We argue that theories of crystallization solely based on translational 
ordering miss important physical constraints coming from excluded volume effects under dense packing 
and/or directional interactions, both of which lead to local bond orientational ordering. 
This is particularly important in the nucleation stage of crystal, where there is no well-developed 
translational order yet. 
We propose to describe crystallization by combining 
two types of orderings, positional 
ordering and bond orientational ordering. 
Densely packed spherical particles with the same size 
usually possesses 12 nearest neighbours in 3D. 
So natural bond orientational order should be associated with face-centred-cubic (fcc), 
hexagonal-closed-packed (hcp), and 
icosahedral (ico) symmetries (see below).  
The fcc and hcp order is represented by a 
combination of $Q_{6m}$ and $Q_{4m}$, whereas ico order is by $Q_{6m}$ 
ordering (involving no $Q_{4m}$ ordering). 

Now we consider couplings between orderings of $\rho$ and 
$\vec{Q}$.  
The lowest order coupling between $\vec{Q}$ and $\rho$ 
should be given by the rotationally and translationally invariant energy \cite{Steinhardt,Jaric}: 
\begin{eqnarray}
F_{\rm int}=\int dq \ \sum_{l,m} \alpha_l(q) 
\int d^2\hat{q} \ Q_{lm} Y_{lm}^\ast (\hat{\vec{q}}) 
\rho (\hat{\vec{q}}) \rho (-\hat{\vec{q}}).  \label{eq:F_int}
\end{eqnarray}
Up to the lowest order, $\rho$ is not coupled linearly to 
$\vec{Q}$, and 
$\rho(\vec{q})\rho(-\vec{q})$ is coupled to it. 
Accordingly, the equilibrium $\rho$ need not have 
the symmetry of the equilibrium $\vec{Q}$. 
This particular type of coupling leads to an asymmetric coupling between 
the orderings. 
If the translational ordering temperature $T_\rho$ is higher than the bond 
orientational ordering temperature $T_Q$ then, because the 
$\vec{Q}$-$\rho$ interaction is linear in $\vec{Q}$, 
the ordering of $\rho$ at $T_\rho$ will necessarily induce an ordering 
in $\vec{Q}$. 
This seems to justify the theory based on the density field alone, but which may not be 
so as we will see later. 
On the other hand, if $T_Q>T_\rho$ then, because 
the $\vec{Q}$-$\rho$ interaction is quadratic in $\rho$, 
the ordering of $\vec{Q}$ at $T_Q$ will have the effect of 
renormalizing the quadratic coupling without necessarily inducing an 
ordering of $\rho$. Jar\'ic proposed that this case of $T_Q>T_\rho$ should  
correspond to quasicrystal formation \cite{Jaric}. 

In ordinary 3D crystallization, $T_\rho>T_Q$. 
The free energy of the equilibrium crystal itself can be expressed by the scalar density (or, translational) 
order parameter alone. However, it does not necessarily mean that bond orientational order 
does not play any role in crystallization. Rather it plays a crucial role in a supercooled state 
as well as in the process of crystal nucleation (see sec. \ref{sec:crystallization}). Below we explain how this becomes possible. 
Due to a rather strong first-order nature of liquid-solid transition, a system can enter 
into a long-lived metastable state, where a liquid-glass transition can 
take place if the temperature is cooled fast enough for a system 
to bypass crystallization. In this branch, thus, we may practically forget the 
ordering of $\rho$ even for $T<T_\rho$ until crystal nucleation starts. 
Because of strong first-order nature of the translational ordering, 
there is little growth of density fluctuations, whose amplitude is basically determined by the isothermal compressibility $K_T$. 
So the only remaining ordering is that of 
$\vec{Q}$. For 3D hard spheres, for example, bond orientational ordering towards fcc/hcp develops 
and competes with that towards ico, which leads to frustration \cite{TanakaMJPCM,TanakaGJNCS,ShintaniNP}. 
A form of the Landau-type free energy associated with tensorial bond orientational ordering with translational and rotational invariance 
can be found, e.g., in refs. \cite{Jaric,nelson1984symmetry,sachdev1985order}. 
For simplicity (see also the speculative reasoning below), 
we consider the following free energy form associated with (scalar-like) 
bond orientational ordering $\vec{Q}$:
\begin{eqnarray}
F_Q=\int d\vec{r} \left( ct \vec{Q}^2 
+I_3(\vec{Q}) +O(\vec{Q}^4) 
+\frac{1}{2} K (|\nabla \vec{Q}|)^2 \right)+\cdots \nonumber
\end{eqnarray}
where $c$ is a positive constant, $t$ is the reduced 
temperature $\tilde{t}=1/\phi-1/\phi_0^b$ (or $\tilde{t}=T-T_0^b$), 
and $\cdots$ represents frustration originating from competing bond 
orientational orderings (e.g., $\mbox{\boldmath$Q$}_{\rm CRY}$ vs $\mbox{\boldmath$Q$}_{\rm LFS}$), 
internal frustration (this is the case for icosahedral order \cite{Steinhardt}), and/or random disorder effects. 
Here $\mbox{\boldmath$Q$}_{\rm CRY}$ is compatible to the symmetry of the equilibrium 
crystal, whereas $\mbox{\boldmath$Q$}_{\rm LFS}$ is incompatible to it, as described above.  
Here $\phi_0^b$ (or $T_0^b$) is the bare transition volume fraction (or temperature). 
Even though the third order invariant $I_3$ is suggestive of the first 
order nature of the transition, the transition 
might be almost continuous. 
Frustration effects originating from competing 
$\mbox{\boldmath$Q$}_{\rm CRY}$ and $\mbox{\boldmath$Q$}_{\rm LFS}$ orderings and/or random disorder effects 
due to polydispersity may change the nature of the transition 
from a continuous (characteristic to a tensorial order parameter) 
to a discrete Ising symmetry (characteristic to a scalar order parameter) 
\cite{TanakaNM}. 
We speculate that renormalization of frustration effects changes 
the symmetry of the transition from the continuous to the discrete Ising 
symmetry and also shifts the critical point from  
$\phi_0^b$ (or $T_0^b$) to $\phi_0$ (or $T_0$), although this should be carefully checked. 
We also emphasize that frustration effects may not only change the type of ordering, but also 
lead to exotic critical phenomena accompanying the growing activation energy towards the hypothetical critical point 
(see sec. \ref{sec:activation}).

The total free energy $F_{\rm total}$ may then be given by the sum of 
translational ordering, local and global orientational ordering, and 
the couplings between them: 
\begin{equation}
F_{\rm total}=F_\rho+F_Q+F_{\rm int}. \label{eq:F_total}
\end{equation}
In the above, however, we need a special care for avoiding double counting. This may be done with 
a proper projection procedure. 

Here we note that there are new important effects 
of bond orientational ordering, which have so far not been considered 
in describing liquid-glass transition: 
(1) thermodynamic effects of short-range bond ordering, which can be considered 
on the basis of the simpler free energy $f_S$ (see eq. (\ref{eq:fS})), 
(2) random field effects of $\mbox{\boldmath$Q$}$ (or, $S(\vec{r})$) on crystallization 
(long-range translational and bond orientational ordering), 
and (3) long-range crystalline (or quasicrystal) ordering 
consistent with the symmetry of $S$ order. 

Later, we consider problems of thermodynamic and kinetic anomalies  
of water-type liquids, liquid-liquid transition, liquid-glass 
transition, and crystallization, focusing on these three effects (1)-(3). 

\subsubsection{Case of hard spheres} 
To illustrate the importance of bond orientational ordering, 
here we briefly discuss why fcc, hcp and ico 
bond orientational orderings are important in a supercooled 
hard-sphere-like liquid state \cite{TanakaNM,Kawasaki3D,KTPNAS,TanakaJSP,TanakaNara}. 
As described above, the driving force of bond orientational ordering 
is the constraint due to dense packing for hard spheres. 
So a minimal structural unit is 
a particle and its 6 nearest neighbour particles for 2D, whereas 
a particle and its 12 nearest neighbour particles for 3D. 
For 2D, it is obvious that hexatic order is the only key bond order 
parameter. For 3D, it is not so obvious as for 2D. 
However, isolated structures composed of one central 
and 12 surrounding particles are only fcc, hcp, and icosahedral packings, 
which are the most probable candidates of 
preferred bond orientational order in a supercooled hard-sphere liquid. 
The fcc and hcp order can extend in space without any internal 
frustration, whereas ico intrinsically suffers from internal geometrical frustration 
\cite{NelsonB,Steinhardt,sadoc1999} 
since icosahedron cannot grow further and fill up the space. 
Furthermore, there is an intrinsic symmetry mismatch between crystal-like 
bond orientational order (fcc and hcp) and icosahedral order, which can be a source of 
frustration against crystallization.  

Here we stress that bcc symmetry provides only 8 
nearest neighbour particles and thus it should not be favoured under 
a constraint of dense packing, although it is favoured by the cubic 
term in the Landau expansion of the density. 
Indeed, bcc crystal is not stable for hard spheres.  
We propose that this is the reason why bcc structures are not seen 
in hard sphere systems, in contrast to the prediction of the 
Alexander-McTague theory \cite{Alexander}. Although hard sphere systems might be out of the range of 
the applicability of this theory, we can say that this feature coming from local symmetry selection due to packing constraint is 
not taken into account properly in theories based on translational ordering alone. 

The constraint from dense packing becomes weaker for systems 
of softer interactions. This may explain why bcc structures 
are more often seen in systems of softer interactions 
(see, e.g., \cite{likos2006soft,likos,Likos2,AuerC}).

\subsection{A few pieces of evidence supporting our scenario of crystallization}
\subsubsection{Importance of bond orientational order in solidification of 
two-dimensional (2D) systems: Its implication on 3D crystallization} 

To illustrate important roles of bond orientational order in crystallization, 
we consider liquid-solid transition in 2D systems. 
For 2D hard disks, it is widely accepted that liquid-solid transition 
sequentially takes place in the order of bond orientational ordering 
and translational ordering upon densification. 
This is known as the Kosteritz-Thoules-Halperin-Nelson-Young scenario 
\cite{NelsonB}. 
The nature of the transition (second order or first order) is still a matter of debate, but this does not 
affect the importance of bond orientational order itself. 
We note that a recent simulation study by Bernard and Krauth \cite{krauth} 
provides the following answer to this long standing problem: 
The transition from the liquid to hexatic phase is actually weak first-order transition whereas 
that from the hexatic to solid phase is continuous.  
Bond orientational order can be expressed by the distribution 
of bonds jointing a particle located at $\vec{r}$ 
to its nearest neighbours \cite{NelsonB}. 
For 2D hard disk systems, 
the relevant bond orientational order is 
the (Mermin) hexatic order parameter, as described above. 

This example of 2D ordering tells us that bond orientational order 
should always play an essential role in a densely packed state, 
in which crystallization usually occurs. 
3D systems should not be exceptional. 

As mentioned above, we argue that in ordinary 3D systems, the ordering point of bond orientational order 
is located below that of translational order and thus the former is hidden behind the latter 
\cite{TanakaJSP,TanakaNara}. 
This is a consequence of the fact that translational ordering automatically accompanies bond orientational ordering. 
However, when we consider structural fluctuations in a liquid and/or the kinetic process of crystal nucleation, bond orientational order always plays 
an important role as will be shown later. 
Quasicrystal may be an exception for the above-mentioned order of the two types of ordering: For this case, the ordering point of bond orientational order 
is located above that of translational order, as mentioned above, and thus long-range bond orientational ordering takes 
place before translational ordering takes place \cite{Jaric,TanakaMJPCM,TanakaJSP,TanakaNara}. 
  
\subsubsection{Plastic crystals, liquid crystals, and quasicrystals}

The importance of (bond) orientational ordering is also supported by the presence 
of plastic crystals and liquid crystals as intermediate phases between a liquid and a crystal phase. 
Plastic crystals have positional order, but without orientational order, whereas 
liquid crystals have orientational order, but without positional order. 
This indicates that the two types of orderings both play crucial roles in the 
ordering from an isotropic liquid to an ordinary crystal.  
So it is natural to consider that both positional and bond orientational ordering 
play crucial roles in a liquid-crystal transition in general.  
As discussed above, the presence of quasicrystals can also be naturally explained by the two order parameter model. 

\subsection{General importance of local structural ordering in liquid} 
Our picture is based on the recognition of the general importance of local structural ordering in liquid. 
The gas phase does not have any structural order. The solid phase has extended translational order, 
for which $k$-space (Fourier-space) analysis and representation are very useful. 
We argue, on the other hand, that local or mesoscopic structural ordering is essential for 
the liquid phase, which has so far not been taken seriously as the important general feature of liquid. 
With a decrease in temperature, starting from a random disorder gas state, 
local order appears in a liquid state and then mesoscopic order develops towards the lower stability limit of the liquid phase.  
Eventually extended translatioal order develops upon crystallization. 
The liquid-crystal transition accompanies the breakdown of both translational and rotational symmetry. 
We note that local and mesoscopic order developing in a liquid is associated with orientational order and not with 
translational order. 

Such local or mesoscopic order is rather difficult to access by scattering measurements because of its local nature 
as well as its orientational (non-translational) nature. 
Although real-space analysis is very powerful, it is not applicable to most of ordinary liquids. 
This may be one of the reasons why the understanding of the liquid phase has been far behind that of the gas and solid 
phase. Direct measurements of bond orientational order in real liquids are highly desirable. 
Recent development of X-ray cross correlation spectroscopy may be promising from this respect \cite{cross1,cross2}.  

\section{Thermodynamic and kinetic anomalies of water-type liquids}

Liquid water exhibits unusual thermodynamic behaviour, which is very much different from that of 
ordinary liquids \cite{Eisenberg,MishimaR,DebenedettiB,DebenedettiR,AngellR,SoperR}. 
The most striking anomaly is the decrease of the density $\rho$ upon 
its freezing at 0 $^\circ$C and the density maximum at 4 $^\circ$C. 
Isothermal compressibility $K_T$ and heat capacity at constant pressure 
$C_P$ also show unusual temperature dependences. Both quantities steeply increase 
on cooling. 
In addition to the thermodynamic anomaly, the viscosity $\eta$ also shows 
anomalous non-Arrhenius behaviour. 
Furthermore, at low temperatures $\eta$ decreases with an increase in 
pressure up to 2 kbar, which is markedly different from the behaviour of ordinary liquids, whose 
$\eta$ monotonically increases with pressure. 

These anomalous thermodynamic and dynamic behaviour of water 
has extensively been studied both experimentally and theoretically 
for a long time. 
Many models of water have been proposed to explain the water's anomaly, 
focusing on the unique features of hydrogen bonding. 
Furthermore, various concepts have been proposed to explain the anomalous 
behaviour of water, focusing on both the thermodynamic anomaly 
and the low-temperature phase behaviour of liquid water 
\cite{Eisenberg,MishimaR,DebenedettiB,DebenedettiR,AngellR,SoperR}: 
(a) a stability-limit conjecture, \cite{SpeedyAngell}, 
(b) a second-critical-point scenario (see e.g., refs. \cite{MishimaR,Ponyatovsky,anisimov2011}), and 
(c) a singularity-free scenario \cite{Stanley0,SastryW}.  
Scenario (a) assumes the existence of a retracting spinodal curve and 
attributes the thermodynamic anomaly to proximity to the spinodal 
curve. Scenario (b), on the other hand,  
assumes the existence of a line of first-order transitions between 
two types of liquid water (low-density and high-density water), 
terminating at a metastable critical point, and attributed 
the thermodynamic anomaly to critical phenomena associated 
with the hidden critical point. Finally, scenario (c) 
predicts that the thermodynamic quantities exhibit extrema but 
no divergence. Scenario (a) and (b) predict the divergence of the thermodynamic 
quantities due to the thermodynamic singularity. 
In scenario (a) and (b), the anomaly of the thermodynamic and 
dynamic quantities has often been analysed with assuming the power-law 
divergence, $\epsilon^{-\gamma}$, where $\epsilon=(T-T_{\rm s})/T_{\rm s}$ 
($T_{\rm s}$: mean-field spinodal temperature) and $\gamma$ is a critical 
exponent, and found to be well described by it. 
However, it should be noted that the critical exponents are often treated as 
adjustable parameters and no hyperscaling relations between the exponents 
have been found so far, unlike the case of the typical critical 
phenomena. Furthermore, we cannot approach to the mean-field spinodal temperature 
$T_{\rm s}$ so closely because homogeneous nucleation of ice crystals takes place far above $T_{\rm s}$.  
Thus, $\epsilon >0.05$ in most cases and there has been no convincing evidence 
of the divergence of the thermodynamic quantities at a critical temperature. 
We note that the thermodynamics of water 
has recently been studied in detail on the basis of critical phenomena \cite{anisimov2011}. 
Finally, scenario (c) predicts no divergence of the physical quantities.  

Focusing on the temperature dependence of hydrogen bonding, 
many rather complicated functional forms have been proposed 
to describe the anomalous behaviour of the thermodynamic quantities.   
Despite these considerable efforts, however, there has so far been 
no consensus on 
which of these three types of scenario is primarily responsible 
for the above-described anomaly of water or whether we need a new scenario 
or not \cite{DebenedettiR}.  
The only consensus is the importance of hydrogen bonding. 
This situation is partly due to the lack of a `intuitively appealing' physical 
model of water, which provides a `simple' analytical prediction for 
the water's anomaly. 
Here we explain how the thermodynamic and kinetic anomalies of water-type liquids can be explained 
in the framework of our two-order-parameter model. 

\subsection{What makes water-type liquids so different from ordinary liquids?}
\begin{figure}
\begin{center}
\includegraphics[width=7cm]{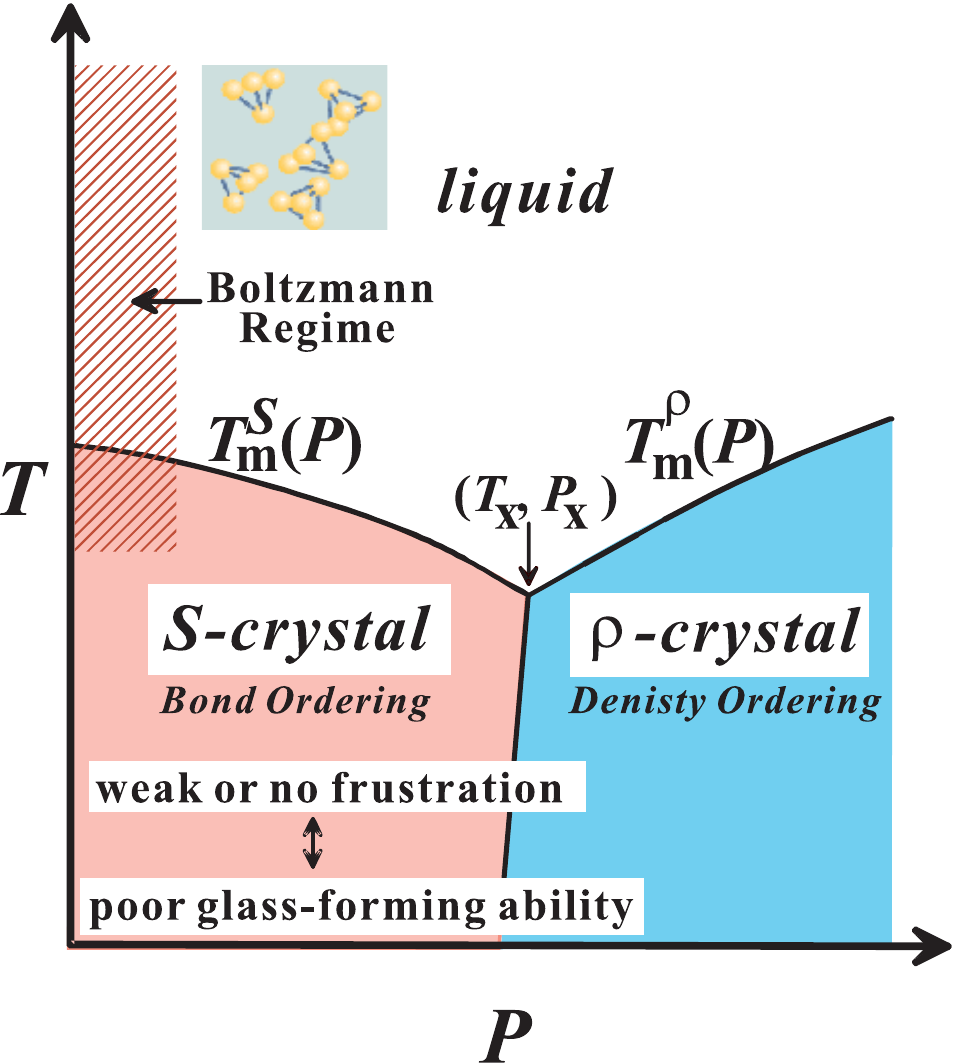}
\end{center}
  \caption{(Colour on-line) $P$-$T$ phase diagram of water-type liquids including 
  water itself and water-type atomic liquids (Si, Ge, Bi, Sb, and Ga). }
  \label{fig:PD}
\end{figure}

    First we consider what makes water so special among `molecular' liquids. 
We pointed out \cite{TanakaWEPL,TanakaWJCP,TanakaWPRB} that water is the only  
molecular liquid, for which local bond orientational ordering is 
compatible with a global crystallographic 
symmetry: The locally favoured tetrahedral structure of water stabilized by hydrogen bonding is 
consistent with the crystallographic symmetry of hexagonal ice Ih and cubic ice Ic. 
It is important to recognize that formation of a tetrahedral structure stabilizes hydrogen bonding with a help 
of local symmetry in a ``cooperative'' manner. 
We argue that all the 
thermodynamic anomalies of water originate from (i) this dominance of bond orientational 
ordering below a crossover pressure $P_{\rm x}$ ($\sim$2 kbar), where the melting 
point of ice crystals has a minimum, and (ii) an unusually large positive value of 
$\Delta v$. Below $P_{\rm x}$, the crystallization 
is due to bond ordering, while above $P_{\rm x}$ it is due to density ordering 
as in ordinary liquids (see fig. \ref{fig:PD}). 
This gives a natural explanation for the unusual pressure dependence 
of the melting point of ice crystals, including its minimum 
around 2 kbar. 
We propose that ice Ih is $S$-crystal, long-range 
ordering of $S$, while high-pressure ices 
are $\rho$-crystals \cite{TanakaWEPL,TanakaWJCP,TanakaWPRB}.  
The V-shaped $T$-$P$ phase diagram of water-type liquids is just a manifestation of 
the Clausius-Clapeyron relation. 

By using this specific shape of the phase diagram as a fingerprint \cite{TanakaWPRB}, 
we classified five elements Si, Ge,  
Sb, Bi, and Ga into water-type atomic liquids. Similarly, some group III-V (e.g., InSb, GaAS, and GaP) and
II-VI compounds (e.g., HgTe, CdTe, and CdSe) can also be classified into water-type liquids. 
As described below, our model provides us with simple analytical predictions for the 
thermodynamic and dynamic anomalies of these water-type liquids \cite{TanakaWPRB}.

\subsection{Thermodynamic anomalies of liquid water}
\begin{figure}
\begin{center}
\includegraphics[width=7cm]{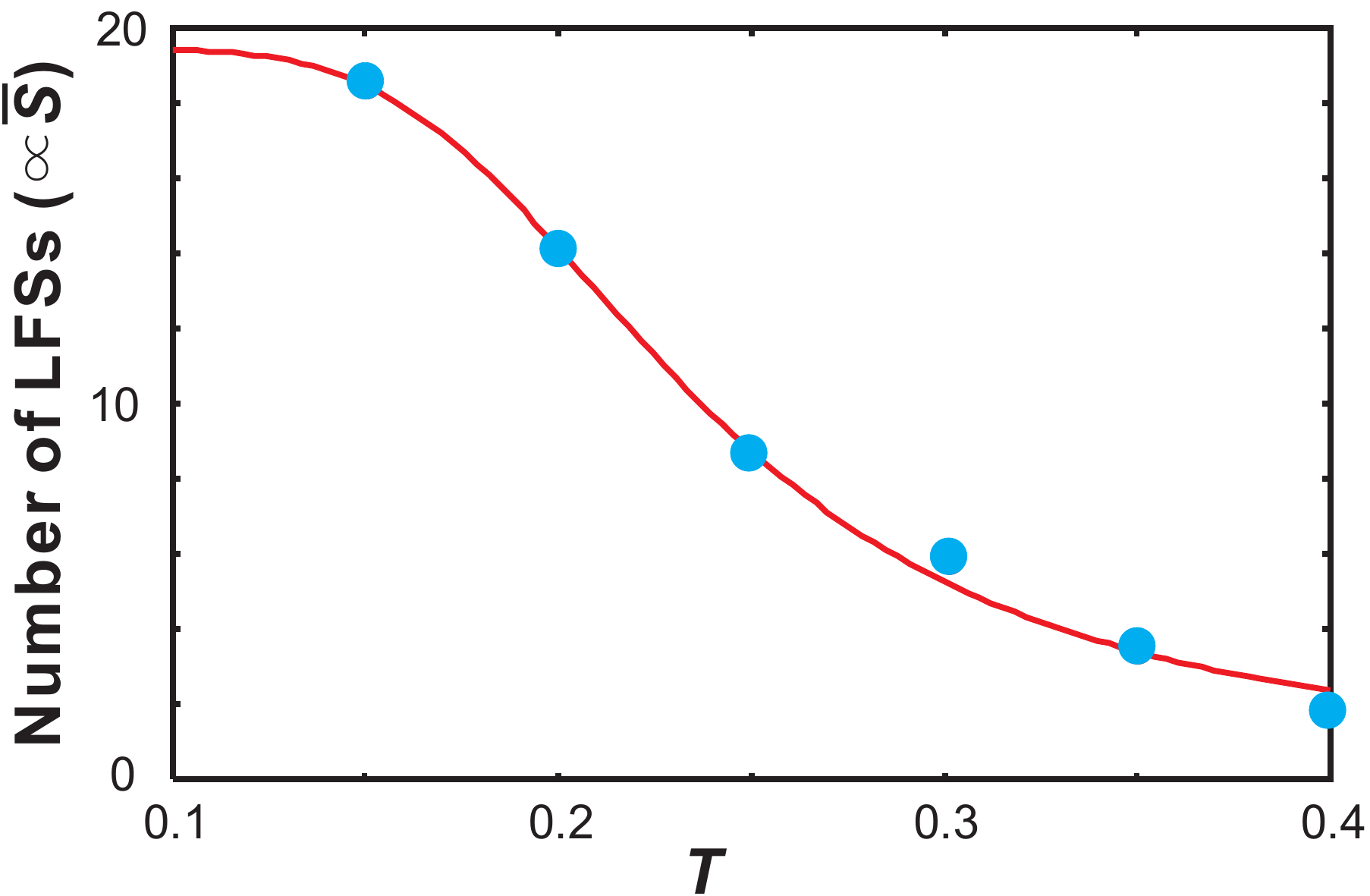}
\end{center}
\caption{(Colour on-line) The number of locally favoured structures (pentagons in fig. 1) as a function of 
the temperature $T$ in our numerical simulations 
 of spherical particles interacting with special anisotropic potential 
 for $\Delta E$=0.2 (on the details, see \cite{ShintaniNP}). 
 We call this liquid ``2D spin liquid''. 
The solid curve is the fitting of eq. (\ref{S_2}). The agreement is satisfactory, 
showing the relevance of the two-state model. 
}
\label{fig:SRO_T}
\end{figure}

Here we consider a simple two-state model of liquid, which corresponds 
to the limit of weak cooperativity in our two-order-parameter model \cite{TanakaWEPL,TanakaWJCP}. 
We first estimate how the average fraction 
of locally favoured structures, $\bar{S}$, increases with 
a decrease in $T$.  
From the condition $\partial f(S)/\partial S=0$ (see eq. (\ref{eq:fS})), 
$\bar{S}$ can be obtained as 
\begin{eqnarray}
\bar{S} = \frac{\frac{g_S}{g_\rho} \exp(\beta (\Delta E-P\Delta v))}{
1+\frac{g_S}{g_\rho} \exp(\beta (\Delta E -P\Delta v))}, \label{S_2}
\end{eqnarray}
where  $\beta=1/k_{\rm B}T$. 
In the above, we assume $J=0$ for simplicity. 
For $J \neq 0$, the cooperativity plays an important role in inducing a  
liquid-liquid transition \cite{TanakaLJPCM,TanakaLLT} (see above). 
Here $\Delta E=E_\rho-E_S$ and $\Delta v=v_S-v_\rho$. 
$E_i$ and $g_i$ are the energy level and the 
number of degenerate states of $i$-type structure, respectively. 
$i=\rho$ corresponds to normal liquid structures of water, while 
$i=S$ to locally favoured structures (see fig. \ref{twostate}). 
The validity of eq. (\ref{S_2}) was confirmed by our numerical simulations 
 of 2D spin liquid (spherical particles interacting with special anisotropic potential 
 favouring pentagon geometry; see fig. \ref{fig:SRO}) 
\cite{ShintaniNP} (see fig. \ref{fig:SRO_T}). 
In this system, $S$ is defined as the number density of the pentagon structures. 

The uniqueness of a locally favoured structure and 
the existence of many possible configurations for normal liquid 
structures lead to the relation $g_\rho \gg g_S$. 
Then, $\bar{S}$ can further be approximated as 
\begin{eqnarray}
\bar{S} \sim \frac{g_S}{g_\rho} \exp[\beta (\Delta E-\Delta v P)], \label{eq:S}
\end{eqnarray}
We stress that this relation should hold even for a non-zero $J$ 
if $\bar{S} \ll 1$ \cite{TanakaWEPL,TanakaWJCP}. 

Here it is worth noting that the above prediction of our two state model on the behaviour of $S$ 
was supported by numerical simulation studies of SPC/E water model by Appignanesi et al. 
\cite{Appignanesi2009,Appignanesi2}. 
It was quite difficult to unambiguously determine both the presence and 
the fraction of each kind of water “species” in terms of numerical simulations. 
They overcome this difficulty by combining a local structure index with potential-energy minimisations. 
Namely, they combined the characterization of tetrahedral structural order parameter  with the identification 
of the inherent structure.  
This allowed them to estimate the fraction of the locally favoured structures as a function 
of temperature.

According to the above picture, the unusual decrease in $\rho$ upon cooling 
below 4 $^\circ$C in water can simply be explained by an 
increase in the number density of locally favoured structures, $\bar{S}$, 
upon cooling. 
The specific volume $v_{sp}$ and the density $\rho$ are, 
respectively, given by 
\begin{eqnarray}
v_{\rm sp}(T,P)&=& v_{\rm sp}^{\rm B}(T,P)+\Delta v \bar{S}, \label{vsp} \\
\rho(T,P) &\sim& \rho_{\rm B}(T,P)-\rho_{\rm B}(T,P) \frac{\Delta v}{v_{\rm sp}} \bar{S}, 
\label{density}
\end{eqnarray}
where $\rho_{\rm B}(T,P)=M/v_{\rm sp}^{\rm B}(T,P)$ ($M$: molar mass). 
Note that the background contributions $v_{\rm sp}^{\rm B}$ and $\rho_{\rm B}$ depend almost linearly on $T$ 
as for those of ordinary liquids.  
Then, $K_T=-\frac{1}{v_{\rm sp}}(\frac{\partial v_{\rm sp}}{\partial P})_T $ 
can straightforwardly be calculated from eq. (\ref{vsp}) as 
\begin{eqnarray}
K_T&=&-\frac{1}{v_{\rm sp}}(\frac{\partial v_{\rm sp}^{\rm B}}{\partial P})_T- 
\frac{1}{v_{\rm sp}}(\frac{\partial \Delta v}{\partial P})_T \bar{S} \nonumber \\ 
&+&\beta \frac{\Delta v^2}{v_{\rm sp}}\frac{ \bar{S}}{1+\frac{g_S}{g_\rho}\exp(\beta(\Delta E-P\Delta v))}.  
\label{KT}
\end{eqnarray}
For a case of $\bar{S} \ll 1$, this relation can further be simplified as 
\begin{equation}
K_T=-\frac{1}{v_{\rm sp}}(\frac{\partial v_{\rm sp}^{\rm B}}{\partial P})_T+ 
\frac{1}{v_{\rm sp}}[-(\frac{\partial \Delta v}{\partial P})_T +\beta \Delta v^2] \bar{S}.  
\end{equation}
The anomalous increase of $K_T$ upon cooling can thus be explained 
by the following two mechanisms: (a) A decrease in $T$ increases 
the population of locally favoured structures, 
which may be softer than normal liquid structures. 
(b) More importantly, the ability (or the degree of freedom) of the transformation 
from locally favoured structures to 
normal liquid structures upon a pressure increase provides softness to a system. 
With an increase in pressure, the anomaly of $K_T$ 
upon cooling becomes weaker, reflecting the decrease in the population 
of locally favoured structures, $\bar{S}$. 

\begin{figure}[t]
\begin{center}
\includegraphics[width=7cm]{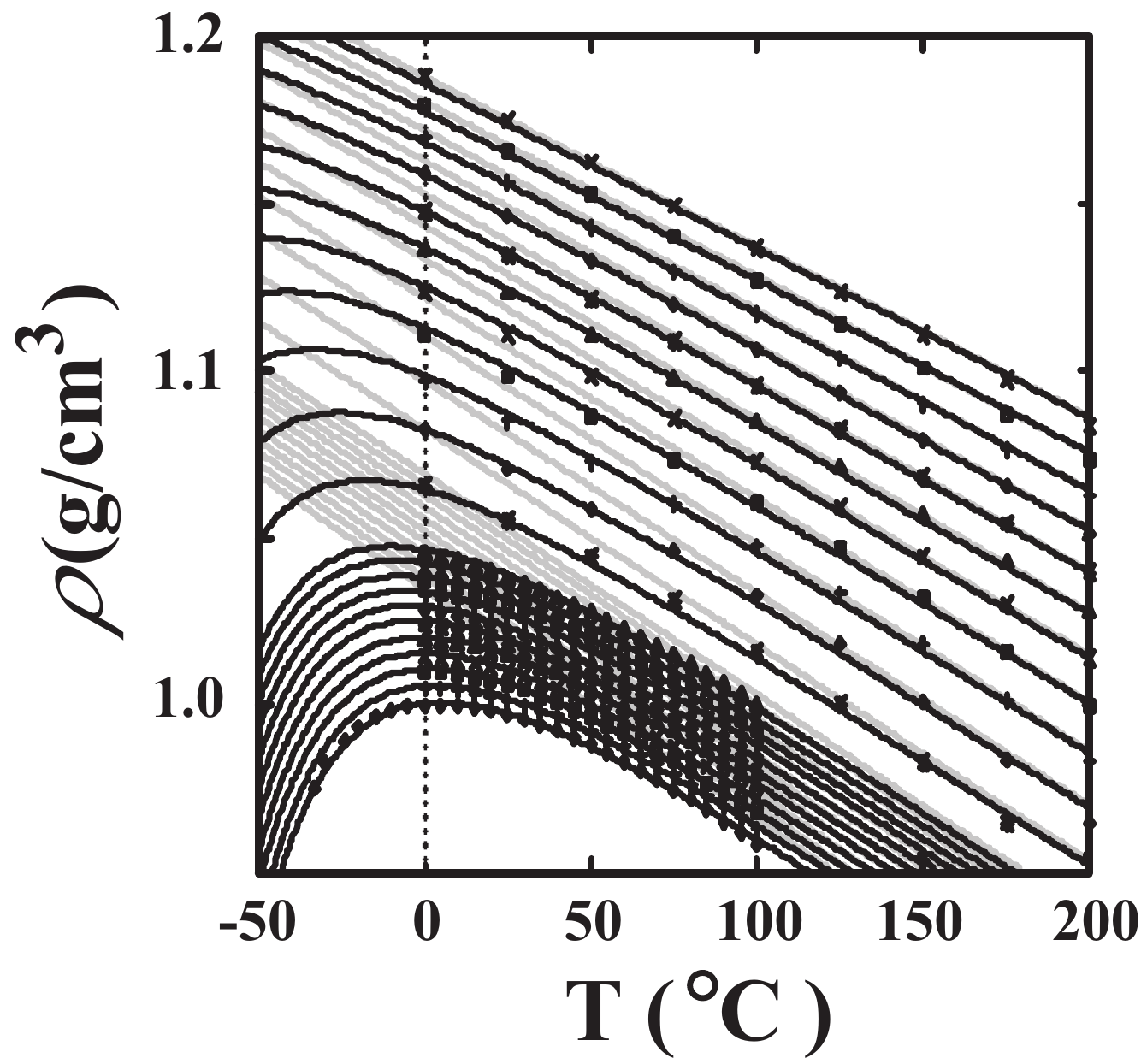}
\end{center}
\caption{
Temperature dependence of density for various pressures. 
The data sets correspond to $P$=1, 100, 200, 300, 400, 500, 
600, 700, 800, 900, 1000, 1500, 2000, 2500, 3000, 3500, 
4000, 4500, 5000, 5500, 6000, and 6500 bar, from bottom to top. 
The solid curves are the theoretical fittings, while 
the gray curves are the background parts.  
This figure is reproduced from fig. 3 of \cite{TanakaWJCP}. 
} 
\label{fig:density}
\end{figure}

\begin{figure}
\begin{center}
\includegraphics[width=7cm]{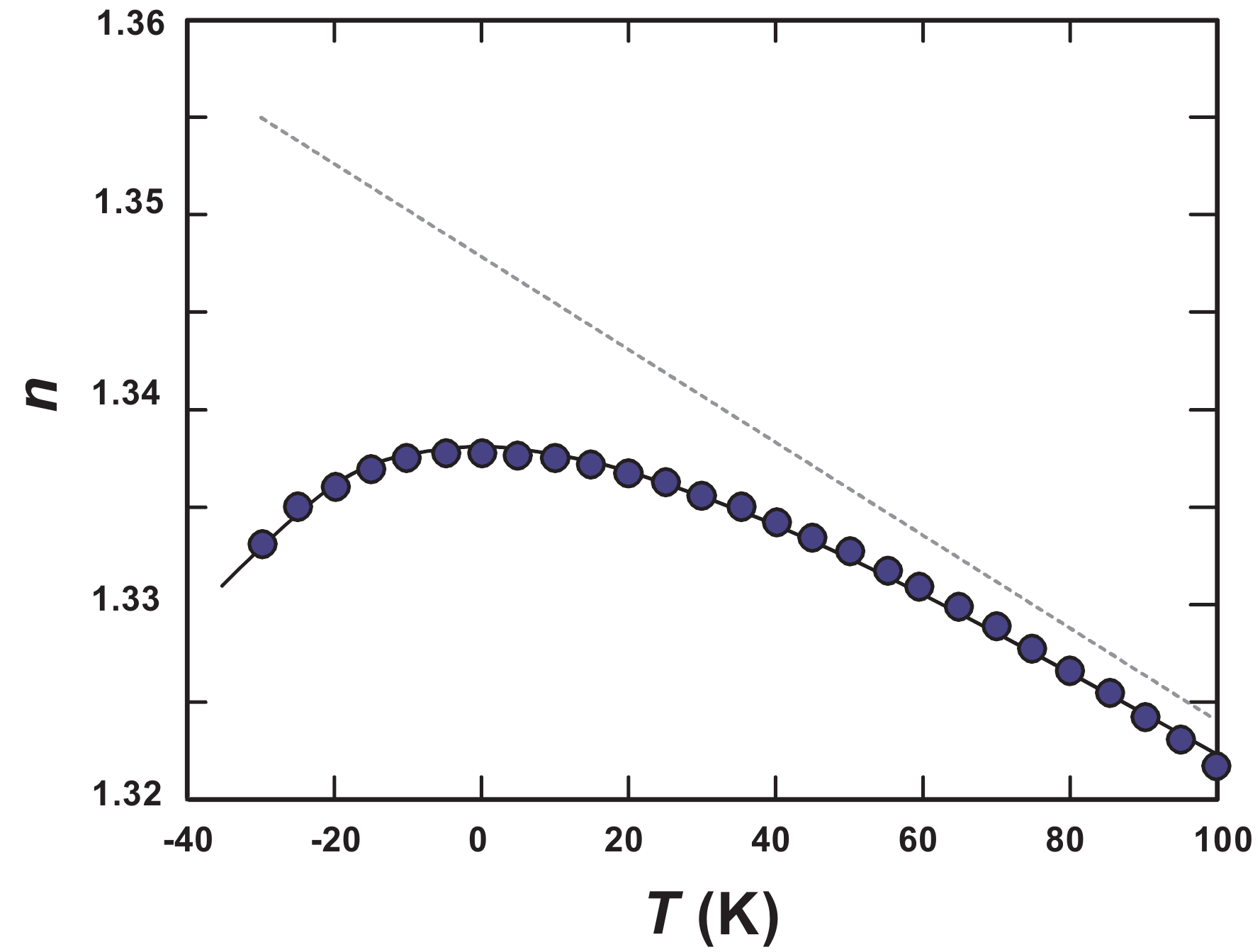}
\end{center}
\caption{The temperature dependence of the refractive index $n$ of water 
at ambient pressure.  The solid curve is composed of the contribution from 
the background part (dotted line) and that of bond ordering given by the 
Boltzmann factor. 
}
\label{fig:n}
\end{figure}

Here we show an example of the fitting 
of eq. (\ref{density}) to the data of $T$, $P$-dependences of density $\rho(T,P)$ 
in fig. \ref{fig:density}. The agreement is satisfactory. We also show a similar fitting 
using the same $\bar{S}$ for the refractive index $n$ in fig. \ref{fig:n}. 

The anomalous increase in $C_P$ upon cooling can also 
be explained as follows. 
The locally favoured structure has a unique 
configuration and the associated degrees of freedom 
are much smaller for it than for the normal liquid structure of water. 
Thus, the entropy $\sigma$ decreases upon cooling, reflecting an increase 
in $\bar{S}$, or short-range tetrahedral ordering: 
\begin{equation}
\sigma=\sigma_{\rm B}(T,P)-\Delta \sigma \bar{S}, \label{eq:sigma}
\end{equation}
where $\sigma_{\rm B}$ is the background part of the entropy associated with 
normal liquid structures. 
Thus, $C_P=T(\partial \sigma/\partial T)_P$ should increase upon cooling as 
\begin{eqnarray}
C_P&=&T(\frac{\partial \sigma_{\rm B}}{\partial T})_P
-T(\frac{\partial \Delta \sigma}{\partial T})_P \bar{S} \nonumber \\
&+& \beta \Delta \sigma (\Delta E-P\Delta v) \frac{\bar{S}}{1+\frac{g_S}{g_\rho}\exp(\beta(\Delta E-P\Delta v))}.  
\qquad \label{Cp} 
\end{eqnarray}
For case of $\bar{S} \ll 1$, this relation can further be simplified as 
\begin{equation}
C_P=T(\frac{\partial \sigma_{\rm B}}{\partial T})_P
+[-T(\frac{\partial \Delta \sigma}{\partial T})_P 
+ \beta \Delta \sigma (\Delta E-P\Delta v)] \bar{S}. 
\end{equation}

The relevance of these predictions was confirmed for water \cite{TanakaWPRL,TanakaWEPL,TanakaWJCP} 
and water-like atomic liquids \cite{TanakaWPRB}. 
The comparisons of eqs. (\ref{KT}) and (\ref{Cp}) with the data of $K_T$ and $C_P$ of water respectively can be found in refs. \cite{TanakaWPRL,TanakaWEPL,TanakaWJCP}. 
The basic physical picture was also supported by numerical simulations by Errington and Debenedetti 
\cite{Errington}, 
which showed how the two order parameters behave as a function of $T$ and $P$ 
for liquid water.  
Recently it was also shown that the steep decrease in the entropy upon cooling 
is due to the development of tetrahedral order \cite{Kumar}. 
This means that local structural ordering is a key to the water anomalies. 
Furthermore, the link between structure, entropy, and diffusivity has also been 
clearly demonstrated for model waters \cite{Agarwal}. 
This work also showed that water anomalies occur at much 
lower pressure than $P_{\rm x}$. In our view this is related to the pressure dependence of 
$\bar{S}(T,P)$ (see eq. (\ref{S_2})). 

\begin{figure}
\begin{center}
\includegraphics[width=7cm]{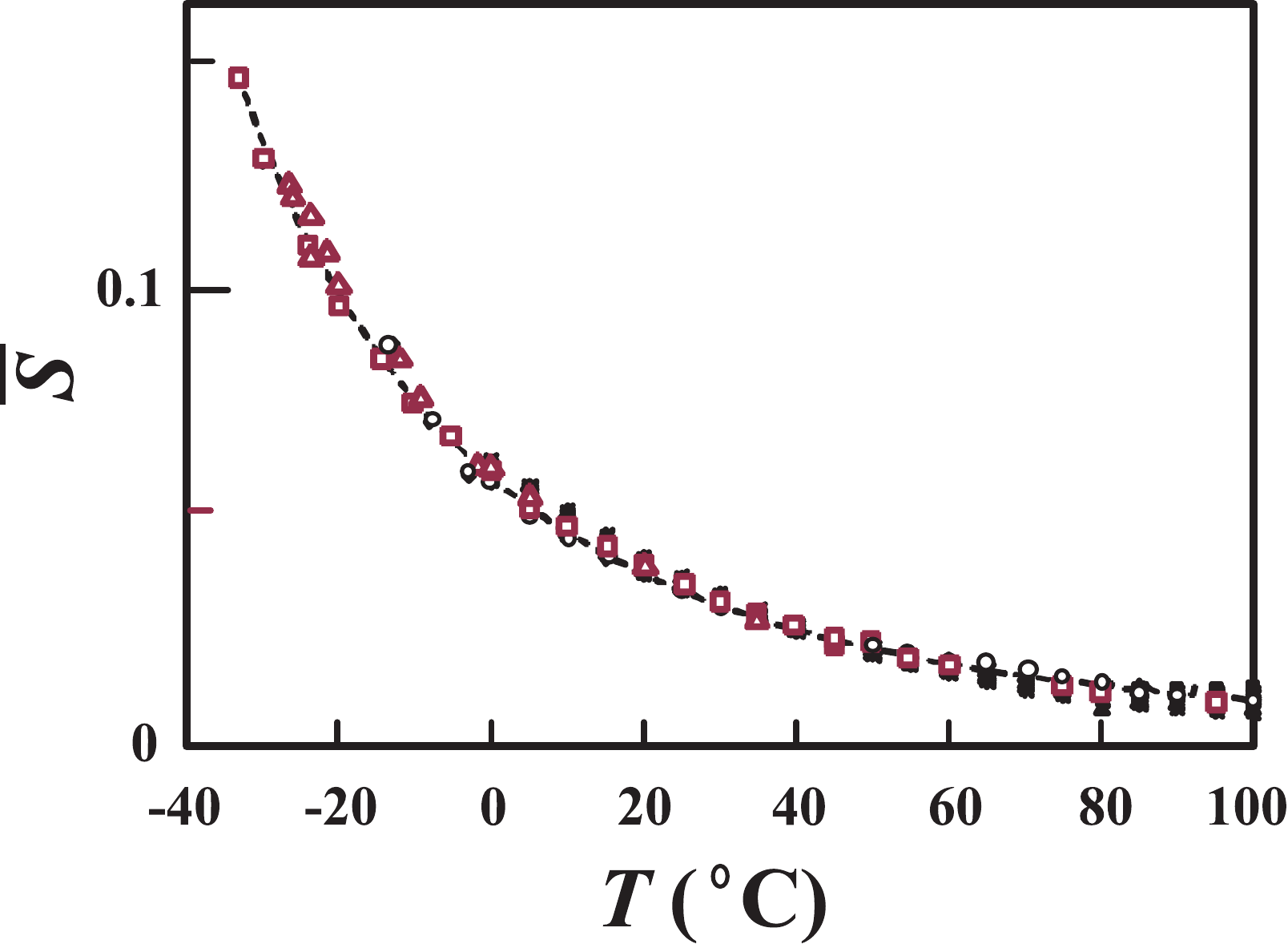}
\end{center}
\caption{Temperature dependence of $\bar{S}$ (see the text on its definition) 
determined by the fitting of our prediction to the experimental data 
of $\rho$, $K_T$, and $C_P$ at various pressures. 
Open squares, triangles, and circles represent, respectively, 
data on $\rho$, $K_T$, and $C_P$ at ambient pressure. 
All the other symbols are data at higher pressures. 
The dashed line is our theoretical prediction for $\bar{S}$. 
The values of $\bar{S}$ determined from the 23 sets of data 
of ``bulk'' liquid water are all collapsed on the master curve, 
which is described by the single Boltzmann factor. 
The figure is reproduced from fig. 1(b) of \cite{TanakaWEPL}. 
See also fig. \ref{fig:STP}, which shows the $T$,$P$-dependence of $S$. 
}
\label{fig:S}
\end{figure}

\subsection{$T$-$P$ dependence of the water anomaly deduced from the above 
two-state model}

\begin{figure}
\begin{center}
\includegraphics[width=7.5cm]{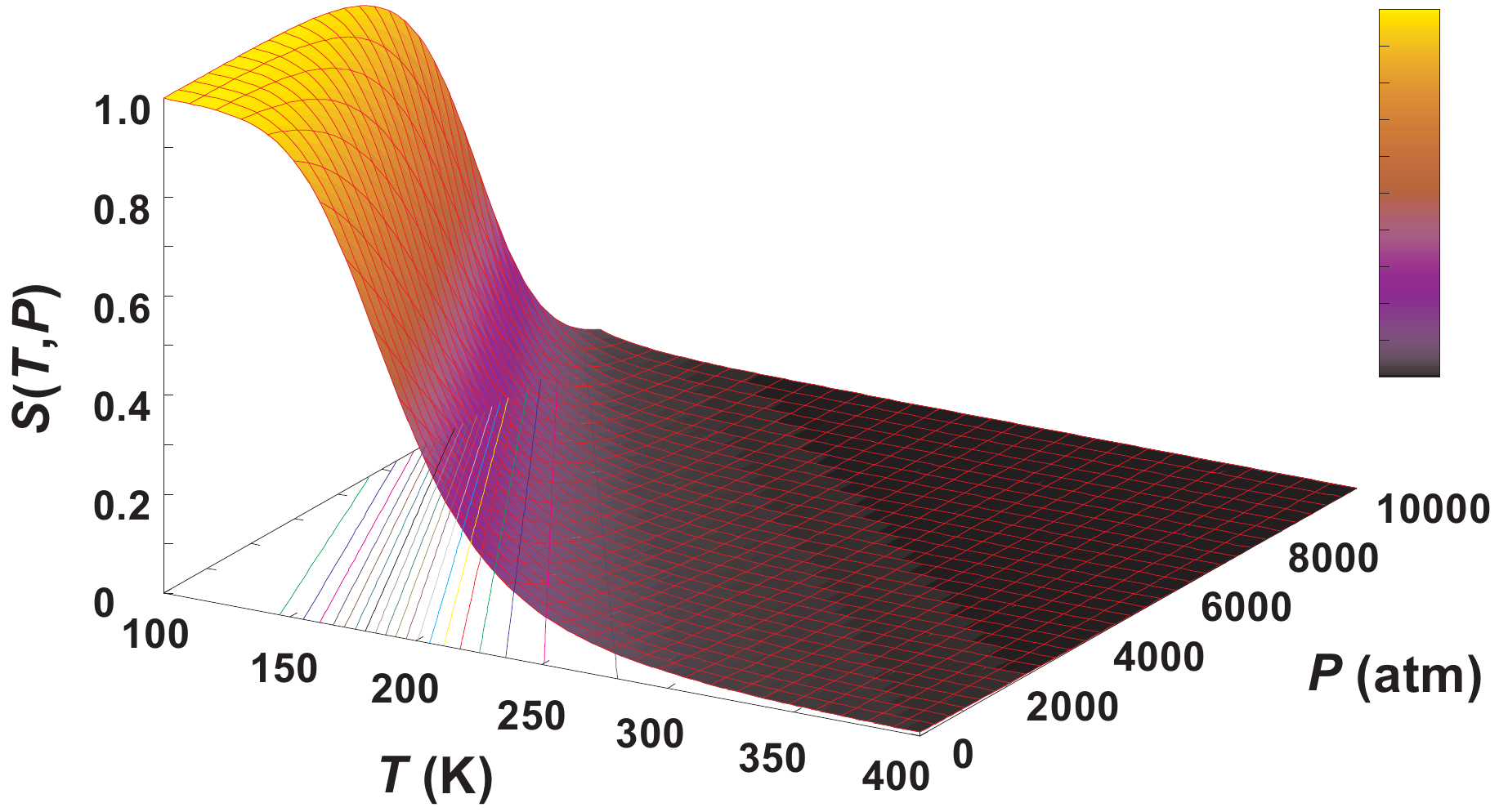}
\end{center}
\caption{(Colour on-line) $T,P$-dependence of $\bar{S}$ for water.  
}
\label{fig:STP}
\end{figure}

\begin{figure}
\begin{center}
\includegraphics[width=7.5cm]{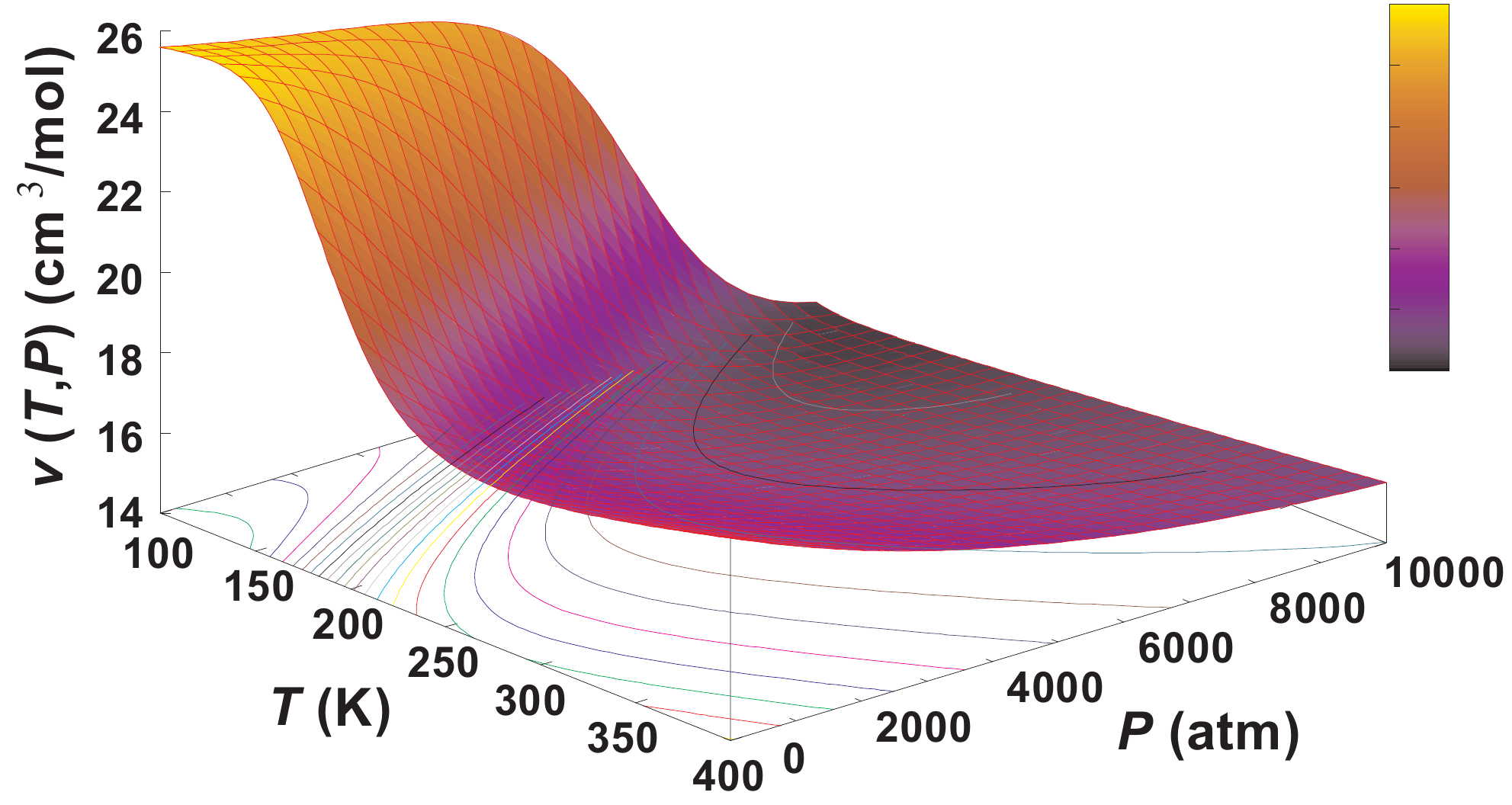}
\end{center}
\caption{(Colour on-line) $T,P$-dependence of $v$ for water.  
}
\label{fig:VTP}
\end{figure}

\begin{figure}
\begin{center}
\includegraphics[width=7.5cm]{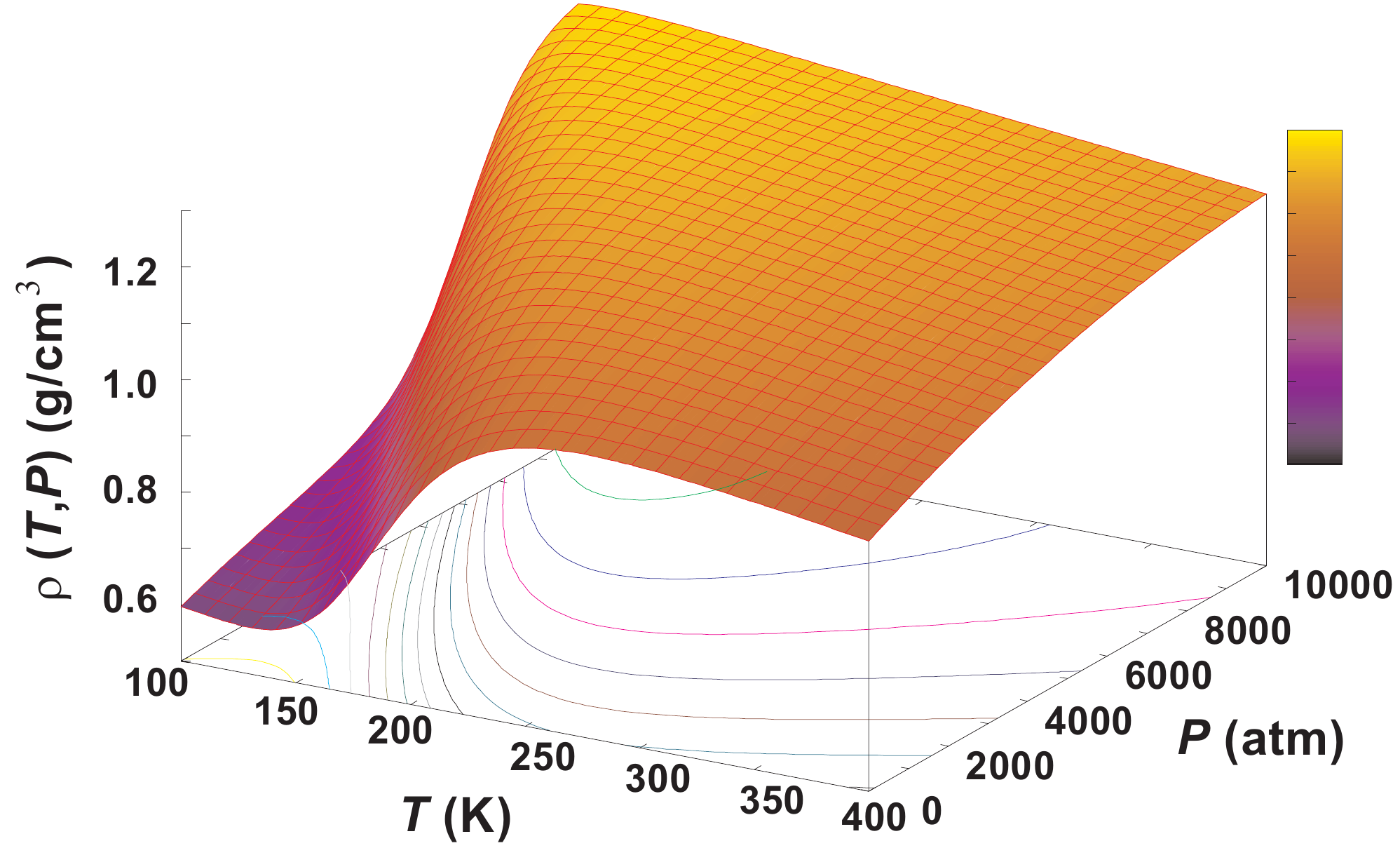}
\end{center}
\caption{(Colour on-line) $T,P$-dependence of $\rho$ for water.  
}
\label{fig:RHOTP}
\end{figure}

\begin{figure}
\begin{center}
\includegraphics[width=7.5cm]{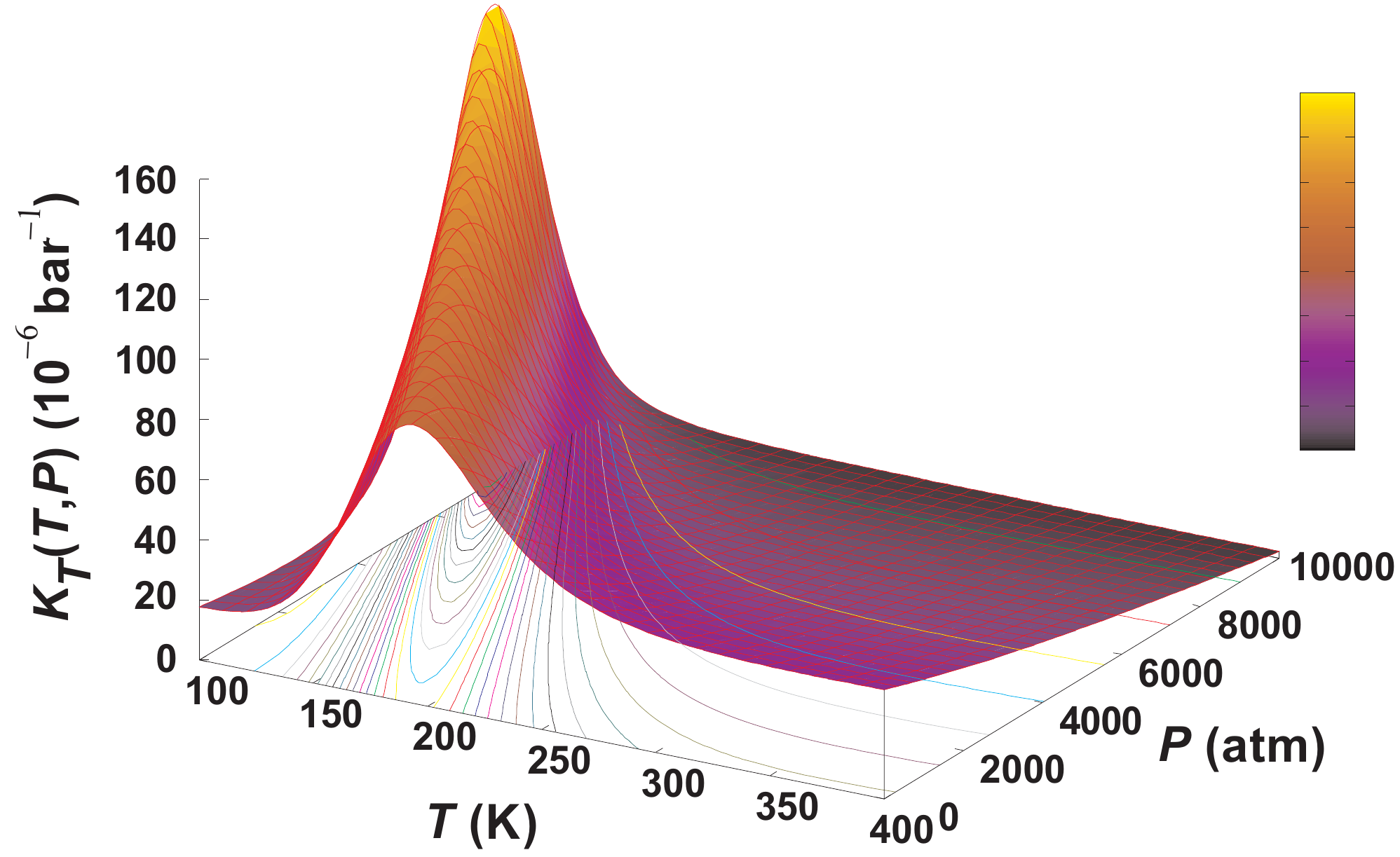}
\end{center}
\caption{(Colour on-line) $T,P$-dependence of $K_T$ for water.  
}
\label{fig:KT}
\end{figure}

In the above, we obtain the $T, P$-dependence of $\bar{S}(T,P)$ (see fig. 8 for the $T$ dependence) 
from the fitting of our model to the experimental data \cite{TanakaWEPL}. Using this result, 
we can calculate the $T,P$-dependence of $\bar{S}$, the specific volume 
$v(T,P)$, the density $\rho(T,P)$, and the isothermal compressibility $K_T(T,P)$ for liquid water. Here we set $J=0$ for simplicity. 
The results are shown in figs. \ref{fig:STP}, \ref{fig:VTP}, \ref{fig:RHOTP}, and \ref{fig:KT}. 
Here it is worth noting that the cooperativity (i.e., non-zero $J$), if it exists, may lead to 
considerable changes of the behaviours of these quantities in a low temperature region. 

\subsection{Thermodynamic anomaly and fluctuations}

Here we consider the origin of the thermodynamic anomaly of water, 
using the anomaly of the isothermal compressibility as an example.  
Using statistical mechanics, macroscopic observables can be expressed 
by molecular-level properties in general.  
Here we express the isothermal compressibility $K_T$ 
in terms of density fluctuations as 
\begin{equation}
k_{\rm B}TK_T/V=\langle (N-\langle N\rangle)^2 \rangle/\langle N \rangle^2, 
\end{equation}
where $N$ is the number of molecules and $V$ is the system volume. 
This can further be expressed as a spatial integral involving the
two-point density-density correlation function as follows \cite{hansenB}:
\begin{equation}
\rho k_{\rm B}TK_T=1+\rho \int \ [g(\mbox{\boldmath$r$})-1] \ d\mbox{\boldmath$r$},
\end{equation}
where $\rho=\langle N \rangle/V$ is the mean number density, and $g(r)$ 
is the molecular pair correlation function. The integral in this last
expression covers all space. 
This relation tells us that the isothermal compressibility 
is linked to deviations of $g(r)$ from its asymptote unity. 
There can be two origins for such deviations: (1) short-range order in
the arrangement of molecules comprised in the liquid and (2) 
long-ranged density fluctuations, which emerge near a critical point. 

In the case of water, we argue that the anomalous increase of $K_T$ primarily 
comes from short-range tetrahedral ordering. 
There are a few reasons supporting this. 
The structure factor $S(k)$ is given by \cite{hansenB}
\begin{equation}
S(k)=1+\rho \int [g(\mbox{\boldmath$r$})-1] \exp(i \mbox{\boldmath$k$} \cdot \mbox{\boldmath$r$})
\ d\mbox{\boldmath$r$}. 
\end{equation}
Although there is not a firm consensus, there is no clear indication of long wavelength 
density fluctuations in small-angle scattering experiments. 
More importantly, the density anomaly itself cannot be explained by long-range density 
fluctuations, but 
can naturally be explained by short-range tetrahedral ordering, as described above. 
These factors seem to support our scenario.

\subsection{Trajectories of the state of water in the two order parameter plane}

\begin{figure}
\begin{center}
\includegraphics[width=7.5cm]{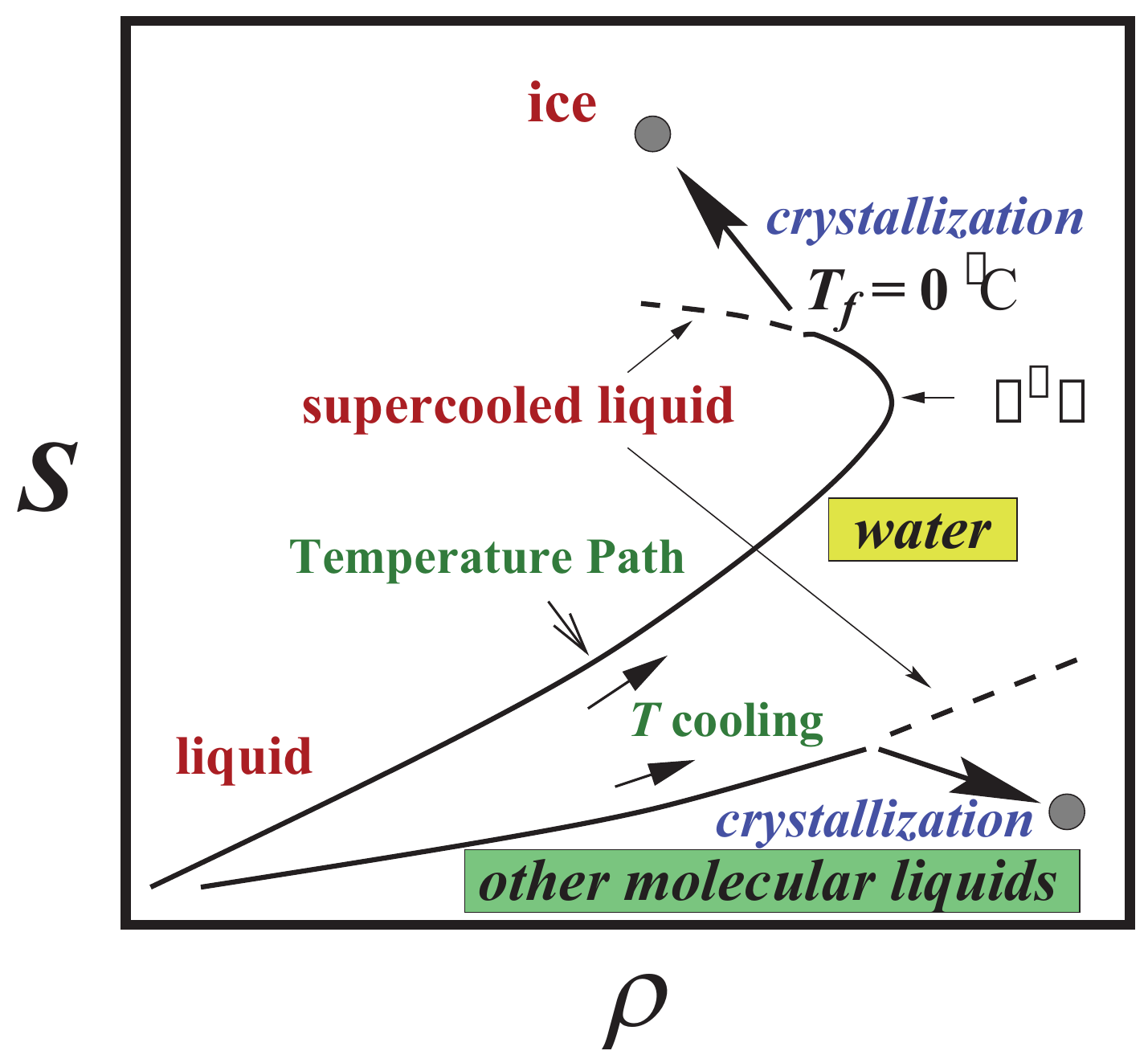}
\end{center}
\caption{(Colour on-line) Schematic figure showing the difference in the liquid-state trajectory on the 
$\rho$-$S$ plane as a function of $T$ 
for water at ambient pressure and an ordinary liquid.   
}
\label{fig:rhoS}
\end{figure}

Here we show a schematic figure representing the trajectories 
of the two averaged order parameters $\rho$ and $S$ for both a water-type liquid and an 
ordinary liquid in fig. \ref{fig:rhoS} (see also ref. \cite{TanakaWPRL}). 
This illustrates the importance of having at least the two order parameters to specify the 
macroscopic state of water properly. 
A similar behaviour was also reported by Errington and Debenedetti \cite{Errington} 
on the basis of numerical simulations. See also ref. \cite{Agarwal} for a recent more detailed 
study.

\subsection{Are there key temperatures associated with water anomalies?}

It is well known that at ambient pressure, the density has its maximum at 4 $^\circ$C, the isothermal compressibility 
has its minimum at 46 $^\circ$C, and the isobaric heat capacity has its minimum at 35 $^\circ$C.
These temperatures are often regarded as important temperatures characterizing the water anomalies. 
According to our scenario, however, the origins of these maxima and minima simply reflect the different temperature dependences of 
the normal background part and the anomalous part proportional to the Boltzmann factor $\bar{S}$. 
Although these temperatures are some measures which can be used to characterize the anomalies, 
their locations are determined by 
the balance between the normal and anomalous part. So we believe that these temperatures 
do not have any significant `physical' meaning.    

\subsection{Relation of our two-state model to so-called mixture and continuum models}

\subsubsection{Relation to the mixture models}
A mixture model was first proposed by R\"ontgen 
\cite{Rontgen} to explain water properties  and then developed by many others 
(e.g., \cite{Eisenberg,Nemethy,BenNaim,BenNaim2,Angell_w,Cho}). 
It was recently applied to 
a water problem by Ponyatovsky et al. \cite{Ponyatovsky}. 
Their model regards water as a mixture of low-density (LDA) and 
high-density amorphous ice (HDA) (see also \cite{Funel}). 
So it may be more appropriate to call this type of model a mixture model 
rather than a two-state model. 

It is worth comparing our model with such a mixture model to 
clarify what physical factors are important to 
determine the scenario relevant to 
water's anomalies. The most crucial difference between our model and the mixture model 
is the value of $\Delta \sigma$. 
We assume that the difference in entropy, or the degeneracy of 
states, between the two states is very large, 
which is a consequence of a disordered 
nature of a normal liquid state and a uniqueness of a locally favoured 
structure. 
We note that normal liquid structures are also made of water molecules temporally hydrogen bonded with neighbouring molecules. 
The important point is that their structural order is still considerably lower than that of locally favoured structures ($g_\rho \gg g_S$). 
On the other hand, it is assumed (see, e.g., \cite{Ponyatovsky}) 
that the difference in the entropy between the two components is small since it is evaluated from the data of 
solid-state amorphous-amorphous (LDA-HDA) transition. 
In other words, it is implicitly assumed that both components have unique structures. 
Considering that a liquid is in a high entropy state, our two-state model approach seems to be more 
reasonable than a mixture model approach. 
This subtle, but important difference leads to a drastic difference in the  
physical picture. In our model, $S$ is very small ($S \ll 1$) at ambient 
temperature and pressure (see fig. \ref{fig:S}), but in most of other models \cite{Angell_w,Cho,Ponyatovsky}  
$S$ (in our terminology) is almost 1/2 or even higher there and  
in some cases the anomaly was ascribed to critical anomaly associated with the second critical point of LLT (see, e.g., \cite{Ponyatovsky}).
In our case, water's anomalies are explained by an increase in $S$ with 
decreasing $T$: The anomalous parts of physical quantities such as 
density are proportional to $S$ and can be described by the Boltzmann 
factor at high temperatures (see eq. (\ref{eq:S})) \cite{TanakaWEPL,TanakaWJCP}. 
However, because detailed microscopic information on hydrogen bondings in water is not 
available, we cannot determine the difference in entropy between the two states in 
a convincing matter. 

Numerical simulations may provide such information, but the current situation is 
still controversial. 
Recent simulation results \cite{Appignanesi2009,Appignanesi2} seem to be consistent with our scenario where $S$ 
is rather small. However, we should also note that 
the estimate of the fraction of the LDL-like component by Cuthbertson and Poole 
is higher \cite{PooleMix}. 
Furthermore, Matsumoto showed that expansion of water upon cooling 
can be explained without invoking any heterogeneity \cite{MatsumotoW}. 
Thus, further studies are necessary to settle this issue (see below). 

Here we mention another reason why we prefer to use ``two-state'' rather than ``mixture''. 
This is because a mixture model gives us an impression that a system is composed 
of A and B component and the order parameter (the fraction of A) is conserved. In reality, however, 
the order parameter should not be conserved: locally favoured structures are created and annihilated without the constraint from its conservation. 
This point is crucial when we consider the nature and the dynamics of water-like anomalies and 
liquid-liquid transition  \cite{TanakaLJPCM,TanakaLLT} (see below). 

We emphasize that the difference in $\Delta \sigma$ leads to the entirely different 
scenarios for water's anomalies, as described above. We believe that this problem, 
which is directly related to the microscopic structural identification of normal and locally favoured structures, 
is a quite important point to understand the physical origin of water's anomalies. 
Thus, careful studies are desired to elucidate which scenario is relevant to water. 
Numerical simulations are obviously very powerful in identifying locally favoured structures. 
Detailed experimental study on the radial distribution function of water may be very useful to settle this issue 
experimentally (see the case of liquid Si described below).

\subsubsection{Relation to the continuum models}

Although we explain water anomalies in terms of the two state model, 
we can also interpret our two-state model as a continuum model \cite{Eisenberg}  
by regarding $\bar{S}$ as the degree of average tetrahedral order instead of the number density 
of locally favoured structures. The same form of the free energy can be used and the expressions 
for various thermodynamic anomalies can be used as they are, if we assume that the volume and the entropy 
of water change in proportion to the degree of average tetrahedral order $\bar{S}$. 
Thus, there is no simple way to distinguish these two scenarios. 
Numerical simulations are the most promising way for this purpose, but there has been no consensus on this 
point so far, as described above \cite{Appignanesi2009,Appignanesi2,PooleMix,MatsumotoW}. 
We note that for 2D spin liquid \cite{ShintaniNP} and hard sphere liquids \cite{MathieuNM}, 
the validity of the two state picture has directly be confirmed (see, e.g., figs. \ref{fig:SRO} and \ref{fig:SRO_T}). 
Recent X-ray absorption spectroscopy also suggests the presence of two distinct structural motifs \cite{Tokushima2008}. 
The two state picture relies on that there are distinct locally favoured structures with a specific symmetry and this symmetry 
is linked to the drastic reduction of the local free energy. This might be related to the possible presence of two types of hydrogen bonding \cite{Tu2009}. 
Thus, this problems lies at the heart of the nature of short-range 
bond orientational ordering in liquids. 

Finally we point out that a continuum model has a difficulty in having a first-order liquid-liquid transition,  
since within its framework, only a continuous transition is allowed. 
This is another reason why we believe the two state picture is more appropriate. 
However, further careful studies are necessary to settle this long-standing issue in a convincing manner. 

\subsection{What is the locally favoured structure of water?}

Suppose that a two state model is a relevant picture of liquid water, a key question is then 
what the locally favoured structure is. The structure should be associated with tetrahedral 
order stabilized by hydrogen bondings. Furthermore, the locally favoured structure should occupy a larger 
specific volume than normal liquid structures. It is reasonable to assume that the former has rather perfect 
tetrahedral order and has an excluded volume in the sense that other water molecules cannot penetrate into its effective volume, 
whereas the latter has low tetrahedral order and other molecules can rather easily access the water molecule at the centre. 
This is suggestive of the importance of the second nearest neighbour shell in determining the locally favoured structures. 
To confirm the presence of such a distinct structure and reveal the exact structural feature, further careful studies 
on water structure by both experiments and simulations are highly desirable.

\subsection{Kinetic anomalies of liquid water} \label{sec:waterdynamics}

\subsubsection{Viscosity anomaly, its characteristics, and previous interpretations}

In addition to the thermodynamic anomaly of water, 
the increase of the activation energy for viscous flow 
or self diffusion upon cooling 
can also be explained by the increase of $\bar{S}$.  
Here we take a standpoint that the dynamic anomaly of water is 
`primarily' neither due to critical phenomena 
nor due to slow dynamics associated with a glass transition, but it is 
due to the existence of locally favoured structures (short-range ordering), 
at least in the temperature region where the bulk water can be supercooled.  

\begin{figure}
\begin{center}
\includegraphics[width=7cm]{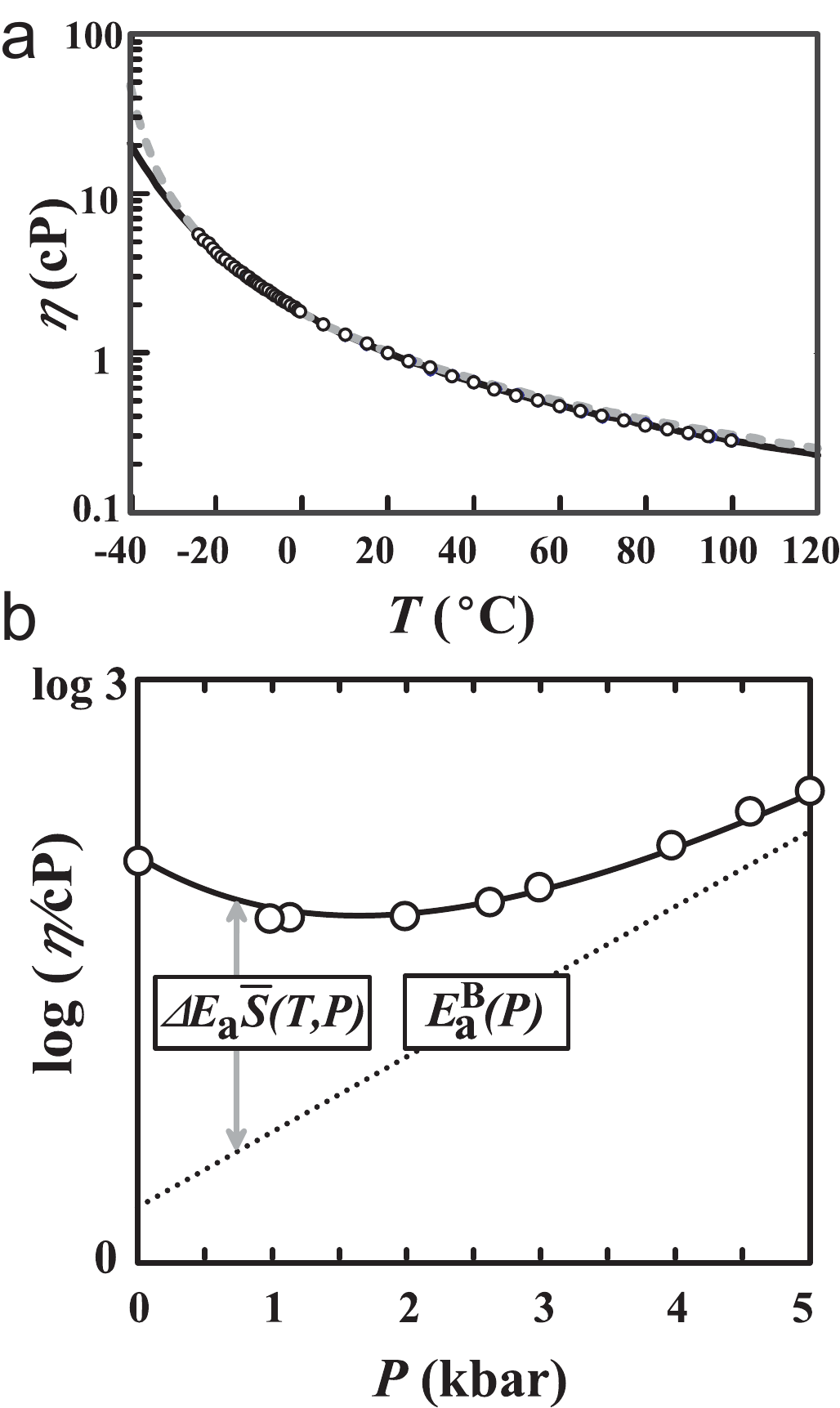}
\end{center}
\caption{
(a) Viscosity anomaly of water at ambient pressure. 
The black solid curve is the prediction of 
eq. (\ref{eq:visco}), 
whereas the grey dashed curve is that of the power law with $T_{\rm s}=228$ K 
$\sim -45$ $^\circ$C and $\nu=1.5$. 
(b) Pressure dependence of viscosity of water 
at $T=-5.0$ $^\circ$C and the fitting curve of our prediction. 
As shown here, our scenario well explains both the $T$ and $P$ dependence 
of viscosity of water. This figure is reproduced from fig. 2 of \cite{TanakaWJPCM}. 
}
\label{fig:viscosity}
\end{figure}

Before proposing a physical picture that can explain this unusual behaviour, 
first we reconsider the statement that water is a fragile liquid near $T_{\rm m}$. 
This statement gives us an impression that water is similar to a typical 
fragile liquid, but it is probably not appropriate from the following reasons. 
\begin{enumerate}
\item[(i)] For ordinary fragile liquids, the temperature distance between $T_{\rm m}$ and $T_{\rm g}$ 
is usually small \cite{TanakaGJPCM,TanakaMJPCM}. 
For water, it is $\sim$140 K, which is unusually larger compared to those for 
typical fragile liquids. 
Note that for typical fragile liquids $T_{\rm m}/T_{\rm g} \sim 1.3-1.5$ whereas for water 
$T_{\rm m}/T_{\rm g} \sim 2.0$. 
\item[(ii)] Bulk water can never be vitrified and always 
crystallizes below $T_{\rm H}$. In other words, water is an extremely 
poor glass former. 
\item[(iii)] More importantly, the viscosity of water first 
decreases with an increase in pressure (or density) \cite{Eisenberg,Bett}, 
and then increases above $\sim$2 kbar [see fig. \ref{fig:viscosity}(b)]. 
This unusual behaviour is markedly 
different from the typical behaviour of ordinary 
liquids that viscosity always increases with increasing pressure (or density). 
Such unusual behaviour cannot be explained by the conventional knowledge 
about supercooled liquids. 
\end{enumerate}

These facts (i)-(iii) cast a serious doubt on the validity 
of the statement that the non-Arrhenius steep increase of the viscosity 
observed around $T_{\rm m}$ is a manifestation of the fragile nature of water, 
which implicitly assumes that water is a glass former in the usual sense. 
We note that glassy slow dynamics of ordinary glass formers always becomes slower 
with an increase in pressure. 
The question that should be answered first is, 
thus, whether the viscosity anomaly of water near $T_{\rm m}$ is caused by slow 
dynamics associated with a glass transition or by other origins. The dynamic 
anomaly of the viscosity $\eta$ and the structural relaxation time 
$\tau$ in water has often been 
explained by the mode-coupling theory (MCT) \cite{DebenedettiB,Starr,Starr1,sciortino2000slow}, 
the model based on the existence 
of a critical-like end point of the hydrogen-bond network formation 
process \cite{Ito,SpeedyAngell},  
or the existence of a retracting spinodal curve \cite{SpeedyAngell}. 
All these models predict the power-law anomaly, 
\begin{equation}
\eta \propto (T-T_{\rm s})^{-\nu},
\end{equation}
where $T_{\rm s}$ is a critical temperature, $\nu$ is the critical exponent,   
and  $T_{\rm s}$ was determined as 228 K at ambient pressure 
\cite{DebenedettiB,Ito,SpeedyAngell}. 
Figure \ref{fig:viscosity}(a) shows such a fitting, which well describes 
the viscosity anomaly. To confirm this type of 
power-law divergence, however, we need to approach very closely to $T_{\rm s}$. 
Since $T_{\rm s}$ is hidden by crystallization in reality [see fig. 
\ref{fig:viscosity}(a)], however, 
this scenario cannot be confirmed in an unambiguous manner. 
More importantly, it should be noted that in a real system 
such a sharp MCT singularity 
is usually smeared out by thermal fluctuation effects, which are considered to activate 
thermally-activated hopping processes. 
For the scenarios based on critical phenomena, we note that the viscosity anomaly associated with critical phenomena is 
usually logarithmic divergence or power-law divergence with 
a very small exponent ($\sim 0.04$) \cite{TanakaWJCP} (see below). 
This is not consistent with the above strong power-law divergence.

We point out some additional problems associated with these scenarios. 
(1) The anomalous pressure dependence of the viscosity 
is ascribed to the pressure dependence of $T_{\rm s}(P)$ and 
$\nu(P)$. This pressure dependence itself is, however, rather difficult to 
explain in a natural manner within their own theoretical frameworks. 
(2) There seems to be no obvious justification for the applicability of the 
mean-field MCT for a system of finite-range interactions. 
Or, why does it work so well only for water? 
It is widely believed that such divergence near the mode-coupling $T_{\rm c}$ 
is not observed in ordinary glass formers, as explained above. 
(3) Furthermore, the absence of the activation process in 
a molecular liquid with hydrogen bonding above $T_{\rm s}$ seems not to be 
reasonable. In other words, there should be a background part 
in the viscosity, but the fitting is usually made while implicitly assuming its absence. 
(4) More importantly, it may be difficult for these scenarios to explain the above facts (i)-(iii). 
Thus, it is worth reconsidering the 
origin of viscosity anomaly from a different standpoint.

\subsubsection{A two-order-parameter description of the viscosity anomaly}
On the basis of our two-order-parameter model of liquid, 
we proposed that the viscosity anomaly of water around $T_{\rm m}$ may be explained as follows 
\cite{TanakaWEPL,TanakaWJCP,TanakaWJPCM}. 
In usual liquids the activation energy required 
for viscous flow or diffusion is associated with 
the creation of a hole, or the disruption of local interactions with 
its neighbouring molecules. 
The existence of the unique activation energy for 
this process is the origin of the Arrhenius behaviour. 
Under the existence of LFS, however, 
an additional activation energy, $\Delta E_{\rm a}$, 
is required for molecules participating these structures to flow. 
Here we note that the lifetime of LFS, which is longer than 
that of NLS, is still quite short ($\ll$ $\mu$s) and 
thus a liquid cannot be regarded as a mixture of stable NLS and LFS. 
Thus, the activation energy should be averaged over all molecules 
participating and not participating LFS, and thus is estimated as   
\begin{eqnarray}
E_{\rm a}(T,P)=E_{\rm a}^{\rm B} (P)+\Delta E_{\rm a} \bar{S}(T,P), 
\end{eqnarray}
where $E_{\rm a}^{\rm B}(P)$ is the background activation energy 
for normal liquid structures without LFS's. 
The $T,P-$dependence of viscosity is thus predicted as 
\cite{TanakaWJCP}
\begin{eqnarray}
\eta(T,P) \propto T^{3/2} \exp[\beta E_{\rm a}(T,P)]. \label{eq:visco}
\end{eqnarray}
We note that this expression may be commonly used in the continuum model if we accept that the extra activation 
energy is proportional to the degree of average tetrahedral order $\bar{S}$, although we believe that the two-state model 
is more appropriate than the continuum model as explained above. 

We made fittings of eq. (\ref{eq:visco}) to the $T-$dependence of viscosity and 
the results are shown in fig. \ref{fig:viscosity}. We obtain 
$E_{\rm a}^{\rm B}(P)=[1832+(0.37-0.0002 \times (T/{\rm K})) \times (P/{\rm bar})]$ K  
and $\Delta E_{\rm a} =2612$ K. 
Here we emphasize that we used the same $\bar{S}$, which we used to describe the thermodynamic anomalies of water. 
Since our prediction and the MCT one equally well describe it, 
we cannot judge solely from this comparison 
which scenario is more reasonable. 
However, we stress that our scenario can explain 
the unusual $P-$dependence of viscosity, or fact (iii), 
in a natural manner. 
As shown in fig. \ref{fig:viscosity}(b), it 
is well explained by the competition between 
the background part, $E_{\rm a}^{\rm B}(P)$, which is a linearly 
increasing function of $P$ as often seen in ordinary liquids, 
and the part related to LFS, $\Delta E_{\rm a} 
\bar{S}(T,P)$, which is an exponentially decreasing function of $P$ 
[see eq. (\ref{eq:S})]. 
In our model, furthermore, all the unusual $T,P-$dependence 
of density, compressibility, heat capacity, and viscosity 
can be described solely by the $T,P-$dependence of 
the common Boltzmann factor, 
$\bar{S}(T,P)$ [see eq. (\ref{eq:S})] in a unified manner  
\cite{TanakaWEPL,TanakaWJCP} (see also fig. \ref{fig:S}). 

It is worth mentioning that the strong correlation between 
structural entropy and diffusion kinetics was established for 
water-type liquids \cite{Truskett,Chakravarty}. 
This suggests the kinetic anomaly can be explained by structural 
ordering and the resulting loss of structural entropy, 
which seems consistent with our scenario. 
We note that the fact that the viscosity anomaly becomes 
less pronounced at higher pressure is not consistent with a scenario 
that the viscosity anomaly is induced by the second critical point, 
provided that the critical point is located at a positive pressure. 

Finally we mention a very recent work by Qvist et al. \cite{Halle} on the rotational dynamics 
of water by combining nuclear magnetic resonance 
measurements with molecular dynamics (MD) simulation. They showed that the origin of the super-Arrhenius temperature dependence of 
the rotational relaxation time cannot be explained by mode-coupling
theory, but rather by the collective dynamics of the fluctuating hydrogen-bond network, i.e., 
thermally induced changes in the tetrahedral hydrogen-bond network, such as the concentration
of over-coordinated (and under-coordinated) ``defects''.  
This looks quite consistent with our two-state model scenario.

\subsubsection{Effects of glass transition}
Finally we mention possible effects of a hidden glass transition point in the 
slowing down of dynamics. 
In the analysis of the $T$-$P$ dependence of the viscosity \cite{TanakaWJCP}, 
we noticed that the low temperature increase of the viscosity cannot be fully explained by 
eq. (\ref{eq:visco}) and we need another source of slow dynamics, which may be linked to glass transition. 
In our study of monodisperse hard spheres \cite{KTPNAS}, we see the growth of the correlation length of bond orientational order 
upon densification, which is a manifestation of glass transition. However, this glass transition is inaccessible as in monodisperse hard spheres, 
because crystallization always takes place before reaching it. 
Such a hidden glass transition, if it exists, should lead to the dynamics characteristic of fragile liquids since 
there is little frustration in this case. 
This example suggests that even a very poor glass former shows behaviour characteristic of fragile glass formers, 
where the tensorial nature of the order parameter plays an important role (see secs. \ref{sec:glass} and \ref{sec:crystallization}). 
Thus, we expect similar behaviour also for water. 
In other words, at lower temperatures bond orientational order starts to extend and may lead to glassy slow dynamics. 
However, we know that hard sphere liquids suffer from competition between crystal-like bond orientational ordering 
and icosahedral ordering. Thus, hard sphere liquids may suffer from stronger frustration effects than water does. 

In contrast to the above expectation, however, the real glass transition of water at 136 K 
exhibits a character of strong liquid. This strong nature of the glass transition may be 
a consequence of strong frustration effects of water locally favoured structures on crystallization into 
a crystal different from the hexagonal ice: 
This crystal which tends to be formed at such a very low temperature may have a symmetry inconsistent with the tetrahedral locally favoured structure of water. 
However, this argument may be too speculative and thus further studies are highly desirable.

\subsection{Anomalies of water-type atomic liquids}

\begin{figure}
\begin{center}
\includegraphics[width=7cm]{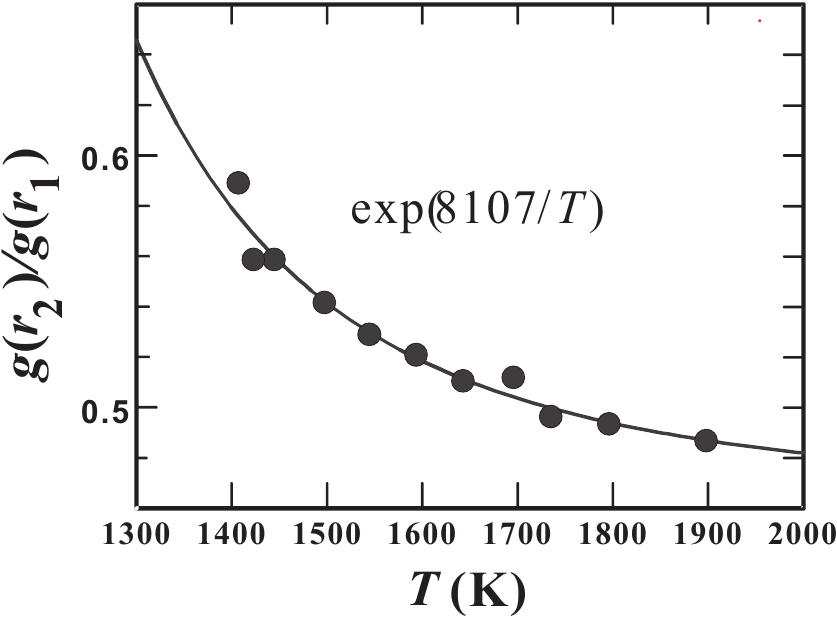}
\end{center}
  \caption{Temperature dependence of $g(r_2)/g(r_1)$ of Si calculated from 
the experimentally measured $g(r)$ \cite{Kimura}. 
The solid curve is our prediction: $g(r_2)/g(r_1)=a+b\bar{S}$, where 
$a$ and $b$ are positive constants \cite{TanakaWPRB}. 
Its anomalous increase upon cooling is very well 
described by the Boltzmann factor, $\exp(8107/T)$ (solid line). 
This figure is reproduced from fig. 2 of \cite{TanakaWPRB}. }
\label{fig:Si}
\end{figure}

For water-type liquids (water, Si, Ge, Ga, $\dots$), the existence of short-range 
bond order with tetrahedral symmetry is evidenced by the shoulder 
in the high wave number ($k$) side of the first peak of 
the structure factor $F(k)$, or the second peak of the radial distribution 
function $g(r)$. 
For Si, for example, the first peak of $g(r)$ is located around 
$r_1$=2.4 \AA, whereas the second one is around $r_2$=3.5 \AA 
\cite{Kimura}. 
The ratio of 3.5/2.4=1.46 is compatible with that 
of the two characteristic interatomic distances of the tetrahedral structures, 
$2 \sqrt{6}/3=1.63$. 
For liquid Si, the temperature dependence of the ratio of the height of 
the second peak to that of 
the first one of $g(r)$, $g(r_2)/g(r_1)$, which may be a direct measure of 
the population of tetrahedral units, is found to be well described 
by $\bar{S}$ with $\Delta E=8107$ K (see fig. \ref{fig:Si}). 
We also found that the anomalies of $\rho$ and $C_P$ 
of liquid Si can also be well explained by our predictions 
[eqs. (\ref{density}) and (\ref{Cp})] with the same $\Delta E$ \cite{TanakaWPRB}. 
Thus, the anomalous thermodynamic behaviour can be well explained 
by our simple scenario. 
Furthermore, the persistence of covalent bonding in metallic liquid silicon above $T_{\rm m}$ 
was recently confirmed experimentally by x-ray Compton scattering \cite{okada2012persistence}. 
We argue that critical phenomena associated with a critical point of LLT may not 
play a primary role in the anomaly \cite{TanakaWEPL,TanakaWJCP,TanakaWPRB,TanakaWJPCM}, as in the case of 
water, even if it exists.

\subsection{On the influence of critical anomaly associated with the second critical point}

One popular scenario of water-type anomalies is based on critical phenomena associated with 
the second critical point. 
Here we mention some difficulties in this scenario. 
(i) Obviously, the anomaly of density cannot be explained 
by the effects of critical fluctuations. 
It is well-known that even at a gas-liquid critical point, 
where the fluctuations 
of density order parameter diverge, the average density 
itself has no critical anomaly. 
In principle, critical fluctuations of bond order parameter $S$ may cause the anomaly 
of the compressibility and the heat capacity, since these quantities 
can be expressed by the correlation function of the order-parameter 
fluctuations. 
For example, the excess anomaly of $C_P$, $\Delta C_P$, should be expressed in terms of critical fluctuations 
of entropy $\delta \sigma$ as \cite{Onuki}  
\begin{eqnarray}
\Delta C_P=k_{\rm B} \bar{\rho} \int \ d\mbox{\boldmath$r$} 
\langle \delta \sigma(\mbox{\boldmath$r$}) \delta \sigma(\mbox{\boldmath$0$}) \rangle. 
\end{eqnarray}
On the other hand, the density anomaly must stem solely from 
the linear coupling of density order parameter to the average 
value of bond order parameter, namely, the increase 
of the average value of bond order parameter itself $\bar{S}$ 
upon cooling, as described above. 
According to our scenario, the density anomaly is expressed by the 
same Boltzmann factor as the anomaly of $K_T$ and $C_P$. 
This strong correlation of the anomaly between 
the critical-anomaly-free quantity (density) and 
the critical-fluctuation-sensitive quantities ($K_T$ and $C_P$) 
indicates the relevance of our scenario.  
(ii) The X-ray scattering measurements of the correlation 
length $\xi$ of fluctuations in supercooled water at ambient pressure shows 
a slight increase \cite{bosio1981,huang2009} in $\xi$ or almost no indication of criticality \cite{xie1993,clark2010}.  
This indicates that the water's anomaly is ``at least primarily'' not due to critical phenomena. 
Note that $\xi$ is the most fundamental 
quantity characterizing the spatial scale of critical fluctuations. 
Without a significant critical enhancement of $\xi$, it is difficult to expect 
strong critical divergence of the physical quantities. 
(iii) It is known that shear viscosity does not show a strong critical anomaly 
even near a critical point. The shear viscosity is given by \cite{Onuki} 
\begin{eqnarray}
\eta=\eta_{\rm b}+\frac{1}{k_{\rm B}T} \int_0^\infty dt \int 
d\mbox{\boldmath$r$} 
\langle \Pi_{xy}(\mbox{\boldmath$r$},t) \Pi_{xy}(\mbox{\boldmath$0$},0) \rangle, 
\end{eqnarray}
where $\eta_{\rm b}$ is the non-critical background. Here $\Pi_{xy}$ 
is the $xy$ component of stress tensor and may be expressed 
by $\Pi_{xy} =C(\partial \phi/\partial x)(\partial \phi/\partial y)$, where $\phi$ is the order parameter.  
In this case, $\phi$ may be $S$. 
Then the viscosity anomaly is calculated by using the decoupling approximation for 
the four body correlation function of the order parameter and 
expressed by the weak logarithmic divergence even near the critical point \cite{Onuki}. 
It means that the critical exponent is effectively zero 
(more exactly, $\sim 0.04$). 
Thus, we cannot expect any strong critical anomaly for the viscosity 
associated with a second critical point, even if it exists. 
Thus, we believe that the viscosity anomaly 
is not primarily caused by critical fluctuations. 
We note that frustration effects may alter this conclusion (see sec. \ref{sec:activation}). 

\subsection{Structural characteristics of liquid water}
Here we mention a recent work by Perera \cite{perera2011water} 
on the structure of liquid water on the basis of radial distribution functions 
obtained by simulations.  
He found that the radial distribution function of water is characterized by a compact
‘three-peaks structure’ over three molecular diameters, which is followed by an apparent loss of the packing
correlations beyond $R_c \sim 9$ \AA. This is in contrast to simple liquids for which the correlations decay continuously
with distance. This indicates the importance of competition between the packing effect and the hydrogen bonding 
interactions, which is consistent with our two-order-parameter scenario. 
It was also suggested that the spatial correlations appear
as a part of a special local structure and not by density fluctuations. 
This indicates that weak enhancement of the structure factor in small angle X-ray scattering 
even at room temperature quite far from any critical
phase transition may be a consequence of special types of short-range correlations 
that are not critical fluctuations. This is again consistent with the above explanation. 

Because of the phenomenological nature of our model, we cannot specify  
what is a locally favoured structure for water, besides its link to tetrahedral structural order. 
It is highly desirable to reveal its very structure, which 
satisfies to be regarded as a symmetry element with a large specific volume.

\subsection{Liquid-liquid transition in water-type liquids} \label{sec:LLTwater}

To explain the water-type anomalies, we do not use the cooperativity 
of short-range bond ordering. That is, we assume $S \ll 1$ (not necessarily $J=0$) 
for simplicity. 
However, it is natural to expect that there is some cooperativity 
in $S$ ordering (i.e., $J \neq 0$). Then there can be a second critical point associated with 
cooperative $S$ ordering. Thus, our model does not preclude 
the existence of LLT, but rather predicts its existence \cite{TanakaLLT}. 
Once we include the cooperativity, it should also influence the thermodynamic 
and kinetic behaviour near the critical point \cite{TanakaWEPL,TanakaWJCP}. 
However, we believe that the water anomalies may be explained without 
invoking such cooperativity because of a long distance from the critical point, even if it exists. 
In our view, the maximum of the isothermal compressibility primarily comes 
from the `non-cooperative' part of the free energy $f(S)$ (see eq. (\ref{eq:fS})), 
which is sometimes called ``Widom line'' \cite{xu_Widom}. 
It is worth mentioning that recently the relationship between various scenarios was studied 
in a general framework \cite{Franzese}. 
We also note that Procaccia and Regev \cite{procaccia2012coarse} recently constructed a simple statistical
mechanics theory which describes the local structure of a generic model of tetrahedral liquids, the Stillinger-Weber model. 
They showed that a finite tendency of a liquid towards tetrahedral symmetry can cause anomalous behaviour with or without
an underlying phase transition, consistent with our scenario, and the presence of three-body interactions is not 
equivalent to the presence of cooperativity leading to LLT.

Since our model is phenomenological in nature, 
we cannot predict the location of the second critical point of water even if 
it exists. The location of the second critical point was predicted 
from experimental results of the 
amorphous (HDA)-amorphous (LDA) transition as well as numerical simulation 
results. The former suffers from elastic effects associated with 
the solid-state volume change accompanied by the transition \cite{TanakaWEPL} (see below), whereas 
the latter suffers from the fact that the location of LLT 
is crucially dependent upon the potentials used \cite{DebenedettiR,PooleN,Brovchenko,TanakaN}. 
Numerical simulations focusing on both glass transition and LLT may shed light in this problem \cite{giovambattista2012interplay}. 
Recently it was suggested by Limmer and Chandler that there is no LLT for water on the basis of numerical simulations 
using the so-called mW and modified ST2 models \cite{limmer2011}, contrary to previous reports. 
This problem is associated with the fundamental question of whether tetrahedral bond orientational order in the supercooled water 
is associated with locally favoured structures characteristic of the second liquid or crystal ice order.  
This is now a matter of active debate (see, e.g., \cite{sciortino2011}). So further careful studies are necessary to settle (a) whether 
LLT really exists in water or not, (b) what is the relation between LLT and crystallization, and (c) where the second critical point is located if it exists. 
See also the next section on this problem. 

The situation is similar in water-type atomic liquids. 
There are many indications of the existence of LLT from numerical simulations, for example, 
in Si \cite{Sastry_Si,Morishita,Widom_Si,Jakse_Si,sastrySi3}. The LLT of Si is also expected to be located 
below the melting point, which is not easy to access experimentally as in the case of water, due to the interference by crystallization. 
Nevertheless, the polyamorphic transition was reported experimentally \cite{McMillan_Si,daisenberger2007}. 
LLT in water-type liquids thus continues to be an interesting topic for future research (see below).

\section{Liquid-liquid transition}

Here we consider the phenomena of liquid-liquid transition in a single-component liquid in the framework of our two 
order parameter model. 

\subsection{Current situations}

Usually it is considered that 
atoms or molecules have random disordered structures in gas and liquid states. 
This leads to a common sense view that any single-component 
substance has only one gas and one liquid state. 
On the other hand, it is widely known that even a single-component 
liquid can have more than two crystal forms, which is known as 
``polymorphism''. The uniqueness of the state is very natural and 
correct for gas, where the kinetic energy dominates. 
However, it is not so obvious for liquid since many body interactions 
come into play, reflecting its high density, as we saw above. 

Recently there has been growing experimental evidence 
that even a single-component 
liquid can have more than two liquid states 
\cite{DebenedettiB,McMillan,McMillan2,AngellR,Poole,harrington1997liquid,Katayama,Katayama2,Monaco,Tsuji,Brazhkin,MishimaR,DebenedettiR}. The transition between these liquid states is called ``liquid-liquid transition (LLT)''. 
There are also experimental indications for the presence of LLT in binary-component liquids such as AsS \cite{brazhkin2008ass,brazhkin2009viscosity}. 
The existence of liquid-liquid transition has also been supported 
by a number of numerical simulations for atomic liquids such as Si \cite{Sastry_Si,Widom_Si,Jakse_Si,sastrySi3} and 
molecular liquids such as water \cite{DebenedettiR,PooleN,Brovchenko,TanakaN}. 
This phenomenon has attracted considerable attention not only 
because of its counter-intuitive nature but also from 
the fundamental importance for our understanding of the liquid state 
of matter. The connection between liquid-liquid transition and polyamorphism 
is also an interesting issue. 

In many cases, however, liquid-liquid transitions exist in a region 
which is difficult to access experimentally, and accordingly its very existence itself is a matter of debate 
and the physical nature and kinetics of the transition remains elusive. 

For example, Katayama et al. discovered the first order LLT in phosphorus 
at high pressure and high temperature with synchrotron X-ray 
scattering \cite{Katayama,Katayama2}. 
They revealed the structure factors for both liquid I and II and 
confirmed the coexistence of liquid I and II during the transition, which suggests the first-order nature of the transition. 
The change in the structure factor suggests that LLT in phosphorus is the 
transformation from tetrahedral to polymeric liquid. 
Such a structural transition was supported by the first principle simulation 
performed by Morishita \cite{Morishita}. 
However, Monaco et al. \cite{Monaco} confirmed that the first-order transition in P 
is between a high-density molecular fluid (not a liquid in the exact sense) 
and a low-density polymeric liquid. 
Thus, the transition is now regarded as a `supercritical fluid'-liquid transition rather than a liquid-liquid transition. 
This explains an unusually large difference in the density between the two states. 
The existence of LLT in liquid Si was
also suggested by high-pressure experiments \cite{McMillan_Si,daisenberger2007,McMillan2} 
and numerical simulations \cite{Sastry_Si,Widom_Si,Jakse_Si}, but the presence of LLT 
still needs to be checked carefully. 
LLT was also reported in yttria-alumina \cite{McMillan,Greaves,McMillan2,greaves2008,greaves2009}. 
However, there are also still 
on-going debates on the composition range over which this
phenomenon occurs and the experimental conditions required
to produce the effect \cite{Skinner} and even on its existence itself \cite{Barnes}. 

For molecular liquids, 
Mishima et al. found an amorphous-amorphous transition in water 
\cite{Mishima}.  
The transition has recently been studied 
in details \cite{DebenedettiR}. 
Computer simulations also suggest the existence of LLT(s) in water 
\cite{DebenedettiR,MishimaR,Poole,TanakaN,Brovchenko,salcedo2011core}. 
On the basis of these findings, the connection 
of amorphous-amorphous transition and LLT in water was suggested 
and actively studied  \cite{MishimaR,DebenedettiR}. 
However, the LLT is hidden by crystallization in water, 
even if it exists. This makes an experimental study 
on the LLT extremely difficult especially for bulk water. 
Even for numerical simulations, difficulties associated with the distinction between LLT and crystallization in a deeply supercooled 
liquid was recently pointed out \cite{limmer2011}. 
It was also pointed out that the 
role of mechanical stress involved in amorphous-amorphous transition 
may make the connection a bit obscure \cite{TanakaWEPL}. 

As briefly reviewed above, LLT is located at high 
pressure and high temperature (e.g., for atomic liquids) or hidden by 
crystallization (e.g., for water) in the above examples. 
This makes detailed experimental studies very difficult. 
This situation has been much improved by recent confirmation 
of LLT at ambient pressure in molecular liquids, triphenyl phosphite (TPP) \cite{TKM,KuriSci} 
and n-butanol \cite{KuriButa}. 
However, this phenomenon was also claimed by Hedoux et al. \cite{Hedoux,Hedoux1,Hedoux2,Hedoux4,HedouxN,HedouxR1,HedouxR2}  
to be induced by the formation of micro-crystallites rather than LLT. 
Recently, a similar claim was also made for n-butanol \cite{Ramos1,Ramos2,krivchikov2011low}. 

So strictly speaking, there has been no firm consensus 
on the existence of LLT for any substance from the experimental side, and 
it remains a matter of debate whether the above-mentioned 
phenomena are the true evidence of LLT or not. 
Theoretically, on the other hand, the generality of LLT, or possible 
existence of LLT in various types of liquids, was recently 
discussed on the basis of phenomenological \cite{TanakaLJPCM,TanakaLLT} 
and analytical models \cite{Lee,Franzese0,Franzese2,Franzese}. 

It is known that LLT usually accompanies a large change in the 
physical properties of liquid even though the component 
is exactly the same. 
This means that the problem of LLT is linked to the 
fundamental question of what physical factors control the 
properties of a liquid. 
Here we discuss the physical origin of liquid-liquid transition 
on the basis of a simple physical picture of local structuring 
of a liquid and its cooperativity. 
We also consider the kinetics of the transition, which was recently observed 
in molecular liquids, triphenyl phosphite and n-butanol. 
The liquid-liquid  transformation kinetics was classified into 
nucleation-growth-type and spinodal-decomposition-type.  
The behaviour is well explained by a scenario that 
liquid-liquid transition is a consequence of 
the cooperative ordering of a non-conserved scalar order parameter, 
which is the number density of locally favoured structures, $S(\mbox{\boldmath$r$})$. 
We also discuss some unsolved problems, which may be at the heart 
of liquid-liquid transition, and alternative scenarios proposed 
on the same phenomena.

\subsection{Thermodynamics}

\begin{figure}
\begin{center}
\includegraphics[width=8cm]{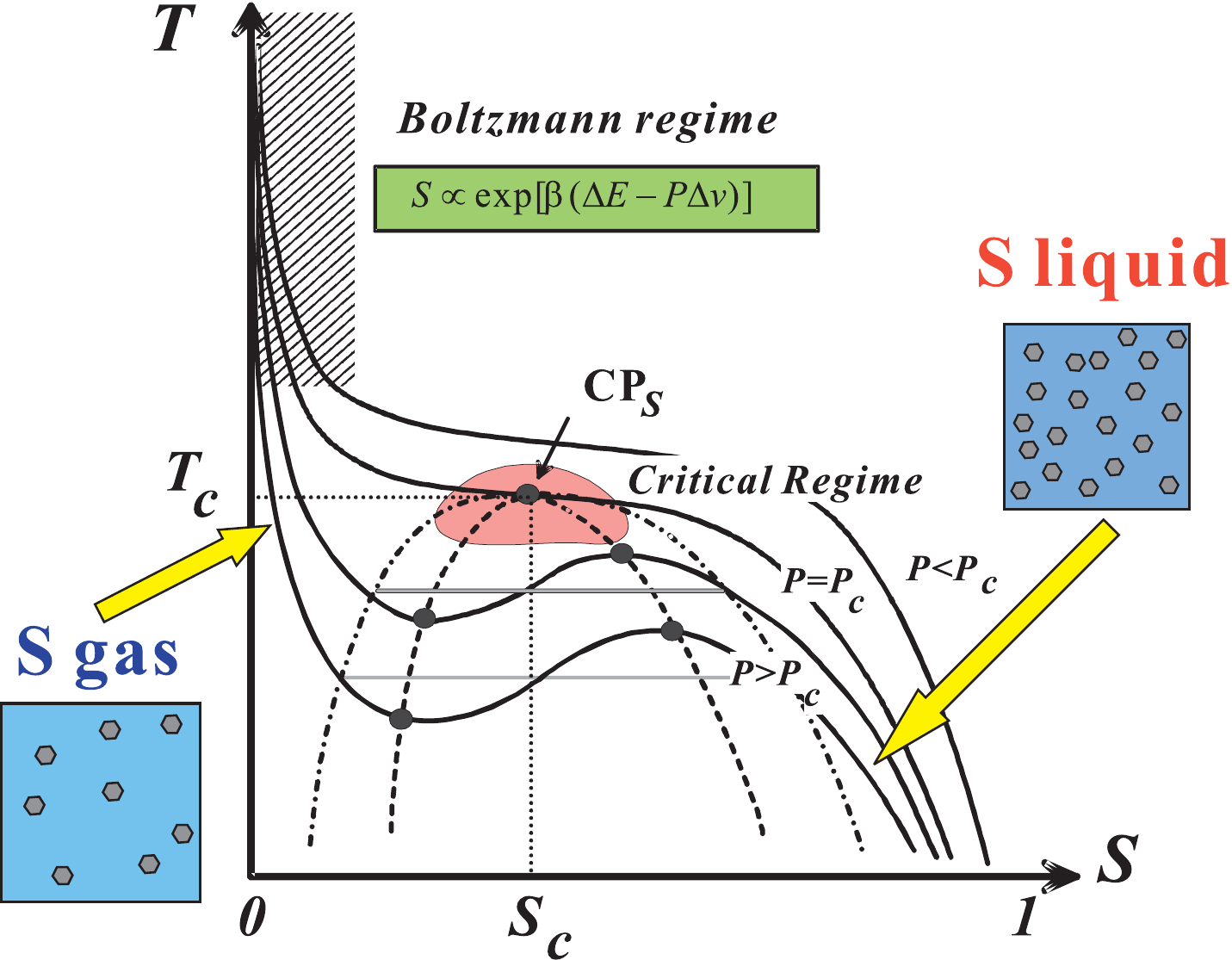}
\end{center}
  \caption{(Colour on-line) Schematic phase diagram of liquid-liquid transition 
  in $T$-$S$ plane \cite{TanakaLLT}. Liquid-liquid transition can be understood as a transition between 
  a $S$-gas state and $S$-liquid state. }
  \label{TS}
\end{figure}

\begin{figure}
\begin{center}
\includegraphics[width=6cm]{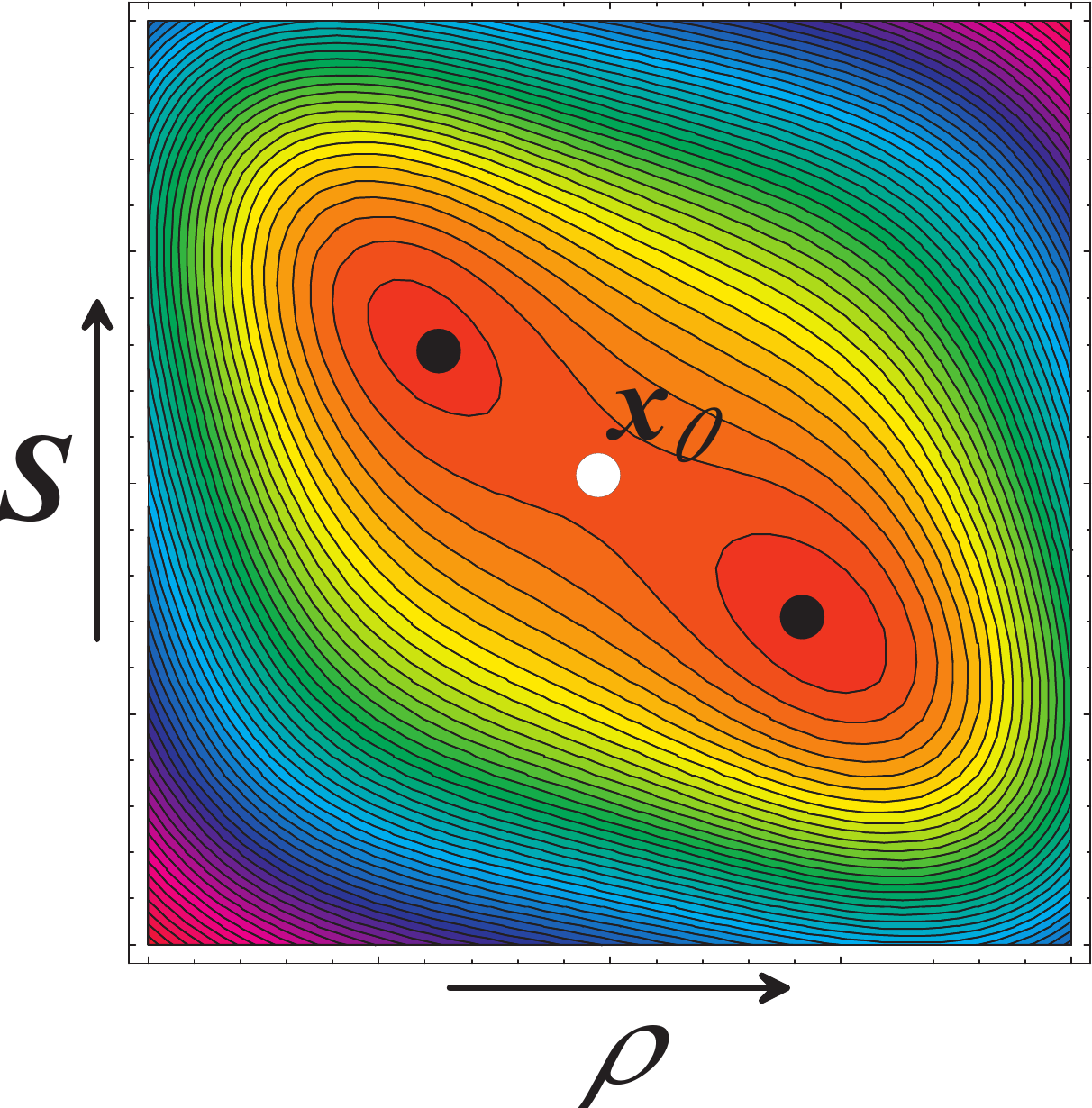}
\end{center}
  \caption{(Colour on-line) Schematic representation of the free energy surface of a system having liquid-liquid transition 
  on the $\rho$-$S$ plane \cite{TanakaLJPCM}. This is a case for $\Delta v>0$. 
  The situation corresponds to the free energy surface at a certain temperature and volume (not pressure), 
  where the two liquid states have the same free energy. 
  This is reproduced from fig. 4 of ref. \cite{TanakaLJPCM}.}
  \label{2Dfree}
\end{figure}

We now consider a liquid-liquid phase transition, or 
{\it cooperative short-range bond ordering} (see fig. \ref{TS}), 
on the basis of the free energy $f(S)$ given by eq. (\ref{eq:fS}) \cite{TanakaLLT}. 
The equilibrium value of $S$ is determined by the condition 
$\partial f(S)/\partial S=0$, or 
\begin{eqnarray}
\beta [-\Delta E+\Delta v P+J(1-2S)]+\ln \frac{g_\rho S}{g_S (1-S)}=0, 
\label{eq:dfds}
\end{eqnarray}
where $\Delta E=E_\rho-E_S>0$, $\Delta v=v_S-v_\rho$, 
and $\beta=1/k_{\rm B}T$. 
It is worth noting that the degeneracy of each state, or the entropy difference between the two states, strongly affects the phase behaviour. 
A critical point is 
determined by the conditions, $f'_S(S_{\rm c})=0$, $f"_S(S_{\rm c})=0$, 
$f^{(3)}_S(S_{\rm c})=0$, and $f^{(4)}_S(S_{\rm c})>0$, as 
\begin{eqnarray}
S_{\rm c}&=&1/2, \\
T_{\rm c}&=&J/(2k_{\rm B}), \\ 
P_{\rm c}&=&[\Delta E-T_{\rm c} \Delta \sigma]/\Delta v. 
\end{eqnarray}
A first-order phase-transition temperature $T_{\rm t}$ is obtained as 
\begin{eqnarray}
T_{\rm t}=(\Delta E-P \Delta v)/\Delta \sigma. 
\end{eqnarray}
Note that a first-order transition occurs only if $T_{\rm t}<T_{\rm c}$. 
For $T_{\rm t}>T_{\rm c}$, this $T_{\rm t}$ is a temperature where $\Delta G=0$ and thus $\bar{S}=1/2$. 
The maximum of $K_T$ is also located near $T_{\rm t}$. 
$\Delta v$ may be positive for liquids such as water and Si, 
but it can also be negative for liquids such as triphenyl phosphite (see below). The sign of $\Delta v$ 
determines the slope of $T_{\rm t}(P)$. 
Liquid I and liquid II are defined as the two possible minima of the liquid-state free energy on 
the $\rho$-$S$ plane (see fig. \ref{2Dfree}).

\subsection{Origin of cooperativity}

The origin of cooperativity in the formation of locally favoured structures is a very fundamental and important 
issue for our understanding of liquid-liquid transition. 
One is the microscopic cooperativity of directional bonding, which is related to the change in the electronic state 
by the formation of locally favoured structures. The second is the modification of the degree of freedoms around 
locally favoured structures due to the local reduction of configurational and vibrational entropy. The third 
is a possible role of long range van der Waals forces, which is due to the difference in density between locally favoured structures 
and normal-liquid structures, $\delta \rho$. The interaction strength may be estimated as 
\begin{equation}
U \sim U_{11} (\delta \rho/\rho)^2 (b/a),  
\end{equation}
where $U_{11}$ is the interaction between basic units (‘atoms’ or ‘molecules’) of size $a$, and $\rho$ is the density, and $b$ is the 
size of locally favoured structures. This interaction might be too weak to cause LLT in the usual situation.  
Since this problem lies at the heart of our understanding of LLT, further careful studies are highly desirable. 
First principle calculations may be a promising way to attack this problem. 

\subsection{Nature of the order parameter governing liquid-liquid transition}

In relation to the above, we briefly discuss the nature of the order parameter governing LLT. So far we put a special focus 
on the local symmetry of locally favoured structures and argue that the order parameter controlling LLT 
should be described by a bond orientational order parameter. 
However, a liquid-liquid transition is also seen in particles interacting with spherically symmetric 
potentials such as the Jagla model \cite{jagla1999core,xu2009monatomic}. 
This indicates that directional interactions are not necessary to have LLT  and the presence of the two length scales 
in the interaction potential is a key to LLT \cite{xu2009monatomic}. 
Nevertheless, we argue that the bond orientational order parameter is still a relevant order 
parameter for this case since it is natural to expect that the presence of the two length scales in the potential 
leads to the selection of a specific local symmetry. The situation is similar to bond orientational ordering in hard spheres. 
Theoretically, there is a merit to have the two order parameters with different characters, density and bond order parameters, rather than to have 
only the density as an order parameter: The presence of the two scalar order parameters allow us to have a gas-liquid-type transition for each order parameter. 
Since this problem is also of fundamental importance, further careful studies are highly desirable.

\subsection{Possible types of liquid-liquid phase transition}

\begin{figure}
\begin{center}
\includegraphics[width=8.5cm]{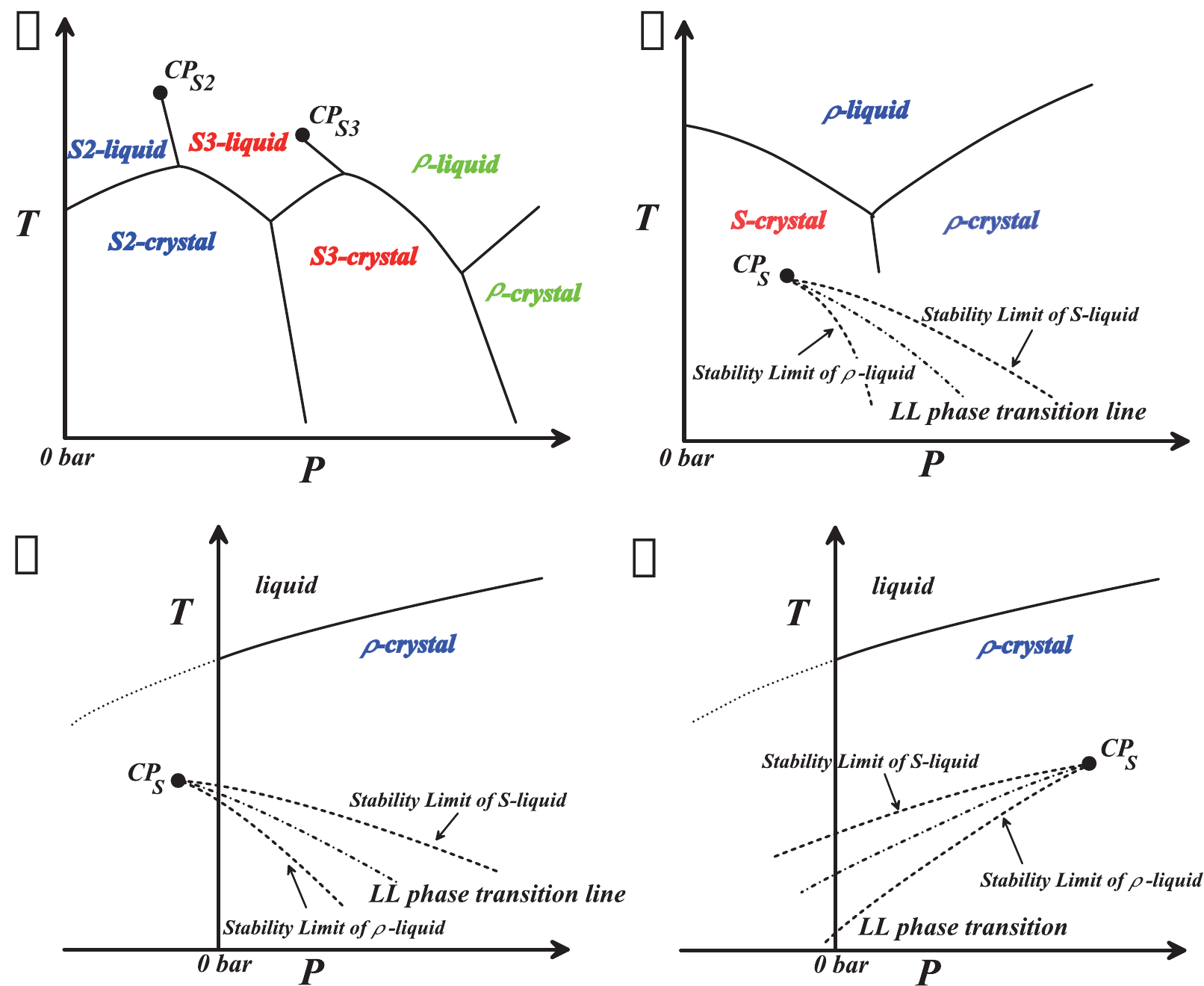}
\end{center}
\caption{(Colour on-line) (a) Schematic $P-T$ phase diagram of a liquid with large $\Delta E$ and $J$ 
such as liquid carbon. In this case, we assume more than two types of locally favoured structures 
in the liquid. The gas-liquid critical point (CP$_{\rho}$) is 
not shown in this figure. (b)
Schematic $P-T$ phase diagram of a liquid with intermediate $\Delta E$ and $J$  such as liquid water. 
CP$_{S}$ is a critical point of $S$ ordering. 
The gas-liquid critical point (CP$_{\rho}$) is not shown in this figure. 
ms stands for ``metastable''. 
The dashed and dot-dashed lines are spinodal and first-order transition lines, 
respectively. 
(c) Schematic $P-T$ phase diagram of a liquid 
with small $\Delta E$ and $J$ for $\Delta v>0$.  
(d) The same for $\Delta v<0$. 
CP$_{S}$ is a critical point of $S$ ordering and 
located at negative pressure. 
The gas-liquid critical point (CP$_{\rho}$) is not shown in this figure. 
ms stands for ``metastable''. 
} 
\label{fig:ph3}
\end{figure}

The examples of possible phase diagrams are shown in figs. \ref{fig:ph3}(a)-(d). 
The type of a phase diagram is classified by the values of 
$J$ and $\Delta E$. The phase diagrams include both 
liquid-solid and liquid-liquid transitions. 
As shown in these figures, we propose that liquid-liquid 
phase transition can, in principle, exist in any liquids 
including even ordinary molecular liquids. 
The necessary conditions are (i) the existence of locally favoured 
structures and (ii) their cooperative excitation ($J>0$): many body interactions and their cooperativity.  
For materials of large $J$ and $\Delta E$, a liquid-liquid transition 
exists in a stable liquid state (see fig. \ref{fig:ph3}(a)), whereas it is hidden by 
crystallization for materials of intermediate $J$ and $\Delta E$ (see fig. \ref{fig:ph3}(b)) 
or it is located in a glassy state for materials of small $J$ and 
$\Delta E$ (see figs. \ref{fig:ph3}(c) and (d)). 
In the last case, there may practically be no direct access to LLT because of the interference due to crystallization, 
as in the case of water.

\subsubsection{Liquid with large $J$ and $\Delta E$}

First we consider the case of large $J$ and $\Delta E$. 
Carbon and phosphorus may be examples of materials having 
large $J$ and $\Delta E$ (see fig. \ref{fig:ph3}(a)). 
Carbon is, for example, known to have a few candidates 
of locally favoured structures, reflecting sp ($S1$), sp$_2$ ($S2$), and 
sp$_3$ ($S3$) bonding. 
Figure \ref{fig:ph3}(a) demonstrates a possible phase diagram of such a liquid, which 
starts from a situation that sp$_2$-type bonding is dominant at ambient 
pressure. There should exist $S1$ liquid in a negative pressure region. 
CP$_{S2}$ and CP$_{S3}$ are critical points associated with 
$S2$ and $S3$ ordering, respectively. 
Above the critical points, the type of liquid changes in a continuous manner. 
In this case, the liquid-liquid transition lines and the associated 
critical points exist in an equilibrium liquid state. 
Note that the relation among the density of each phase is as follows: 
$S2$ liquid $<$ $S2$ crystal $<$ $S3$ liquid $<$ $S3$ crystal $<$ 
$\rho$ liquid $<$ $\rho$ crystal. 
For carbon, $S2$ crystal is graphite and $S3$ crystal is diamond. 
The sign of the slope of a melting line 
is determined by the Clausius-Clapeyron relation, 
$d T_{\rm m}/dP =\Delta v_{\rm m}/\Delta \sigma_{\rm m}$, 
where $T_{\rm m}$ is the melting point, and 
$\Delta \sigma_{\rm m}$ and $\Delta v_{\rm m}$ are the changes 
in entropy and volume upon melting, respectively. 
Since $\Delta \sigma_m>0$, the sign of $dT_m/dP$ is determined solely 
by $\Delta v_m$. The melting lines in fig. \ref{fig:ph3}(a) are drawn by using this fact together with 
the above relation among the density of each phase. 
The phase diagram shown in fig. \ref{fig:ph3}(a) is basically consistent 
with that of liquid carbon  
obtained by experiments \cite{Togaya} and simulations (see, e.g., 
fig. 2 of ref. \cite{Ree}) in a low pressure region. 
More quantitative comparisons require the information 
on physical quantities such as $\Delta E$, $\Delta v$, $\Delta \sigma$, 
and $J$. Our model predicts the existence of an additional 
critical point (CP$_{S3}$) 
at a high pressure region of the phase diagram. 
Experimental studies in this high pressure region are highly desirable.

\subsubsection{Liquid with intermediate $J$ and $\Delta E$}
Water-type liquids may be examples of materials having intermediate $J$ and 
$\Delta E$ (see fig. \ref{fig:ph3}(b)). In this case, the liquid-liquid 
transition line and the associated critical point exist 
in a metastable state below the melting line \cite{Poole}. 
Actually, recent experimental \cite{MishimaS} and molecular dynamics 
simulations \cite{PooleN,Harrington} studies have indicated evidence of 
a first-order liquid-liquid transition in a metastable state 
of water. 
The presence of multiple LLTs was also suggested for water 
\cite{Brovchenko}. 
For water, for example, ice Ih is identified as $S$ crystal, whereas  
ices III, V, $\cdots$ are identified as $\rho$ crystal. 
The liquid density should be higher than $S$ crystal, 
but lower than $\rho$ crystal, which is consistent with 
what is known for the real water. 

The situation of liquid Si is quite similar to that of water. Recent simulations 
indicate the existence of LLT in liquid Si \cite{Sastry_Si,Sastry2004,miranda2004,MorishitaSi,jakse2007,ganesh2009,sastrySi3}. 
There is also experimental evidence for it \cite{McMillan_Si}. 
On the prediction of LLT of Si on the basis of numerical simulations, it was also pointed out that 
the details of the potential could affect strongly the nature and even the existence of the liquid-liquid phase transition \cite{beaucage2005}: 
the liquid-liquid transition disappears when the three-body term of the potential is strengthened by as little as 5 \%.

\subsubsection{Liquid with small $J$ and $\Delta E$}
Finally, we argue that even an ordinary liquid, which has 
small $J$ and $\Delta E$, may have a 
liquid-liquid transition (see figs. \ref{fig:ph3}(c) and (d)). 
For this case, $P_c$ may also be either negative or positive. 
As mentioned above, we recently found LLT in molecular liquids, TPP and n-butanol 
\cite{TKM,KuriSci,KuriButa}. 
The state diagrams of these molecular liquids correspond to the case of fig. \ref{fig:ph3}(d), 
namely, the locally favoured structures are denser than the normal liquid structures 
($\Delta v<0$). This was supported by high pressure measurements \cite{kivelson2001,Paulch_TPP}. 
These systems may be ideal for the study of LLT in the sense that it is 
very easy to access LLT experimentally. We will discuss LLT in these liquids in detail later. 

\subsubsection{Fischer clusters}

The above picture for liquid with small $J$ and $\Delta E$ might provide us with a possible 
scenario of ``Fischer clusters'' \cite{Fischer}. 
Fischer and his coworkers found that some supercooled molecular and polymeric liquids exhibit strong excess scattering, 
which indicates mesoscopic-lengthscale fluctuations of the refractive index.   
While approaching the mean-field spinodal line, 
there may be the critical enhancement of $S$ fluctuations, 
which causes the excess scattering. 
This can happen at ambient pressure if a critical point of $S$ ordering 
is located at a negative pressure and $\Delta v>0$ (see fig. \ref{fig:ph3}(c)) 
or if it is located at a positive pressure and $\Delta v<0$ (see fig. \ref{fig:ph3}(d)). 
Thus, ``Fischer clusters'' might be viewed as 
critical-like fluctuations of $S$ near 
a hidden mean-field spinodal of a gas-liquid-like phase 
transition of locally favoured structures ($S$ ordering). 
The possible existence of LLT in any liquid is a natural consequence of our picture that 
cooperative short-range ordering exists in any liquid. 
Dynamic anomaly associated with Fischer clusters can also be reasonably 
explained by our model \cite{TanakaLJPCM,TanakaLLT}, although 
this should be checked carefully. 
If this scenario is correct, these phenomena can be used to probe LLT.  

The following predictions can be made on the basis of our model: (i) Liquids 
exhibiting ``Fischer clusters'' may have a liquid-liquid phase transition at 
a lower temperature. This transition may be hidden by a liquid-glass 
transition. 
(ii) Applying a higher pressure at the same temperature 
should weaken the critical-like anomaly for $\Delta v>0$, while 
strengthen it for $\Delta v<0$. 
Here it is worth noting that our discussion is based on the mean-field 
approximation. 
The critical-like anomaly near a spinodal line exists only in the mean-field 
limit. This may be consistent with the fact that 
``Fischer clusters'' are characterized by a long bare correlation length 
and are commonly observed in many polymeric glass formers 
\cite{Fischer,Kanaya}, on noting that 
the Ginzburg criterion is safely satisfied in a system with a long-range 
interaction as in polymer systems. 

However, we should note that there is no consensus on the presence of Fischer clusters itself 
and the situation is very controversial. 
Some people think that Fischer clusters are associated with impurities in a sample, 
since any organic liquid inevitably contains some impurities. 
This is a possible scenario, however, it is unclear how impurities can induce ``critical-like'' excess scattering. 
There is also a possibility that the phenomena are caused by an unknown mechanism. 
Thus, we need further experimental studies on this problem.

\subsection{Remark on the relationship between 
melting-point maximum and LLT}

The melting-point maximum is often regarded as a signature of liquid-liquid transition 
and it was a motivation of the mixture model developed by Rapoport \cite{Rapoport}. 
However, the presence of the melting point maximum does not necessarily 
means the presence of LLT. Recently this problem was discussed by Imre and Rzoska \cite{Rzoska} 
and by Makov and Yahel \cite{Makov}. 

In relation to this, it may be worth mentioning that particles interacting with soft potentials such as the Gaussian Core model (GCM)
show the maximum melting point in the $T$-$P$ phase diagram. 
Because the pair potential between GCM particles is bounded, it cannot maintain a crystalline
order at high enough pressures, causing a re-entrant melting of the solid phase at high densities \cite{stillinger}. 
This leads to a maximum melting temperature for the GCM, but without LLT. 
The presence of a maximum melting temperature is one of the striking features of systems of particles with a bounded 
repulsive interaction \cite{likos,likos2006soft,malescio2007}.  
At high temperature and/or pressure the core is unable to give rise to the excluded volume
effects responsible for crystallization, and thus the stable phase is the liquid. 
Furthermore, the GCM shows a solid-solid phase transition from the fcc to the bcc structure, where the
fcc phase is stable at low temperature and pressure. In fact, at such conditions the repulsion 
is strong enough to avoid penetration and the fcc structure is favoured, whereas at high 
pressure the core is more easily penetrable and the bcc phase becomes the
stable one. 
In relation to this, it was recently reported \cite{Parrinello} that the re-entrant
behaviour in Na results from the screening of interionic interactions by conduction electrons, which
at high pressure induces a softening in the short-range repulsion, and such an effect plays an important role in governing 
the behaviour of a wide range of metals and alloys. 
These examples also suggest that the presence of a melting point maximum is not necessarily linked to the presence 
of LLT. 

\begin{figure}
\begin{center}
\includegraphics[width=8.5cm]{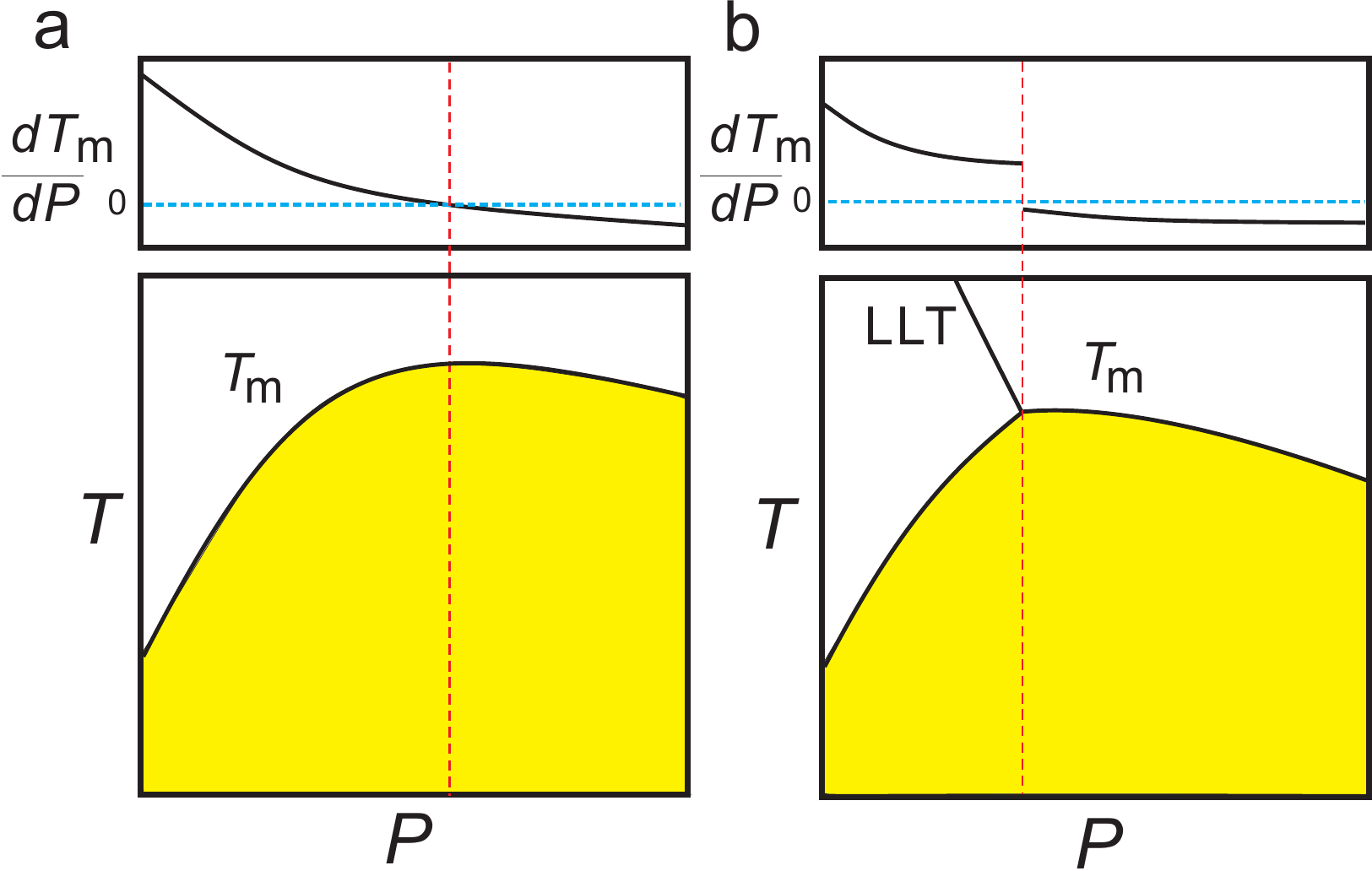}
\end{center}
\caption{(Colour on-line) (a) Schematic $P-T$ phase diagram of a liquid. 
(a) A system without a kink in the melting curve. 
$dT_{\rm m}/dP$ is also schematically shown. Our two-order-parameter model 
predicts the exponential-like functional shape, reflecting the pressure dependence 
of $\bar{S}$.  
 (b) 
A system with a kink in the melting curve, reflecting the presence of LLT. 
In this case, $dT_{\rm m}/dP$ has a jump, reflecting the kink in the pressure 
dependence of the melting point $T_{\rm}$. 
} 
\label{fig:melting_max}
\end{figure}

Here we point out on the basis of our two-order-parameter model that a melting point maximum can also be induced by the change in the liquid side:  
The pressure dependence of the melting point $T_m$ may exhibit a non-monotonic 
$P$-dependence, reflecting the $P$-dependence of $\bar{S}$. 
Our model predicts that 
the shape of the melting point curve is approximated by the straight line minus 
the term proportional to the Boltzmann factor, as shown below (see also fig. \ref{fig:melting_max}(a)): 
\begin{equation}
T_{\rm m}(P) \sim T_{\rm m}^{\rm b}(P)-\Delta T \bar{S}(P), 
\end{equation}
where $T_{\rm m}^{\rm b}(P)$ is the background part of the pressure dependence of $T_{\rm m}$ 
and $\Delta T$ is the amplitude of the effect of local bond orientational ($S$) ordering on the 
melting point. 
The Clausius-Clapeyron relation tells us that $d T_{\rm m}/dP=\Delta v_{\rm m}/\Delta \sigma_{\rm m}$, 
where $\Delta v_{\rm m}$ and $\Delta \sigma_{\rm m}$ are respectively the change in the specific volume and entropy upon a liquid-solid transition. 
We note that the volume and entropy of a crystal have monotonic pressure dependences besides the above-mentioned systems interacting 
soft potentials. 
On the other hand, the volume and entropy of a liquid depends on $P$ as expected from eqs. (\ref{vsp}) and (\ref{eq:sigma}), respectively. 
Depending upon the balance of these two effects, thus, the melting point can have 
a broad maximum as a function of $P$ even without 
liquid-liquid transition. LLT should be associated with a discontinuity 
of $dT_{\rm m}/dP$, or a sharp kink in the melting curve  (see fig. \ref{fig:melting_max}(b)). 
So a special care is necessary to seek LLT by using the shape of the melting curve. 

\subsection{Kinetics of LLT} 

In LLT, we argue that the bond order parameter $S$ 
plays essential roles, as explained above, and the density order parameter 
$\rho$ is slaved by $S$. 
Using $\delta S=S-\bar{S}$, we introduce the following minimal 
Landau-type free energy density by expanding $f(S)$ in terms of $\delta S$, 
which governs $S$ fluctuations near a gas-liquid-like critical point or 
mean-field spinodal lines of bond ordering, where $S=S_{SD}$: 
\begin{eqnarray}
\frac{f(\delta S)}{k_{\rm B}T}=\frac{\kappa}{2} \delta S^2 
+\frac{b_3}{3} \delta S^3
+\frac{b_4}{4} \delta S^4 +h \delta S, \label{eq:fe}
\end{eqnarray}
where $\kappa = \frac{1}{S_{SD}(1-S_{SD})} \Theta$, 
$b_3 = -\frac{1}{2}\left [ \frac{1}{S_{SD}^2} 
- \frac{1}{(1-S_{SD})^2} \right ]$, 
$b_4 = \frac{1}{3}\left [\frac{1}{S_{SD}^3}+\frac{1}{(1-S_{SD})^3}
\right ]$, and 
$h = \left [ \ln \frac{g_\rho}{g_S} + \ln \frac{S_{SD}}{1-S_{SD}}\right ]
\Theta$. 
In the last relation, we use $\partial f/\partial S=\partial^2 
f/\partial S^2=0$ at $T=T_{\rm SD}^\ast$. 
In the above, $\Theta$ is the scaled temperature: $\Theta=(T-T_{\rm SD}^\ast)/T$, 
where $T^\ast_{\rm SD}$ is a critical or spinodal temperature 
of bond ordering without the coupling to $\rho$, and $b_2$ and $b_4$ are 
positive constants. By further including the gradient term, we obtain the following 
Hamiltonian that we believe is relevant 
to the physical description of liquid near a gas-liquid-like transition of 
locally favoured structures \cite{TanakaLLT}: 
\begin{eqnarray}
\beta H_{S}=\int d\vec{r}\ [f(\delta S)
+\frac{K_S}{2} |\nabla \delta S|^2]. \label{eq:H0} 
\end{eqnarray}  
For simplicity, we assume the density $\rho$ is given as a function 
of $S$ as follows: $\rho(\mbox{\boldmath$r$})=\rho_N(1-S(\mbox{\boldmath$r$})) 
+\rho_S S(\mbox{\boldmath$r$})$, where $\rho_N$ is the density of 
the normal-liquid structure and $\rho_S$ is the density of the 
locally favoured structure. 
For $\Delta v<0$, which is a case of TPP, an increase in $S$ leads to 
an increase in $\rho$. 

In our previous papers \cite{TanakaLJPCM,TanakaLLT}, 
we employed a more complex coupling between $\rho$ and $S$, 
which leads to the constraint for the global density. 
However, real experiments are performed at constant pressure 
and there is no constraint for the total density. 
Indeed, our preliminary light scattering experiments show that 
the scattering intensity at $k \rightarrow 0$ grows upon LLT, which is indicative 
of the evolution of the structure factor of a non-conserved order parameter. 
Although we need a more complete description, which takes into account 
the couplings to the density (mass conservation), velocity fields (momentum conservation), and 
temperature fields (energy conservation), we here stick to the 
simplest version of the kinetic theory. 
Here we note that Takae and Onuki recently investigated the roles of latent heat 
on LLT. This might also play an important role \cite{Takae} if local heating is allowed due to 
a weak thermal contact between the sample and the temperature bath. 

The kinetic equation describing the time evolution of the 
non-conserved scalar order parameter $S$ is then given by \cite{TanakaLLT} 
\begin{eqnarray}
\frac{\partial \delta S}{\partial t}
&=&-L_S \left[ -K_S \nabla^2 \delta S+\frac{\partial f(\delta S)}
{\partial \delta S} \right] \nonumber \\ 
 &=& -L_S 
 [-K_S\nabla^2\delta S + h + 
\kappa\delta S + b_3\delta S^2 \nonumber \\
&& + b_4\delta S^3+
C\delta{\rho} ] + \zeta_S',
\label{eq2:S}
\end{eqnarray}
where $L_S$ is a kinetic coefficient and $\zeta'_S$ is  
normalized thermal noise, which satisfies the following 
fluctuation-dissipation relation: 
\begin{eqnarray}
\langle \zeta'_S(\mbox{\boldmath$r$}_1,t_1) 
\zeta'_S(\mbox{\boldmath$r$}_2,t_2) \rangle=2 (L_S \xi_\rho^2 / L_\rho) 
\delta(\mbox{\boldmath$r$}_1-\mbox{\boldmath$r$}_2) \delta(t_1-t_2). \nonumber
\end{eqnarray} 
We stress that the non-conserved nature of the order parameter $S$ originates from the fact that locally favoured 
structures can be created and annihilated without the constraint of its conservation.

\subsubsection{Nucleation-growth(NG)-type LLT}

When we quench and anneal TPP or n-butanol in the metastable region 
with respect to LLT, droplets of liquid II are randomly nucleated in both space and time in liquid I 
and the domain size $R$ grows 
with a constant interface velocity as $R \propto t$ \cite{TKM,KuriSci,KuriButa}.  
In the late stage, droplets of liquid II collide, coalesce, and further grow. 
This behaviour is characteristic of the NG behaviour.
Then, the new phase covers the entire region and eventually the boundary between droplets  
tends to disappear; and, thus, liquid I almost transforms 
to homogeneous liquid II (see fig. \ref{fig:NGSD}(a)). 
This is a consequence of the non-conserved nature of $S$ and the off-symmetric quench. 
If the LLT were governed by a conserved order parameter, 
the diameter would grow in proportion to 
$t^{1/3}$ and the system would never become homogeneous again \cite{Onuki}. 

The temporal evolution of the average order parameter 
$S$ during the transformation, which is directly related to  
the temporal evolution of the enthalpy $H(t)$ measured in experiments. 
$H(t)$ can be estimated 
from the latent heat flux $dH/dt$ released 
during the transformation from liquid I to liquid II \cite{KuriPRB}, 
which was measured by differential scanning calorimetry (DSC). 
Microscopically, the latent heat is a consequence of the formation of locally favoured structures. 
We found that the time evolution of $H(t)$ obeys 
the Avrami-Kolmogorov equation \cite{Onuki}, $ H(t) / H_{m} = 1-\exp(-Kt^n)$, 
where $ H_{m}$ is the final value of $H(t)$.  
We obtained $n=4$ from the fitting, which means 
that liquid II is nucleated homogeneously and grows isotropically 
with a constant rate in 3D. This is consistent with the above result of microscopy observation. 
Note that $n=d+1$ ($d$: the spatial dimensionality) 
for the case of homogeneous nucleation and isotropic linear growth 
\cite{Onuki}. 

The NG-type transformation is characterized by 
the temporal change of the 
probability distribution function of the bond order parameter $S$, 
$P(\delta S)$: 
$P(\delta S)$ changes from a single Gaussian shape at $t=0$ to 
another Gaussian 
($t \rightarrow \infty$) through a double-peaked shape for NG 
\cite{TanakaDIA,TanakaProb}.
This reflects the fact that the nuclei have already the final value of 
the order parameter when they appear.

\subsubsection{Spinodal-decomposition(SD)-type LLT}

Next we consider SD-type LLT, which occurs when 
a liquid is quenched into an unstable region 
below $T_{\rm SD}$ \cite{TKM,KuriSci,KuriButa}. 
First we summarize our experimental findings. 
We studied pattern evolution during LLT. 
The initial stage is reminiscent of the Cahn's linear regime \cite{Onuki}. 
In the beginning, the amplitude of fluctuations exponentially grows 
with time and thus the contrast increases. 
Then, the domain size and the contrast both increase. 
Later the liquid becomes more homogeneous, which leads to the decrease in 
the contrast.
Finally, the liquid becomes homogeneous liquid II.

We also followed the temporal evolution of the heat released upon the 
transformation, $H(t)$ (see fig. \ref{DSC}), 
which is expected to be proportional 
to the bond order parameter $S(t)$, provided that 
the heat is released upon the formation of locally favoured structures. 
It is well described by the SD-type evolution of 
a non-conserved order parameter. 
For the ordering of a non-conserved order parameter, 
the interface motion is described by the Allen-Cahn equation: 
$v = dR/dt = -L \Upsilon$, where $v$
is the interface velocity, $L$ is the kinetic coefficient, and 
$\Upsilon$ is the mean curvature of the interface ($\sim 1/R$). 
This relation yields the domain coarsening law of $R \sim \sqrt{Lt}$.
However, this is a case for a symmetric quench where there is a symmetry 
against the fluctuations of the order parameter and a sharp interface between the two phases is formed. 
For an off-symmetric case, a sharp interface is never formed and the scaling argument is not firmly justified, which makes 
the coarsening law less obvious.

The evolution of the average value of $S$ for a critical quench 
can be described by the theoretical prediction for the SD-type evolution 
of a non-conserved order parameter:  
\begin{eqnarray}
\frac{S(t)}{S_{m}} = \frac{S_{m}[1+ ((\frac{S_{m}}{S(0)})^2-1) 
e^{-2\gamma t}]^{-0.5}-S(0)}{S_{m}-S(0)}, \nonumber
\end{eqnarray}
where $\gamma$ is the growth rate of the order parameter 
\cite{Onuki}. 
This relation is obtained from  the kinetic equation 
by ignoring the gradient term, 
the thermal noise, and the couplings of the two order parameters there.  
In the above, the linear and cubic terms in the free energy eq. (\ref{eq:fe}) are also neglected, which 
may yet play an important role in many cases. 

Finally we mention the temporal change of the 
probability distribution function of the bond order parameter $S$, 
$P(\delta S)$.
For SD-type LLT, the distribution transiently becomes broader.  
Reflecting the non-conserved nature of $S$, the mean value of 
$S$, $\bar{S}$, increases with time in a discontinuous manner 
for NG, whereas in a continuous manner for SD \cite{KuriSci,TanakaProb}.

\subsubsection{Behaviour of the isothermal compressibility}

Next we consider the behaviour of the isothermal compressibility. 
This problem is related to a liquid exhibiting excess scattering from Fischer clusters 
(see \cite{Fischer}). 
A liquid with Fischer clusters exhibit strong excess scattering. 
However, the isothermal compressibility does not exhibits any anomaly. 
This cannot be explained if we consider that the excess scattering comes directly from density fluctuations. 
This apparently strange phenomenon can be naturally explained by accepting an additional 
order parameter, which is coupled with density. 
Then the isothermal compressibility should be given by 
\begin{eqnarray}
K_T=\frac{1}{k_{\rm B}T \rho^2}\int \ d\mbox{\boldmath$r$} 
\langle \rho^r(\mbox{\boldmath$r$})\rho^r(\mbox{\boldmath$0$}) \rangle, 
\end{eqnarray}
where $\delta \rho^r=\delta \rho^\ast+c \delta S^\ast$ 
is the real fluctuation of density under the coupling to bond ordering. 
For a small value of $|c|$, the major contribution 
to $K_T$ comes from the direct density-density correlation. 
On the other hand, the scattering intensity mainly comes from 
fluctuations of bond order parameters, 
which may be dominant near the mean-field spinodal of $S$ ordering. 
Thus, we suggest that there is a possibility of the apparent violation of the compressibility 
sum rule, which may be due to (a) the existence of an additional hidden order parameter, 
namely, bond order parameter, which has critical-like fluctuations, 
and (b) its direct coupling to the refractive index $n$. 
However, near the second critical point, the contribution from fluctuations of $S$ 
may finally become dominant even in the isothermal compressibility. 
We should note that in the above scenario we cannot deny a possibility that the hidden order parameter 
is associated with the concentration fluctuations of impurities. 
As in the case of water, $K_T$ may also exhibit an anomalous increase, 
due to the increase in the number density of locally favoured structures, $S$, itself (not due to 
its fluctuations).  

Finally, it is worth noting that critical-like fluctuations and the steep increase in the susceptibility, 
which might be due to LLT,   
were recently reported for supercooled yttria-alumina melts 
\cite{greaves2008,greaves2009,greaves2010,greaves2011}.

\subsection{Comparison of our model with experiments in molecular liquids}

\subsubsection{LLT in pure molecular liquids}

Kivelson and coworkers found the following unusual phenomena 
in their study of a supercooled state of TPP \cite{Ha,Cohen,Kivelson2}. 
If TPP is cooled rapidly enough, it first enters into 
a supercooled liquid state below the melting point $T_m$ as usual 
liquids, and then into a glassy state, which we call glass I. 
This supercooled liquid (liquid I) behaves as a typical fragile glass former. 
On the other hand, if TPP is quenched to a certain temperature 
between 213 K and 223 K and then anneal it at that temperature, 
a new apparently amorphous phase (the so-called glacial phase) is nucleated 
in a supercooled liquid and grows with time. 
Eventually, the entire system transforms into the glacial phase. 
Surprisingly, the glacial phase is apparently an optically transparent 
homogeneous amorphous phase, but it is obviously different from 
ordinary liquid (liquid I) and glass (glass I). 

This finding stimulated intensive experimental researches on this 
unusual phenomenon. However, the nature and origin of the glacial phase 
has been a matter of debate and many different, 
even controversial, explanations have been proposed on it. 
The glacial phase was thought to be a new amorphous phase 
\cite{Cohen,Rossler,Senker,Senker2} or 
a highly correlated liquid \cite{Oguni}. 
However, the most researches have shown that the glacial phase has 
some crystallinity or anisotropy. Hence the newly formed glacial phase 
appears to be neither a standard glass nor liquid. 
It is this that has led researchers to conclude that the glacial phase 
is actually some type of defect-ordered crystals 
(orientationally disordered or modulated crystal) 
\cite{Cohen,Kivelson2,Alba}, liquid crystal \cite{Johari}, 
plastic crystal \cite{Johari,Senker2}, 
aborted crystallization 
\cite{Hedoux,Hedoux1,Hedoux2,Hedoux4,HedouxN,HedouxR1,HedouxR2}, 
or nano-clustering \cite{Yarger}.

We recently succeeded in directly observing the process of liquid-liquid transition with optical microscopy 
for two pure organic liquids, triphenyl phosphite (TPP) \cite{TKM,KuriSci} and n-butanol \cite{KuriButa}. 
TPP has been known to show an anomalous transition from a supercooled liquid to the 
so-called glacial phase \cite{Cohen}. 
On the basis of our experimental results, we concluded that this transformation is actually 
a transition from a supercooled state of liquid I to a glassy state of liquid II. 
The situation is complicated by the presence of micro- or nano-crystallites, which are formed during the transformation. 
Indeed, this same phenomenon was interpreted as aborted crystallization by Hedoux et al. \cite{Hedoux}, as mentioned 
in the introduction of this section. We confirmed that LLT accompanies the formation of micro- or nano-crystallites at rather high temperatures 
(above 214 K), 
but at low temperatures (e.g., at 212 K) there is no indication of crystallization and only LLT takes place. 

In the case of n-butanol, on the other hand, crystallization always occurs and the situation is a bit more complicated. 
For example, Ramos and his coworkers recently claimed that the phenomena observed in n-butanol is aborted 
crystallization and not LLT \cite{Ramos1,Ramos2}, but we argue that it is LLT on the basis of the kinetic features 
of the transformation process. 

\begin{figure}
\begin{center}
\includegraphics[width=7cm]{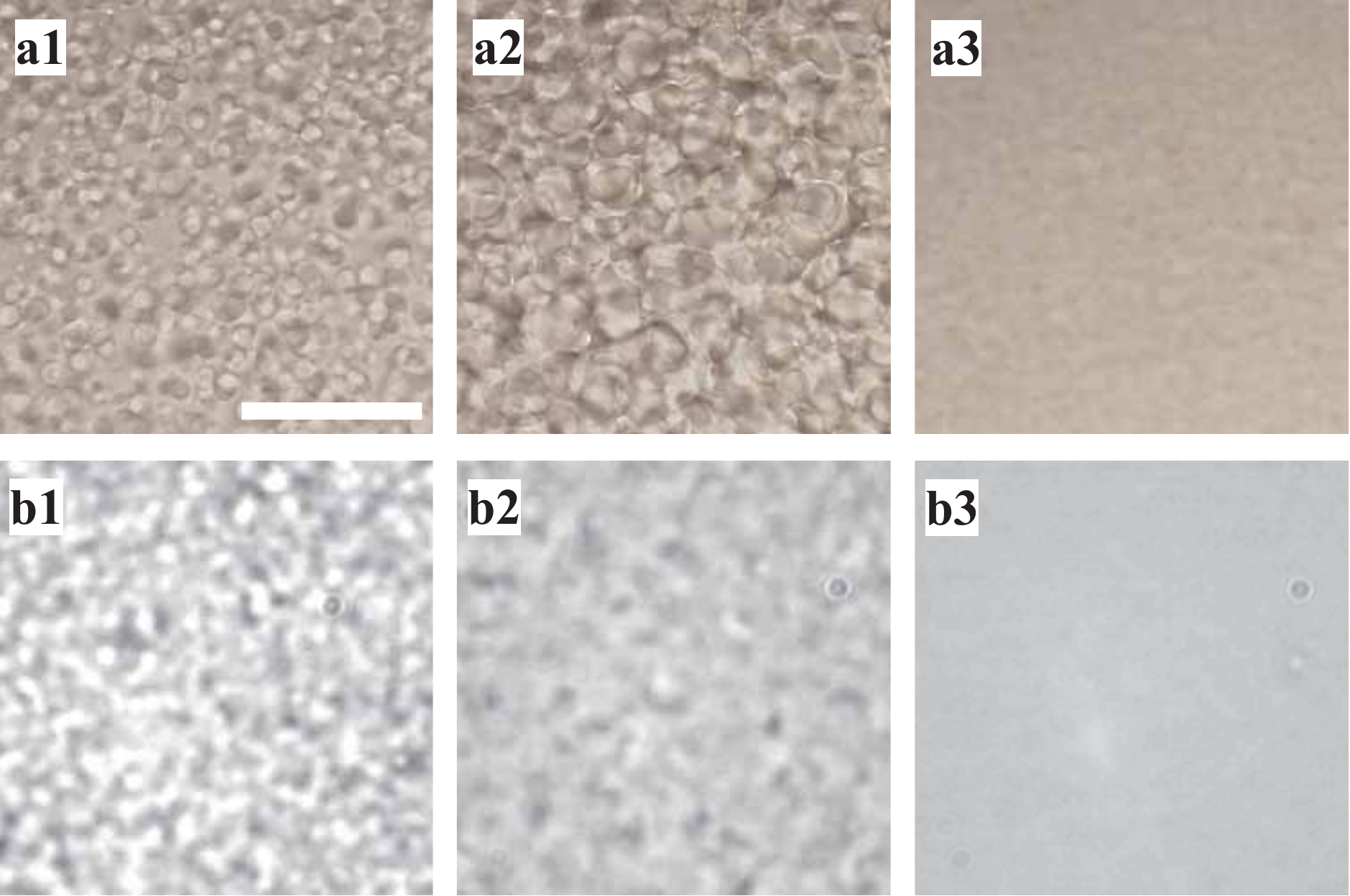}
\end{center}
\caption{(Colour on-line) 
Pattern evolution observed 
during the annealing of a supercooled liquid at $T_a$. 
(a1)-(a3) are observed with normal microscopy 
at $T_a=220$ K at the annealing time 
$t_a=60$ min, 120 min, and 240 min, respectively. 
(b1)-(b3) are observed with phase-contrast microscopy at $T_a=213$ K 
at $t_a=120$ min, 240 min, and 360 min, respectively. 
The length of the white bar in (a1) corresponds 
to 100 $\mu$m for (a1)-(a3), 
while to 20 $\mu$m for (b1)-(b3). 
The sample thickness was 100 $\mu$m for (a), while 
20 $\mu$m for (b). This figure is reproduced from fig. 1 of \cite{TKM}. 
}
\label{fig:NGSD}
\end{figure}

\begin{figure}
\begin{center}
\includegraphics[width=7cm]{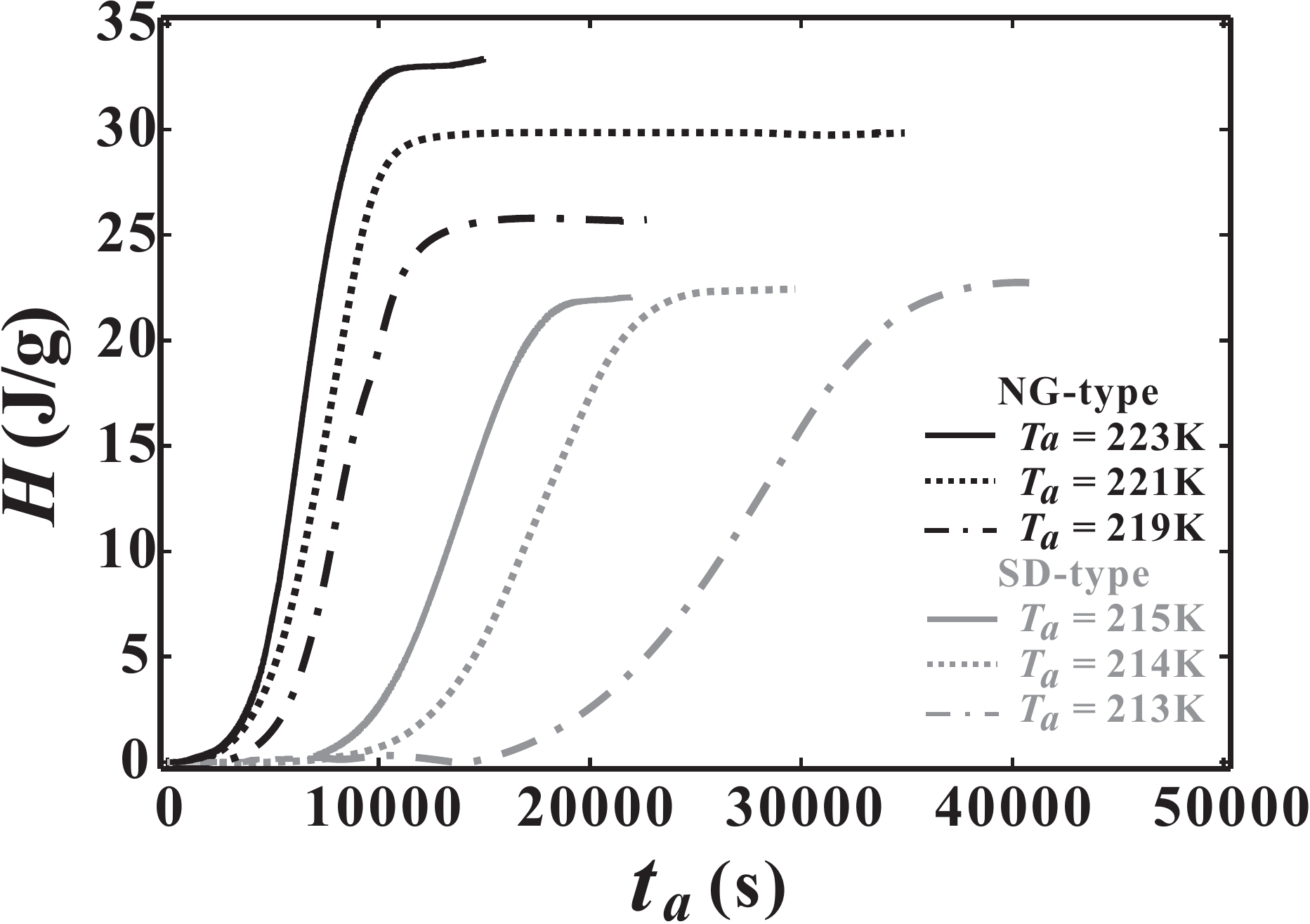}
\end{center}
\caption{
The annealing time dependence of the heat released during LLT, $H(t)$, at 
$T_a$=213 K, 214 K, 215 K, 219 K, 221 K, and 223 K. 
NG-type transformation occurs at  $T_a=$219 K, 221 K, and 223 K, 
while SD-type one occurs at $T_a$=213 K, 214 K, and 215 K. 
This figure is reproduced from fig. 2 of ref. \cite{KuriPRB}. }
\label{DSC}
\end{figure}

Here we show the typical kinetic processes of LLT observed in TPP in fig. \ref{fig:NGSD}: NG-type 
and SD-type LLT.  
The process is basically consistent with our simple kinetic theory. 
The heat evolution is also measured during LLT (see fig. \ref{DSC}), which is also consistent with our model 
that assumes the formation of locally favoured structures with a lower local free energy. 
According to our model, the heat evolution is proportional to the development of the bond order 
parameter $S$, since the heat is released in the process of the formation of locally favoured structures. 
This was supported by the structural study of the process of LLT by X-ray scattering \cite{KuritaXray}.

\subsubsection{Glass transition and the fragility of liquid I and II}

\begin{figure}
\begin{center}
\includegraphics[width=7cm]{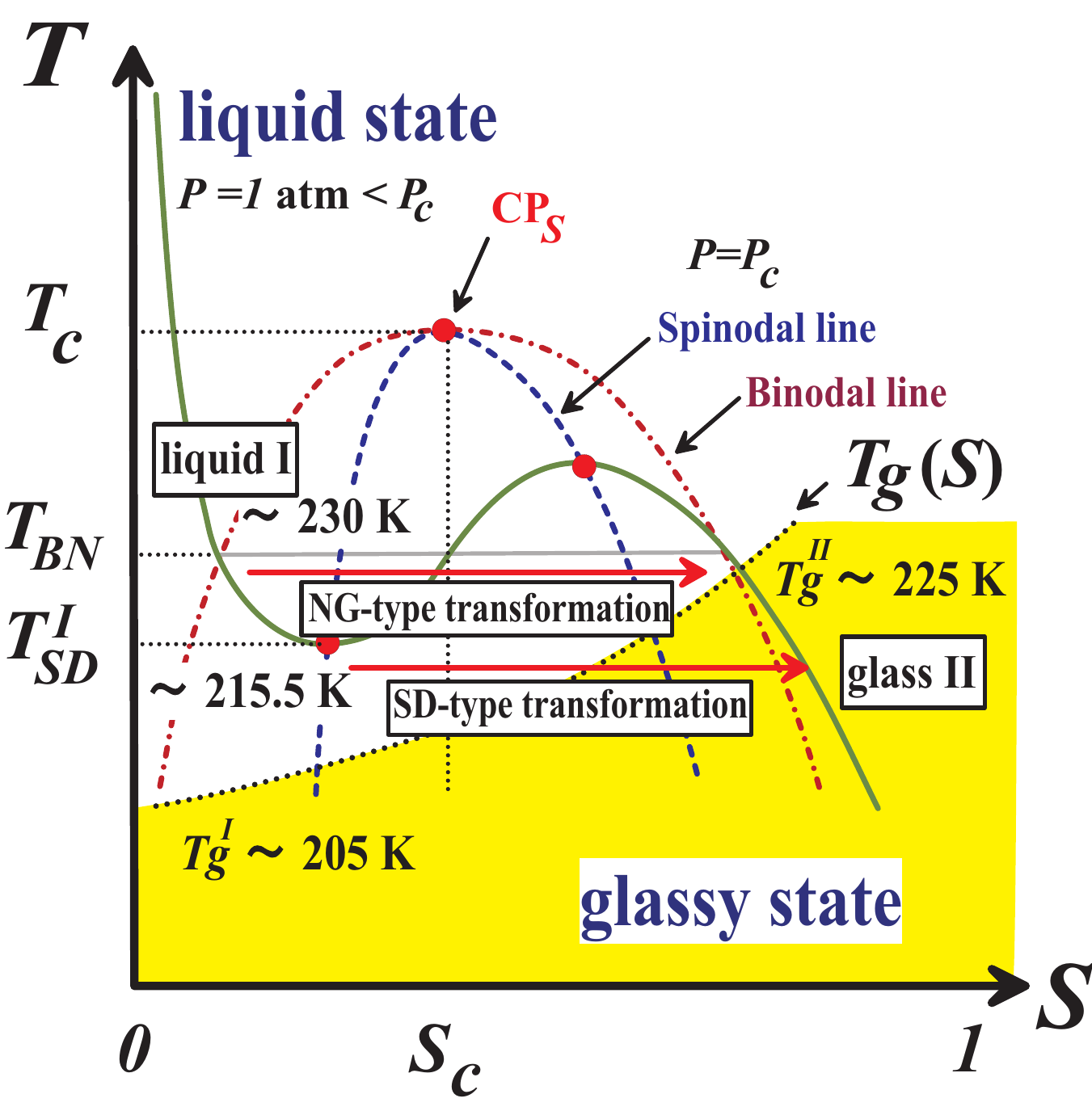}
\end{center}
\caption{(Colour on-line) Schematic $T-S$ phase diagram of TPP. 
The dashed and dot-dashed lines are spinodal and binodal lines, 
respectively. CP$_S$ is a gas-liquid-like critical point 
of the bond order parameter $S$, which might exist at a high pressure. 
A dotted curve is the glass transition line $T_{\rm g}(S)$ and the gray 
region corresponds to a glassy state. $T_{\rm BN}$ and $T_{\rm SD}^I$ 
represent the binodal temperature and the lower spinodal temperature 
at atmospheric pressure, respectively. 
Note that the liquid I $\rightarrow$ liquid II (glass II) transition 
inevitably 
accompanies vitrification, which makes a glass II state non-equilibrium 
in the sense that both $\rho$ and $S$ cannot reach their equilibrium 
values. 
}
\label{fig:figS1}
\end{figure}

\begin{figure}
\begin{center}
\includegraphics[width=8cm]{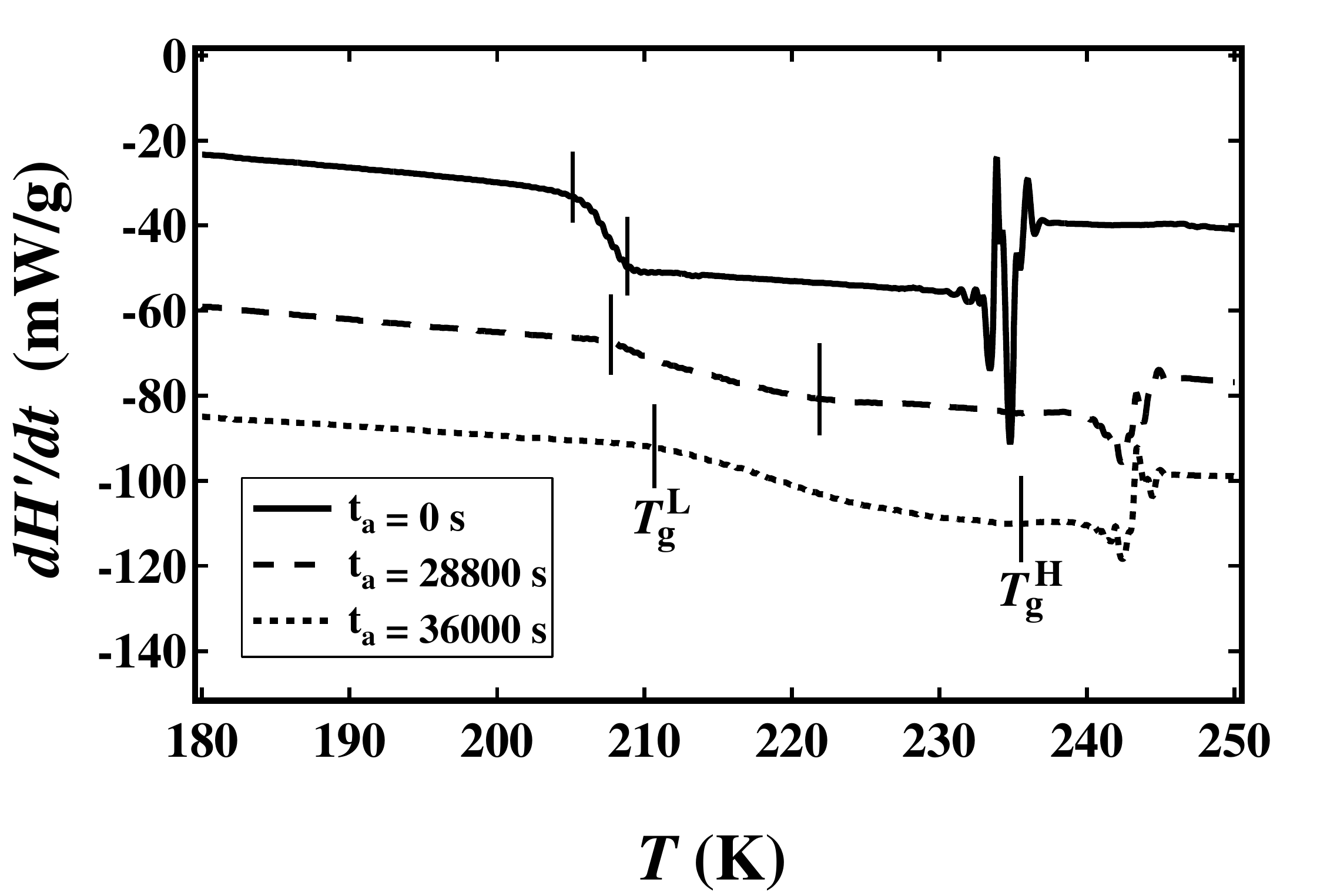}
\end{center}
\caption{The reversible part of the heat flow ($dH'/dt$) 
during the heating process of 
TPP annealed at 213K for $t_a=$ 0 s, 28800 s, and 36000 s. 
The measurements were made by AC calorimetry. 
A sample was heated with an alternating rate, whose average was 
2 K/min and whose period and amplitude were 30 s and 0.16 K, respectively. 
A different value of $t_a$ means a different value of $\tilde{S}$: 
For $t_a=0$ s, 28800 s, and 36000 s, $\tilde{S}$ is 0, 0.66, and 0.95, 
respectively. 
We can clearly see that the steplike change near $T_{\rm g}$ 
becomes broader with an increase in $t_a$: $\Delta T_{\rm g}=T_{\rm g}^H-T_{\rm g}^L$ 
increases with $t_a$. This figure is reproduced from fig. 2 of 
\cite{Kurifra}. 
}
\label{fig:fra}
\end{figure}

We confirmed that in both TPP and n-butanol, liquid I transforms into a glassy state of liquid II. 
The situation can be understood in the schematic state diagram of LLT for TPP on the $T$-$S$ plane 
(see fig. \ref{fig:figS1}), 
which includes the $S$-dependent glass transition line.  
We measured the glass transition temperature of liquid I and II. 
Figure \ref{fig:fra} shows the glass transition behaviour of liquid I, liquid II, and liquid 
during the SD-type liquid-liquid transformation. 
We can see that liquid I has a lower glass transition temperature $T_{\rm g}$  
than liquid II \cite{TKM} and liquid I is more fragile than liquid II \cite{TKM,Kurifra}. 
The width of the glass transition range of glass II ($\sim 23$ K) 
is much broader than that of glass I ($\sim 4$ K) \cite{TKM}. 
We can see the fragility monotonically decreases 
with the transformation, indicating the fragility is negatively correlated with the 
number density of locally favoured structures $S$ \cite{Kurifra}. 
In other words, the fragility is controlled by the same order parameter 
controlling LLT. These findings are difficult to explain by the scenario of aborted crystallization. 

Liquid I is more fragile than OTP, whereas liquid II is stronger 
than B$_2$O$_3$. 
This provides us with information on the nature of two liquids 
associated with their glass transitions: Liquid II 
is stronger than liquid I. This conclusion is consistent with 
the fact \cite{Rossler,Senker,Oguni} that 
the temperature dependence of the structural relaxation time $\tau_\alpha$ 
is super-Arrhenius (typical for fragile liquids) for liquid I, 
while it is almost Arrhenius (typical for strong liquids) for 
liquid II. 
A similar difference in the fragility between two liquid 
states of liquid-liquid transition was also reported 
for other materials \cite{AngellR,McMillan3,Sastry,McMillan2}. 
Thus, this may be a common feature of liquid-liquid transition. 

With the above-mentioned transition map of the structural relaxation time 
for liquid I and II, 
we can also naturally explain the observed temperature 
dependence of the complex dielectric constant 
\cite{Oguni,Johari} and its temporal change 
during the transformation from liquid I 
(normal liquid) to glass II (the glacial phase) \cite{Hedoux3}. 
During the transformation from liquid I to glass II, 
the structural relaxation time enormously slows down 
and its distribution becomes broader \cite{Hedoux3}. 
Upon the transformation of liquid I to glass II, 
the real part of the complex dielectric constant 
should decrease if the structural relaxation time 
becomes slower than the measurement frequency. 
Such behaviour was indeed observed \cite{Oguni,Johari}. 
One remaining question is why the distribution 
of the structural relaxation is so broad for liquid II. 
For ordinary glass formers, it is well established 
that the distribution of the structural relaxation time 
is narrower for a stronger liquid. The extremely broad 
distribution for strong liquid II is thus unusual \cite{Rossler}. 
This unusual behaviour might be related to the presence of locally favoured structures, but needs further studies. 

The dependence of the fragility on $S$ is consistent with our scenario 
of glass transition (see below), provided that locally favoured structures formed in TPP 
are not consistent with the symmetry of the equilibrium crystal and disturb crystallization. 
Our study clearly indicates that the fragility is not a material-specific 
quantity; namely, it is not related to the type of the 
interparticle potential in a direct manner. 
This result may shed new light on a fundamental question 
of what physical factor controls the fragility of liquid 
and contribute to our deeper understanding of liquid-glass transition.

\subsubsection{LLT in a mixture of TPP with other liquids}

It is interesting to consider how LLT of a liquid affects the phase behaviour, or the miscibility, of its mixture with another liquid. 
This problem was discussed experimentally \cite{AngellMIX}, theoretically \cite{DebeMIX} and numerically \cite{Le}. 
Recently we found that LLT can be observed in mixtures of TPP with other liquids, and in some cases 
LLT induces phase separation \cite{Kurimiscibility}. This indicates that the miscibility of liquid I with other liquids 
is different from that of liquid II with them. 

\begin{figure}
\begin{center}
\includegraphics[width=8.5cm]{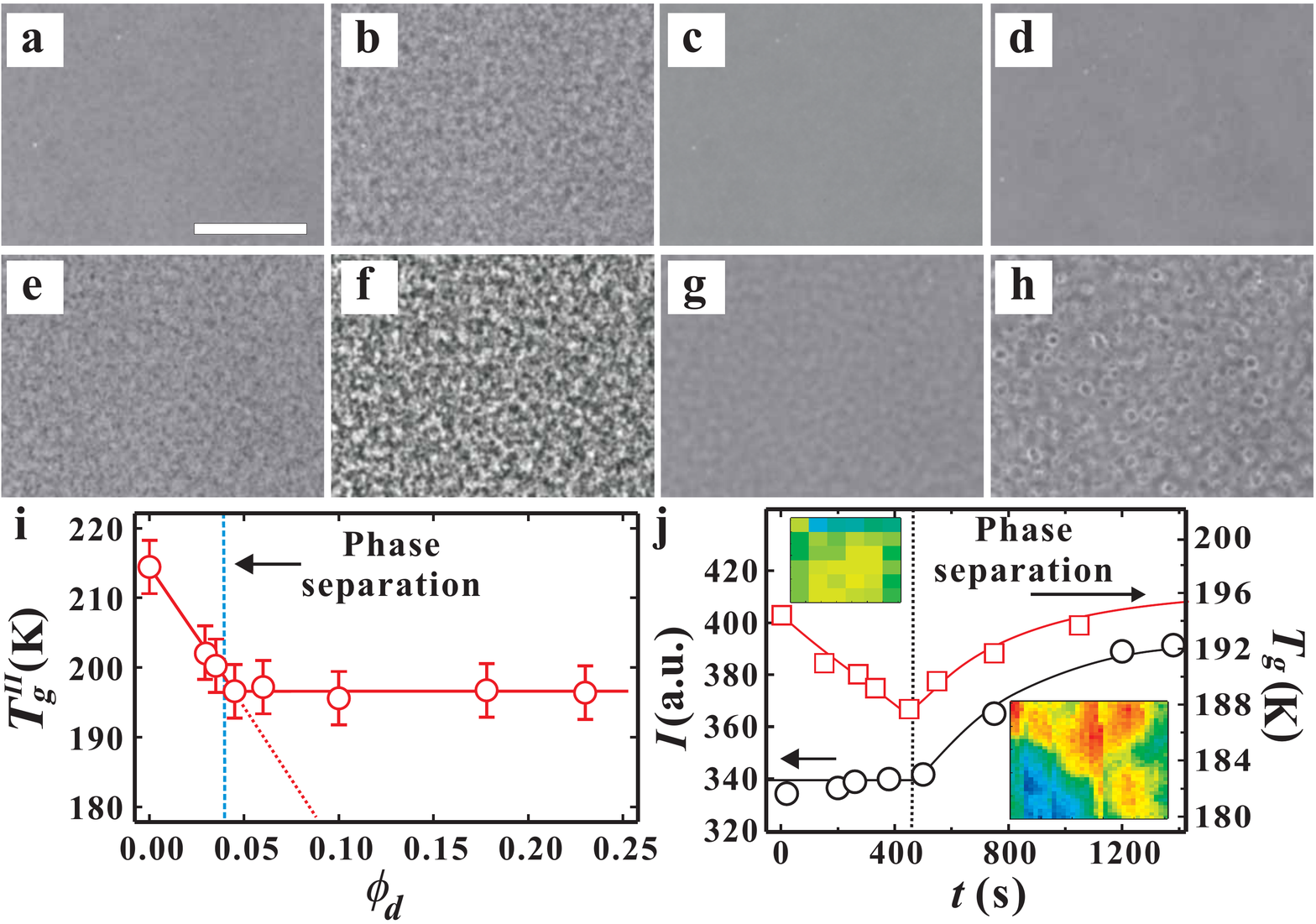}
\end{center}
\caption{(Colour on-line) Demixing induced by LLT in TPP-diethyl ether mixtures. 
(a-d) Pattern evolution  
during the transformation at 209 K for $\phi_d$= 2.98 \%. 
(a) $t=$80 min, (b) 100 min, 
(c) 150 min, and (d) 400 min. 
(e-h) Pattern evolution during the transformation at 209 K 
for $\phi_d$= 4.45 \%.
(e) $t=$20 min, (f) 25 min, 
(g) 70 min, and (h) 112 min.
The scale bar corresponds to 100 $\mu$m. 
\textbf{i,} $\phi_d$-dependence of $T_{\rm g}^{II}$.
$T_{\rm g}^{II}$ decreases monotonically with an increase in $\phi_d$ 
below 4\%, 
whereas $T_{\rm g}^{II}$ keeps almost constant for $\phi_d \geq 4$ \%. 
This indicates that phase separation occurs for $\phi_d \geq 4$ \%, 
which is consistent with microscopic observation ((a)-(h)). 
\textbf{j,} Temporal change of $T_{\rm g}$ (squares) 
and the intensity of Raman peak at 3068 cm$^{-1}$ (circles) 
for $\phi_d$ = 8 \% and at $T$ = 209 K. 
Note that this peak comes from TPP and does not from diethyl ether.   
The insets of (j)  
are the intensity maps for the peak at 3068 cm$^{-1}$ 
measured at $t$ = 200 s (left) and at $t$=1800 s (right). 
The area size and the spatial resolution of the left inset are 
36 $\mu$m $\times$ 36 $\mu$m and 6 $\mu$m $\times$ 6 $\mu$m, respectively.
On the other hand, those of the right inset are 100 $\mu$m $\times$ 
100 $\mu$m and 3 $\mu$m $\times$ 3 $\mu$m, respectively. 
At $t$ = 200 s, the speed of LLT is still fast and thus 
we needed to measure the 2D intensity map very quickly. 
This is the reason why the image size is smaller and the 
spatial resolution is poorer there.  
More reddish (bluish) colour means higher (lower) 
intensity, i.e., a higher (lower) fraction of TPP. 
This figure is reproduced from fig. 2 of ref. \cite{Kurimiscibility}
}
\label{fig:mix}
\end{figure}

Figures \ref{fig:mix}(a)-(d) and \ref{fig:mix}(e)-(h) show pattern evolution 
in a mixture of TPP and diethyl ether 
at 209 K for $\phi_d$ = 2.98 \% and 4.45 \%, respectively, 
where $\phi_d$ is the volume fraction of diethyl ether. 
The early-stage pattern evolution looks the same between the two cases. 
The amplitude of fluctuations grow with time while keeping 
the characteristic length constant in the early stage 
of the transformation, which is characteristic 
of the Cahn's linear regime of SD-type LLT (see figs. \ref{fig:mix}(a) and (e)) 
\cite{KuriSci}. 
However, the analysis of the process may require a special care since an image obtained by phase contrast microscopy 
generally loses long-wavelength components of the spatial fluctuations of the refractive index and thus has to be 
corrected by the proper optical transfer function.  
In the later stage, the domain size coarsens with time (see figs. \ref{fig:mix}(b) and (f)) 
and afterwards the system becomes homogeneous again 
(figs.  \ref{fig:mix}(c) and (g)), which is 
typical of SD-type LLT \cite{KuriSci,KuriPRB,KuriSimu}. 
However, unusual behaviour is observed for $\phi_d$ = 4.45 \%: 
Reappearance of droplets after the homogenization. 
We found that such behaviour is observed 
only for $\phi_d  \geq $ 4 \%, and not for 
$\phi_d  \leq $ 3 \%, and the final volume fraction of droplets 
(fig.  \ref{fig:mix}(h)) increases 
with an increase in $\phi_d$.
This can be explained by a scenario that phase separation 
occurs for $\phi_d \geq 4$ \%, accompanying the emergence 
of diethyl-ether-rich droplets.

To confirm this scenario, 
we annealed samples at 209 K for 10 hours 
to complete the process of LLT and measured the glass transition behaviour. 
We found that $T_{\rm g}^{II}(\phi_d)$ decreases with an increase in $\phi_d$ 
below $\phi_d$ = 4 $\%$, whereas it does not depend on $\phi_d$ 
for $\phi_d \geq 4$ \% (fig.  \ref{fig:mix}(i)). 
This indicates that phase separation indeed takes place for $\phi_d \geq 4$ \%, 
resulting in the formation of TPP-rich and diethyl-ether-rich phases. 
Each phase should have its own $T_{\rm g}$, but 
$T_{\rm g}$ of the diethyl-ether-rich phase 
may be located at too low a temperature, which is out of the range 
of our DSC measurements. 
Thus, only $T_{\rm g}$ of the TPP-rich phase was measured. 

Furthermore, we investigated the spatial distribution of the 
two components of the mixture 
by using micro Raman spectroscopy measurements, whose 
spatial resolution is 1 $\mu$m $\times$ 1 $\mu$m. 
We used the Raman peak at 3068 cm$^{-1}$ as a fingerprint, 
since this peak exists only 
for TPP (not for diethyl ether) and is not affected by LLT \cite{Hedoux}.
We followed the temporal evolution of the peak intensity 
at a fixed point of the sample of $\phi_d$ = 8 \% at 209 K. 
The results are shown in fig.  \ref{fig:mix}(j). 
The intensity is almost constant before $t$ = 450 s, 
indicating that the fraction of TPP at this point is constant with time. 
Then, the intensity increases continuously. 
We also measured a peak at 3020 cm$^{-1}$, which belongs to 
diether ether, at the same point. 
The intensity of this peak does not change until 450 s, 
but decreases after 450 s. 
Both facts suggest that phase separation starts to take place at $t$ = 450 s 
and our measurement spot belongs to a TPP-rich domain 
after the initiation of the phase separation. 
Using the fact that the peak intensity is proportional to the molecular fraction, 
we can estimate the local volume fraction of diethyl ether, $\phi_d$, 
after the phase separation. 
This result tells us that $\phi_d$ changes from the initial value, 8 \%, to 
the final one, 5 \%, which coincides well with the above-mentioned threshold $\phi_d$, 
above which phase separation is induced by LLT. 
We also performed real-time 2D Raman microscopy measurements 
(the insets of fig.  \ref{fig:mix}(j)). 
We can see that the system is homogeneous at $t$ = 200 s, but 
phase separates into TPP-rich and TPP-poor regions at $t$ = 1800 s.  
This confirms the occurrence of LLT-induced phase separation. 
The temporal change of $T_{\rm g}$ 
during this process (see the squares of fig. \ref{fig:mix}(j)) 
also supports our scenario. 

This LLT-induced phase separation may provide a new route for triggering phase separation 
of a binary liquid mixture. 
We also revealed that toluene is miscible with both liquid I and II of TPP and adding a small amount of toluene to TPP 
transforms the LLT of pure TPP, where liquid II is in a glassy state, to a true LLT in the sense that both liquids I and II are in a liquid state  
with fluidity. 
This provides an example of a `true' liquid-to-liquid transition 
in molecular liquids. 
This transition accompanies a drastic change in the 
transport properties (viscosity and diffusivity) by many orders of 
magnitude. Thus, it may be used as a liquid whose transport properties can be changed drastically. 

LLT has so far attracted attention purely from a scientific viewpoint, 
but the above finding may open new possibilities for the control of fluidity, 
which governs flow, diffusion, and chemical reaction rate, 
and the control of chemical miscibility and reactivity of a liquid 
without modifying its chemical structure. 

It is also worth noting that mixing a liquid with another liquid possessing LLT 
will add considerable variety to liquids exhibiting LLT, 
which may be important for future applications of LLT. 
Furthermore, LLT in a mixture can be used to seek LLT in a single-component liquid,  
which is difficult to access experimentally, for example, due to crystallization or glass transition, as in water. 
Mixing another component may reveal LLT hidden by these phenomena (see below).

\subsubsection{LLT in a mixture of water and glycerol}

One of the important remaining questions is whether LLT exists in pure water or not, as discussed in sec. \ref{sec:LLTwater}.
Unlike the cases of TPP and n-butanol, experimental verification of LLT in water is quite difficult due to 
the interference by instantaneous crystallization. 
There may still be a few routes to access a hidden LLT in water, if it exists. 
One strategy is to use water confined in nm-size pores 
\cite{Liu_conf,Mallamace} or to use protein-hydration water \cite{Doster}. 
These works show evidence suggestive of a fragile-to-strong liquid transition and LLT. 
However, these experiments inevitably suffer from criticisms  
that water confined into a nm-scale space surrounded by a wall is intrinsically different from 
bulk water because of the presence of water-wall interactions 
and the reduced dimensionality \cite{Mancinelli,Morineau,Findenegg}. 
It was shown \cite{Mancinelli}, for example, that (i) without a special care 
we cannot conclude even whether water inside a nm-size pore is liquid or 
solid or amorphous or crystalline and (ii) the interaction with the wall 
makes the confined state inhomogeneous. 
Since the presence of the amorphous-amorphous transition in bulk does not 
prove the presence of LLT either (see above), it may be fair to say  
that we do not have any firm experimental evidence 
for LLT in water yet, although there are many implications. 

\begin{figure}[h]
\begin{center}
\includegraphics[width=8cm]{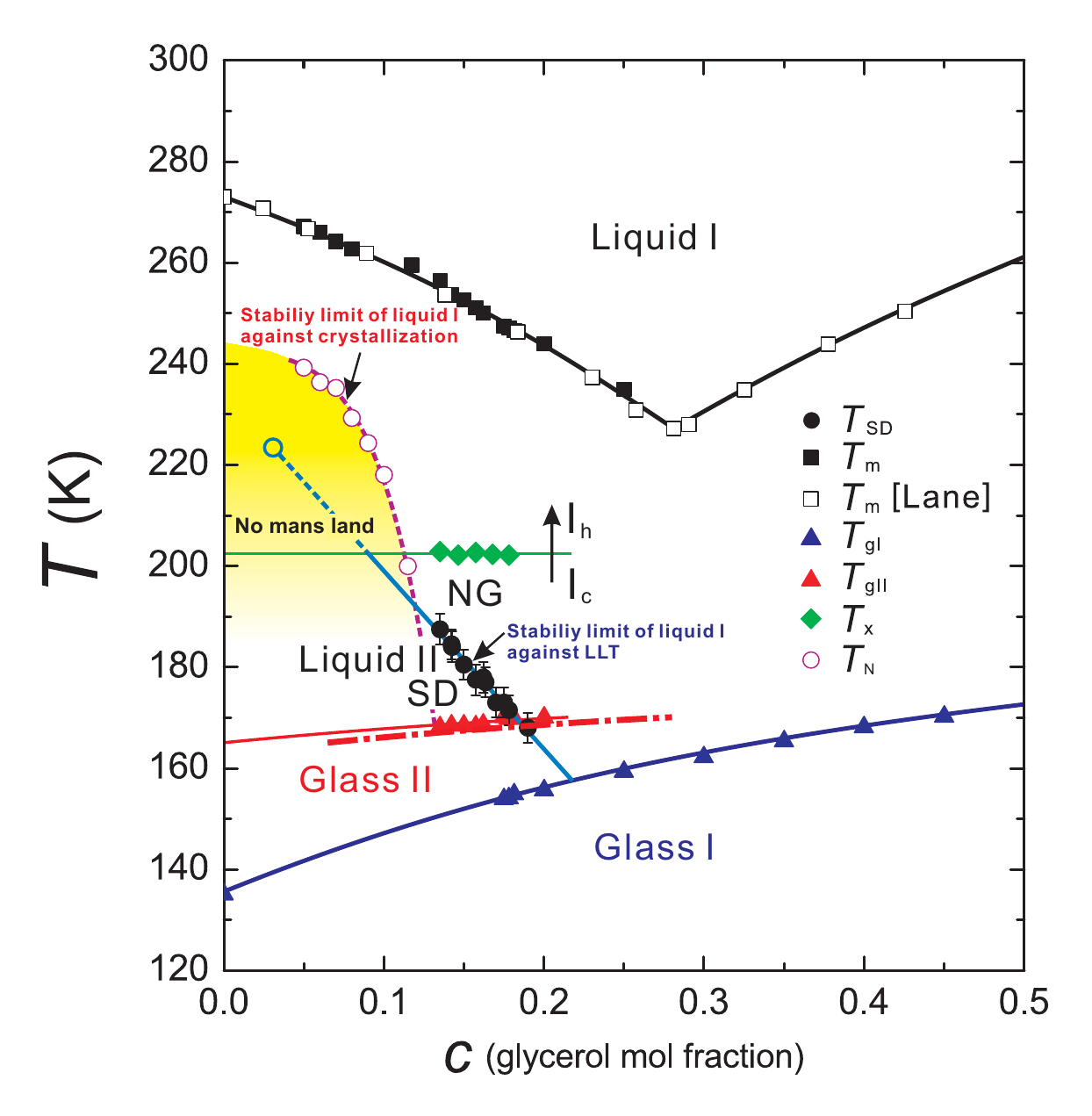} 
\end{center}
\caption{(Colour on-line) Glycerol concentration vs temperature ($c-T$) state diagram of water/glycerol.
$T_{\rm SD}$: LLT spinodal temperature (black filled circles); $T_{\rm gI}$: the glass transition temperature of liquid I 
(blue filled triangles). For pure water ($c=0$), we use the widely accepted value of 136 K \cite{DebenedettiR} for $T_{\rm gI}$; 
$T_{\rm gII}$: the glass transition temperature of liquid II (red filled triangles). Dot-dashed line indicates $T_{\rm gII}$ of pure liquid II without ice I$_{\rm c}$, provided that liquid II contains $\phi_{\rm c}=$ 17 \% of ice I$_{\rm c}$, which should result in the increase in the glycerol mole fraction of liquid II by 6.4 \% (see SI for the related discussion); 
$T_{\rm H}$: the homogeneous nucleation temperature (violet open circles) measured for the cooling rate of 100 K/min 
(see SI for the standard $T_{\rm H}$ measured for an emulsified sample);
$T_{\rm X}$: the transition temperature from ice I$_{\rm c}$ to I$_{\rm h}$, which was determined by microscopy observation (green filled diamonds); 
$T_{\rm m}$: the melting (liquidus) temperature (black filled squares: our data; open squares: the data of Lane \cite{Lane}). 
We make a linear extrapolation of $T_{\rm SD}$ to estimate the position of a hypothetical critical point (CP) (light blue 
open circle), since 
we cannot access $T_{\rm SD}$ for $c < 0.13$ due to rapid nucleation of ice I$_{\rm h}$ before reaching the final target temperature 
in the quench process. For $c>0.19$, on the other hand, the kinetics of LLT drastically 
slows down, which also prevents us from accessing LLT during the observation time. 
Finally we note that the $T_{\rm SD}$ we measured is the stability limit of liquid I, 
and we could access neither the binodal line nor the stability limit of liquid II 
because of interference by ice crystallization. This figure is reproduced from fig. 4 of ref. \cite{murataNM}.}
\label{fig:watergly}
\end{figure}

Recently we took a different strategy: mixing water with glycerol to avoid crystallization 
of water. Note that glycerol is a well-known non-crystallizable liquid and can cause 
strong frustration against water crystallization. 
In an aqueous glycerol solution we found the direct experimental evidence for 
genuine (isocompositional) LLT without accompanying phase separation \cite{murataNM}. 
We confirmed that liquid I transforms via the two types of kinetics characteristic of 
the first-order transition of a non-conserved order parameter, NG and SD, 
towards homogeneous liquid II. The processes are essentially the same as those observed in TPP. 
The state diagram of water/glycerol mixtures is shown in fig. \ref{fig:watergly}. 
The liquid-solid phase diagram of water/glycerol mixtures is very similar to the $T$-$P$ phase diagram of pure water, which also has a $V$-shape. 
We found that 
liquid I and II, differ in the density, the refractive index, the structure, the hydrogen bonding state, 
the glass transition temperature, and the fragility. 
We revealed that this transition is mainly driven by local structuring of water rather than glycerol, suggesting 
a possible link to LLT in pure water. 
In relation to this, it was recently pointed out by Towey and Dougan \cite{towey2012structural} that 
glycerol molecules act to ``pressurize'' water. This further suggests a link between a water/glycerol mixture 
and pure water. 
However, further study is necessary to clarify whether water has LLT without glycerol or not.

\subsubsection{LLT and wetting phenomena}

LLT of a liquid may also affect its wetting properties to a substrate. 
We found that the wettability of liquid I is different from that of liquid II \cite{Murata_wetting}. 
These results show that liquid I and liquid II differ in the density, the refractive index, 
the glass transition temperature, the fragility, the miscibility with other liquids, 
and the wettability to a substrate. 

\begin{figure}
\begin{center}
\includegraphics[width=8.5cm]{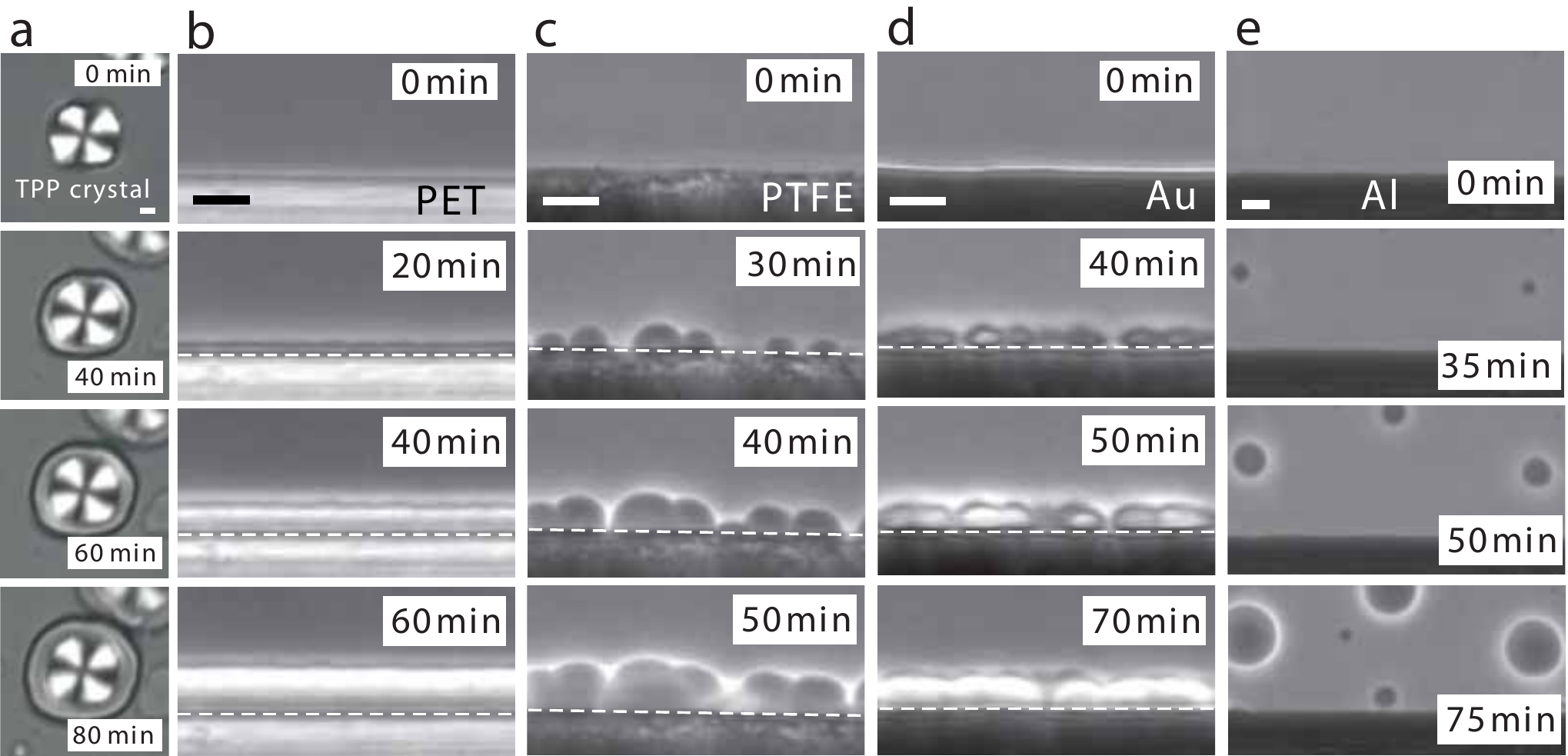}
\end{center}
\caption{(Colour on-line) Heterogeneous nucleation of liquid II on various solid 
surfaces at 220 K. 
(a) Time evolution of LLT in the presence of a TPP crystal, 
which was observed with polarizing microscopy under the 
crossed Nicols condition.
In the beginning (0 min), we observe only the TPP crystalline spherulite  
with the Maltese cross pattern.  
Then, the layer of liquid II, which has no birefringence, 
is formed on the surface of the TPP spherulite (40 min)
and its thickness linearly grows with time (60 min, 80 min) (see (f)). 
(b) Time evolution of LLT in the presence of a PET 
(poly(ethylene terephthalate)) surface. 
In this case, LLT proceeds while forming a thin film of liquid II: 
compete wetting. White dashed lines indicate the location of the surface. 
(c),(d) Time evolution of LLT in the presence of 
PTFE (poly(tetrafluoroethylene)) (c) and Au (d) 
surface.
Nuclei of liquid II is preferentially formed on the substrates 
with a finite contact angle ($\theta \le 90^{\circ} $): 
partial wetting.
(e) Time evolution of LLT in the presence of an Al surface.
Nuclei of liquid II do not have any contact to the Al surface 
and normal NG-type droplet growth in bulk was observed: non-wetting.
The scale bars correspond to 10 $\mu$m. 
This figure is reproduced from fig. 1 of ref. \cite{Murata_wetting}. 
}
\label{fig:wet}
\end{figure}

Here we show NG-type LLT ($T_{\rm SD}<T<T_{\rm BN}$, where $T_{\rm BN}$ is the binodal line of LLT, below which liquid I 
becomes metastable to liquid II) for a case of complete wetting. Figure \ref{fig:wet}(a) shows pattern evolution 
during heterogeneous nucleation on a TPP crystalline surface at 220 K. 
Before the temperature quench ($t<0$), the spherulite of TPP crystal 
with optical birefringence (manifested by the Maltese cross) 
grows in the homogeneous liquid I at 235 K. 
Immediately after a temperature quench to 220 K, 
LLT is initiated and the layer of liquid II is formed preferentially 
on the surface of the TPP crystalline spherulite ($t=40$ min). 
Then its thickness linearly grows with time ($t=60$ min, 80 min) (see below). 
Figure \ref{fig:wet}(b) shows the process of heterogeneous nucleation of liquid II 
on a poly(ethylene terephthalate) (PET) surface at 220 K. 
Unlike usual NG-type transformation, 
LLT proceeds while accompanying the formation of a thin film 
of liquid II on the solid surface for both TPP crystal and PET case: 
complete wetting. 
In these cases, nucleation occurs preferentially on the solid surfaces. 
Figures \ref{fig:wet}(c) and (d) show the processes of heterogeneous nucleation of liquid II 
on poly(tetrafluoroethylene) (PTFE) and Au surfaces at 220 K, 
respectively. 
In these cases, we observe partial wetting behaviour. 
We can see that the nucleation rate is much higher on 
the surface than in bulk. 
This behaviour is consistent with the typical heterogeneous nucleation 
behaviour on a wettable substrate, which is known for other types of phase transitions such as 
crystallization and phase separation \cite{DebenedettiB}. 
Finally, for Al we observe non-wetting behaviour (see fig. \ref{fig:wet}(e)). 

We revealed that significant surface wetting effects 
on NG-type LLT are induced by specific interactions 
(weak hydrogen bonding) 
between substrates and TPP molecules \cite{Murata_wetting}. 
This `bottom-up' wetting mechanism of a microscopic nature 
is markedly different from the ordinary macroscopic mechanism 
responsible for wetting effects on phase separation \cite{tanakawet}, where 
dispersion forces play a crucial role. This has an interesting implication on the roles of specific interactions 
in bond orientational ordering. 
We also demonstrate critical point wetting, i.e., 
a partial-to-complete wetting transition when approaching $T_{\rm SD}$. 
This critical-point-wetting-like behaviour \cite{cahn1977} may be regarded as evidence 
for the criticality associated with LLT, although the extrapolation is too large to draw 
a definite conclusion. 
The interfacial tension between liquid I and II decreases in a manner 
consistent with the mean-field criticality \cite{KuriSci}. 
This further supports that the transition observed in TPP is truly 
``liquid-liquid transition''. 

Unlike the significant change of the kinetics for NG-type LLT by a substrate, 
we reveal that SD-type LLT is not affected by surface wetting effects. 
This is markedly different from wetting behaviour observed in spinodal decomposition 
of a system of a conserved order parameter \cite{tanakawet}, and can be 
explained by the non-conserved nature of the order parameter governing LLT. 

These findings may have a significant implication not only for 
the mechanism of LLT itself but also for applications of LLT. 

Our results show that we can use solid substrates or particles as 
a catalyst to promote LLT in a metastable region. 
This may open up novel possibilities not only 
of spatial patterning of liquid I and liquid II 
using chemically or topologically patterned surfaces 
but also of controlling the kinetics of LLT. 
The generality of the bottom-up wetting mechanism for LLT in atomic liquids, 
oxides, and chalcogenides  
is an interesting topic of the future study.

\subsubsection{LLT or other phenomena: the nature of the glacial phase in TPP}

As discussed above, there have been various proposals for the nature 
of a new amorphous state of TPP. 
Similarly, the phenomena observed in n-butanol were recently interpreted as 
a consequence of the formation of micro- or nano-crystallites, contrary to our scenario 
\cite{Hedoux_buta,Ramos1,Ramos2}.

As shown above, we confirmed \cite {TKM} that the glacial phase formed below 
212 K is a homogeneous glassy state of liquid II, namely, pure glass II, 
whereas that formed above 213 K is a mixture of glass II and 
micro-crystallites. This absence of micro- or nano-crystallites 
was confirmed particularly for liquid II made below 212 K. 
The fraction of micro-crystallites decreases 
if an annealing temperature 
$T_a$ approaches to 212 K, 
and becomes almost zero (undetectable) below 212 K \cite{TKM}. 
This lower bound temperature for nano-crystallites formation seems to have sample-dependence. 
In some cases the temperature is located around 215 K, whereas in other cases it is around 212 K 
for TPP. So this temperature may be determined independently from LLT, e.g., by impurities in a sample. 
In the case of n-butanol, micro-crystallites were observed even in 
the lowest temperature below $T_{\rm SD}$. 
In n-butanol, we cannot access a state free from crystallization, probably because 
$T_{\rm SD}$ is located only slightly above $T_{\rm g}$. 
Nonetheless, 
we believe that the observed phenomena cannot be explained by 
crystallization alone. 

The peculiar crystallization behaviour accompanied by LLT 
may be a consequence of three factors: The first factor is a lower surface 
tension between crystal and liquid II than that between crystal and liquid II 
(see fig. \ref{fig:wet}(a)) \cite{Murata_wetting}. This should lower a barrier for crystal nucleation and lead to 
the large increase in the crystal nucleation frequency. 
The second factor is a stronger frustration effect of locally favoured structures 
against crystallization in liquid II than in liquid I. 
The third factor is that the transformation of liquid I to glass II prevents further growth of micro-crystallites. 
These three factors may be responsible for the formation of micro- or nano-crystallites, 
which should be confirmed in the future. 
  
Thus, we argue that all the confusions concerning the nature 
of the glacial phase may originate from that 
the glacial phase prepared above 213 K, which is actually 
a mixture of glass II and micro-crystallites, has been misinterpreted 
as either a homogeneous phase with anisotropy 
or a mixture of liquid I and micro-crystallites or nano-clusters. 
For example, the signatures indicative of 
nano-crystals or defect-ordered crystals may arise 
from the micro-crystallites embedded in the glass II. 
H\'edoux et al. \cite{Hedoux} reported that the Raman spectra 
can be decomposed into those of the supercooled liquid and 
crystals and they concluded that 
a glacial phase is composed of nano- or micro-crystallites possibly 
mixed with a fraction of untransformed supercooled liquid (liquid I 
in our terminology), depending upon the annealing temperature. 
This conclusion is partly consistent with our picture, 
but we argue that it should be decomposed into those 
of glass II and crystals and not into those of liquid I and 
crystals. 
We stress that this scenario reasonably 
explains the results of x-ray scattering \cite{Cohen,Hedoux} 
and neutron scattering \cite{Alba,Yarger,Hedoux1,HedouxN} and 
those of Raman scattering \cite{Hedoux1,Hedoux4,HedouxR1,HedouxR2} for TPP 
and similar results for n-butanol \cite{Hedoux_buta,Ramos1,Ramos2}.  

Finally we consider spontaneous heat evolution during 
annealing, which is observed by 
isothermal differential scanning calorimetry measurements 
\cite{Oguni,Hedoux2,Hedoux3}. 
This should reflect both liquid-liquid transformation and 
crystallization above 213 K, according to our scenario. 
To extract the information on the kinetics of LLT from such data, 
we need to subtract the contribution of crystallization. 
Once this is properly done, we can deduce the time evolution of 
the bond order parameter $S$. 
Results of such analyses and detailed comparison of them 
with the kinetics of liquid-liquid transition deduced from 
pattern evolution will be reported in the future. 

Thus, we conclude at this moment that the existing data on the glacial phase 
of TPP can be naturally explained by our scenario of 
liquid-liquid transition, although further careful studies are still necessary 
to derive a definite conclusion. 

\subsection{Link between LLT and polyamorphism}

\subsubsection{Elastic effects on a polyamorphic transition}

Convincing experimental evidence for polyamorphism has been reported for many liquids including 
water \cite{Mishima,MishimaR,MishimaS}, Si \cite{McMillan_Si,daisenberger2007}, silica \cite{lacks2000,silica2004_1,silica2004_2}, 
and metallic glasses \cite{sheng2007}. 
Typically there are low-density and high-density amorphous states. 
The low-density one is formed at ambient pressure, whereas the high-density one 
is formed by applying a pressure to either an amorphous or a crystal state prepared at 
ambient pressure. 
When pressure is reduced, a high-density amorphous state transforms back to a low-density 
one at a certain pressure. This hysteresis behaviour is interpreted as a manifestation of 
the first-order nature of the transition. 
However, we point out that this transition between the two non-ergodic states 
is far more complicated than liquid-liquid transition from the following reasons. 
It is not easy to figure out the role of the cooperativity ($J$) 
in such a solid-state transition because of the non-ergodic nature of the transition. 
Furthermore, the link between the transition and the underlying free energy is obscured by 
the effects of mechanical stress inevitably involved upon the transition. 
It is not easy to separate the thermodynamic factors and the mechanical ones for pressure-induced phase transitions 
in a non-ergodic state. 

Here we consider this problem in more detail.  
The standard elastic theory of isotropic matter tells us 
\cite{Onuki} that 
the elastic energy is given by 
\begin{eqnarray}
f_{\rm el}=\int d \mbox{\boldmath$r$} [\frac{B}{2} (\mbox{\boldmath$\nabla$} 
\cdot \mbox{\boldmath$u$})^2 
+\frac{\mu}{4} \Sigma_{ij}(\frac{\partial u_j}{\partial x_i}
+\frac{\partial u_i}{\partial x_j}
-\frac{d}{2}\delta_{ij}\mbox{\boldmath$\nabla$} \cdot 
\mbox{\boldmath$u$})^2], \nonumber
\end{eqnarray} 
where $B$ is a bulk modulus, $\mu$ is a shear modulus, $\mbox{\boldmath$u$}$ 
is the deformation vector, and $d$ is the dimensionality. Note that there is 
a coupling between the bond order parameter ($S$) and 
$\mbox{\boldmath$\nabla$} 
\cdot \mbox{\boldmath$u$}$, which is given by the following free energy 
in the lowest order: 
\begin{eqnarray}
f_{\rm int}=\int \ d \mbox{\boldmath$r$}\ \alpha (S \mbox{\boldmath$\nabla$} 
\cdot \mbox{\boldmath$u$}), \nonumber
\end{eqnarray}
where $\alpha$ is the coupling constant. For water, $\alpha$ should be 
negative. 
The total free energy should include these elastic contributions.
This causes a marked difference between solid-state and liquid-state 
transitions. In addition to the possible differences in 
the values of $\Delta E$, $\Delta v$, and $\Delta \sigma$, and $J$ 
between these two cases, thus, we need to include 
the above-described elastic terms into our free energy for a solid case. 
As the result, for example, there should be an elastic energy barrier in 
addition to a barrier coming from the interfacial energy, for 
a phase-transformation (nucleation) process in a solid state. 
Its importance can, for example, be recognized from the fact that 
HDA of water breaks in pieces upon the transformation into LDA while releasing mechanical stress 
\cite{MishimaR,DebenedettiR}. 
Since the elastic energy is proportional to the volume, the elastic terms 
should significantly increase the nucleation barrier, which leads to 
a large difference in the location of the apparent phase-transition 
lines between solid-state thermomechanical and liquid-state thermodynamic transitions.

\subsubsection{Case of water as an example}
As an example, we consider the relationship between a possible liquid-liquid transition 
of liquid water and the corresponding amorphous-amorphous transition 
of solid amorphous water. 
Spinodal-like lines associated with the stability limits of LDA and HDA 
are often used to determine the location of a 
liquid-state transition and the corresponding critical point. 
In relation to the difference between 
solid-state and liquid-state transitions, we point out the 
following fundamental problems: (i) Amorphous ices are not in an equilibrium 
state, but in a non-equilibrium glassy state. Thus, we cannot apply an 
equilibrium liquid-state theory to predict the solid-state 
phase transition between these non-equilibrium 
amorphous ices, as described above. 
A solid amorphous state may even depend upon the history of sample 
preparation because of the non-equilibrium nature of the state. 
(ii) The entropy difference 
between normal-liquid structures and locally favoured structures in a 
liquid state should be significantly larger than that in a glassy state, 
as described above, 
since in a glassy state a free volume, or the translational and rotational 
degrees of freedom, is very small for both types of amorphous ices (i.e., 
small $\Delta \sigma$). 
Thus, there may be no direct connection between a liquid-state transition 
and a solid-state transition even if we assume the equilibrium nature for a 
solid-state transition. (iii) More importantly, the solid-state phase 
transformation with a volume change inevitably accompanies the elastic 
deformation \cite{Onuki}, as discussed above. 

The existence of two amorphous ices might be a manifestation of the existence 
of a second-critical point \cite{MishimaR}. 
However, the above consideration suggests that the experimental data of 
an amorphous-amorphous transition 
may not necessarily be useful for determining the location of a liquid-state 
critical point in a straightforward manner. 
Then, how can we determine its location experimentally? 
A search of a kink in the melting curve 
of ice crystals \cite{MishimaS} is a very promising way, since it reflects 
the location of a ``liquid-liquid'' transition line (see fig. \ref{fig:melting_max}). 
Further careful studies are required to prove the existence 
of a hidden second-critical point and specify 
its location.

\subsubsection{Possibility of amorphous-amorphous transition without cooperativity}

Here we consider a fundamental problem on the nature of amorphous-amorphous transition, 
which takes place in a non-ergodic state. 
Amorphous-amorphous transition is often believed to be linked to thermodynamic 
liquid-liquid transition in a supercooled state. 
The first-order-like discontinuous nature of the transition is regarded as 
evidence supporting this picture. 
In relation to this, it is worth considering a situation that 
there are two states of liquid structures, but without any cooperativity (i.e., $J=0$ in eq. (\ref{eq:fS})). 
In a liquid state, it is rather easy to see whether a transition is of cooperative nature or not. 
In a non-ergodic state, however, it is rather difficult to see whether the transition 
induced by pressure is linked to the thermodynamic first-order transition with cooperativity  
or the mechanically induced transition between the two states without cooperativity. 
This implies that the first-order-like transition between two amorphous states 
may not necessarily be linked to the thermodynamic LLT. The consequence of the cooperativity 
in a solid-state amorphous-amorphous transition is an interesting issue for further investigation.

\section{Glass transition}
\label{sec:glass}

When crystallization is kinetically avoided, a liquid increases its viscosity 
steeply upon cooling and eventually becomes elastic or plastic when passing through 
the glass transition temperature $T_{\rm g}$. So a liquid gradually transforms to a solid-like state at $T_{\rm g}$. 
On the other hand, a discontinuous first-order liquid-solid transition takes place upon crystallization. 
In this case, the elasticity originates from the long-range translational order with periodicity. 
In a crystal, thus, a motion of a single particle must inevitably accompany all the other particles, if we assume 
perfect order. 
In a glass, on the other hand, the situation is far less obvious and the origin of elasticity is not so clear 
(see sec. \ref{sec:transient} and \cite{barrat2011,Tanguy2011review} for review). 
Furthermore, the elasticity appears rather gradually in glass transition, accompanying the continuous slowing down of the mechanical (or structural) relaxation time $\tau_\alpha$. 
The transition is from a liquid to a solid via a viscoelastic state. 
There is also ageing (slow temporal change of the physical properties) in a glassy state, reflecting its intrinsically non-equilibrium nature. 
Unlike a liquid-crystal transition, there is no evident change in the liquid structure through $T_{\rm g}$: 
The structural relaxation time $\tau_\alpha$, or the viscosity $\eta$, increases 
by more than ten orders of magnitude while accompanying little change in the liquid structure 
probed by the scalar density field (e.g., by the static two-point density correlator). 
The origin of slow dynamics associated with glass transition has been a long-standing fundamental problem in 
condensed matter physics. 

\begin{figure}
\begin{center}
\includegraphics[width=8.5cm]{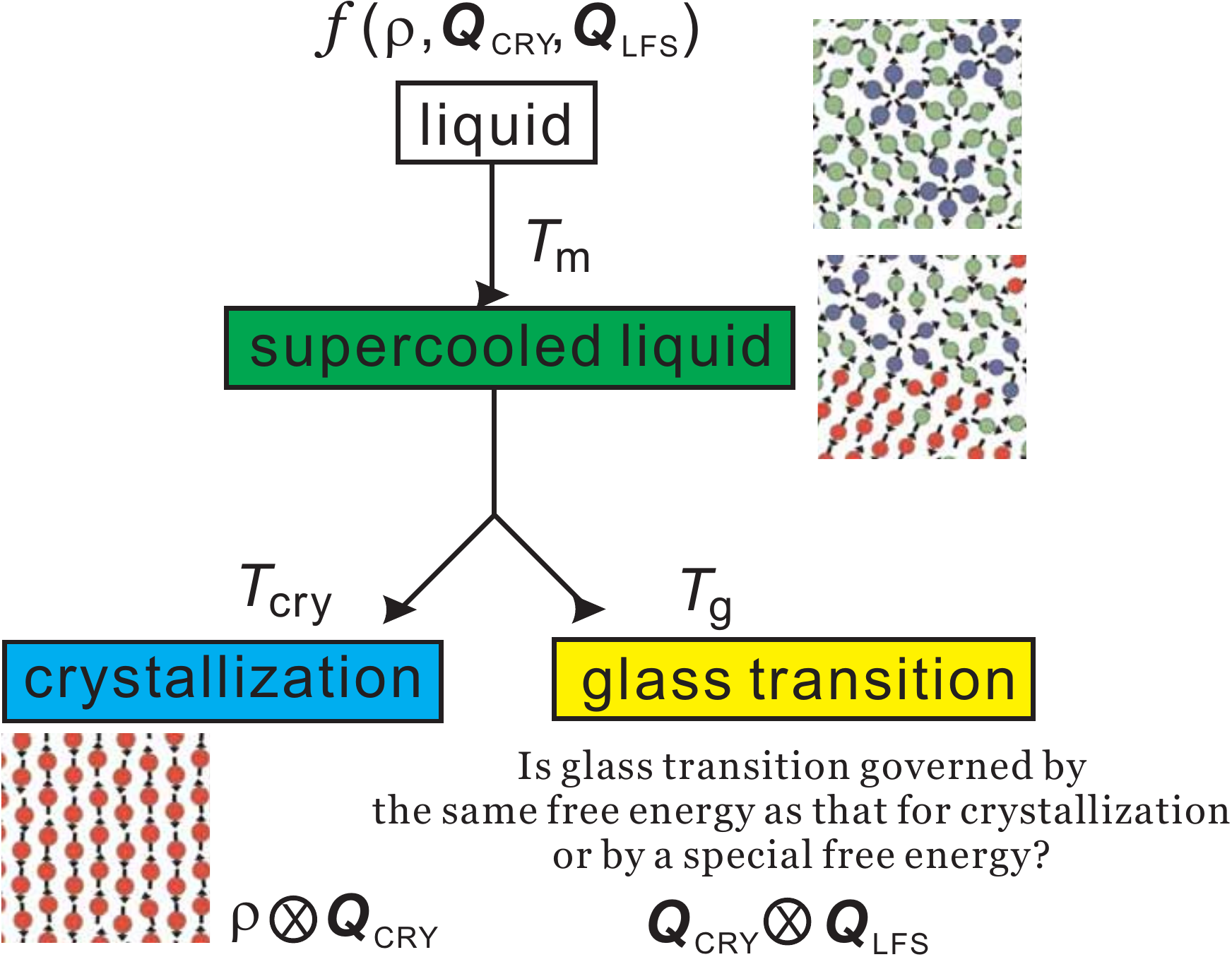}
\end{center}
\caption{(Colour on-line) Schematic figure explaining the behaviour of liquid upon cooling. 
The liquid becomes a metastable supercooled state below the melting point $T_{\rm m}$ 
and further cooling leads either to crystallization or to glass transition. 
The former takes place at the crystallization temperature $T_{\rm CRY}$, whereas 
the latter at the glass transition temperature $T_{\rm g}$. 
The former is a thermodynamic phase transition, but the latter is a kinetic transition. 
The key fundamental question here is whether the glass transition behaviour is controlled by the 
same free energy as that for crystallization or a special free energy?  
}
\label{fig:cry_glass}
\end{figure}

In general, when a liquid is cooled, it is either crystallized or vitrified. 
Except liquids with quenched disorder, such as atactic polymers or polydisperse colloids, 
a single-component liquid can in principle crystallize below the melting point $T_{\rm m}$ without accompanying inhomogeneization (phase separation). 
Glass transition is thus observed only when crystallization is `kinetically' avoided. 
This is suggestive of a deep link between crystallization and vitrification. 
However, most of previous approaches did not consider crystallization to be important for the physical description of 
vitrification itself. In these approaches, either a purely kinetic origin for dynamic arrest is sought or 
the special free energy describing the vitrification branch is 
newly introduced. In both cases, the crystallization branch is simply ignored, as we will see 
below. This is probably because people who are interested in glass transition 
are not interested in the crystallization branch but only in the glass transition branch (see fig. \ref{fig:cry_glass}). 
Another reason may come from our intuition linked to a different, but related phenomenon, jamming transition. 
When we consider slowing down of motion of people in a packed train, we do not care about crystallization. 
This is also related to the fundamental question concerning the link between glass transition and jamming transition. 

About a decade ago, however, we proposed \cite{TanakaGJPCM,TanakaGJCP1,TanakaGJCP2,TanakaI,TanakaII,TanakaIII} 
that crystallization should be regarded as the basis for understanding glass transition, more specifically, 
vitrification can be regarded as frustration on the way to crystallization. 
We argued that frustration against crystallization is a key to understanding the physical origin of 
glass transition: a supercooled liquid is controlled by the same free energy of the system, which leads to crystallization 
under the influence of frustration effects (see fig. \ref{fig:cry_glass}).
In this sense, jamming transition (without being driven mechanically) may be essentially different from 
glass transition, although there are many similarities.  
The absence of thermal noises makes jamming transition of `mechanical' character essentially different from glass transition 
of `thermodynamic' character in our opinion.  Below we explain our two-order-parameter model of liquid-glass transition. 

\subsection{Background}

\subsubsection{Major previous approaches to glass transition}

Our approach to the problem of glass transition based on the two-order-parameter 
model was already reviewed briefly in sec. 2. 
Here we mention other popular theoretical approaches (on the details of these approaches, please refer 
\cite{DebenedettiB,DebenedettiN,CavagnaR,Parisi2010,BerthierR,binder2011glassy,das2011statistical}). 
Here we classify previous theoretical approaches to glass transition into the following five types: 
\begin{enumerate}
\item[(A)] Approaches such as 
free-volume theory \cite{Free}, mode coupling theory \cite{GotzeB}, and trap model \cite{monthus1996} 
focus on the slowing down of the dynamics while approaching $T_{\rm g}$. 
According to the mode-coupling theory (MCT), the density-density correlation function is the main mode for slow relaxation 
in a glass-forming liquid. The free-volume theories also suggest that the transport in a supercooled liquid is 
controlled by the slow relaxation of the density fluctuations. 
The geometrical constraints due to dense packing on molecular motion have been expressed in terms of a few different concepts, 
such as ‘free volume’ and ‘caging-induced memory effects’. 
Among these theories, MCT provides a scenario which directly 
bridges macroscopic and microscopic structural relaxation  \cite{GotzeB,das2004mode}. According to MCT, 
the blocking of particle motion by the neighbouring particles, which is called ``caging'', is the essential origin of the slowing down 
of `macroscopic' structural relaxation. 
Thus, density fluctuations whose characteristic lengthscale is the cage size ($\sim$ particle size $d$) slows down and is eventually frozen. 
Because of this microscopic nature of the mechanism of slowing down, 
MCT has been considered to be a microscopic or first-principles theory that can explain 
various key aspects of glass transition. It was also shown that MCT has a deep link to the trap model 
\cite{bouchaud2008anomalous}. 

\item[(B)] In the Adam-Gibbs theory \cite{Adam}, on the other hand, it is assumed that under a dense packing condition 
the motion of a particle becomes possible only when particles in its surrounding region move in a cooperative manner. 
A number of particles required for a collective rearrangement is a manifestation of the 
growing degree of dynamical correlation in a supercooled liquid. This is expressed by the concept of ‘cooperatively rearranging regions’. 

\item[(C)] There are also approaches putting a focus on local relaxation mechanisms. 
Chain-like excitation \cite{Langer}, single-particle barrier hopping \cite{schweizer}, and elastically constrained motion \cite{DyreR}. 
In these models, localized relaxation events become more and more difficult with a decrease in temperature 
and the activation energy for the relaxation becomes larger and larger.  
For example, in the shoving model \cite{DyreR} the activation energy is associated with elastic cost for shoving the surrounding particles.  

\item[(D)] Kinetically constrained model focuses 
on the role of mobility defects and can thus be regarded as a purely kinetic model (see, e.g, \cite{ritort2003glassy,fredrickson1984,Chandler1,chandler2009}). 
Chandler and Garrahan and their coworkers proposed that in contrast to equilibrium phase transitions, which occur in a configuration space, the glass transition 
occurs in a trajectory space, and it is controlled by variables that drive the system out of equilibrium. 
Thus, they connected the glass transition to an exotic time-domain phase transition between active and inactive states. 
This scenario relies on the recognition that the glass transition is the freezing of a liquid into a solid state ``without structural ordering''. 
This type of model, thus, shows trivial thermodynamics but non-trivial dynamics and is able to explain dynamic features such as 
super-Arrhenius behaviour, dynamic heterogeneity, and the violation of the Stokes-Einstein relation. 
However, this model inevitably results in a complete decoupling between dynamics and thermodynamics, which may be 
its weakness \cite{biroli2005defect}.  

\item[(E)] Approaches such as frustration model and spin-glass-type model focus on geometrical frustration effects, 
and apply the knowledge of (a) a ‘frustrated system’ \cite{Steinhardt,Tarjus} 
to the problem of glass transition or 
(b) ‘spin glass’ \cite{Kirkpatrick,Parisi2010,Xia,lubchenko2007}, whose glassy behaviour is much more deeply understood than that of structural glass. 
The random-first-order transition scenario \cite{Kirkpatrick,Xia,lubchenko2007} belongs to this category. 
Here we note that the energy landscape picture \cite{goldstein1969} has a close connection to spin-glass-type approach.  
Unfortunately, however, these approaches do not provide us with a molecular-level explanation for the origin 
of frustration and the nature of an underlying ordering transition. 
We note that unlike approaches (A)-(D), approach (E)-(a) puts focus on the symmetry or structure rather than the density field. 
Interestingly, however, there is a deep connection between MCT and some mean-field spin glass models. So there is a link between 
MCT and approach (E). The link between approach (B) and (E)-(b) was also discussed \cite{bouchaud2004adam}. 
\end{enumerate}

These approaches (A)-(E) are essentially different in the basic physical interpretation of the liquid-glass transition with each other; 
for example, approaches (A)-(D) presuppose that glass transition is not associated with any ordering transition, 
while approach (E) puts more emphasis on frustration effects on an underlying thermodynamic ordering transition, 
which however is not related to crystallization. 
However, the connection between MCT and spin-glass models, which implies a link between dynamical memory effects and geometrical frustration, 
makes the situation a bit obscure. Furthermore, we may say that all these models 
have some common features at least on an intuitive level. 
For example, the free volume theory and the Adam-Gibbs theory look closely related with each other, although the languages used are quite different:  
We can apparently connect them by saying that a minimum ‘free volume’ required for motion of a particle is shared 
by many particles in the ‘cooperatively rearranging region’ in a supercooled state. 
Such apparent similarities between different approaches, which reflect the fact that each approach captures some important feature 
of glass transition, make it very difficult to answer which model is most relevant to the phenomena. 
We may say that we do not yet have a clear answer for what is the physical origin of slow dynamics associated with 
the liquid-glass transition.

Here it is worth noting that the free-volume theory and Adam-Gibbs theory are consistent with the presence of dynamic heterogeneity, while 
the original MCT, which deals with a two-body density correlator in the mean-field level, cannot explain it. 
The extension of MCT is, however, now in progress \cite{schweizer,biroli4,berthier2007general,berthier2007spontaneous}. 
However, in this scenario the dynamic heterogeneity is diverging towards $T_{\rm c}$, which is located far above $T_{\rm g}$ and $T_0$. 
How smearing of a sharp transition by thermal fluctuations affects MCT criticality is also not so clear. 
We point out that these models do not seem to provide a clear answer for what physical parameter controls 
the fragility and glass-forming ability of liquids, although there are some efforts towards this direction. 

We believe that besides the drastic slowing down accompanied by glass transition, on which all the above theories 
put a focus, we should also consider another important problem, namely, why some molecules do crystallize without vitrification 
and the others can easily form glasses without crystallization. This is because `not to crystallize on cooling' is 
the same as `to vitrify'. Since most glass-forming molecules except for systems such as atactic polymers can 
crystallize under a certain condition, it is important for any physical model describing the glass transition 
to explain why crystallization is easily avoided in the so-called good glass formers. All approaches (A)-(E) 
presuppose the avoidance of crystallization. This is partly due to a wide and basically correct belief that crystallization is 
avoided kinetically in vitrification in the spirit of Classical Nucleation Theory (CNT). Later, we will show that this is not 
necessarily the whole story and a thermodynamic 
factor associated with frustration on crystallization (more strictly, crystal-like bond orientational ordering) 
may also play a crucial role in the avoidance of crystallization. 

As explained above, previous models of glass transition themselves cannot provide any direct information on the glass-forming 
ability of liquids. Unlike these approaches, we propose that geometrical or energetic frustration against 
crystallization plays a key role in the avoidance of crystallization (or vitrification) in addition to the kinetic factor,  
and controls the nature of the liquid-glass transition and the glass-forming ability.

\subsubsection{Our approach}
Crystallization is suppressed by kinetic and thermodynamic factors such as high viscosity, 
large crystal-liquid interfacial energy, and small liquid-crystal free energy difference. 
We propose that these factors are affected by 
geometrical or energetic frustration effects against crystallization, which also control the nature of liquid-glass transition. 
Here we emphasize that we are not claiming that crystallization is avoided by energetic frustrations alone. 
In our model, crystallization can in principle occur in a metastable branch of a glassy state. 
Our standpoints are (i) to choose a crystalline state as a reference state of a glassy state and (ii) to clarify 
the physical origin of frustration against crystallization hidden in glass formers that apparently have no intrinsic quenched disorder. 
Contrary to the common belief that the density is the only order parameter required for the physical description of 
liquids, we proposed that it is necessary to consider bond orientational order parameters, which represent 
medium-range bond ordering towards a crystal as well as 
short-range bond ordering towards the formation of locally favoured structures (see sec. \ref{sec:crystallization}). 
This has already been emphasized throughout this article. 
We believe that this approach should be relevant at least to single-component glass-forming liquids. 
We note that many molecular glass formers are made of a single component.  
As will be discussed later, when a system suffers from strong disorder effects, e.g., as in binary and multicomponent systems, 
we may need to consider other types of structural order parameters.  

Locally favoured structures act as random disorder effects and ‘symmetry-breaking fields’ 
against the long-range bond orientational and  density ordering (crystallization) 
in much the same way as in spin systems \cite{TanakaGJPCM,TanakaGJCP1,TanakaI}. 
To our knowledge, this was the first approach to the problem of the liquid-glass transition 
directly focusing on crystallization (long-range density and bond orientational ordering). 
According to our model, a liquid below $T_m$ is in an unusual metastable state. 
Its free energy has extensive numbers of local minima due to the frustration effects of short-range bond ordering 
(or random disorder effects) in addition to the deep minimum of a stable crystalline state. 
Based on our recent numerical and experimental studies \cite{ShintaniNP,STNM,KAT,WT,TanakaNM,Kawasaki3D,KawasakiJPCM}, 
we proposed \cite{TanakaJSP,TanakaNara} that positional ordering is avoided by frustration rather easily, 
but bond orientational order linked to crystallization (MRCO) still survives and grows continuously upon cooling. 
This may be because upon crystallization translational ordering comes only after bond orientational order is sufficiently developed (see sec. \ref{sec:crystallization}).  
Such crystalline bond orientational ordering (MRCO) competes with bond orientational ordering towards locally favoured structures in a supercooled state. 
Under such competing orderings, the correlation length of MRCO still diverges towards the ideal glass transition temperature $T_0$, 
which is the origin of dynamic heterogeneity in this type of system (see below for the details) and the growing activation energy (see sec. \ref{sec:activation}). 

\subsubsection{Concept of geometrical frustration}

Here we consider models of glass transition which put focus on geometrical frustration \cite{NelsonB,sadoc1999} in a liquid. 
Geometrical frustration is canonically illustrated by considering the packing of equal-size spherical particles. 
This problem is directly linked to the spatial arrangement of regular simplices (triangles in 2D and tetrahedra in 3D), which 
are the most well-packed densest possible local packing of spheres. 
In 2D Euclidean space, triangles can assemble into the regular triangular lattice. Frustration can still be induced by 
introducing a curvature in the 2D surface \cite{NelsonB}. 
For 3D Euclidean space, on the other hand, tetrahedra cannot fill up the space without defects. 
Thus a regular tiling of simplices inevitably leads to strong geometrical frustration. 
This is the basis of many frustration-based models of glass transition \cite{Kirkpatrick,Parisi2010,Xia,lubchenko2007}. 
However, such purely geometric consideration may not be appropriate when considering 
the structure of liquid, which is in a thermodynamic state where entropy plays a crucial role. 
For example, the concept of defects is not well-defined in a liquid, and may be better to be interpreted 
as structural fluctuations. 
As will be shown later, in addition to icosahedral packings, fcc and hcp packings are locally favoured in a thermodynamic system 
of hard spheres. Below we consider approaches based on two different types of frustration, 
`internal geometrical frustration' and `frustration against crystallization' and compare them.

\subsubsection{Comparison between our model and other frustration models}
Here we compare our two-order-parameter model with frustration or spin-glass models, focusing on the difference in 
the underlying ordering phenomenon behind vitrification between them. In particular, we focus on the following facts:
\begin{itemize}
\item[(a)] Stereo-irregularity in polymer structures prevents crystallization and leads to vitrification for atactic polymers. 
\item[(b)] Mixing of different-size particles prevents crystallization for simulations using hard and soft spheres and Lennard-Jones particles 
\cite{Andersen}. 
\item[(c)] There are a number of examples of glass-forming mixtures, both of whose component molecules themselves 
are very poor glass formers. 
\item[(d)] Good glass formers can often be made by mixing many component atoms in metallic glasses. Glass formation is helped not only by random mixing effects, but also by chemical and topological short-range ordering towards icosahedral structures (see, e.g., refs. \cite{TanakaMJPCM,TanakaGJNCS} and the references therein). 
These glass formers are now widely known as multi-component ‘bulk’ metallic glass forming alloys \cite{WangR,Ma_review}.
\item[(e)] It is also known that conformational disorder of molecules helps vitrification. For example, ortho-terphenyl 
(OTP) has significant bond-angle and out-of-plane distortions of the phenyl-phenyl bonds. Such structural 
irregularities may explain why OTP can be undercooled far easier than m- or p-terphenyl. Furthermore, 
it was also demonstrated by Torre and coworkers \cite{Torre} 
that o- and m-toluidine have better glass-forming ability than p-toluidine. They showed that this difference 
arises from the fact that o- and m-toluidine tend to form clusters, which prevent crystallization, whereas p-toluidine 
does not. This is quite consistent with our scenario.
\item[(f)] It is pointed out by van Megen and his coworkers \cite{vanMegen,vanMegenX} 
that the size distribution (polydispersity) of colloids is essential for the retardation of crystallization and the resulting vitrification of colloidal suspensions.
\end{itemize}

All these examples strongly indicate that for vitrification of liquids, or for preventing `crystallization', 
frustration or disorder effects are always required. 
More explicitly, disorder effects on crystallization are enough to cause vitrification and thus slow glassy dynamics. 
This fact, which is well known but not properly recognized, supports our 
physical picture that frustration or disorder effects on crystallization is crucial for vitrification of any 
liquids, including simple one-component liquids, which apparently look free from such disorder effects. 
In all the above cases, to make a liquid a `good glass former' is the same as to introduce frustration or 
disorder effects against crystallization. 
These examples clearly indicate that crystallization is avoided not only by a 
kinetic reason, but also by energetic and/or entropic frustration. Our model is based on a picture 
that `to vitrify' is essentially the same as `not to crystallize'. 

We believe that our picture that a liquid always tends to crystallize into the equilibrium crystal even under 
frustration effects is more natural than the conventional frustration or spin-glass-type picture that a liquid tends to be ordered 
into a special hypothetical ordered state or a special glass structure, which are different from the 
equilibrium crystal. In the latter picture, for example, we need to assume an unconventional 
type of ordering, whose nature is generally not clear, in a liquid branch, in addition to the ordering into a crystal. 
The possible candidates of such an ordered state are a hypothetical ordered state of locally favoured structures 
(e.g., quasicrystal for icosahedral structures \cite{NelsonB,sadoc1999} and an exotic amorphous order \cite{lubchenko2007}). 
In this scenario, a locally favoured structure is not necessarily localized and tends to extend further 
towards its long-range ordering, and it is this ordering that is the driving force of vitrification. 
However, it is not clear whether real liquids tend to be ordered into such a hypothetical ordered state 
(e.g., stacking tetrahedra) in a liquid branch or not. More importantly, disorder effects in the above examples 
(a)-(f) are introduced purely to avoid crystallization, and neither to help the hypothetical ordering nor to 
increase disorder effects on it. These considerations seem to suggest that our model is physically more natural 
than frustration or spin-glass models, although further careful studies are necessary to draw such a conclusion.
We note that this view was supported by recent simulation studies \cite{ShintaniNP,KAT,TanakaNM,Kawasaki3D,Coslovich}, 
as will be shown later. 

\subsubsection{Dynamic heterogeneity: Length scale relevant for slow dynamics}

Recently there have been growing evidence that slow dynamics in a supercooled liquid is spatially heterogeneous and 
the length scale of this heterogeneity grows when approaching $T_{\rm g}$. 
This raises a fundamental question on the origin of slow glassy dynamics. Many approaches such as MCT ascribe slow dynamics 
to the local blocking of particle motion in a densely packed situation. On the other hand, strong spatial correlation of particles 
over a mesoscopic length scale may be the origin of slow dynamics. The physical picture behind this has a link to dynamic 
critical phenomena, although there is an essential difference between them, as will be discussed later. 
This question on the relevant length scale, microscopic or mesoscopic, lies at the heart of the origin of slow dynamics 
associated with glass transition and will be the central topic below. 

\subsection{Roles of locally favoured structures in vitrification}
\subsubsection{Frustration effects}
In the above, we demonstrate the importance of locally favoured structures 
in the problem of liquid-state thermodynamic anomaly and liquid-liquid 
transition. 
For example, spherical particles interacting with the Lennard-Jones potential are known to form 
icosahedral structures, whose energy is even lower 
than the corresponding fcc, hcp or bcc crystals locally \cite{Frank}. 
This is also the case for hard spheres \cite{MathieuNM,karayiannis2011,karayiannis2011s}. 
Unlike water, the symmetry of a locally favoured structure 
is not consistent with any crystallographic symmetry for these cases \cite{TanakaMJPCM}. 
This energetic frustration hidden in 
the interaction potential causes the frustration effects on 
long-range  bond orientational and density ordering (crystallization). 
The situation is very similar to that in fig. \ref{fig:KTTP}. 

Here we consider how such local bond ordering affects 
glass-forming ability. 
Conventional models of glass transition cannot answer the question 
of what controls the glass-forming ability, since they do not 
put focus on crystallization itself. 
The explanation is then given by the classical nucleation theory (CNT) alone without considering 
frustration effects. 
Of course, the CNT is always the basis when considering the glass-forming ability. 
However, this may not necessarily be the whole story. Significant modifications may be required (see sec. \ref{sec:crystallization}).  
In our model of liquid-glass transition, the glass formability is related to the strength 
of frustration between long-range crystal ordering and short-range 
bond ordering \cite{TanakaGPRL,TanakaMJPCM,TanakaGJNCS} in addition to the 
physical factors considered in the classical nucleation theory. 

By modifying the classical nucleation theory \cite{Turnbull} while  
including the effect of translational-rotational decoupling \cite{TanakaK}, 
the nucleation frequency $I$ is given by 
\begin{equation}
I=k_n D_T \exp[-\Delta F^c/k_{\rm B}T],
\end{equation}
where $k_n$ is a constant and $D_T$ is the translational diffusion 
constant. $\Delta F^c$ 
is the free-energy barrier for nucleation of a critical nucleus, 
which is estimated as 
\begin{eqnarray}
\Delta F^c=16 \pi \gamma_{l-c}^3/(3 \delta \mu)^2, \nonumber 
\end{eqnarray}
where $\delta \mu$ is the Gibbs free energy 
of a supercooled liquid over the crystal per unit volume and $\gamma_{l-c}$ is the interface tension between liquid and crystal. 
Usually, it is assumed that 
\begin{eqnarray}
\delta \mu=\Delta H_f(1-T/T_m), \nonumber
\end{eqnarray}
where $\Delta H_f$ is the enthalpy of fusion. 
This assumption should be valid as far as the degree of supercooling is not so large. 
According to our model, however, 
it should be modified due to the existence of locally favoured structures 
as follows: 
\begin{eqnarray}
\delta \mu \cong \Delta H_f(1-T/T_m)
+\Delta G(T_m) \bar{S}(T_m) 
-\Delta G(T) \bar{S}(T). \nonumber
\end{eqnarray}
This reflects the lowering of the liquid free energy due to local structural ordering. 
The downward deviation of $\delta \mu$ from the linear temperature dependence 
is indeed observed for various metallic glass formers \cite{Cp,Lu}. 
Furthermore, this deviation is larger for a 
stronger (better) glass former \cite{Cp,Lu}.  
According to our model, a stronger glass former may have 
larger $\bar{S}$, provided that the local symmetry of $S$ is not consistent with that of the crystal. Thus, the above observation is quite 
consistent with our model. 
We also note that $\gamma_{l-c}$ should be larger for a liquid with larger $\bar{S}$. 
These factors should increase the nucleation barrier for a system suffering from strong frustration 
(i.e., a system of large $\bar{S}$). 

As will be discussed in sec. \ref{sec:crystallization}, crystallization is initiated by 
the enhancement of the coherency of crystal-like bond orientational order. Thus, frustration effects on crystalline bond orientational ordering 
leads to the increase of the barrier height for nucleation.   
From these considerations, we conclude that the better glass formability 
is at least partially due to smaller $\delta \mu$ and larger $\gamma_{l-c}$, 
which are induced by a stronger tendency of short-range bond ordering 
(larger $\bar{S}$) or random disorder effects for a stronger liquid with larger $D$. 
Here $D$ is the so-called fragility index. Note that the smaller $D$ means the larger fragility of the liquid, i.e., 
the steeper increase of the viscosity near $T_{\rm g}$.  
Thus, our model suggests that a better glass former likely suffers from 
stronger frustration effects, provided that the values of $\Delta H_f$ and the bare interfacial tension are about the same between different liquids. 
For a single-component liquid, thus, we expect that a stronger tendency of short-range bond ordering not only 
makes the liquid stronger (i.e., larger $D$), but also enhances the glass-forming ability.  
Glass-forming ability is often characterized by the critical cooling rate $R_c$, which is the 
slowest cooling rate to form a glassy state from a supercooled liquid without crystallization. 
We note that the positive correlation between $R_c$ and $D$ in metallic glass formers may naturally 
be explained by our model \cite{TanakaMJPCM,TanakaGJNCS}. 
In the future, we need to improve a theory of crystal nucleation on the basis of 
the microscopic physical picture described in sec. \ref{sec:crystallization}.

\subsubsection{Apparent disparity between $T_0$ and $T_{\rm K}$}
\begin{figure}
\begin{center}
\includegraphics[width=6cm]{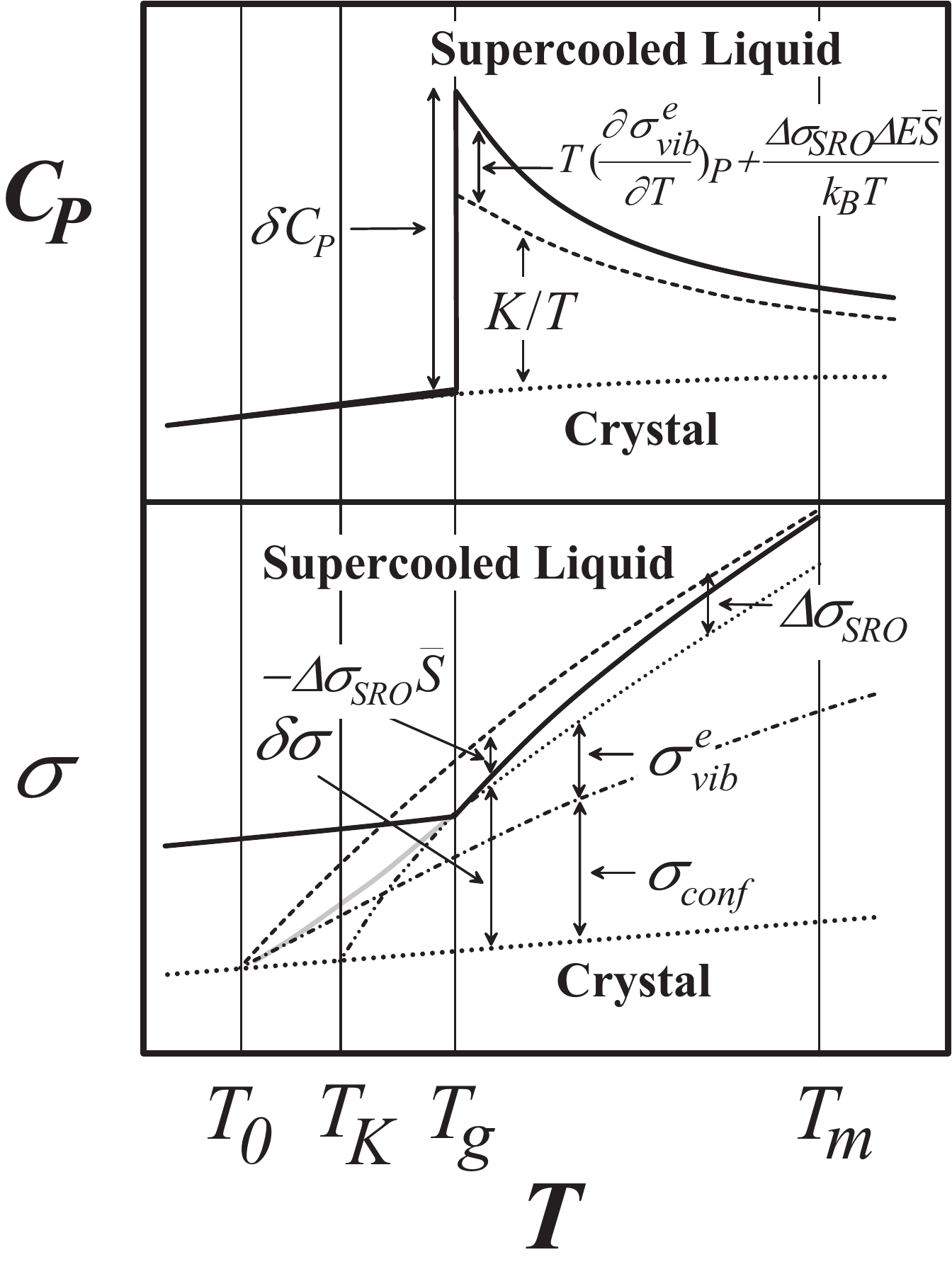}
\end{center}
\caption{Temperature dependence of the heat capacity 
$C_P$ (top) and the entropy $\sigma$ (bottom). 
Short-range ordering results in the deviation of $T_{\rm K}$ from $T_0$ 
toward the high-temperature side and the extra 
contribution to the heat capacity jump at $T_{\rm g}$, $\delta C_P$. 
The latter may explain unusually large $\delta C_P(T_{\rm g})$ for 
alcohols and metallic glass formers \cite{donth}. 
This figure is reproduced from fig. 2 of ref. \cite{TanakaGPRL}} 
\label{fig:t0tk}
\end{figure}

We also showed that this short-range bond ordering affects 
the conventional picture even on a qualitative level. 
For example, the excess entropy of liquid over the crystal 
should be modified, 
reflecting the extra entropy decrease due to short-range bond ordering. 
This leads to the apparent violation of the well-known 
relation $T_0=T_{\rm K}$ ($T_0$: the Vogel-Fulcher-Tammann temperature; $T_{\rm K}$: the 
Kauzmann temperature). 
The configurational entropy is supposed to vanish at the 
Vogel-Fulcher-Tammann temperature $T_0$, namely, $T_{\rm K}=T_0$.  
This decrease in the configurational entropy upon cooling is linked to the growth of dynamical correlation length, 
or underlying static correlation length $\xi$, as will be shown below. 
Upon cooling, however, a system also loses entropy associated with the formation of locally favoured structures. 
This leads to an extra decrease 
of the entropy by $\Delta \sigma_{\rm SRO} \bar{S}$ (see eq. (\ref{eq:sigma})).  
The situation is schematically shown in fig. \ref{fig:t0tk}. 
The deviation of $T_{\rm K}$ from $T_0$ 
should be larger for a stronger liquid having a stronger tendency of local bond ordering, which has recently 
been confirmed for a wide variety of liquids \cite{TanakaGPRL}. 
However, we emphasize that this may be only an apparent deviation originating from 
the large extrapolation of the entropy change (see fig. \ref{fig:t0tk}). 
We also note that as shown later, the Vogel-Fulcher-Tammann temperature $T_0$ and the 
Kauzmann temperature $T_{\rm K}$ can never be accessed since crystallization should take place before 
(quasi-)equilibrating a supercooled liquid state \cite{TanakaK}. This also provides a natural resolution of the 
Kauzmann paradox (see sec. \ref{sec:Kauzmann}).

\subsubsection{Quasicrystal formation}
Here we consider an interesting problem of the relationship among 
local icosahedral ordering, glass formability, and quasicrystal formation 
in bulk metallic glass formers \cite{TanakaMJPCM,TanakaGJNCS}. 
For example, Chen et al. \cite{Chen} recently reported the structural similarity 
between a supercooled liquid and an icosahedral phase in 
Zr$_{65}$Al$_{7.5}$Ni$_{10}$Cu$_{12.5}$Ag$_{5}$. 
They found that (i) the effective activation energy of transition 
from a supercooled 
liquid to an icosahedral quasicrystalline phase is much lower 
than that from a supercooled liquid to eutectic crystalline 
phases and (ii) the activation energy of transition from an icosahedral 
to a crystalline phase is almost the same as 
that from a supercooled liquid to a crystalline phase. 
These facts strongly suggest the similarity of the 
local atomic structure between the supercooled and the icosahedral phase. 
Our model provides us with a natural scenario for 
the close relationship among the degree of local icosahedral 
ordering in liquid, 
glass formability, and quasicrystal formability \cite{TanakaMJPCM,TanakaGJNCS} (see also \ref{sec:metallic}).

\subsection{Our scenario of glass transition: Critical-like 
glassy structural ordering and slow dynamics} 

We recently found critical-like behaviour of glassy structural ordering 
in several glass-forming liquids, which indicates that liquid is almost homogeneous 
if we look it through density order parameter, more specifically, two-body density correlation, but it is quite heterogeneous if 
we look it through a relevant glassy structural order parameter such as a bond orientational order parameter.  
This means that contrary to the common belief, glass transition actually involves a significant structural change, 
although this cannot be seen by the two-point density correlator.  
The critical-like fluctuations apparently diverges towards the hypothetical 
critical point, which seems to be located at the ideal glass transition point $T_0$. 
We interpret this phenomenon as critical-like phenomena associated with 
glassy structural ordering under frustration. The critical behaviour observed 
at least apparently belongs to the Ising universality class, despite that in some cases (e.g., polydisperse 
hard sphere systems) the order parameter is 
a bond orientational order parameter, which is tensorial and not scalar \cite{TanakaNM}.   
Our finding suggests an intimate link between such critical-like fluctuations, dynamic heterogeneity, and glassy slow dynamics. 
Below we describe more details about this and discuss related problems. 

\subsubsection{Glassy structural order under frustration: Competing bond orientational ordering and/or random disorder effects}

In our scenario, bond orientational ordering which comes from a part of the 
crystallization Hamiltonian (see eq. (\ref{eq:F_total})) tends to grow in a metastable supercooled 
liquid state. This may be regarded as a shadow of crystallization. 
As will be shown in sec. \ref{sec:crystallization}, translational order develops, or crystallization takes place, only when 
bond orientational order and its coherency grow beyond a certain critical threshold. 
The growth of the phase coherency of bond orientational order is rather easily suppressed or destroyed by frustration effects, 
although the degree of bond orientational order itself can grow even under frustration.   
The sources of such frustration can be (i) competing bond orientational 
orderings which have symmetries inconsistent with each other and/or (ii) 
random disorder effects such as polydispersity of colloidal particles,  
stereo-irregularity of polymers, and mixing of more than two components. 
As we discussed above, for case (i) glassy structural order can be represented by 
bond orientational order. This is also the case for case (ii) if disorder effects are not so strong. For 2D polydisperse colloids it is hexatic order parameter $\Psi_6$, 
whereas for 3D polydisperse colloids it is a combination of six-fold and four-fold 
bond order parameters, $Q_6$ and $Q_4$ (more precisely, crystal-like (fcc-like) bond orientational order parameter) 
\cite{MathieuNM,russo2011}.

For 2D polydisperse system, we used the hexatic order parameter averaged over a certain period of time 
to remove vibrational distortion ($\leq \tau_\alpha$). 
For 3D polydisperse systems, on the other hand, we employ the following coarse-grained bond order parameter \cite{TanakaNM}, 
which is obtained by combining the time average and the spatial average which was developed by Lechner and Dellago~\cite{lechner}. 
The time-averaged $l$-th order bond orientational order of particle $k$ is calculated as 
\begin{eqnarray}
\bar{Q}^k_l=\frac{1}{\tau_{\alpha}}\int_{t_0}^{t_0+
\tau_{\alpha}}dt\left(\frac{4\pi}{2l+1}
\sum^{l}_{m=-l}|Q^k_{lm}|^2\right)^{1/2}. \nonumber 
\end{eqnarray}
Here 
\begin{eqnarray}
Q^k_{lm}=1/N^k_b\sum_{j=0}^{N^k_b}q_{lm}(\vec{r}_{kj}), \nonumber
\end{eqnarray}
where the sum from $j=0$ to $N^k_b$
runs over all neighbours of particle $j$ plus particle $k$ itself \cite{lechner} (see also sec. \ref{sec:defineBOO}). 
We note that a coarse-graining of the order parameter in space and/or time is crucial to extract 
static structural order in a supercooled liquid: 
This coarse-graining and/or temporal averaging added to the standard 
Steinhardt bond orientational parameter \cite{Steinhardt} is useful in 
detecting local structural ordering explicitly. 
However, the time averaging may mix up static and dynamic information and thus we need a special care. 
The meaning of the coarse-graining operation is rather clear. 
There are two types of bond orientational order: extendable (fcc and hcp) and non-extendable (icosahedral) order. 
If we calculate the spatial correlation without distinguishing these two types of order, 
the correlation tends to decay rather quickly due to the absence of the spatial correlation between them. 
The coarse-graining procedure allows us to pick up only the growing extendable order. 
However, this problem may also be related to the fragility of the definition of the bond orientational order parameter 
to microscopic motion of particles. Thus, further careful studies are necessary to settle this issue. 

Strong frustration effects on crystallization as in 
binary hard sphere mixtures 
lead to decoupling between `glassy structure' 
and crystalline bond orientational order and thus make an identification 
of glassy structural order quite difficult. 
This is a natural consequence of the fact that in this type of systems phase separation 
is required for crystallization to take place. 
This leads to a decoupling between glassy structural order and bond orientational order linked to the 
crystal symmetry. 
Thus, we take a different approach to elucidate structural 
features of a supercooled state of 2D binary colloid mixture (2DBC). 
For a system interacting with hard-sphere-like repulsive interactions, 
the free energy is dominated by entropy. 
For example, the ordering in a monodisperse hard sphere system 
takes place to increase the total entropy by gaining the correlational 
(vibrational) 
entropy while sacrificing the configurational (or structural) entropy. 
This essential physics should also be the same for glass transition: 
the system tends to lower the free energy by gaining the correlational 
entropy. 
This motivates us to make the estimation of so-called `pair structural entropy', or 
the two-body translational correlation contribution 
to the excess entropy, $s_2$ \cite{Green,Mountain,Evans}:
\begin{eqnarray}
s_2=-\frac{\rho}{2} \int d\vec{r}\ [g(\vec{r}) \ln g(\vec{r})
-(g(\vec{r})-1)]. \nonumber 
\end{eqnarray}
It was shown that this entropy has a direct link to the 
transport in various liquids 
\cite{Rosenfeld,Rosenfeld2,DzugutovL,Charu,Truskett}. 
For 2DBC, the partial excess entropy for particle type $\alpha$ 
($\alpha=A$ or B) is given by \cite{Ghosh}
\begin{eqnarray}
s_2^\alpha=-\frac{1}{2}\sum_\nu  \rho_\nu \int d\vec{r} \ 
[g_{\nu \alpha}(\vec{r}) \ln g_{\nu \alpha}(\vec{r})
-(g_{\nu \alpha}(\vec{r})-1)], \nonumber 
\end{eqnarray}
where $\rho_\nu$ is the individual component of the total number density 
$\rho$. 
The local version of $s_2$ for particle $j$, $s_2^j$ (or, 
$s_2(\vec{r})$) can measure 
local structuring around particle $j$ located at $\vec{r}$, 
even if there is no obvious structural order. 
To do so, we calculate (time-averaged) $g(r)$ for particle $j$ and 
then $s_2^j$. 
It is worth noting that this local $s_2$ might be used as a  
structural indicator of glassy structural order for a variety of systems covering 
from polydisperse to bidisperse systems. However, since this does not involve many body 
correlations in a direct manner, its validity is not so clear at this moment (see below). 

In a system where bond orientational order plays an important role, the coupling between bond order parameter 
and density is given by eq. (\ref{eq:F_int}). 
Since there is no direct coupling to density, bond orientational order does not necessarily accompany a local density 
change, but may accompany 
an increase in the density correlation. If there is a similar coupling between glassy structural order and density 
in general, $s_2$ may be a useful measure of glassy structural order.  

In relation to this, we note that both bond orientational order and structural entropy $s_2$ reflect the degree of short-range 
structural ordering. 
The latter can be applied even to cases where there is no specific local symmetry around a particle. 
The analysis of structural entropy $s_2$ is motivated by the successful description of the transport anomaly 
of liquids including water in terms of $s_2$ \cite{Rosenfeld,Rosenfeld2,DzugutovL,Charu,Truskett,Chakravarty}, 
which suggests that a link between local tetrahedral ordering and structural entropy $s_2$. 
Both are linked to low configurational entropy, more generally, low local free energy, which leads to low fluidity, or high solidity. 
However, $s_2$ is basically linked to the two-body correlation function, but not `directly' to 
many-body correlations, although a hidden connection between $s_2$ and bond orientational 
order may be expected as mentioned above. The long time average to obtain the local version of $s_2$ may have mixed up the static and dynamical 
heterogeneity \cite{furunote}. Thus, it still remains elusive whether structural entropy at a pair level can be 
used to pick up static structural order or not. This point is now under investigation.  

In the above, we consider two types of glassy order parameters: One is the bond orientational order parameter, 
and the other is the structural entropy. The former is a tensorial order parameter, whereas the latter is a 
scalar order parameter. 
As described later, the tensorial nature of the glassy order parameter may be essential for crystallization. 
We note that these structures with low configurational entropy can support stress transiently.

In relation to this, here we mention recent works by \"Ottinger and his coworkers \cite{ottinger2006,del2008,mosayebi2010}. 
They estimated the static correlation length $\xi$ in a binary Lennard-Jones liquid by using coarse-grained non-affine 
displacement field to small shear deformation \cite{ottinger2006,barrat2011,Tanguy2011review}. 
The typical linear size of the regions where non-affine deformation takes place is roughly the particle size 
at a high temperature, but grows upon approaching $T_{\rm g}$. 
It was shown that the correlation length diverges towards $T_0$ with the same power law as ours (see below). 
The regions whose inherent structures show more non-affine deformation should have more disorder. 
This non-affine nature upon deformation persists in a glass state and plays a significant role 
in the non-affine and nonlinear response of glasses to large deformation and the mechanical fracture \cite{barrat2011,Tanguy2011review}. 
This implies that regions of high glassy structural order is linked to regions whose inherent structures exhibit more affine deformation. This 
is consistent with our scenario that regions of high glassy structural order has a long structural lifetime, i.e., has 
(transient) stress-bearing solid-like character.  
In relation to this, a causal link between the localized low-frequency
normal modes of a configuration in a supercooled liquid and the irreversible structural reorganization
of the particles within this configuration was also pointed out recently \cite{widmer2008}. 
Such a correlation was also observed experimentally in a colloidal glass \cite{ghosh2011}.

\subsubsection{Critical-like behaviour of glassy structural order}
\label{sec:Critical}
\begin{figure}
\begin{center}
\includegraphics[width=7cm]{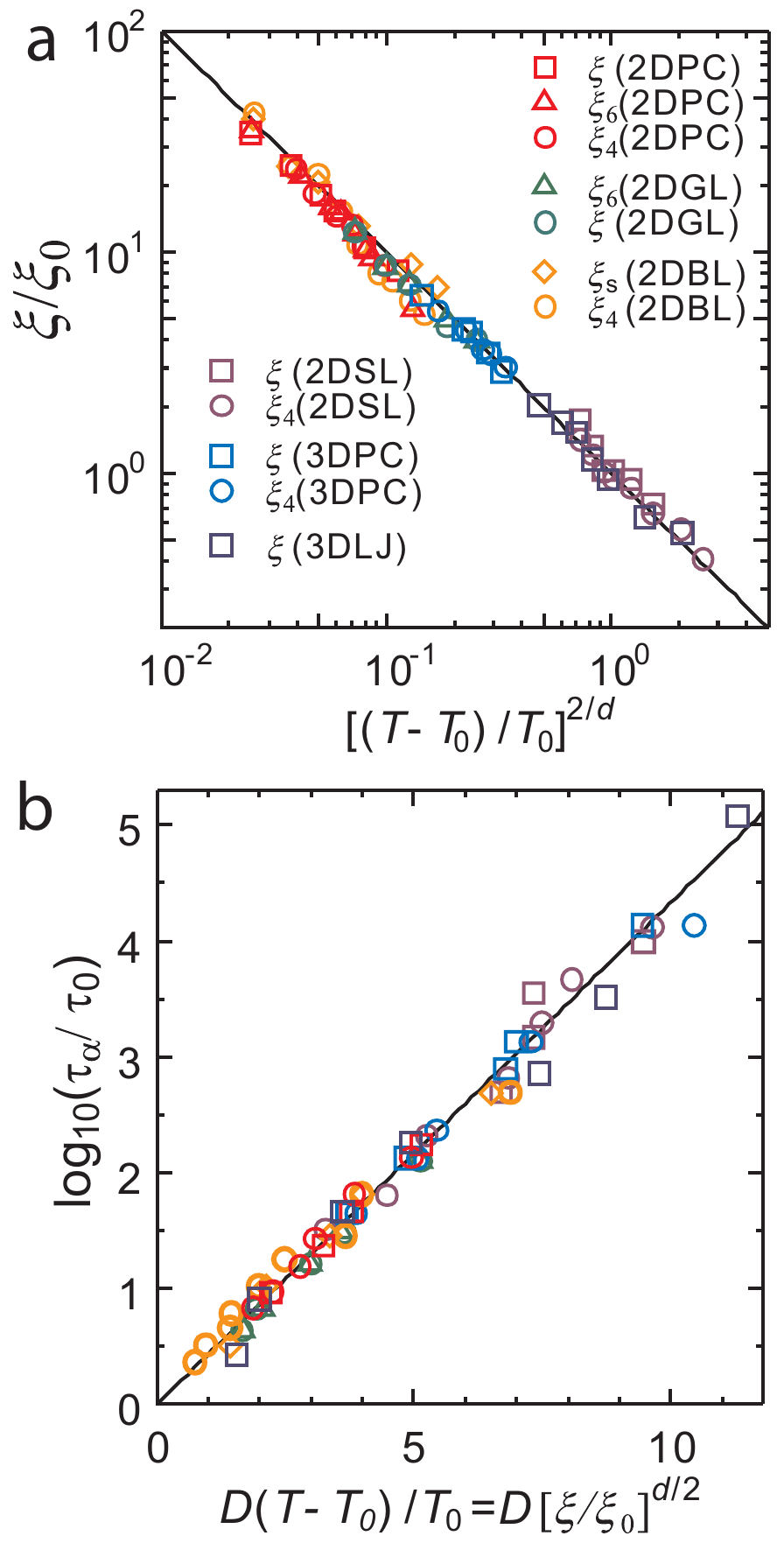}
\end{center}
\caption{(Colour on-line) The growth of medium-range crystalline order 
and its relation to the slow dynamics.  
(a) Relation between $\xi/\xi_0$ and $t^{\nu}$ 
for all the systems: 2DPC ($\Delta_{2DPC}=9$ \%) \cite{KAT}, 
2D granular liquid (2DGL) ($\Delta_{2DGL}=10.7$ \%) \cite{WT}, 
3DPC ($\Delta_{3DPC}=6$ \%), 
3D polydisperse Lennard-Jones liquid (3DLJ) 
($\Delta_{3DLJ}=6$ \% and the density $\rho=1.2$), 
2D spin liquid (2DSL) ($\Delta_{2DSL}=0.6$) \cite{ShintaniNP}, and 
2D binary soft sphere liquid (2DBL). 
Note that $t=(T-T_0)/T_0$. 
The fitted line has a slope of -1, indicating the relation 
$\xi/\xi_0=t^{-2/d}$. 
(b) Relation between $\log_{10}(\tau_\alpha/\tau_0)$ and 
$DT_0/(T-T_0)=D(\xi/\xi_0)^{d/2}$ for all the systems. 
Symbols are the same as in {\bf a}. 
The solid line is the relation of 
$\log_{10}(\tau_\alpha/\tau_0)=(\log_{10} e)[DT_0/(T-T_0)]
=(\log_{10} e)[D(\xi/\xi_0)^{d/2}]$. 
$D$=0.24, 0.41, 7.3, 0.16, 0.78, 3.85 
for 2DPC, 2DGL, 2DSL, 2DBL, 3DPC, and 3DLJ, respectively. 
The fact that the fragility index $D$ is not a universal constant 
suggests that the relation between a 
diverging lengthscale (static criticality)  
and slow dynamics (or, viscosity) may not be universal and depends 
upon the degree of frustration. 
This figure is reproduced from fig. 4 of ref. \cite{TanakaNM}.
}
\label{fig:critical}
\end{figure}

We recently found in several model systems that glassy structural order (e.g., 
bond orientational ordering) exhibits Ising-like critical anomaly towards 
the ideal glass transition point $T_0$ in a supercooled state \cite{TanakaNM}. 
The correlation length of bond orientational order $\xi$ grows 
as 
\begin{equation}
\xi=\xi_0 t^{-\nu},
\end{equation}
where $t=(T-T_0)/T_0$ or $t=(\phi_0-\phi)/\phi$ (see sec. 7.4 of ref. \cite{TanakaNara}), 
and its susceptibility diverges 
as 
\begin{equation}
\chi=\chi_0 t^{-\gamma}.
\end{equation}
We found that $\nu \sim 2/d$ (see fig. \ref{fig:critical}(a)) and $\gamma \sim2 \nu$, which 
are consistent with the $d$-dimensional Ising universality class \cite{TanakaNM}. 
The criticality was checked \cite{TanakaNM} by using the finite-size scaling analysis \cite{berthier2003finite,karmakar2009growing}. 
The similarity between dynamic heterogeneity and critical fluctuations has also been pointed 
out by many other researchers (see, e.g., \cite{mountain1995,mountain1998,yamamoto1998,yamamoto1998dynamics,whitelam2004,mosayebi2010}). 
We speculate that this Ising-like 
criticality is a consequence of frustration caused by either competing 
bond orientational orderings or random disorder effects on structural ordering. 
We note that the bond orientational order parameter with continuous symmetry should exhibit 
a (fluctuation-induced) 
first-order phase transition because it has a continuous symmetry. 
Nevertheless, we observe Ising-like criticality characteristic 
of a discrete ($Z_2$) symmetry \cite{TanakaNM} (see below on its possible origin).

Although we see that the power-law divergence of $\xi$ at least practically describes 
the observed change of $\xi$ in the accessible $t$ region, it is not clear at this moment 
whether $\xi$ really diverges at $T_0$ or not. The limited range of $\xi$ (at most a decade) makes it difficult 
to derive a definite conclusion. This originates from the present ability of numerical simulations, 
but more essentially from the intrinsic inaccessibility to the hidden hypothetical critical point because of 
the extremely steep slowing down of the structural relaxation time $\tau_\alpha$ towards it.  

The minimum reduced temperature, $t=(T-T_0)/T_0$, we can access is still larger than $10^{-2}$, 
which is markedly different from the case of ordinary critical phenomena, where we can easily realize $t \sim 10^{-5}$. 
In relation to this, there is a possibility that there is no singularity above $T=0$  K (see below). 
In our scenario, this problem may be viewed as whether the ordering transition is second-order, rounded, or 
weakly first-order.  As shown below, crystallization must take place before reaching $T_0$: 
Resolution of the Kauzmann paradox \cite{TanakaK}. 
Thus, the ideal glass transition point, $T_0$, is intrinsically an inaccessible critical point.  
We also note that recently even a non-monotonic temperature dependence of $\xi$ was reported \cite{kob2011non}. 
In this work, however, the way to extract the correlation length is different from ours (see sec. \ref{sec:wall}), which might have 
an influence on the non-monotonic behaviour of $\xi$.  
Further careful studies are necessary for revealing the origin of the criticality or even its presence.

\subsubsection{Crossover between critical (low temperature) and non-critical 
(high temperature) behaviour} \label{sec:cross}

Here we consider a possible crossover between critical (low temperature) 
and non-critical (high temperature) behaviour. 
Our scenario of a hidden critical point at the ideal glass transition point 
$\phi_0$ suggests that the glass transition 
volume fraction $\phi_g$ is located far below $\phi_0$. 
This implies that we are almost always quite far from the 
critical point and the accessible reduced temperature $t$ is rather large, as described above. 
Recently, we found for gas-liquid critical phenomena of a colloid-polymer 
mixture that the crossover of the correlation length from a critical to a 
non-critical, classical regime can be expressed by replacing 
the ordinary $t=(T-T_c)/T_c$ by $t=(T-T_c)/T$ ($T_c$: the critical point) 
\cite{Paddy}. 
This expression avoids an unphysical behaviour that $\xi$ goes to zero 
for $T \rightarrow \infty$, and guarantees $\xi \rightarrow \xi_0$, 
where $\xi_0$ is the bare correlation length reflecting 
a characteristic length of microscopic interactions, 
for $T \rightarrow \infty$. 
This expression was also theoretically proposed for magnetic systems 
\cite{Campbell}. So we should be able to describe the crossover 
from a critical to a non-critical regime by replacing $t=(\phi_0-\phi)/\phi$ 
by $t=(\phi_0-\phi)/\phi_0$ in a natural manner. 
In hard sphere colloids, it is known that 
for $\phi \leq 0.45$ the relaxation time is almost independent 
of $\phi$ \cite{vanMegenX}, which implies that there is no 
cooperativity for that $\phi$ range. 
For ordinary molecular liquids, this relation may describe a crossover 
from a high-temperature Arrhenius to a low-temperature super-Arrhenius 
behaviour. This point needs further studies. 
This crossover marks the onset of the criticality, which 
induces all sorts of characteristic glassy behaviours such as 
dynamic heterogeneity, translational-rotational decoupling, the violation of Stokes-Einstein relation, 
and the emergence of Johari-Goldstein slow $\beta$ process \cite{ngai2011,tanaka2004origin}.

\subsubsection{Link between the correlation length and the structural relaxation}
\label{sec:Link}

What physical mechanism connects the growing length scale 
with slow dynamics also remains an important open question, which lies at the heart of the origin of 
the glass transition (see also \ref{sec:activation}). 
Here we consider this problem, namely, the relation between the correlation length of glassy structural order $\xi$ and the 
structural relaxation time $\tau_\alpha$. 
We find the following empirical relation between $\tau_\alpha$ and $\xi$  (see fig. \ref{fig:critical}(b)): 
\begin{equation}
\tau_\alpha=\tau_\alpha^0 \exp(D (\xi/\xi_0)^{d/2}),  \label{eq:tau_D}
\end{equation}
where $\tau_\alpha^0$ is the microscopic time, 
$D$ is the fragility index, and $d$ is the spatial dimensionality. 
Here it should be mentioned that similar relations for $\xi$ and $\tau_\alpha$ were recently reported on the basis 
of numerical simulations, but with slightly different functional forms or exponents 
\cite{karmakar2009growing,szamel2010,mosayebi2010,kob2011non,karmakar2011direct}. Thus, we need further careful studies 
to settle this issue. 
We note that the above relation reduces to the following Vogel-Fulcher-Tammann (VFT) relation 
by inserting $\xi=\xi_0 t^{-2/d}$:
\begin{equation}
\tau_\alpha=\tau_\alpha^0 \exp(D T_0/(T-T_0)).  \label{eq:VFT}
\end{equation}
Whether the relation (\ref{eq:tau_D}) implies the direct importance of the growing length scale in slow dynamics or merely 
the indirect relation via, e.g., the number density of defects remains a subject for future study. 
However, we argue that the former may be the case: 
The relation (\ref{eq:tau_D}) implies that the structural relaxation involves the activation process 
whose energy scales as $\xi^{d/2}$, which can be understood as a consequence of cooperativity, or the coherency of particle motion 
over $\xi$. 

\begin{figure}
\begin{center}
\includegraphics[width=6.5cm]{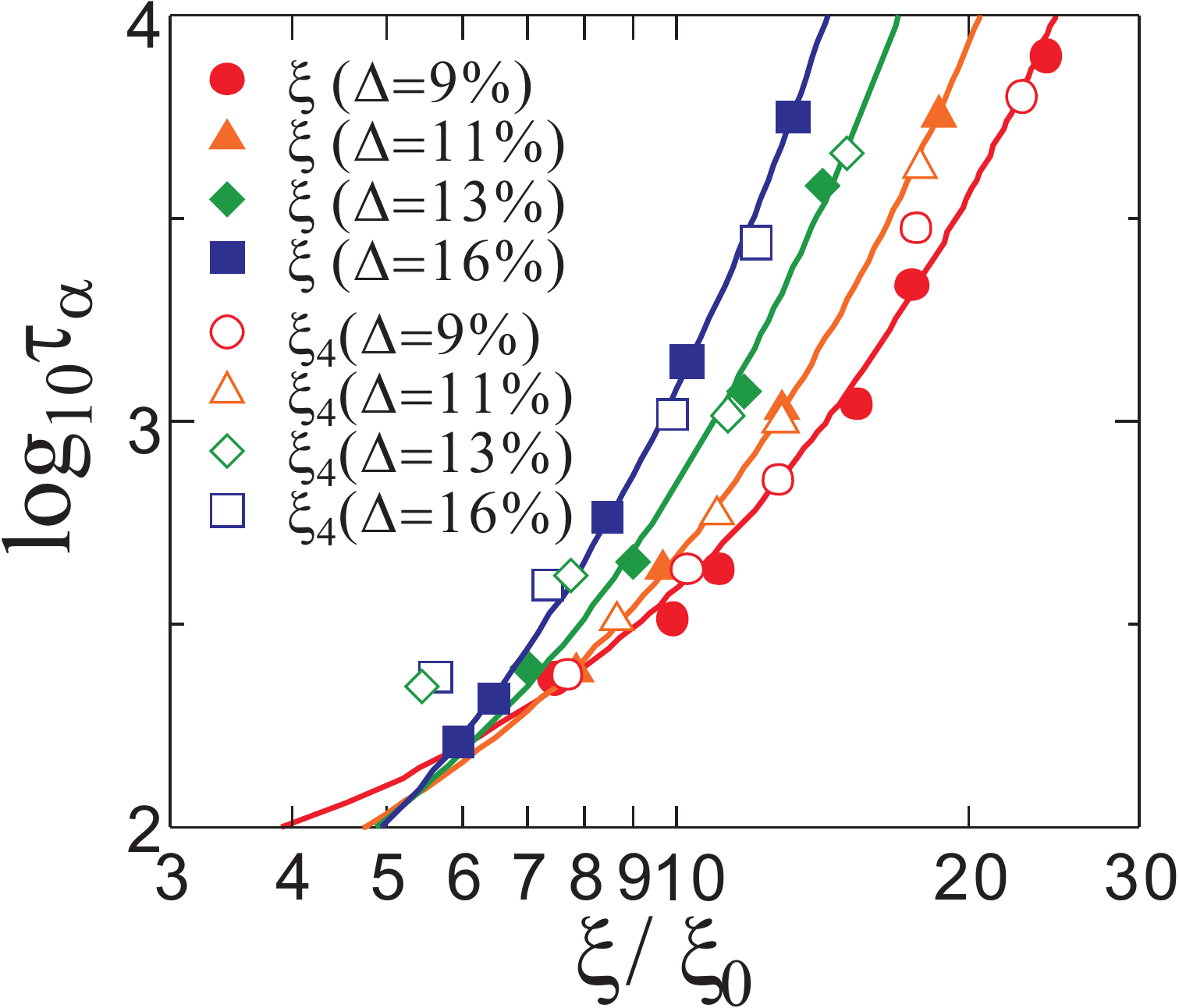}
\end{center}
\caption{(Colour on-line) Log-log plot of $\tau_\alpha$ against $\xi$ for 3DPC. 
Solid lines are the fitting of eq. (\ref{eq:tau_D}). }
\label{fig:tau_xi_log}
\end{figure}

A different type of relation, the power law,  
\begin{equation}
\tau_\alpha \propto \xi^z, \label{eq:power_xi}
\end{equation}
was also proposed, relying on the analogy to critical phenomena \cite{yamamoto1998,yamamoto1998dynamics}. 
This power-law form has widely been used in the analysis of simulation results, but not so popularly used in in the analysis of experimental results. 
We also note that the following form which takes into the activated dynamics into account was also proposed 
for a low temperature region where cooperativity plays a role \cite{erwin2002temperature}:
\begin{equation}
\tau_\alpha \propto (\xi/\xi_0)^z \exp(\Delta_a/k_{\rm B}T), \label{eq:power_A},
\end{equation}
where $\Delta_a$ is the microscopic activation energy. 
For 3D polydisperse spheres, however, the relation (\ref{eq:tau_D}) better fits the data than the relation (\ref{eq:power_xi}) 
(see fig. \ref{fig:tau_xi_log}). 
See also ref. \cite{szamel2010} on the relation between $\tau_\alpha$ and $\xi$. 
A possibility of a zero temperature dynamical critical point ($T_0=0$ K) was also pointed out \cite{whitelam2004}.

\begin{figure}
\begin{center}
\includegraphics[width=8.0cm]{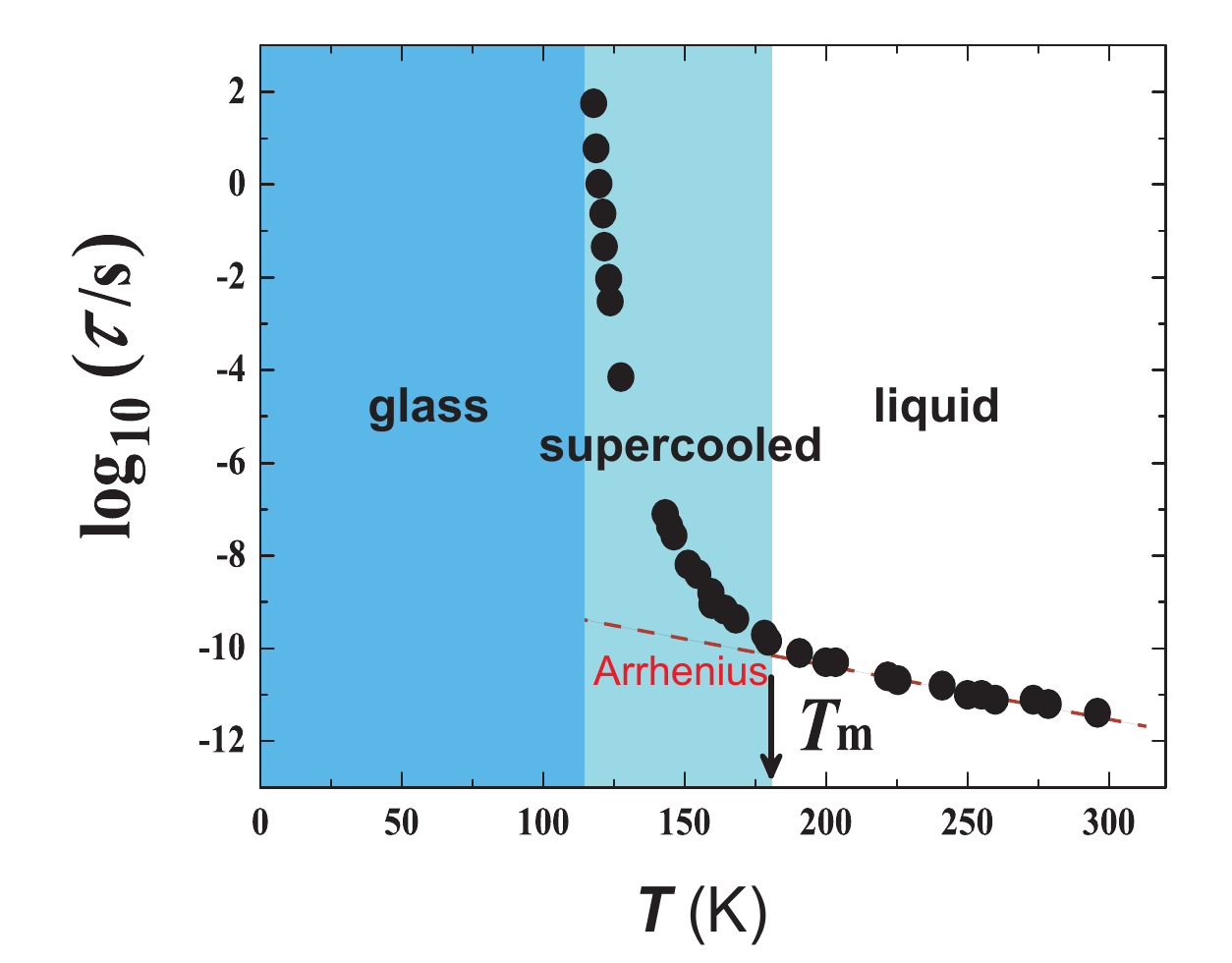}
\end{center}
\caption{(Colour on-line) The temperature dependence of the structural relaxation time $\tau_\alpha$ for toluene. 
The data were taken from ref. \cite{wiedersich2000}. 
The dashed red line is an extrapolation of the high temperature Arrhenius behaviour. }
\label{fig:toluene}
\end{figure}

Here we point out a difficulty in the power-law type relation (\ref{eq:power_xi}). 
The power exponent $z$ was reported to be in the range between 2 and 4 on the basis of results of numerical simulations. 
In the range that can be covered by simulations, the difference between the power law and the exponential law is not so significant. 
In experiments, on the other hand, the situation is very different. As such an example, we show in fig. \ref{fig:toluene} the temperature dependence of $\tau_\alpha$ for toluene, 
which is a typical fragile liquid. As can be seen there, $\tau_\alpha$ starts to deviate from the high temperature Arrhenius behaviour 
around the melting point $T_m$ and then increases by 12 orders of magnitude during a temperature interval of several tens K upon cooling. 
This means that even for $z=4$, $\xi$ has to increase by a factor of 1000. However, such a long-range correlation (1000 times of a molecular size) 
has never been reported. This seems to be a weakness of the power law scenario.  
However, since the range covered by numerical simulations, which can provide information on $\xi$, is rather narrow and there might also 
be a crossover from microscopic to mesoscopic dynamics, 
further careful studies are necessary to reveal the definite relation between $\tau_\alpha$ and $\xi$.

\begin{figure}
\begin{center}
\includegraphics[width=8cm]{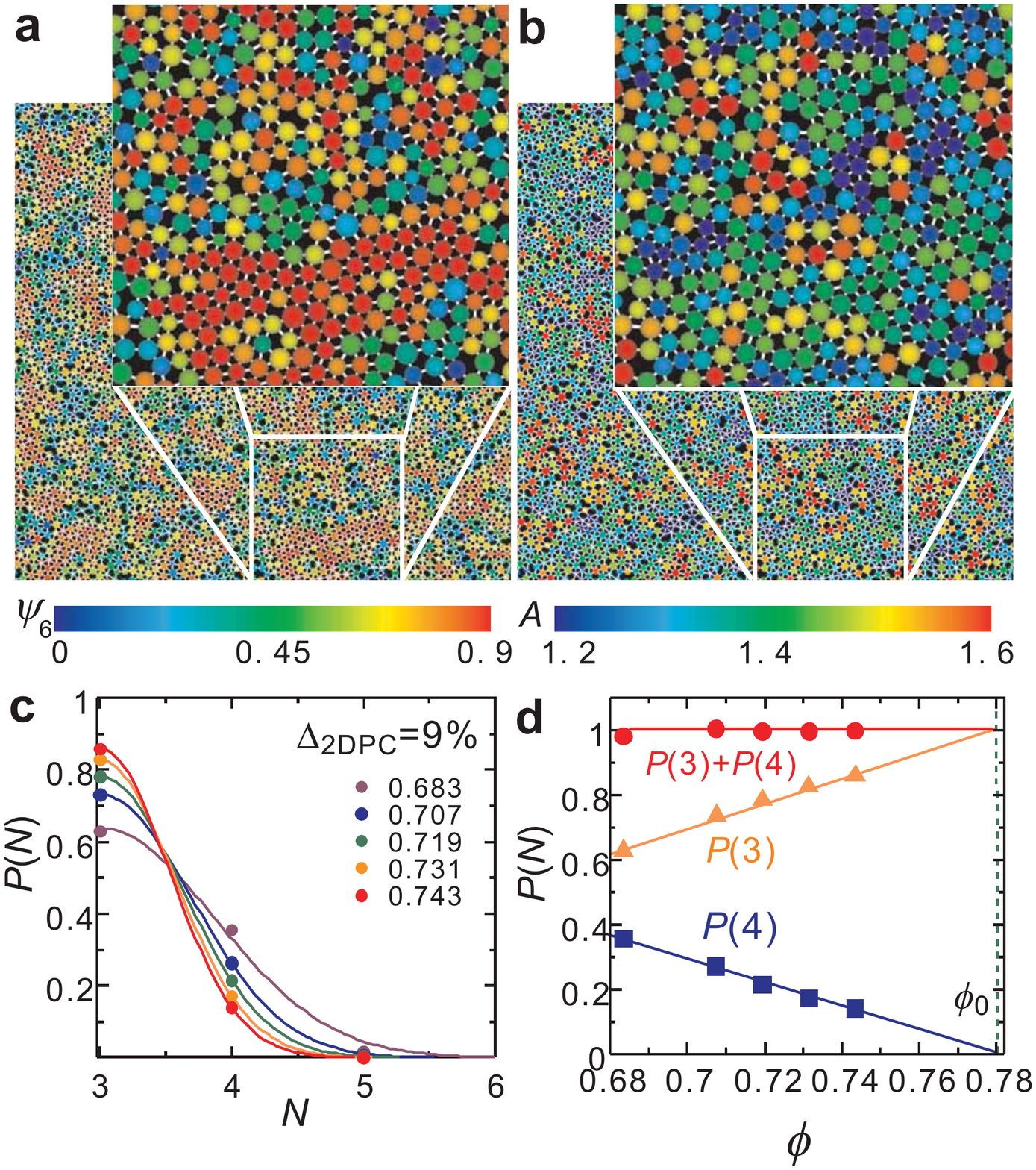}
\end{center}
\caption{(Colour on-line) Geometrical tiling patterns of 2DPC ($\Delta_{2DPC}=9$\% 
and $\phi=0.73$) and its $\phi$ dependence.  
(a) Correlation of the (instantaneous, or not time averaged) 
order parameter, $\psi_6$, to 
tiling units (triangles, squares, pentagons, $\cdots$), 
or geometrical defects for 2DPC.  
Bonds are shown as thin white lines. 
$\psi_6$ is evidently anti-correlated 
with the number density of geometrical defects (voids). 
(b) Correlation of a local volume (area) per particle, $A$, 
calculated from the area of a Voronoi polygon to 
a tiling unit (triangles, squares, pentagons, $\cdots$), or 
geometrical defects, for 2DPC. 
Geometrical defects, or voids (red particles), 
accompany densely packed triangles (blue particles) nearby, and thus 
density fluctuations are suppressed over a long range. 
Note that particles with large $\psi$ have green colour, indicating 
they have intermediate $A$. 
See the colour bars for the meaning of the particle colour. 
{\bf c,} Distribution of polygons with $N$ sides. 
Polygons are predominantly composed of triangles ($N=3$) and squares ($N=4)$. 
{\bf d,} $\phi$-dependence of the fraction of triangles and squares, 
which are averaged over $10 \tau_\alpha$. 
Squares (geometrical defects) decrease, or transform to triangles, 
with an increase in $\phi$ and tend to completely disappear around $\phi_0$. 
This figure is reproduced from fig. 2 of ref. \cite{TanakaNM}.
}
\label{fig:defect}
\end{figure}

\subsubsection{Speculation on the nature of the hypothetical underlying transition at $T_0$}
\label{sec:Speculation}
Here we discuss the nature of 
the underlying ordering, taking 2DPC as an example. 
For 2DPC, we take an order parameter as $\psi_6$, 
which is a non-conserved complex hexatic order parameter. 
The question here is why the complex order parameter of a 
continuous symmetry exhibits discontinuous Ising ($Z_2$) critical 
behaviour. This may be a consequence of the fact that 
the spatial coherence of the phase component $\theta$ 
of the bond orientational order parameter  
$\psi_6=|\psi_6|e^{i \theta}$ is easily disrupted by disorder effects. 
The loss of long-range phase coherence due to disorder effects, more specifically, polydispersity, (see below) 
may prohibit gapless long-wavelength excitations 
in the ordered phase and transform the symmetry of the ordered 
phase from continuous to discrete Ising ($Z_2$) symmetry. 

Physically, the order parameter is correlated with 
the degree of triangular tiling and anti-correlated 
with the number density of geometrical defects 
(squares, pentagons, $\cdots$). 
This feature can be clearly seen in fig. \ref{fig:defect}(d): The population 
of squares $P(4)$ (defects) monotonically decreases with $\phi$ and its extrapolation 
goes to zero around $\phi_0$, suggesting 
the transformation of squares to triangles with an increase in $\phi$. 
The hypothetical ordered state supposed to be attained at $\phi_0$ 
may be a state of global triangular tiling, but with `distortion' 
due to polydispersity-induced geometrical disorder. 
This state has no (or little) configurational entropy. Even in this hypothetical ideal non-ergodic state, 
correlational entropy, or local vibrational degrees of freedom, remains; 
that is, particles dress correlation volumes around them. This feature 
may make thermal glass transition distinct from athermal jamming transition. 
Geometrical defects tend to disappear completely at $\phi_0$ 
(see fig. \ref{fig:defect}(d)), 
which results in the disappearance of the configurational entropy 
and the resulting freezing of a particle configuration, or the loss of fluidity. 
The importance of defects was also emphasized by Aharonov et al. 
\cite{Schupper} in their study of a 2D binary soft-sphere mixture. 
They considered that some defects must be excited at a finite temperature 
and thus any phase transition does not exist (the absence of $T_0$). 
This may further imply that there is a finite probability to excite defects even below $T_0$, 
ultimately indicating the absence of the ideal glass with static elasticity. 
As will be shown later (see sec. \ref{sec:Kauzmann}), the system may eventually crystallize at the lower metastable limit $T_{\rm LML}$ before reaching 
$T_0$ \cite{TanakaK} and thus it may not practically be meaningful to consider the exact nature of this 
hypothetical ideal glass transition point, despite its conceptual importance. 

This link between the degree of bond orientational order and the degree of triangular tiling suggests 
the link between `glassy structural order' and `solidity'. 
This latter link may be a generic feature of glassy structural order, which is formed to lower 
the free energy of the system locally. This further suggests that the lowering of the free energy must be associated with the increase in solidity, or slow dynamics. 

The degree of distortion increases with an increase in the degree of polydispersity, $\Delta$. 
This means that a denser packing is required to attain a state of global 
triangular tiling, which explains the increase of $\phi_0$ and $D$, 
i.e., the decrease of the fragility, with an increase in $\Delta$. 
We find that the correlation lengths estimated from the spatial correlations of the complex and scalar (rotationally invariant) 
bond orientational order parameters both diverge 
towards $\phi_0$, following the same power law $\xi/\xi_0=(\phi_0/\phi-1)^{-1}$. 
This scalar nature of the order parameter `correlation' also suggests that 
the system is to belong to the Ising universality class, 
whose lower critical dimension $d_L$ is 1   
\cite{Onuki,Lubensky}, consistent with our observation. 
However, we stress that the structural order parameter 
itself is not scalar, but complex (or tensorial). 

Such transformation of the phase ordering 
from (Heisenberg-type) continuous to Ising ($Z_2$) symmetry due to frustration and random disorder effects 
has also been known 
for spin systems \cite{Chandra,Weber}, implying the generality of 
frustration and random disorder effects on the nature of the ordering. 
Very recently, the liquid-to-hexatic transition in 2D was revealed to be weakly first-order \cite{krauth}. 
The introduction of the polydispersity might also alter the nature of the transition. 

These facts lead us to the following conjecture: The introduction of any frustration or random disorder effects disturb 
the phase coherence of the direction of the vector or tensorial order parameter. This may eventually transform the nature of the transition from the continuous to the discrete Ising symmetry. 
So we speculate that the Ising-like criticality associated with 
bond orientational ordering may be a manifestation of the underlying frustration against the bond orientational ordering. 
Finally, we emphasize that the most important characteristics of this transition is that the order parameter is not only linked to 
local structural symmetry, but also to high solidity, or low fluidity. This may be the most crucial feature of glass transition.

\subsubsection{Critical-like slow dynamics of bond orientational order parameter in 2DPC}
\label{sec:longlifetime}

Here we consider critical-like slowing down of hexatic order parameter $\psi$ observed in 2DPC. 
The kinetic equation describing the dynamics of 
the non-conserved complex order parameter $\psi$ 
should be expressed as 
\begin{eqnarray}
\frac{\partial \psi}{\partial t}=-L_{\rm R} \frac{\delta F(\psi)}
{\delta \psi^\ast}, 
\end{eqnarray}
where $L_{\rm R}$ is the renormalized transport coefficient. 
Although $L_{\rm R}$ may depend on $\psi$, here we neglect such a dependence for simplicity. 
Here $F(\psi)$ is the Landau-type free energy functional of the order 
parameter $\psi$: 
\begin{eqnarray}
F(\psi)=\int d \mbox{\boldmath$r$} \ 
\left[ \frac{\tau}{2} |\psi|^2 +\frac{u}{4} |\psi|^4 
+\frac{K}{2} |\nabla \psi|^2+\cdots \right],   \label{freep}
\end{eqnarray}
where $\tau=a(T-T_0)$ and $a$, $u$ and $K$ are positive constants 
at the mean-field level.  
However, $\tau$, $u$, and $K$ can be replaced 
by the renormalized ones for a more accurate 
description including fluctuation effects \cite{Onuki}. 
Here $T_0$ is the hypothetical ordering point of $\psi$ 
under an influence of frustration. 
Equation (\ref{freep}) may be regarded as a part of the free energy 
describing crystallization, in which density (positional) ordering 
is prevented by frustration.  
This kinetic model belongs to ``model A'' in the Hohenberg-Halperin classification of dynamical 
critical phenomena \cite{Onuki,Lubensky}. 
Here we neglect possible couplings of $\psi$ 
to the velocity and stress fields for simplicity. 
Then, the order parameter correlation function $S(k,t)$ 
is predicted to decay as a function of time $t$ as 
\begin{eqnarray}
S(k,t) \equiv \langle \psi_k(t)\psi_{-k}(0) \rangle=\chi_k \exp(-\Gamma_k t), \label{eq:Sqt}
\end{eqnarray}
where $\chi_k (\equiv S(k,0)=\langle \psi_k \psi_{-k} \rangle)$ 
is the $k$-dependent susceptibility and $\Gamma_k$ 
is the decay rate of the order parameter. 
Here $\Gamma_k$ is given by 
\begin{eqnarray}
\Gamma_k=L_{\rm R}/\chi_k \cong 
(L_{\rm R}/\chi_0)t^\gamma [1+(k \xi)^{2-\eta}].   \label{eq:Gammaq} 
\end{eqnarray}

\begin{figure}[h]
\begin{center}
\includegraphics[width=8cm]{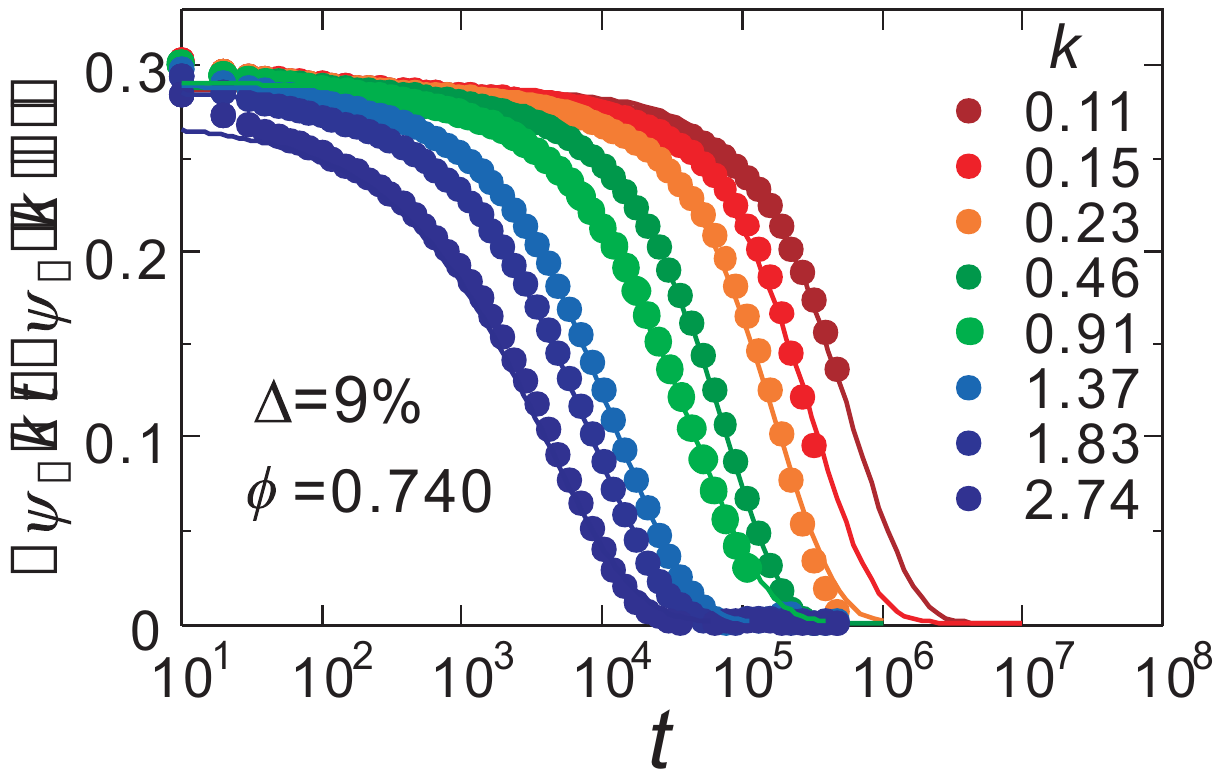}
\end{center}
\caption{(Colour on-line) $k$-dependence of $S_k(t)$ ($=\langle \psi_6(k,t)\psi_{6}(-k,0) \rangle$) 
for $\phi=0.740$ and $\Delta=9$ \%. The curves are the fitted 
functions. }
\label{fig:Gamma}
\end{figure}

Here we show the order parameter correlation function 
$S_k(t)$ for different values of $k \xi$ in fig. \ref{fig:Gamma}. 
For $k a \leq 1$ (note that the average particle radius $a=1$), 
$S_k(t)$ can be well fitted by 
the single exponential function, which is typical of critical 
fluctuation dynamics, whereas 
for $k a > 1$  (note that $a=1$) the function becomes stretched, reflecting 
the influence of microscopic structural relaxation. 
This crossover from the low $k$ to high $k$ behaviour of the 
order parameter correlation function 
around $ka \sim 1$ is reasonable 
and can be regarded as the crossover from the critical-fluctuation-dominated 
to the microscopic-relaxation-dominated regime.

\begin{figure}
\begin{center}
\includegraphics[width=8.5cm]{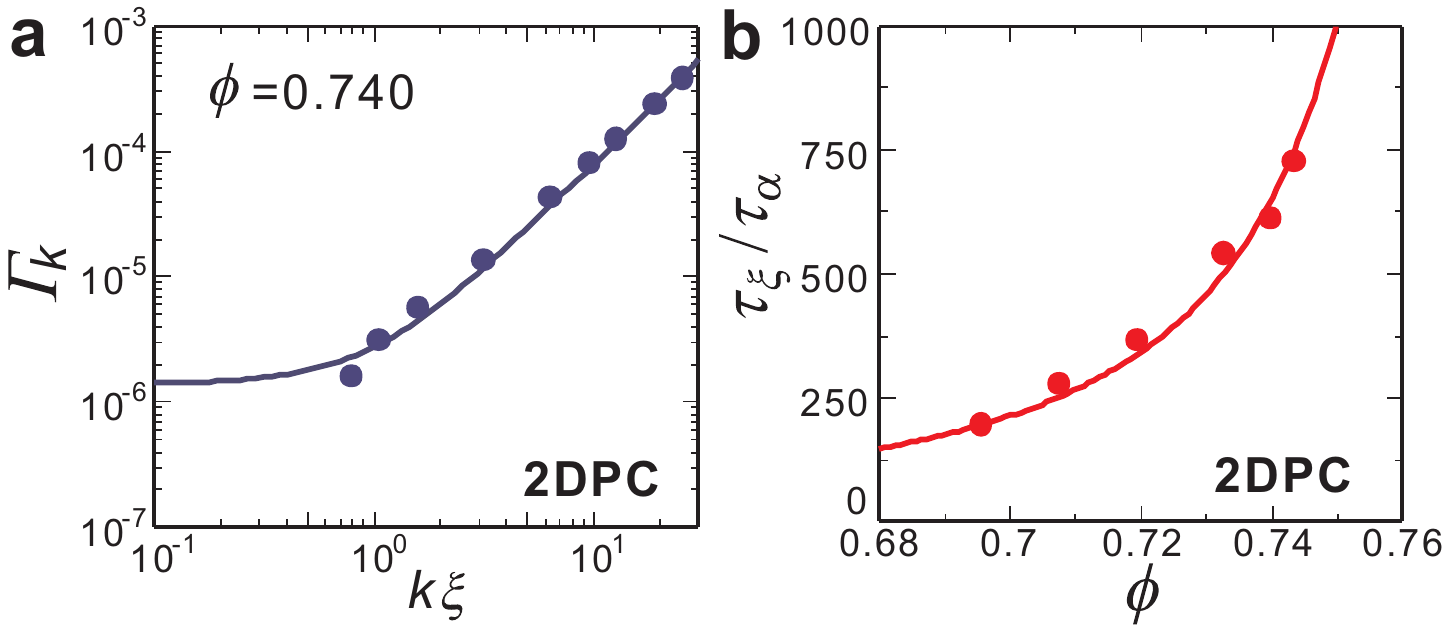}
\end{center}
\caption{(Colour on-line) Critical-like behaviour of dynamics correlation 
in 2DPC ($\Delta_{2DPC}=9$ \%).  
(a) The $k$ dependence of the decay rate $\Gamma_k$ for 2DPC. 
The fitted curve is $\Gamma_k \propto [1+(k \xi)^{7/4}]$. 
Here $\xi$ was determined independently by the cluster size (see text). 
(b) The $\phi$ dependence of $\tau_\xi/\tau_\alpha$ for 2DPC. 
The fitted curve is $\tau_\xi/\tau_\alpha \propto t^{-7/4} 
\propto (\phi_0-\phi)^{-7/4}$. 
$\phi_0$ was determined by the independent fitting for $\tau_\alpha$ 
as 0.78. This figure was reproduced from fig. 6 of ref. \cite{TanakaNM}. 
}
\label{fig:Gamma_q}
\end{figure}

Next we compare the simulation results of 2DPC  
with the theoretical prediction 
of the above phenomenological model for 2DPC ($d=2$). 
Because of the Ising ($Z_2$) symmetry of the underlying ordering, 
we expect that this system belongs to the 2D Ising universality 
\cite{Lubensky}, where 
the values of the critical exponents are: $\gamma=7/4$ (susceptibility), 
$\alpha=0$ (heat capacity), $\beta=1/8$ (spontaneous magnetization), and 
$\nu=1$ (correlation length), and $\eta_{\rm F}=1/4$ 
(Fisher correction) \cite{Onuki,Lubensky}. 
This means that $\Gamma_k \propto [1+(k\xi)^{7/4}]$ 
and $\Gamma_{k=1/\xi} \equiv 1/\tau_\xi =(2 L_{\rm R}/\chi_0) 
t^{7/4}$ for 2D (see eq. (\ref{eq:Gammaq})). 
Figure \ref{fig:Gamma_q}(a) plots the $k$-dependence of $\Gamma_k$ 
obtained by the fitting of eq. (\ref{eq:Sqt}) to $S(k,t)$ (see fig. \ref{fig:Gamma}), 
which is consistent with the above prediction based on model A, although we have few data of $\Gamma_k$ for $k \xi <1$. 
Figure \ref{fig:Gamma_q}(b), on the other hand, plots 
$\tau_\xi/\tau_\alpha=(\chi_0/2L_{\rm R} \tau_\alpha)t^{-7/4}$ 
against $\phi$, suggesting $\tau_\xi/\tau_\alpha \propto t^{-7/4}$. 
This means that $L_{\rm R} \propto 1/\tau_\alpha \propto 1/\eta$, 
which needs to be explained and justified 
on a fundamental level in the future (see also \ref{sec:activation}). 
These results suggest that the behaviour of glass-forming liquids (2DPC) 
can be explained by a critical-phenomena-like scenario in which the 
correlation length of glassy structural order $\xi$ diverges 
towards the hypothetical ideal glass transition temperature $T_0$. 
We note that $\tau_\xi$ should be the lifetime of crystal-like bond orientational order, 
which may be the slowest timescale of the system. 

\subsubsection{Brief review on the analysis based on the four body density correlator}

The first attempt to seek a growing correlation length in computer simulations 
was made by Dasgputa et al. \cite{dasgupta1991there} and Ernst et al. \cite{ernst1991search}, but both 
of which failed in finding such evidence. However, after the discovery of its experimental evidence \cite{spiess}, 
many studies have revealed the presence of growing dynamical correlation length with a help of the increase in computer power 
and a consensus on it has now been established firmly (see, e.g., \cite{harrowell2011}).  
Here we describe standard methods analysing dynamic heterogeneity; namely, the particle trajectory analysis and the four-time density 
correlation function analysis. Since the former is straightforward, here we explain only the latter. 

Density fluctuations play an important role in the description of liquid as spin fluctuations in magnets. 
The time correlation of density fluctuations $\delta \Psi(\vec{r},t)=\langle \delta \rho(\vec{r},t)\delta \rho (\vec{r},0) \rangle$ 
has been thought to be a key order parameter of glass transition, which is often called ``Edwards-Anderson'' or non-ergodicity order parameter, since 
it is zero in an ergodic liquid state and becomes non-zero in a nonergodic glassy state. 
Then the natural correlation function for this non-ergodicity order parameter is given by the following four-point correlation function: 
\begin{eqnarray}
G_4(\vec{r},t)=\langle  \delta \rho(\vec{0},t)\delta \rho (\vec{0},0)  \delta \rho(\vec{r},t)\delta \rho (\vec{r},0) \rangle  \nonumber \\
-\langle  \delta \rho(\vec{0},t)\delta \rho (\vec{0},0) \rangle \langle  \delta \rho(\vec{r},t)\delta \rho (\vec{r},0) \rangle.   
\end{eqnarray} 
Then the susceptibility of this order parameter, a four-point susceptibility, is given by 
\begin{eqnarray}
\chi_4=\int d\vec{r} G_4(\vec{r},t). 
\end{eqnarray}

Although the above approach is theoretically more appealing, 
practically a mobility correlation is easier to handle. Here we explain 
this following ref. \cite{Glotzer}. 
First a time-dependent order parameter that measures 
the number of overlapping particles in two configurations 
separated by a time interval $t$ is defined 
as 
\begin{eqnarray}
Q(t)= \Sigma_{i=1}^{N} \Sigma_{j=1}^{N} 
w(|\mbox{\boldmath$r$}_i(0)-\mbox{\boldmath$r$}_j(t)|), \nonumber
\end{eqnarray}
where $w(r)=1$ for $r \leq b$ whereas $w(r)=0$ for $r>b$. 
$b$ is a typical amplitude of vibrational motion and set to $0.3-0.4a$.  
The structural relaxation can then be characterized by the 
variance of $Q(t)$ as 
\begin{eqnarray}
\chi_4(t)=\frac{V}{k_{\rm B}TN^2} \left[ \left<Q(t)^2 \right>
- \left<Q(t) \right>^2 \right]. \nonumber
\end{eqnarray} 
The spatial correlation of the overlapping function $w$ at a time $\chi_4(t)$ has a peak provides us with 
the characteristic length of dynamic heterogeneity, $\xi_4$.

\subsubsection{Relationship between the structural relaxation time $\tau_\alpha$, 
the lifetime of dynamic heterogeneity $\tau_{\rm DH}$, and that of glassy structural ordering $\tau_\xi$} \label{sec:decoupling}

Next we consider the relation of critical-like dynamics of the bond orientational order parameter 
to structural relaxation dynamics and its dynamic heterogeneity. 
We have confirmed that the characteristic length scale of dynamic heterogeneity 
$\xi_4$ determined by the four-point 
density correlator is almost equivalent to the characteristic length scale of bond 
orientational order parameter $\xi$ \cite{ShintaniNP,KAT,WT,TanakaNM,Kawasaki3D,KawasakiJPCM} 
(see also fig. \ref{fig:critical}). 
This is further supported by a one-to-one correspondence between the degree of bond orientational order and the 
slowness of particle dynamics (see fig. \ref{fig:compare}) \cite{TanakaNM}. 
However, there is a large difference in the lifetime of dynamic heterogeneity and that of the 
bond orientational order parameter, as shown in fig. \ref{fig:Gamma_q}(b). 
The former, which is characterized by the time when the susceptibility $\chi_4(t)$ 
has a peak, is the order of the structural relaxation time $\tau_\alpha$. 
On the other hand, the latter can be a few orders of magnitude times longer than $\tau_{\alpha}$ 
for 2DPC (see fig. \ref{fig:Gamma_q}(b)). 
For 3D systems, on the other hand, this difference becomes much smaller 
and less than a factor of 10, although this is dependent on the degree of supercooling.

\begin{figure}
\begin{center}
\includegraphics[width=8.5cm]{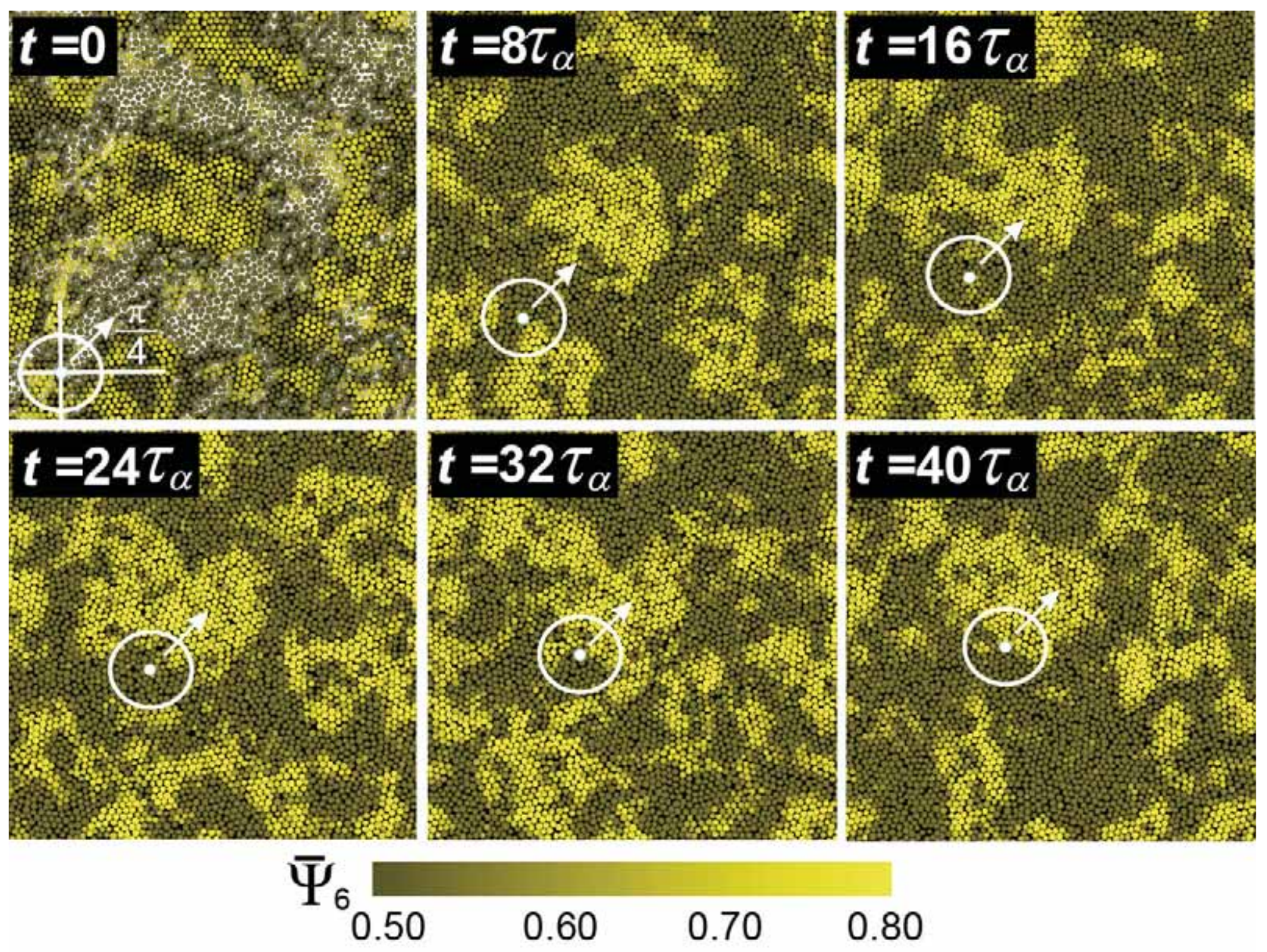} 
\end{center}
\caption{(Colour on-line) The position of a tagged particle for 
$t=0, 8\tau_{\alpha},16\tau_{\alpha}, 24\tau_{\alpha}, 
32\tau_{\alpha}$, and $40\tau_{\alpha}$.  
Here $\phi=0.617$, $\Delta=9\%$ and $F_{\rm ex}=0.014$.
The tagged particle is coloured in white, whereas 
other particles are coloured 
from dark yellow to light yellow, reflecting 
the value of $\bar{\Psi}_6$ (see the color bar). 
The whitish background in the image ($t=0)$ represents 
the extremely fast moving particles (hoppers \cite{Douglas}), 
whose local mean squared 
displacements $\langle \Delta r^2(10\tau_{\alpha}) \rangle>2.5$. 
This threshold is the ``hop'' distance introduced in \cite{Douglas}.
This figure is reproduced from fig. 2 of ref. \cite{KTviolation}.  
}
\label{fig:FDT}
\end{figure}

Because of this disparity of the two lifetimes, 
one might cast a doubt on the relevance of bond orientational order in slow glassy dynamics. 
So we consider the reasons for this discrepancy. 
First of all, dynamic heterogeneity is defined by local particle dynamics, whose relevant wavenumber 
is the order of the inverse of the particle size $a$. 
The four-body density correlation apparently deals with the wavenumber of $1/\xi$ 
and seems to pick up mesoscopic dynamics. 
Although both reflect mesoscopic dynamics, there is still a crucial difference. 
This is natural since the change in local particle configurations is enough for local structural relaxation to take place. 
On the other hand, the lifetime of bond orientational order parameter fluctuations 
is measured for the wavenumber of $1/\xi$ 
(or, $k \xi \sim 1$) and thus it is not associated with local relaxation, but 
with mesoscopic relaxation. 
We can see a steep dependence of $\Gamma_k$ on the wavenumber $k$ (see fig. \ref{fig:Gamma_q}), 
as in critical phenomena \cite{Onuki}.

Furthermore, by definition, the overlapping function $w$ becomes zero if a particle inside 
a cage moves together with the cage itself over $b$. On the other hand, bond orientational order 
does not decay if all the particles move translationally or rotationally together. 
In such a case, the dynamic heterogeneity detected by $w$ should disappear 
within a time scale of $6\pi \eta b^2 \xi/k_{\rm B}T$, which is comparable to $6 \pi \eta a^3/k_{\rm B}T$. 
Thus, it is natural to expect that the lifetime of dynamic heterogeneity is shorter than 
that of bond orientational order fluctuations, which may be given by $\tau_\xi=6 \pi \eta \xi^3/k_{\rm B}T$. 
Furthermore, bond orientational order is already not a single-particle quantity, and involves at least 6 particles 
for 2D. Moreover, the medium-range nature of bond orientational order imposes the coherency of particle motion over its size 
and its lifetime $\tau_\xi$. 
Since structural and stress relaxation takes place locally, the lifetime of bond orientational order 
fluctuations does not affect the ordinary relaxation dynamics, but structural relaxation may be dominated 
spatially by medium-range bond orientational order, which is a measure of the coherence length of particle motion, 
as described above. 
However, since the connection of the correlation length of the bond orientational order parameter to the structural 
or stress relaxation is not clear yet, further studies are necessary to establish the importance of the 
static glassy structural order in slow dynamics (see also sec. \ref{sec:lengthdynamics}).  

Here it may be worth noting that Shiba et al. \cite{shiba2012relationship} compared the lifetime of dynamic heterogeneity obtained 
from the four-point density correlator and that from the bond-breakage lifetime introduced by 
Yamamoto and Onuki \cite{yamamoto1998dynamics,yamamoto1998}. They showed that the former is fragile against low-frequency 
vibrations but the latter is robust. We speculate that the correlation length of our bond orientational order, which is of static origin, may have a link to that of long bond-lifetime particles, which is of dynamic origin, for polydisperse colloid systems.

Finally, we note that critical-like fluctuations are characterized by its fractal nature. 
Fluctuations of high bond orientational order, whose correlation length is the order of $\xi$, 
contains shorter lengthscale fluctuations and thus local structural relaxation can take place 
much faster than the lifetime of long wavelength fluctuations, as we can see in the $k$-dependence 
of $\Gamma_k$ (see fig. \ref{fig:Gamma_q}). 
Note that both structural and stress relaxation are accomplished
by average particle displacements of less than half a
particle diameter \cite{Perera1999b}. 

Although the lifetime of MRCO is different from the structural relaxation time, 
we note that this long lifetime of bond orientational order can have physical significance. 
For example, it leads to an apparent violation of the fluctuation-dissipation theorem for 2DPC 
\cite{KTviolation}. 
As shown in fig. \ref{fig:FDT}, when we drive a particle by using an external force, its mobility 
largely fluctuates and depends on the degree of local bond orientational order. 
When a particle encounters a high bond orientational region, particle cannot penetrate 
into that region and have to wait for its transformation to a state of low bond orientational order. 
This is a consequence of transient elasticity of ordered regions due to the slow dynamics. 
This intermittency is the origin of the apparent transient violation of the 
fluctuation-dissipation theorem. 
This violation is absent in the time scale longer than $\tau_\xi$, as it should be for an ergodic system 
\cite{KTviolation}. 

We also note that the lifetime of glassy structural order parameter fluctuations $\tau_\xi$ 
might play a role in nonlinear shear thinning behaviour of a supercooled liquid. 
Furukawa et al. \cite{furukawa_shear} demonstrated that shear thinning is not associated with the structural relaxation time $\tau$, but with a time scale much slower than $\tau$ by a few orders of magnitude in 2D soft sphere mixtures.     
It was suggested that the critical shear rate characterizing shear thinning behaviour $\dot{\gamma}_c$ might be given by 
$\dot{\gamma}_c \sim 1/\tau_\xi$, as in the case of a coupling of shear flow to critical concentration fluctuations 
in a critical binary mixture \cite{Onuki}. Further careful consideration is necessary on this problem, including 
other possible mechanisms leading to shear thinning behaviour. 

Finally we note that the lifetime of dynamic heterogeneity was directly estimated experimentally (see, e.g., \cite{spiess,spiess2,cicerone1995,wang2000lifetime,ediger2000R,richert2002}) 
and also discussed on the basis of numerical simulations (see, e.g., \cite{leonard2005lifetime,kim2010multi,harrowell2011}). 
The link between these observations and the lifetime of glassy structural order (e.g., bond orientational order) is an interesting issue since 
at least there is a decoupling between the dynamic heterogeneity estimated from the four-point density correlation function and 
the lifetime of bond orientational order, as described above.

\subsubsection{Critical phenomena with growing activation energy} \label{sec:activation}

In the above, we show a possibility that 
the correlation length of the order parameter, $\xi$, exhibits a power law divergence, whereas the activation barrier height, 
not the relaxation time, exhibits a power law divergence in proportion to $\xi^{d/2}$ towards the ideal glass transition point $T_0$. 
Here we consider such unconventional critical phenomena with diverging barrier heights \cite{villain1985,fisher1986,fisher1987activated,Sethna,bulbul2004}. 
Below we follow the argument given in ref. \cite{fisher1987activated}. 

In conventional critical phenomena, there are no barriers which diverge with the correlation length and the dynamics is controlled 
by the diffusive motion \cite{Onuki}. The transport coefficient $L$ itself does not exhibit any significant anomaly.  
This is because the characteristic scale of the singular part of the free energy in volume $\xi^d$ is of the order of $k_{\rm B}T$, which yields the 
hyperscaling relation $d \nu=2-\alpha$, where $\alpha$ is the critical exponent for the heat capacity. 
Thus, no activated dependence of times on length scale is expected. The critical dynamics does not involve motion over large barriers. 
This is the consequence of the fact that the competition which causes the critical point is between energy and entropy 
and the contribution of the latter to the free energy is always of $k_{\rm B}T$.  

In certain random systems, on the other hand, the long distance behaviour is controlled by competition 
between two types of competing energy, rather than between energy and entropy. 
There is an ordinary critical divergence for the static correlation length $\xi$: 
\begin{eqnarray}
\xi =\xi_0 \epsilon ^{-\nu}. \label{eq:xi1}
\end{eqnarray}
In these cases, however, it is expected that the system should be controlled by a zero temperature fixed point at which 
the important parts of the free energy grow as a positive power of the length scale: $F \sim \xi^\theta$, 
where $\theta$ is the exponent. Note that in the above we assume $\theta \cong d/2$. 
Thus, it is natural to consider that the free-energy barrier $B$ is scaled as 
$\xi^\theta$ (for simplicity, we do not consider a possibility of a different exponent). 
Since every quantity on the length scale $\xi$ experiences fluctuations due to randomness, 
a distribution of barrier heights likely again scales as $\xi^\theta$. 
This implies an extremely broad distribution of the associated times $\tau_\alpha \sim \exp(B/k_{\rm B}T)$. 
Thus, the natural valuable is not time, but the logarithm of time. 
Thus, we expect that the structural relaxation time scales as 
\begin{eqnarray}
\ln (\tau_\alpha/\tau_\alpha^0) \sim (\xi/\xi_0)^\theta (\Delta_{\rm a}/k_{\rm B}T), \label{eq:exp}
\end{eqnarray}
where $\tau_0$, $\xi_0$, and $\Delta_{\rm a}$ are the appropriate microscopic time, length, and energy scales, respectively. 
By comparing this with eq. (\ref{eq:VFT}), we find 
\begin{eqnarray}
\Delta_{\rm a}/k_{\rm B}T=D.
\end{eqnarray}
This relation implies that the energy scale of the interaction which leads to the order determines the fragility 
of a liquid ($D$). This is quite consistent with our picture (see, e.g., ref. \cite{ShintaniNP}). 
We note that $\Delta_{\rm a}$ provides the natural energy scale of frustration. 
The configurationally averaged dynamic correlation of a variable $A$ for $\ln t$ and $\xi$ then scales as 
\begin{eqnarray}
C_A(r,t,\epsilon) \sim \xi^{-2x_A} \Gamma_A[r/\xi,(\ln t)/\xi^\theta], 
\end{eqnarray}
where $x_A$ is the scaling index for the operator $A$. 
This is the basic picture drawn for random field magnets, which exhibits glassy slow dynamics.

\subsubsection{Relevance of the above picture to structural glass transition}

This scenario of activated dynamic scaling has been expected to be valid for other frustrated systems 
\cite{fisher1987activated} and even for structural glasses \cite{Sethna}. 
In our scenario, the relevant order parameter is glassy structural order, more specifically, 
bond orientational order in many cases. 
The glassy structural order has a link to high solidity, or low fluidity. 
Here we emphasize that high solidity, or low fluidity, means a long lifetime of configuration 
and not high elastic modulus. Note that the distinction between liquid and solid is made just by comparing 
the rheological relaxation time with the observation time.  
This link between the glassy structural order parameter and solidity looks natural since in our model of structural glass transition 
the underlying ordering 
without frustration is crystallization, which accompanies the emergence of static elasticity. 
The lack of translational order is an origin of the absence of the static elasticity in a glassy state. 
This basic link between the structural order and the coherence of motion may be preserved even for the absence of 
long-range order, which is a consequence of frustration effects. 
The above analogy implies that the activation energy for rotational motion of 
solid-like regions of size $\xi$, which is directly linked to high bond orientational order regions, 
is roughly given by $(\xi/\xi_0)^\theta$ times of the activation energy 
for particle-level motion, which is expressed by $\Delta_a$. 
Furthermore, the distribution of barrier heights is expected to be proportional to $(\xi/\xi_0)^\theta$, 
which is consistent with the decrease of the stretching exponent $\beta$ upon cooling. 
These analogies provide a useful and important guide to understand the activated nature of glassy slow dynamics. 

In this scenario the fragility of liquid is linked to the strength of frustration, which is positively 
correlated to $\Delta_a$. In the fragile limit or without frustration, there would be no growing 
activation energy and the transition just accompanies ordinary non-activated dynamics. 
However, we note that the complete absence of frustration to crystallization is unlikely. 
In the strong limit, bond orientational order cannot grow due to strong frustration and thus 
the structural relaxation should obey the Arrhenius-like law. 
Between the two limits, the growing activation energy is given by $\Delta_a (\xi/\xi_0)^\theta$. 
This is basically consistent with our simulation results \cite{ShintaniNP,KAT,TanakaNM,Kawasaki3D,KawasakiJPCM}. 

In structural glasses, however, the order parameter should have a deep link to the flow and viscoelastic properties and 
thus the order parameter should be coupled to the velocity and/or stress fields. These features are absent in random spin systems 
where spins are located on a lattice. 
Thus, the above simplified argument may still not be enough to elucidate the true microscopic origin of the slow dynamics, and further 
studies are certainly necessary. In relation to this, it may be worth mentioning the work by Dattagupta and Turski \cite{dattagupta1993interplay}, 
which studied how the kinetic effects associated with bond orientational ordering in a supercooled liquid
lead to enhancement of the fluid viscosity and the elastic moduli of the system on the mean-field level 
as the glass transition is approached from above.

\subsubsection{Other simulation works relating to our scenario of crystal-like structural ordering}

Here we mention numerical simulation works, which also indicated the presence of growing 
crystal-like structural order in a supercooled state and the resulting dynamic heterogeneity. 
Some time ago Coslovich and Pastore studied the dynamics and structure of supercooled liquid states of 
binary mixtures of Lennard-Jones particles and Wahnstr\"om mixtures 
by numerical simulations \cite{coslovich1,coslovich2}. 
They found locally favoured structures whose number density increases upon cooling. In the Wahnstr\"om mixture, 
they found that icosahedral structures are locally favoured structures and grow upon cooling. 
This is apparently at odds with our scenario and looks consistent with other scenarios 
such as the frustration limited domain theory and spin glass theory. 
However, it was found later \cite{pedersen2007,pedersen2010} 
that the Wahnstr\"om mixture crystallizes into MgZn$_2$ consisting of tetrahedral network of 
Frank-Kasper bonds. Thus, icosahedral structures are identified as 
key ingredients of Frank-Kasper clusters \cite{pedersen2010}. 
Coslovich \cite{Coslovich} showed that the growth of domains formed by interconnected locally preferred structures signals 
the onset of the slow-dynamics regime and allows the rationalization of the different dynamic behaviours of the models. 
He also showed that these growing structures have a link to the symmetry of the crystal structure, which is 
consistent with our scenario. However, this type of liquid behaves differently from the systems we 
described above (see below). 

In 2D polydisperse systems, we found the growth of hexatic order and concluded that this 
is a manifestation towards crystal-like bond orientational ordering, since the underlying crystal has hexatic order \cite{KAT}. 
However, this order can also be interpreted as a 2D version of icosahedral order \cite{Sausset}. 
This was a matter of discussion \cite{KATPRLC1,KATPRLC2}. 
However, it was recently found in 3D polydisperse colloidal systems 
that the key structure is characterized by fcc-type crystal-like bond order parameter \cite{TanakaNM,Kawasaki3D}. 
Although icosahedral structures are formed in hard-sphere liquids, slow dynamics is mainly due to fcc-like bond orientational 
order and not primarily due to icosahedral order \cite{MathieuNM}. 
This supports our scenario that slowing down is more closely linked to extendable bond orientational ordering.  

Finally we mention that a recent study showed that a modified Kob-Andersen binary Lennard-Jones mixture crystallizes 
in lengthy simulations by forming pure fcc crystals of the majority component 
\cite{toxvaerd2009}. This indicates that crystallization is important even in a system 
which has been believed to be an ideal glass former free from crystallization. 
However, we note that in a system where fractionation, or demixing, is prerequisite for crystallization, there may be 
no direct link between glassy structural order and the crystal structure (see below). 

\subsection{Counter examples for our scenario of critical-like scaling and classification 
of glass-forming liquids}

\subsubsection{Counter examples for the scaling relation}
So far we showed that the static and dynamic correlation length coherently grow 
when approaching the glass transition as eq. (\ref{eq:xi1}) and the structural relaxation time slows down 
with the increase in the correlation length as eq. (\ref{eq:exp}).  
For the systems we studied we find $\nu=2/d$ and $\theta=1/\nu=d/2$ within errors. 
However, a recent careful study by Szamel and his coworkers \cite{flenner2011analysis} showed 
that for a 3D binary hard spheres the dynamic correlation length $\xi_4 \cong \xi_{40} \epsilon^{-2}$ (or $\epsilon^{-1}$) and 
$\theta \cong 1$. This result is different from ours, which questions the universality 
of a critical-like scenario. 

In relation to these behaviours, we note that Montanari and Semerjian \cite{Montanari2006} recently showed 
that a static length scale grows with 
relaxation time, but the physical relevance of that length scale may vary with the specific relaxation regime considered 
and/or the type of glassy dynamics. Two recent studies have shown that the two types of length scales (static \& dynamical) can have very different behaviour as the system becomes sluggish. For example, Charbonneau et al. \cite{charbonneau2012} 
showed that the growth of the two lengths can be decoupled, which is in agreement with both a facilitation-based description and the random first order transition (RFOT) theory, but not geometrical frustration. 
A similar strong decoupling was also reported by Hocky et al. \cite{reichmann2012} for three model supercooled liquids 
which have similar static pair correlations. 
On the other hand, Kob et al. \cite{kob2011non} reported that the dynamical length can behave non-monotonically 
as the static length steadily grows, which is reminiscent of a critical-like behaviour and is only consistent with
RFOT theory. 

Here we mention some controversies on the presence or absence of a growing static lengthscale.   
In the case of binary Lennard-Jones (Kob-Andersen) mixtures, no clear growing structural lengthscales have been identified by a point-to-set analysis \cite{reichmann2012}, 
as mentioned above, whereas Mosayebi at al. \cite{mosayebi2010} found a diverging static lengthscale by analysing the response of the inherent structure to static perturbation. Similarly, for binary hard and soft sphere mixtures, no local ordering, nor structural lengthscale has been found to accompany the growing dynamical lengthscale \cite{widmercooper2006,kob2011non,charbonneau2012,dunleavy2012}, but again Mosayebi et al. found the opposite for the latter mixture \cite{mosayebi2012}. 
The presence of a growing static length in Kob-Andersen and repulsive binary mixtures has also been reported by Karmakar and Procaccia \cite{karmakar2012}.  
These results suggesting a growing static order are consistent with our scenario, but further careful studies are necessary to settle this controversial situation.  

We also note that our recent study on the Wahnst\"om mixture also 
indicates the deviation from the scaling behaviour we found for polydisperse hard-sphere liquids 
\cite{malins2012identification}. 
This problem is directly linked to the fundamental question of whether the growth of the static correlation length 
is responsible for slow glassy dynamics or just a subsidiary effect and of whether its answer depends on the type of systems or not. 
This issue is a cornerstone of the debate between geometrical frustration-based descriptions, facilitation-based descriptions, the RFOT theory, and our scenario based on frustration against crystal-like bond orientational ordering. 
Below we consider a possible scenario accounting for this problem.

\subsubsection{Classification of glass-forming liquids}

On the basis of the above-mentioned non-universality of the scaling behaviour, 
here we propose to classify glass-forming liquids into three types on the basis 
of the type and degree of frustration against crystallization. 
This is based on a physical picture that a liquid has a general tendency to form 
local or mesoscopic structures which lowers the free energy locally. 
In this scenario, structural ordering acts to minimize the free energy locally and dominates the regions of slow dynamics.  
In other words, glassy slow dynamics is linked to the formation of local structures of low free energy, 
which can support stress for a long time. 

Tentatively we group glass-forming liquids into three types \cite{malins2012identification}: 
\begin{description}
\item{\bf Type I glass-forming liquids:} 
Quasi-one-component systems (monodisperse or weakly polydisperse systems) in which crystal-like bond orientational order grows upon cooling  (see, e.g., \cite{ShintaniNP,TanakaNM}). 
This type of system has a direct link in symmetry between local free-energy minimum structures in a liquid 
and its crystal structure, 
and thus bond orientational order can be a relevant order parameter to describe glass transition. 
\item{\bf Type II glass-forming liquids:} Systems in which non-space-filling amorphous order develops as a part of (quasi-)long range order upon cooling (see, e.g., \cite{Dzugutov1992,coslovich1,coslovich2,pedersen2010,Coslovich}). 
Type II glass formers have rather large crystal unit cells, and in a liquid state structures having a link to 
only a part of the crystal structure are formed. This type of frustration against crystallization 
is called ``entropic frustration''~\cite{pedersen2010}. 
Entropic frustration may also be relevant for a liquid in a higher dimension, such as 4D liquid, 
for which a barrier for crystal nucleation is higher \cite{van2009geometrical}.  
In the case of the Wahnst\"om mixture, the local free-energy minimum structure has a link to icosahedral 
order. However, the local nature, or non-expandability of this structure, leads to a peculiar relation 
between static order and dynamics. 
In type II liquids, bond orientational order is still a useful order parameter for characterizing 
a local free-energy minimum structure. However, its link to the crystal orientational symmetry may not be so direct. 
The way of the growth of a static lengthscale may also be different from type I liquids.  
On this point, further careful studies are highly desirable. 
\item{\bf Type III glass-forming liquids:} Multi-component systems in which no specific structural order has been detected (e.g. some binary mixtures and highly polydisperse systems), but dynamical correlations exhibit a growing lengthscale  \cite{hurley,kob1997dynamical,yamamoto1998,yamamoto1998dynamics,hiwatari1998structural,Perera1999a,Perera1999b,BiroliN,flenner2011analysis}. 
Type III glass formers involve phase separation or fractionation when they crystallize. 
This leads to decoupling between low-free-energy local structures in a supercooled liquid and crystalline structures. 
So far we have controversial reports: For example, indications of static (translational) structuring in 
a binary Lennard-Jones system were reported in ref. \cite{Phillies2005extended}, whereas 
indications of neither translational nor bond orientational ordering were reported in ref. \cite{ernst1991search}.  
Although relevant low free-energy structures have not been identified for binary mixtures in a clear manner yet, 
we believe that there are static structures responsible for slow dynamics. 
This may be supported by the finding of Widmer-Cooper et al.  \cite{Widmer2004} for this type of mixtures. 
Furthermore,  Mosayebi et al. \cite{mosayebi2012} demonstrated that there is a static growing length for both Kob-Andersen 
and binary soft sphere mixtures, as mentioned above. 
At this moment, however, we cannot rule out a possibility that slow dynamics is purely a consequence of 
dynamical correlation. This is one of the most important problems, which lies at the heart of the nature of 
glass transition and, thus, further careful studies are highly desirable. 
\end{description}

Here we consider a possible unification of the above different types of glass formers. 
The common scenario for all types of glass formers may be as follows. 
Liquid has a tendency to form locally or mesoscopically stable structures, which lower the free energy locally. 
The nature of such structures may depend on the type of glass formers, but it is common that 
these structures have a long lifetime and thus 
low fluidity, reflecting lower local free energy and large barrier. Growth of these long-lived structures in a supercooled 
liquid may be a general origin of glassy slow dynamics.

\subsection{Relation between `mesoscopic' glassy structural order and `microscopic' caging} 

Here we consider the relationship between mesoscopic bond orientational order and microscopic caging in detail, since this 
problem is linked to a fundamental question on the very origin of slow dynamics. 

\begin{figure}
\begin{center}
\includegraphics[width=8.0cm]{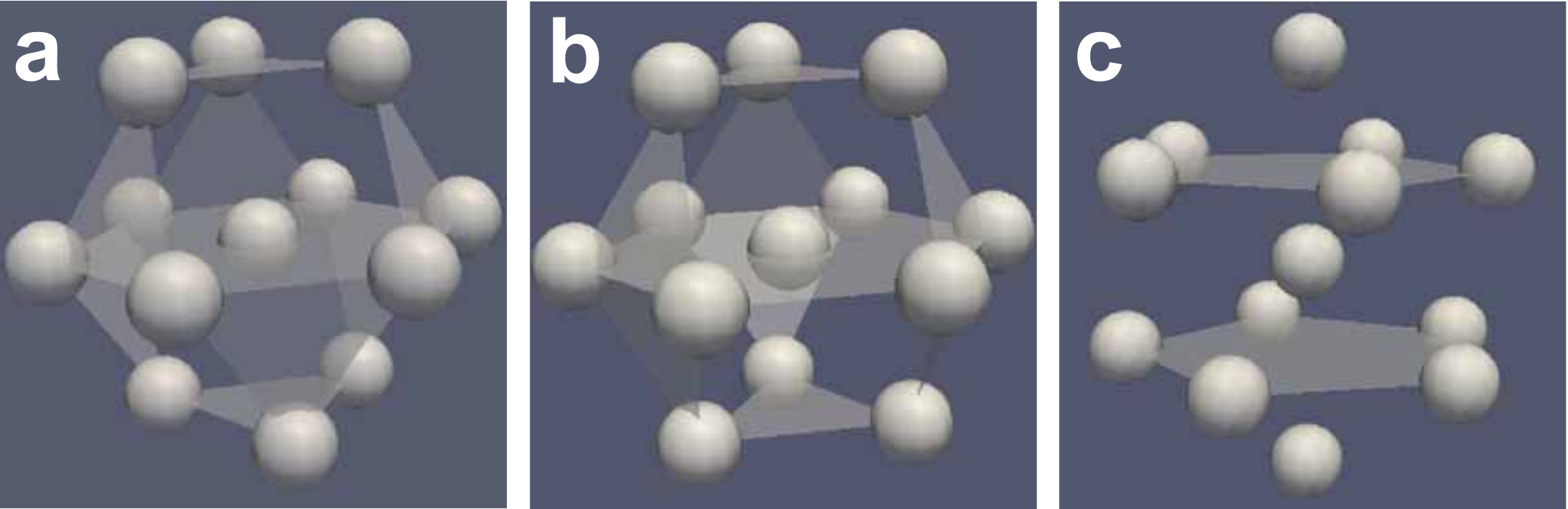}
\end{center}
\caption{(Colour on-line) Schematic figure representing the densely packed structures made of 13 particles, 
which has fcc, hcp, and icosahedral bond orientational order. 
(a) fcc, (b) hcp, and (c) icosahedron. 
} 
\label{fig:fcc}
\end{figure}

First we consider local bond orientational ordering and its link to caging for a case of hard-sphere-like system. 
We emphasize that local bond orientational ordering is a direct consequence of packing effects 
for hard spheres.
The degree of caging is related to the hight of the local radial distribution function $g(r)$, on which 
the mode-coupling theory relies. 
This peak height can be regarded as a manifestation of short-range (bond orientational) ordering. 
In relation to this, we note that bond orientational order is coupled to $\rho(\mbox{\boldmath$k$})\rho(-\mbox{\boldmath$k$})$, 
as shown in eq. (\ref{eq:F_int}). 
Thus, we argue that the caging is closely linked to bond orientational order since 
bond orientational order in hard-sphere-like liquid is a consequence of dense packing. 
The natural number of nearest neighbour particles are 12 for 3D systems if the polydispersity 
is not so large. 
There are three relevant bond order parameters linked to the densest packing 
consisting of a central particle and 12 nearest neighbour particles: fcc, hcp and icosahedral 
structure (see fig. \ref{fig:fcc}). 
It seems that this fact has been overlooked so far, and 
only caging represented by the pair distribution function has been considered. 
The fact that bond orientational order, or more generally, glassy structural 
order, is responsible for particle caging is related to the fundamental feature of caging, 
which is many-body effects and cannot perfectly be captured by the two-body density correlator alone. 
It may be worth noting that as discussed above, bond orientational ordering is 
anti-correlated with the number density of defective structures (voids), which are necessary 
for particles or solid-like regions to move. 
We note that bond orientational order is induced to lower the free energy by gaining correlational entropy for hard spheres. 

For 3D hard spheres, similarly to chalcogenide glasses (see sec. \ref{sec:chalcogenide}), 
we may regard particles surrounded by less than 12 particles 
as a floppy state and those surrounded by more than 13 particles as an over-constrained state. 
When the number of nearest neighbours is 12, the central particle is most efficiently caged.   
For hard spheres of large polydispersity, this optimum number of neighbouring particles depend upon 
the size of a central particle and those of its neighbouring particles. 
Again this fact should be linked to the coupling between $\mbox{\boldmath$Q$}_{\rm cry}$ and $\rho(\mbox{\boldmath$k$})\rho(-\mbox{\boldmath$k$})$. 
This is very difficult to quantize, but may still be correlated with some type of correlational entropy. 
Thus, glassy structural order (e.g., bond orientational order) is linked to local solidity, 
but caging is `not'.  

Next we consider local bond orientational ordering for a case of directional energetic bonding. 
In liquids like silica, silicon, and water, atoms or molecules tend to form tetrahedral order by directional 
covalent or hydrogen bonding. For these cases, local (tetrahedral) bond orientational order becomes a good measure 
of energetic caging. 
For energetic caging, there is also a particular bond orientational order, which lowers the local free energy 
by forming directional bonding. There is an optimal condition to form locally stable bonded structures: 
For chalcogenide glasses which are mixtures of a few elements, this condition may be realized in the so-called 
intermediate phase (see sec. \ref{sec:chalcogenide}). The role of glassy structural order should be 
exactly the same as in the above case of hard spheres. 

Now we consider which of glassy structural order parameter or caging is more relevant to glassy slow dynamics. 
Our bond order parameter analysis shows its extended or coarse-grained nature is essential for 
a direct correlation between static order and dynamic heterogeneity \cite{KAT,WT,TanakaNM}, as mentioned above. 
Berthier and Jack also suggest that a connection between the static and dynamic properties of glass formers 
at the particle level is not so clear, but such a connection does exist on larger length scales. 
This is further confirmed by our recent experiments on colloidal glasses \cite{MathieuNM} as well as a recent work 
by Candelier \cite{candelier}.  For example, icosahedral order in hard spheres cannot contribute to glassy slow dynamics 
in a direct manner because of its isolated localized nature \cite{MathieuNM}. 
We argue that this conclusion is a natural consequence of `mesoscopic' caging due to bond orientational ordering. 
A single cage is also represented by local bond orientational order, however, the environment of the cage 
plays a crucial role in determining the lifetime of this particular cage. 
We emphasize that local rotation of bonds can easily lead to the loss of memory, or the decay of the overlap function $w$. 
Thus, we may say that isolated local clusters alone cannot cause slow dynamics. 
The cage concept seems to miss the importance of such rotational motion in stress (or structural) relaxation 
(see sec. \ref{sec:rotation}). 
If the bond orientational order is high around the cage, this means that there are few defects or voids 
and the cage will live for a long time because of constraint to both translational and rotational motion. 
We stress that bond orientational order represents a constraint on bond orientations and thus on bond rotations, which 
are key to structural relaxation. 
It is medium-range correlation that leads to the coherency in the particle motion (both translational and rotational motion) there, or the dynamic heterogeneity. 
Our study shows that the delocalized nature of cages, or medium-range structural correlation, is 
essential for glassy structural order, which is responsible for slow dynamics and local solidity. 
The pair structural entropy $s_2$ is calculated by a two-body correlation, but it sees not only the nearest neighbour particles independently but also their correlations 
as well as the second and third nearest neighbours. However, there should be a much better general 
indicator for low free-energy configurations, which reflects many-body correlation effects.  

Thus, we may conclude that glassy structural ordering  
has an apparent connection to the other concepts such as caging and cooperative rearranging region, but at the same time 
there is the above mentioned essential difference. 
For example, glassy structural order is linked to the coherency of particle motion as a consequence of low fluidity, whereas 
a cooperative rearranging region puts focus on cooperative motion required for the motion of a single particle.  

Finally, we consider a type of local particle motion associated with escape of a particle from its cage. 
Cage escaping motion is often expressed by hopping motion.  
However, we speculate that local circulative motion of particles (or string-like motion) provides a much easier way for a particle 
to escape from its cage (see also below) since this type of transverse (or rotational) motion involves excitation of little free volume.

\subsubsection{Relation between glassy structural order and inherent structures}

About 30 years ago Stillinger and Weber \cite{stillinger1982hidden} proposed a very interesting idea that 
hidden structures in liquid may be revealed by classifying particle configurations according to multi-dimensional 
potential energy minima that can be reached by steepest-descent paths (``quenches''). 
In a two-dimensional Gaussian core model, they found that a remarkable degree of polycrystalline order is hidden 
in a liquid by vibrational distortion. This method allows us to separate structures into a vibrational part and 
an inherent structural part. 
Since this seminal work, the structure of the system at its local potential energy minimum is called ``inherent structure''. 
This idea has also been used to identify important structures hidden in liquids. 
For example, in a Lennard-Jones liquid, icosahedral structures were assigned to be inherent structures \cite{andersen1988icosahedral}. 
The application of this idea to glass transition was also suggested, 
and has contributed to its deeper understanding \cite{stillinger1995topographic}. 
The slowing down of the dynamics has been connected to the presence of basins in the configuration space, or the potential energy landscape 
(see, e.g., refs. \cite{sastry1998signatures,sciortino1999inherent,sciortino2005potential}). 
The short time dynamics (fast $\beta$ mode) was related 
to the process of exploring a finite region of phase space around a local potential energy minimum, whereas  
the long time dynamics was connected to the transition between different local potential energy minima. In 
this picture, upon cooling the intrabasin motion becomes more and more separated in time from the slow (and 
strongly $T$ dependent) interbasin motion. The decrease of the entropy of supercooled liquids on cooling was associated with the 
progressive ordering of the system in the configuration space, i.e., in the progressive population of 
basins with deeper energy but of lower degeneracy. 

So it is natural to expect that inherent structures have a close relation to our glassy structural order. 
For example, the similarity between our results for static growing length and those obtained by Mosayebi et al. 
\cite{mosayebi2012} suggests such a relation. 
There is certainly a connection, but here we mention a few important difference. 
First inherent structures are linked to local potential energy minimum, but not local free-energy minimum, 
since it is obtained by quenching to the zero temperature and throw away the entropic contribution. 
For example, in hard spheres structural order is a consequence of maximizing the entropy. 
The inherent structures in the hard-sphere limit was studied by Stillinger and Weber \cite{stillinger1985inherent} 
and suggested to be randomly packed configurations. This is very different from our conclusion that 
the key structural motif in a supercooled hard sphere liquid is fcc\&hcp-like bond orientational order. 
We speculate that inherent structures for a polydisperse hard sphere liquid are random packed structures 
only at low volume fractions above the freezing volume fraction whereas structures with crystal-like (fcc+hcp) 
bond orientational order at high volume fractions. This point needs to be checked. 
Another important difference is that quenching in searching the inherent structures may lead to local translational ordering, 
which is absent in a liquid state with thermal fluctuations.

\subsubsection{What is the relevant order parameter describing glass transition in general}

In relation to the above, we consider what is the relevant order parameter which can pick up a structural signature 
of glass transition. 
For systems without random disorder, the order parameter governing glass transition 
is the same as a part of order parameters governing crystallization, as described above. 
That is, it is the bond orientational order parameter, which has a link to the 
local rotational symmetry of the equilibrium crystal.  

However, the order parameter for a system suffering from quenched disorder, such 
as bidisperse colloids and atactic polymers, is less obvious. 
A difficulty arises from the fact that we do not have a mathematical 
means to extract such `configurational structural order' 
besides bond orientational order. 
Nevertheless, `configurational structural order' 
can be identified as a state of low local free energy, more specifically, a state of less voids or less configurational entropy for 
hard spheres. In such a state, the correlation volume, or free volume as a source of correlational 
or vibrational entropy is locally homogeneously shared among surrounding particles. 
Bond orientational order satisfies this condition. 

Glass transition can in principle take place in any systems if we can avoid crystallization. The glassy structural order for 
particles with irregular shapes is, for example, far from obvious. In our view, however, it must still be correlated with 
configurations of low local free energy. In relation to this, we mention that to seek an unknown amorphous order (including bond orientational order) 
and its correlation length, a cavity method or an estimation of `overlap' is believed to be a powerful 
means with applicability to various types of unknown structures \cite{CavagnaR,BiroliN,kob2011non,Cavagna2012}. 
The so-called point-to-set length is estimated by freezing the position of a set of particles in an equilibrium configuration and performing 
sampling in the presence of this additional constraint. 
Although this is certainly an attractive method, the applicability of such point-to-set correlations for an off-lattice liquid system does not look so obvious 
and may need to be carefully checked.

\subsubsection{A link between glassy structure ordering and slow glassy dynamics: observation}  \label{sec:lengthdynamics}

The above physical picture leads us to the following scenario of slow glassy dynamics. 
Upon cooling, liquid enters into a metastable state where long-range orientational and positional ordering is prohibited 
by frustration effects. However, bond orientational order still 
develops towards the ideal glass transition point $T_0$ to lower the free energy of the system. 
Bond orientational ordering is a manifestation of many-body correlations 
among strongly correlated neighbouring particles around a particle. For a strongly disordered system, such unique bond orientational 
order linked to the symmetry of the equilibrium crystal no longer exists, however, we expect that strong correlations may still be represented by 
some structural signature, which has a link to configurations of long lifetime, namely, low fluidity. 

\begin{figure}[h]
\begin{center}
\includegraphics[width=8cm]{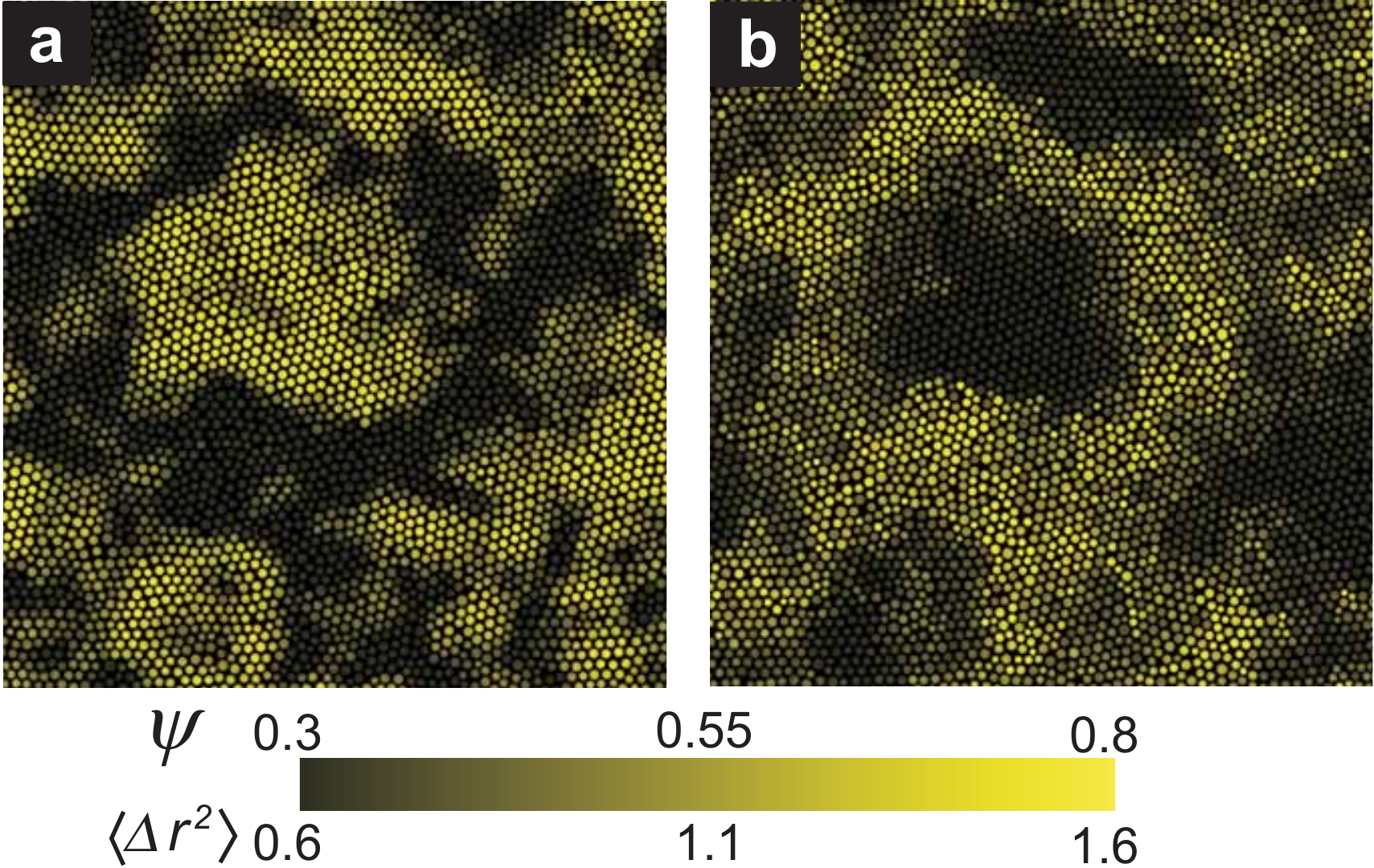}
\end{center}
\caption{(Colour on-line) Relationship between glassy structural order and local mobility in 2DPC ($\phi=0.740$ and the polydispersity $\Delta=9$ \%). 
(a) The spatial distribution of the coarse-grained hexatic order parameter $\psi$. (b) The spatial distribution of the mean-square displacement 
over 10$\tau_\alpha$. We can see the almost one-to-one correspondence between highly ordered regions and regions of low mobility (or low fluidity). 
This figure is reproduced, using a part of fig. 1 of ref. \cite{TanakaNM}. }
\label{fig:compare}
\end{figure}

\begin{figure}[h]
\includegraphics[width=8.5cm]{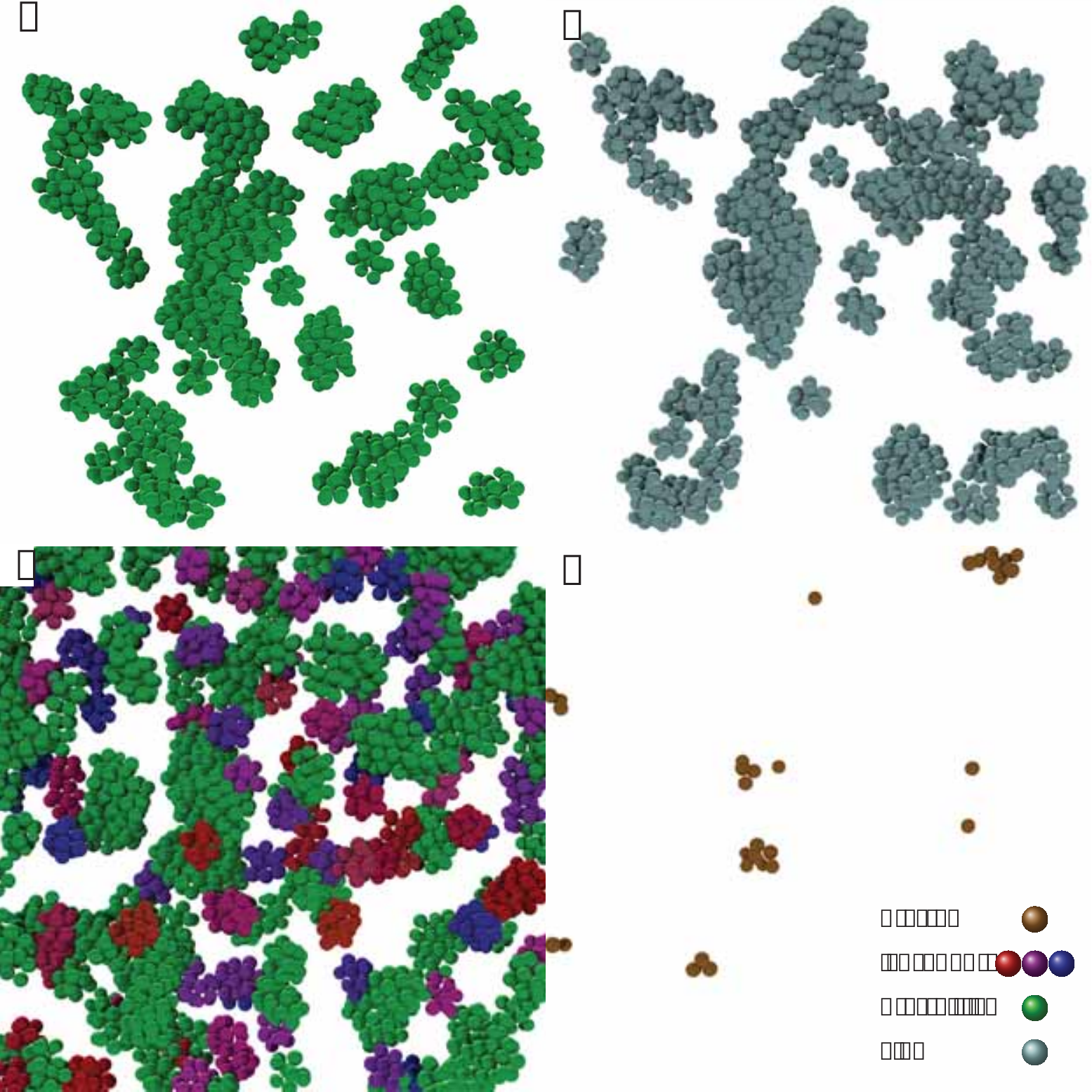}
\caption{(Colour on-line) Computer reconstruction from confocal microscopy coordinates for a polydisperse colloidal suspension ($\phi=0.575$). Only particles of interest and their neighbours are displayed. The depth of the image of 12 times of diameters. Each particle is plotted with its real radius. (a) Particles having high crystal-like bond orientational order alone (the order parameter was averaged over the order of the structural relaxation time). (b) Slow particles with respect to the coarse-grained displacement. Due to particles going in and out of the field of view, assignment of particles located very near the edges of (a) and (b) were not accurate and have been removed. (c) A typical configuration of bond ordered particles. Icosahedral particles are shown in the same colour if they belong to the same cluster. If a particle is neighbouring both crystal-like and icosahedral structures, it is displayed as icosahedral. 
(d) Particles with more than $7$ crystalline bonds. Note that these crystal nuclei (still smaller than the critical nucleus size) are located only in regions of high crystal-like bond orientational order. This figure is reproduced from fig. 4 of ref. \cite{MathieuNM}.}
\label{fig:compare3D}
\end{figure}

Hereafter, we consider a case in which bond orientational order is relevant, just for simplicity. 
In regions of high bond orientational order, particle motion is on average slow since only the coherent motion while keeping bond orientational order 
is allowed. 
A distinct correlation between glassy structural order and slowness of particle motion can be clearly seen for a 2D polydisperse colloidal simulation in fig. \ref{fig:compare}. 
We can see a similar structure-dynamics correlation for a 3D polydisperse colloid experiment in fig. \ref{fig:compare3D}.  
As mentioned above, the length scale of the structural order, or the coherency of particle motion, is a key to the slowness of dynamics. 
This may also be the origin of dynamic heterogeneity. 
Here it may be worth noting that a possible difference in the dynamic and static correlation length. In fig. \ref{fig:compare}, we can see 
almost the one-to-one correspondence between static order and mobility. 
However, this visual comparison is affected by color codes we employ. 
There is no proportionality between the static glassy structural order and the local dynamics, which is clear from a strongly 
nonlinear relation between them (see eq. (\ref{eq:exp})). Thus, the bare correlation length can be different between the static and dynamic ones.

The average bond orientational order is anti-correlated with the defect density and controls 
the average structural relaxation, or the degree of fluidity. Its extrapolated value at $T_0$ may depend 
upon the degree of random disorder or frustration in a system. 
Its critical-like spatio-temporal fluctuations control the dynamical heterogeneity of 
a supercooled liquid. As discussed above, they determine the spatial scale 
of dynamical heterogeneity measured by the four-point density correlator, 
but their lifetime may be decoupled from the lifetime of the dynamic heterogeneity. 

The increase in glassy structural order, or the decrease in defective structures, leads to 
the decrease in the fluidity and thus to the slowing down of structural relaxation. 
Such a direct link between glassy structural ordering and slow structural relaxation dynamics 
is a characteristic nature of glass transition and 
is absent in ordinary critical phenomena (see sec. \ref{sec:activation}). In the latter, the slowing down of the 
dynamics is linked to the characteristic size of the order parameter fluctuations alone, 
but not directly to the transport coefficient such as viscosity. 
For example, in a critical binary mixture, viscosity exhibits a very weak logarithmic 
divergence towards a critical point \cite{Onuki}. 
Figure \ref{fig:defect} shows defective structures or voids in two-dimensional 
polydisperse colloids (2DPC). As can be seen in fig. \ref{fig:defect}(a) and (b), voids or defective structures 
are located in regions of low bond orientational order, as expected. 
The number density of voids decreases with an increase in the volume fraction $\phi$ 
and tends to become zero at the ideal glass transition volume fraction $\phi_0$. 
This implies that the ideal glass is a state of no voids, or a state of high 
bond orientational order with random distortion. This decrease of voids as a consequence 
of bond orientational ordering may be responsible for slowing down of dynamics 
and for glass transition that accompanies the emergence of `quasi-static' elasticity. 
The remaining question here is thus which of the average and the spatial heterogeneity is more essential 
for glassy slow dynamics. 

To address this issue, we consider what lengths control the structural relaxation dynamics. 
There are two candidates: (i) a microscopic cage size and (ii) a mesoscopic length scale associated with dynamic heterogeneity. 
Such a mesoscopic length may be of static or dynamic origin. 
Scenario (i) is based on the physical picture that slow dynamics is due to the local caging of particles 
without any relevant length scale beyond the interparticle distance $a$. This is the schematic MCT scenario of glassy slow dynamics. 
Other models based on a single particle picture such as hopping dynamics, energy landscape, and free volume concept belong to this 
category \cite{berthier2004}.   
On the other hand, scenario (ii) is based on the physical picture that slow dynamics is a consequence of 
cooperative phenomena where single particle dynamics is coherent over the length scale $\xi$ larger than $a$. 
For example, Berthier et al. \cite{berthier2004,berthier2005} showed that the length scale marking a 
crossover from persistent to Fickian diffusion motion has a significant connection to slow dynamics, i.e., 
dynamic heterogeneity is a central aspect of the dynamics of supercooled liquids in that time and length scales are 
intimately connected. This is consistent with our scenario. 
A recent study by Furukawa and Tanaka \cite{furukawa2009,furukawa2011,furukawa2012} has also clearly shown that the viscous dissipation in a supercooled liquid takes place predominantly 
in the length scale over $\xi$ (see also \cite{kim2005,puscasu2010}), which indicates the intrinsic importance of the growing 
mesoscopic correlation length. Although it remains an unsolved problem whether it has a static or a kinetic origin, 
this study clearly indicates that the mean-field (or microscopic) mechanism may not be relevant, but the mesoscopic spatial 
correlation is essential to glassy slow dynamics.  

The next question is then whether the mesoscopic spatial correlation is of purely kinetic or static origin. 
One possible scenario is based on the kinetically constrained model (see, e.g., \cite{berthier2005}), which does not involve 
any static correlation but still exhibits strong dynamic correlation. 
Another scenario is based on the presence of static spatial correlation 
\cite{Widmer2004,Widmer-Cooper2005,KAT,WT,TanakaNM,Kawasaki3D,KawasakiJPCM,Sausset,widmer2008,widmer2009localized,mosayebi2010,Coslovich}. 
The low fluidity of our glassy structural order indicates that slow dynamics may be due to the growing static correlation 
over $\xi$, which can explain the above-mentioned crossover from persistent to diffusional motion quite naturally. 
The discussion in sec. \ref{sec:activation} also supports this conclusion. 

Recently Karmakar et al. \cite{karmakar2009growing} showed in their study of a binary mixture of Lennard-Jones particles 
that the variation of the dynamic susceptibility and the structural relaxation time with respect to the system size 
have an opposite sign, contrary to the expectations of finite-size scaling. 
This indicates that the dynamic susceptibility $\chi_4$ does not contain all of the information about the collective relaxation 
process in the liquid necessary to establish the relaxation time. The authors found that relaxation times are instead determined by configurational entropy. 
On noting that our glassy structural order is linked to structural entropy, which further has a direct connection to configurational entropy, 
this finding may be consistent with our scenario. 
In relation to this, we note that static order (bond orientational order) is not `physically' equivalent to dynamic heterogeneity, although 
they are closely related to each other. 
For example, their lifetimes can be very different (see sec. \ref{sec:decoupling}). 

\subsection{Strongly correlated liquid: a possible mechanism behind a link between glassy structural order and slow dynamics} \label{sec:rotation}

Now we come to a central question of what is responsible for glassy slow dynamics. 
As we discussed above, we have evidence for the importance of crystal-like extendable bond orientational order 
in slow dynamics for quasi-one-component type I liquids. 
Historically bond orientational order had attracted a considerable attention since the seminal work of Frank \cite{Frank}. 
In a series of pioneering papers Nelson and coworkers \cite{Steinhardt,nelson1984symmetry,sachdev1985order} 
proposed that the characteristic of supercooled liquids is the growth of icosahedral bond orientational order but 
geometrical frustration prevents the occurrence of infinite-range icosahedral order and leads to a finite number of defects in the ground state. 
The similar conclusion was independently derived by Sadoc and coworkers \cite{sadoc1999}. 
Slow viscous relaxation was also explained on the basis of this idea \cite{sachdev1986viscous}. 
We have shown \cite{ShintaniNP,KAT,WT,TanakaNM,Kawasaki3D,MathieuNM} that extendable crystal-like bond orientational order plays a much more essential role 
in causing slow dynamics than non-extendable pentagonal or icosahedral order does. 
This indicates that a growing static correlation length is responsible for super-Arrhenius slowing down of the dynamics. 
However, we note that this order parameter is not relevant to slow dynamics in binary mixtures, which has been shown by many researchers 
(see, e.g., \cite{mountain1987molecular,ernst1991search}). 

One possible mechanism for slow dynamics is a modified MCT scenario which takes into account a novel coupling between two-body density correlator 
and bond orientational coupling (see eq. (\ref{eq:F_int})). 
This may lead to a situation that memory effects are more significant in regions of higher bond orientational order. 
We speculate that this may result in dynamic heterogeneity whose correlation length is determined by 
that of bond orientational order. We may say that density correlation is slaved by bond orientational order. 
However, as discussed below, this mechanism itself may still miss the important constraint on particle motion originating from 
many-body correlations. We note that many-body correlations are included in bond orientational order, but real dynamics is still controlled 
in the level of the two-body density correlator in this scheme. 

\begin{figure}
\begin{center}
\includegraphics[width=7cm]{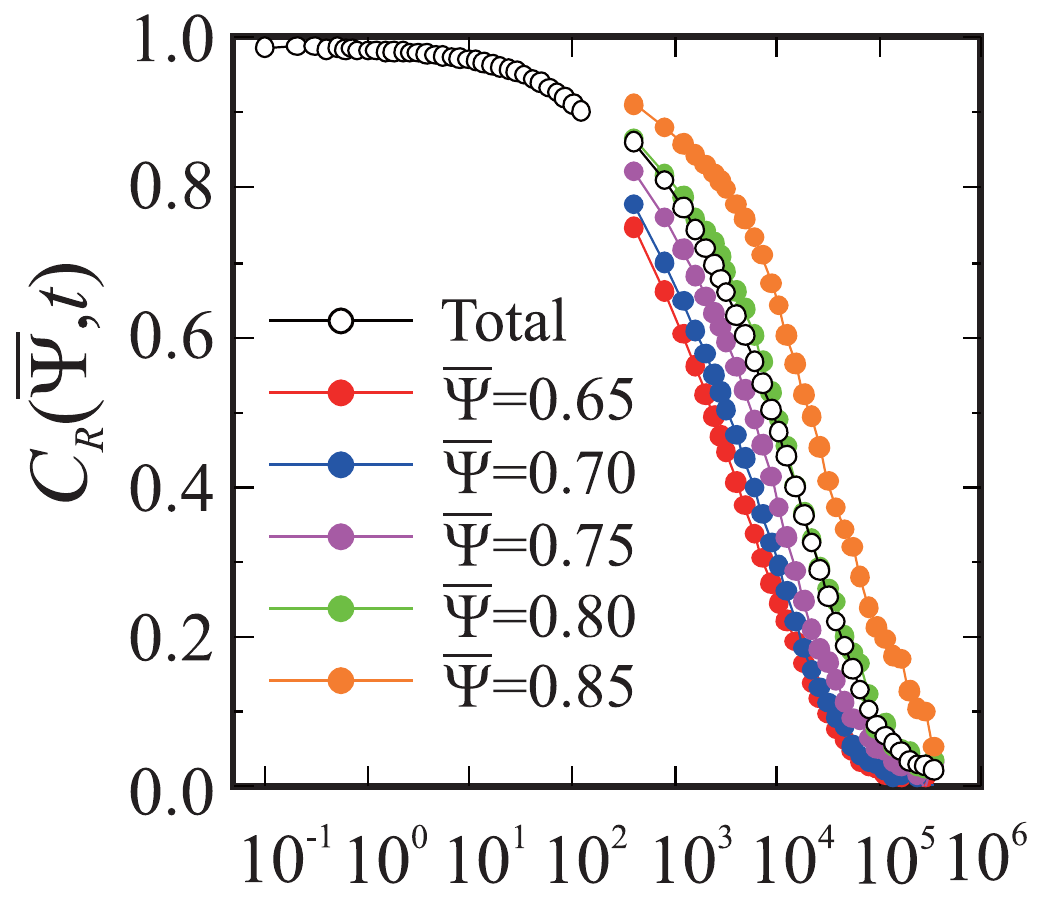}
\end{center}
\caption{(Colour on-line) Rotational correlation function $<C_R(\bar{\Psi},t)>$ for $T=0.17$ and $\Delta =0.6$. 
Here $\bar{\Psi}$ is the degree of antiferromagnetic order. 
Note that the dynamic heterogeneity smears out for large $t$. 
Here we can see particles having high orientational order decays more slowly, which is the origin of the 
stretched exponential decay and is responsible for dynamic heterogeneity. 
This figure is reproduced from fig. 5a in \cite{ShintaniNP}.}
\label{fig:spinrotation}
\end{figure}

We have a feeling that the above type of microscopic description based on two-body density correlation 
may not be able to explain glassy slow dynamics. 
For example, if we see a movie of particle motion in 2DSL \cite{ShintaniNP}, 
particle rotation very rarely takes place inside red regions with high crystal-like antiferromagnetic bond orientaional order (see fig. 3). 
However, red regions themselves translationally migrate slowly with time. 
We confirmed that particles having higher bond orientational order relax more slowly (see fig. \ref{fig:spinrotation}) \cite{ShintaniNP}. 
That is, the local structural relaxation is an monotonically increasing function of the degree of 
antiferromagnetic bond orientaional order, or many-body correlations.  
This implies that the structural relaxation reflects motional coherency in ordered regions (see below).  
 
Such many-body correlations and the resulting slow dynamics can phenomenologically be expressed 
by a scaling argument described in sec. \ref{sec:activation}, on noting 
a link between more extended order in frustrated systems and a higher activation energy for local structural relaxation. 
Here we seek a more microscopic mechanism.  
In a series of interesting papers, Mountain and Thirumalai \cite{thirumalai1987relaxation,mountain1987molecular,mountain1990dynamical} 
sought a local order parameter of glass transition. They suggested \cite{mountain1987molecular} that the orientation of near-neighbour bonds might be a useful means 
of characterizing local order and a function $\psi(t)$ which describes a possibility of the bond rotation provides a time scale for the loss of local 
order. They also showed that neither bond orientational order nor its time correlation function is a useful indicator of a binary glass. 
They also considered its link to the stress relaxation \cite{mountain1990dynamical}. 
The stress tensor is a second rank tensor, characterized by three principal axes and their orientation. 
Thus, the local stress tensor, associated with a finite number of particles, relaxes either because of the magnitudes of 
the principle axes change or because their orientations change. 

Here we connect the above notion and our observation. 
Bond orientational order is directly linked to bond orientation as its name stands. 
The rotation of bonds is not blocked by localized bond orientational order such as icosahedral order. 
However, it is transiently blocked if there is extended bond orientational order. 
Note that bond orientational order is the order stabilizing bond orientation. 
This also has an interesting connection to the idea described in sec. \ref{sec:Speculation}, 
where the degree of triangular tiling is linked to the slow dynamics. This concept of space tiling may be relevant 
not only to type I liquids, but also to type II and III liquids. 
This link between bond order and slow dynamics can be considered to be a consequence of the fact 
that orientational correlation is directly linked to higher order particle distribution 
functions beyond the two-body correlation, which may be the essential feature of glassy slow dynamics. 
In other words, the two-point density correlation function may not be enough or even not relevant to 
cooperative structural rearrangements, which are key to slow dynamics. 
The relevant motion for the structural relaxation may be more of rotational character rather than of translational one. 
The decay of the two-point density correlator might be a consequence of rotational relaxation. 
On noting that the viscosity is related to stress-stress correlation, this picture seems physically natural for 
the mechanism of slow stress relaxation and structural relaxation.

Finally we consider the link between static glassy order and slow dynamics on the basis of the above picture. 
At high temperatures, bonds can rotate almost independently and their dynamics is described by the Arrhenius 
law whose activation energy $\Delta_a$ is determined by the strength of bonds that constrain bond rotation. 
With a decrease in temperature, bond orientational order, for example, locally better triangular tiling 
and tetrahedral tiling respectively for 2D and 3D hard spheres, develops its spatial correlation. 
This spatial correlation may have a long lifetime, which can be longer than the structural relaxation time. 
This spatial correlation over $\xi$ puts a constraint on the rotational motion of bonds, which leads to slow local structural relaxation time. We note that 
in frustrated systems more extended higher order regions suffer from stronger frustration effects 
and require a higher activation energy for reconfiguration. 
Thus, the distribution of bond orientational order parameter leads to the broad distribution of the relaxation time 
(see fig. \ref{fig:spinrotation}) \cite{ShintaniNP}: Particles with higher antiferromagnetic order $\bar{\Psi}$ have longer 
rotational correlation time. This may be an origin of strong structure-dynamics correlation (see fig. \ref{fig:compare}). 
The major stress (or structural) relaxation may be associated with local bond rotation under the constraint from spatial 
coherence. 
The activation energy of such local structural relaxation may be determined by the coherence length $\xi$: Under frustration, 
it may scale as $\Delta_a(\xi/\xi_0)^\theta$ (see sec. \ref{sec:activation}). 
For example, this scenario can also explain the fact that long-lived bond orientational order fluctuations can block 
the translational motion of a tagged particle, but at the same time allows much faster structural relaxation. 
We note that the structural relaxation time is an average of the local relaxation times (see fig. \ref{fig:spinrotation}), which results in a very broad distribution. 
We believe that long-lived low local free-energy configurations such as configurations of high bond orientation order leads to 
slow structural relaxation due to spatial coherence of structural relaxation, which forms the slow relaxation part of the relaxation spectrum. 
In this sense, a supercooled liquid with glassy slow dynamics may be called ``strongly correlated liquid''. 

We also note that regions of high bond orientational order can translationally 
migrate coherently, which leads to the decay of the overlap function, or the 
four-point density correlation. This provides an intuitive explanation for a translational-rotational decoupling. 
Since this argument is speculative, however, further detailed 
studies are necessary but this provides a plausible explanation on the very origin of slow relaxation near the glass transition.  
A clearer physical picture for the violation of the Stokes-Einstein law has recently been 
provided by Furukawa and Tanaka \cite{furukawa2012}.

\subsection{Relation between MRCO, softness of structures, and the excess vibrational density of states}

The physical properties of a topologically disordered amorphous material (glass), such as heat capacity and thermal conductivity, 
are known to be markedly different from those of its ordered crystalline counterpart. 
The understanding of these phenomena is a notoriously complex problem. One of the universal features of disordered glasses is the so called `boson peak', 
which is observed in neutron and Raman scattering experiments. The boson peak is typically ascribed to the excess density of vibrational states. 
Recently we discovered evidence suggestive of the equality of the boson peak frequency to the Ioffe-Regel limit for `transverse' phonons, above which transverse phonons no longer propagate \cite{STNM}. 

In this study, we argued that boson peak is associated with (quasi-)local transverse vibrational modes, whose frequency 
is located around the boson peak frequency. We confirmed the absence of scattering of longitudinal and transverse phonons 
due to inhomogeneities of elasticity up to the Ioffe-Regel frequency. This is consistent with the absence of 
the wavenumber dependence of the shear elastic modulus in a mesoscopic and macroscopic lengthscale revealed by Furukawa and Tanaka 
\cite{furukawa2009,furukawa2011,furukawa2012}.   

\begin{figure}
\begin{center}
\includegraphics[width=8cm]{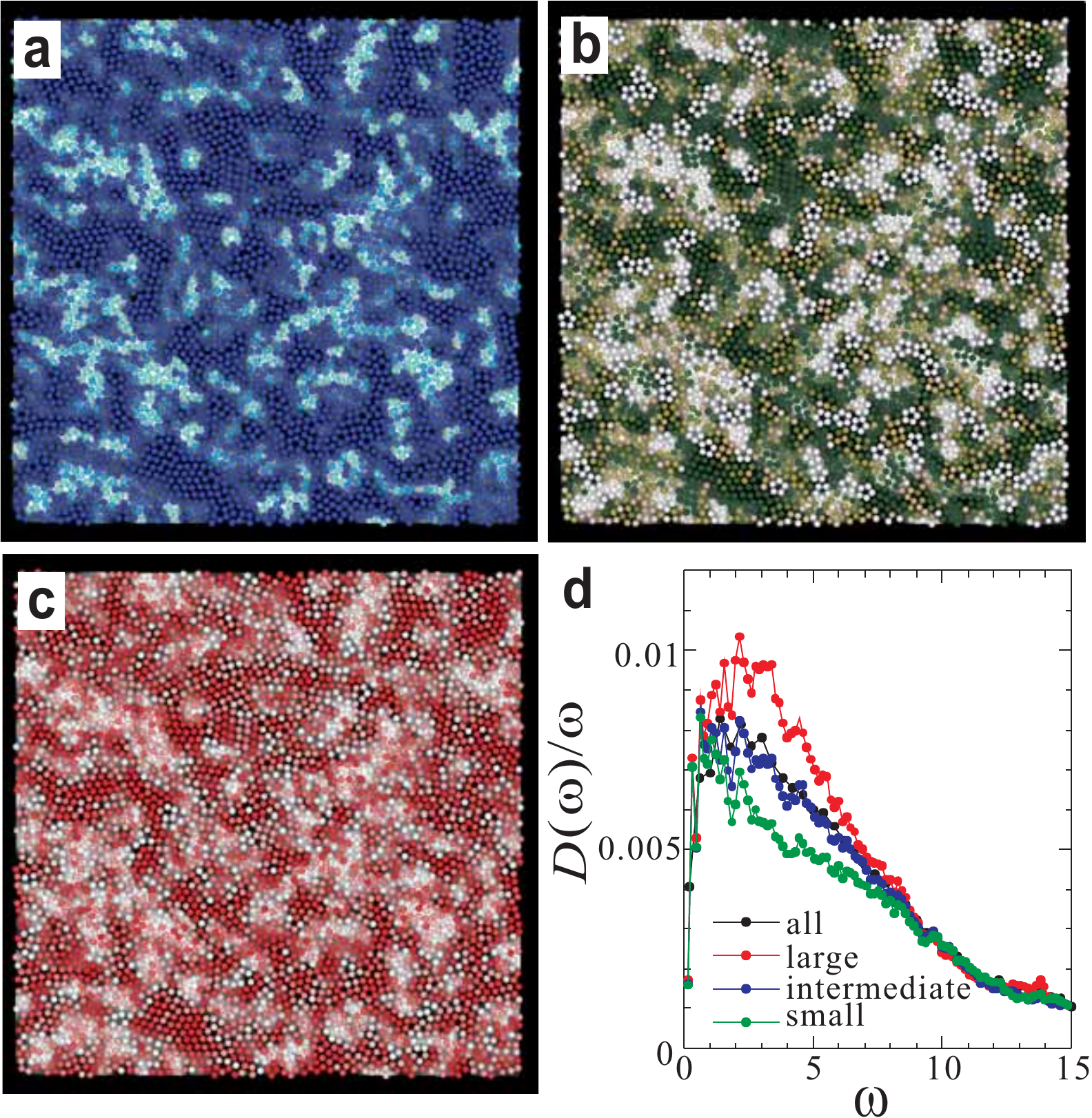}
\end{center}
\caption{(Colour on-line) Structural origin of the boson peak. 
(a) Spatial distribution of $D_i(\omega_{\rm BP})$, which is the amplitude of 
the vibration modes with frequencies around the boson peak frequency. 
Particles with less boson peak intensity are more blue. 
The background is coloured the same way (darker for less boson peak 
intensity), and this same background is repeated in (b) 
and (c), to see the correlation of 
$D_i(\omega_{\rm BP})$ with local order and local volume, 
respectively. 
Note that we equilibrated the liquid of $\Delta$=0.6 at $T=$0.18 
and $P$=0.5 and quenched it to $T$=0.02 to prepare this sample. 
Thus, MRCO is more developed and the number density of LFSs is higher than in 
samples prepared by a rapid quench.
(b) Spatial distribution of the order parameter $\bar{\Psi}_i({\bf r})$ 
(see \cite{ShintaniNP} for its definition). 
Dark green, green, and white particles represent MRCO, 
normal liquids, and LFSs, respectively. 
(c) Spatial distribution of local volume (the area of Voronoi polygon 
per particle). 
Brighter particles have larger local volume (less local density). 
(d) $D(\omega)/\omega$ per particle 
averaged over the one-third of particles having 
large local volume (``large''), small local volume (``small''), 
intermediate local volume (``intermediate''), and all the particles.  
This figure is reproduced from ref. \cite{STNM}. 
} 
\label{boson}
\end{figure}

Since the very origin of the boson peak is out of scope of this article, 
here we discuss the relationship between the degree of bond orientational order and the vibrational degrees of freedom of structures, 
which is related to the structural origin of the boson peak. 
Here we consider this problem, using our study on 2DSL \cite{STNM}. 
In our 2DSL, there are three structural candidates 
which may give rise to quasi-localized vibrational modes 
with the characteristic frequency $\sim \omega_{\rm BP}$ and 
couple with transverse phonons: 
medium-range crystalline (antiferromagnetic) order (MRCO) 
(dark green particles in fig. \ref{boson}(b)), 
locally favoured structures (white particles forming pentagons in fig. \ref{boson}(b)), 
and low-density defective structures (white particles in fig. \ref{boson}(c)) 
in the normal liquid structure. 
In figs. \ref{boson}(a)-(c), the background (outside a particle) colour is brighter 
for particles with high excess vibrational density of states, $D_i(\omega_{\rm BP})$.  

Comparison of the colour between particles and their background in fig. \ref{boson}(b) 
tells us that the vibrational amplitude 
is small in regions of high MRCO, which indicates that particles belonging to  
MRCO are not responsible for the boson peak. 
This is natural in the sense that the excess vibrational state should be linked to structural disorder. 
The vibrational density of states excess over those of the crystal should be smaller for more crystal-like structures. 
On the other hand, locally favoured structures (pentagons) 
at least partly contribute to the boson peak since white particles in fig. \ref{boson}(b) 
often have large $D_i(\omega_{\rm BP})$. 
We see the most distinct (positive) correlation between the local boson peak 
intensity and the local (free) volume, or the low-density defective structures (see fig. \ref{boson}(c)). 
This relation can also be clearly seen in fig. \ref{boson}(d), where 
the boson peak intensity for particles with large, intermediate, 
and small free volume is shown. Particles with larger free volume 
mainly contribute to the boson peak over the Debye value. 
These defective structures are linked to floppy modes because of their 
softness against shear deformation: Structural disorder allows particles to have 
rather isolated `transverse' vibrational modes since the number of constraints may be smaller than 
the number of the degrees of freedom for the defective structures. 
Thus, we argue that transverse vibrational modes associated with 
low-density defective structures are responsible for the excess vibrational density of states, or the boson peak. 
This conclusion was recently supported by experiments in a two-dimensional colloidal system \cite{tan2012understanding} 
and also by simulations of a realistic metallic glass \cite{jakse2012boson}. 
Our results also suggest that these (quasi-)localized transverse vibrational modes are responsible for sound absorption, 
or dissipation of transverse phonons, over a wide frequency range below $\omega_{\rm BP}$. The mechanism behind this needs to be 
clarified in the future.

\subsection{Confinement effects on the structure and dynamics of a supercooled liquid} \label{sec:wall}

\begin{figure}
\begin{center}
\includegraphics[width=8.5cm]{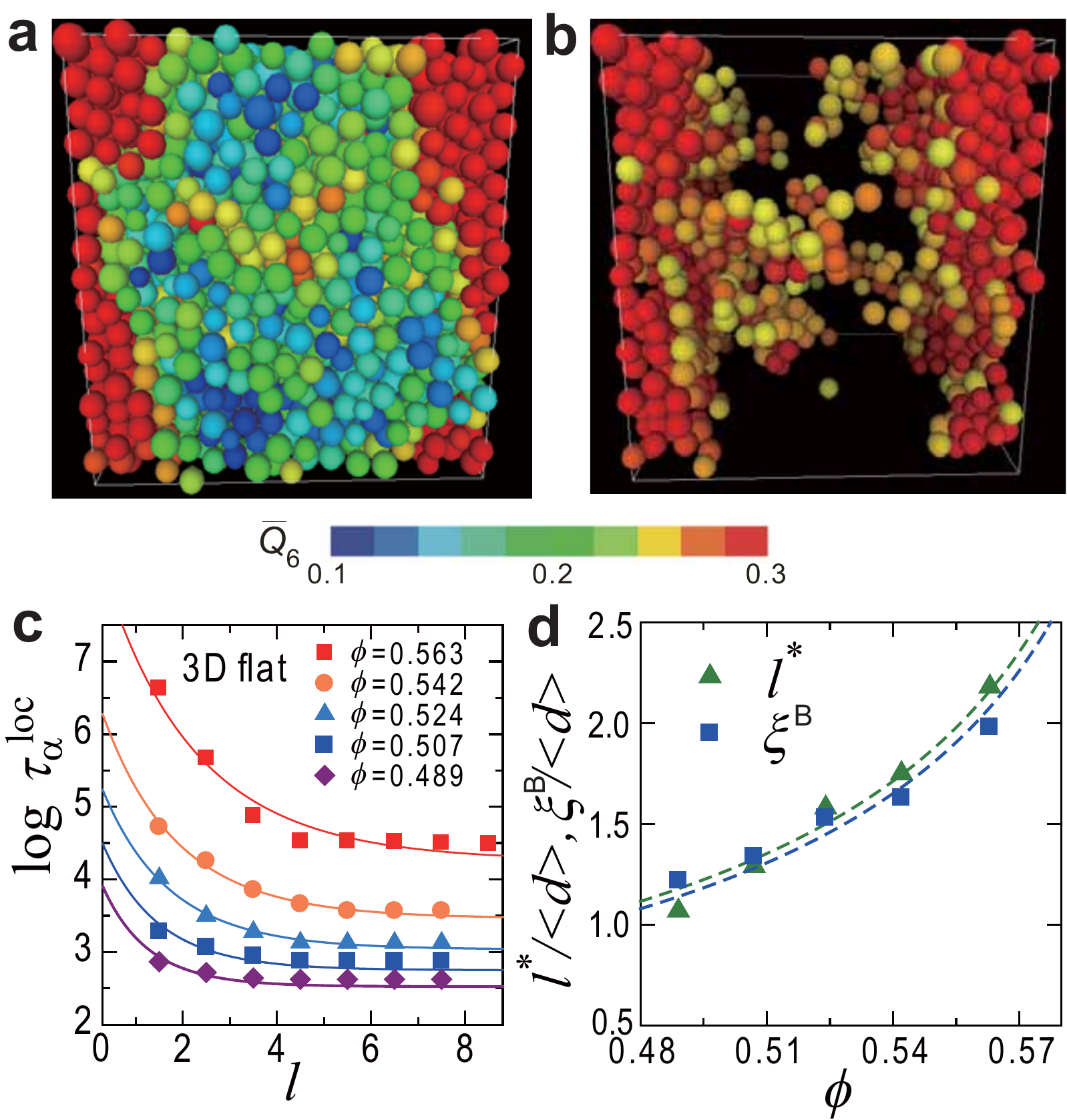}
\end{center}
\caption{(Colour on-line) 3D polydisperse colloidal liquid (3DPL) confined between two smooth flat walls.  
(a) Spatial distribution of $\bar{Q}_6$ for all particles 
at the volume fraction $\phi=0.563$ and $\Delta=8\%$. 
(b) Particles having $\bar{Q}_6>0.25$ (the same as {\bf a}). 
We can clearly see that 
particles near the walls have higher hcp-like bond orientational order.  
(c) The $\phi$-dependence of $\tau_{\alpha}^\mathrm{loc}(\phi, l)$ 
for 3DPL ($\Delta=8$ \%) confined between two flat walls. 
Solid curves are fits to data with $\xi_0^{\rm W}=0.45$. 
(d) The $\phi$-dependence of $l^\ast$. 
We can see that $l^\ast \propto \xi_B$. 
The curve is $\l^\ast =l_0 [(\phi_0^{\rm B}-\phi)/\phi]^{-2/3}$, where 
$l_0=0.507 \cong \xi_0^{\rm B}=0.50$. 
This figure is reproduced from fig. 4 of ref. \cite{watanabe2011}. }
\label{fig:confine}
\end{figure}

Spatial confinement is known to induce a drastic change in the viscosity, relaxation times, and 
flow profile of liquids near the glass (or jamming) transition point \cite{McKenna,Teboul,Goyon,Kob4}. The essential underlying question is 
how the presence of a solid wall affects the dynamics of densely packed
systems. We recently studied this problem \cite{watanabe2011}, using experiments on a driven granular hard-sphere liquid \cite{WT} 
and numerical simulations of polydisperse and bidisperse colloidal liquids. 
The nearly hard-core nature of the particle-wall interaction
provides an ideal opportunity to study purely geometrical confinement effects. We revealed that the slower dynamics 
near a wall is induced by wall-induced enhancement of ‘glassy structural order’, which is a manifestation 
of strong interparticle correlations. By generalizing the structure-dynamics relation for bulk systems (see eq. (\ref{eq:tau_D})), we 
find a quantitative relation between the structural relaxation time at a certain distance from a wall and 
the correlation length of glassy structural order there. 
That is, the structure-dynamics correlation found in bulk, 
$\tau_{\alpha}^{\rm B}=\tau_0^{\rm B} \exp[D^{\rm B}(\xi^{\rm B}/\xi_0^{\rm B})^{d/2}]$  (see eq. (\ref{eq:tau_D})) 
\cite{ShintaniNP,KAT,WT,TanakaNM,Kawasaki3D} can be extended 
to confined systems as follows: 
\begin{eqnarray}
\tau_{\alpha}^\mathrm{loc} (\phi, l)&=&\tau_0^\mathrm{B} 
\exp [D^\mathrm{B}(\xi^\mathrm{loc}(\phi, l)/\xi_0^\mathrm{B})^{d/2}], 
\label{tau} \\
\xi^\mathrm{loc}(\phi, l)&=&\xi_0^\mathrm{loc}(l) 
\left( \frac{\phi}{\phi_0^\mathrm{B}-\phi} \right)^{2/d}, 
\label{xi} \\
\xi_0^\mathrm{loc}(l)&=&\xi_0^\mathrm{B}+\xi_0^\mathrm{W} 
\exp(-l/l^*). 
\label{xi_0}
\end{eqnarray}
An example of the fitting of these relations to the results of a 3D polydisperse colloidal liquid 
confined between two flat smooth walls is shown in fig. \ref{fig:confine}.  

Here we note that the above eqs.~(\ref{tau})-(\ref{xi_0}) have significant physical 
implications. 
First, eq.~(\ref{tau}) suggests that
the relation between the structural relaxation time, $\tau_{\alpha}^{\rm B}$, 
and the correlation length of glassy structural order, $\xi^{\rm B}$, found in the bulk systems 
\cite{ShintaniNP,KAT,WT,TanakaNM,Kawasaki3D} 
also holds for confined systems, except that the $l$-dependence 
of the local correlation length $\xi^{\rm loc}$ must further be taken into account. 
Furthermore, eq.~(\ref{xi}) indicates that $\xi^\mathrm{loc}$ obeys 
the same power law for $\phi$ as the bulk correlation length $\xi^{\rm B}$.
All the wall effects are expressed by eq.~(\ref{xi_0}): 
the $l$-dependent bare correlation length, $\xi_0^\mathrm{loc}(l)$. 
That is, a wall influences {\it only} the bare correlation length. 
This is a reflection of the fact that near a wall, motion is more correlated even in a dilute state. 
Note that the bare correlation length is the correlation length at a dilute state far from $\phi_0$. 
Thus, the effects are expressed by a new term $\xi_0^\mathrm{W}$, 
which represents the extra bare glassy structural ordering induced 
by the presence of the wall. 
Our finding further supports the scenario that static glassy structural ordering is the origin of the slow glassy 
dynamics of a supercooled liquid. 

Here it should be noted that the functional form of $\tau_\alpha(l)$ we employed is different from that was used in refs. \cite{kob2011non,Kob1,Kob2,Kob3,Kob4}: 
\begin{eqnarray}
\ln \left[ \frac{\tau_\alpha^{\rm loc}(l)}{\tau_\alpha^{\rm B}} \right]=A(T) \exp \left[ -\frac{l}{l^\ast} \right], \nonumber 
\end{eqnarray} 
where $A(T)$ is a function of $T$. 
Our relation is expressed in the following form, using eqs. (\ref{tau})-(\ref{xi_0}) : 
\begin{eqnarray}
\ln \left[ \frac{\tau_\alpha^{\rm loc}(l)}{\tau_\alpha^{\rm B}} \right]=\frac{DT_0}{T-T_0} \left[ \left( 1+\frac{\xi_0^W}{\xi_0^B} \exp(-l/l^\ast) \right)^{d/2} 
-1 \right]. \nonumber
\end{eqnarray} 
These relations are almost the same for 2D ($d=2$) except that in our case $A(T)=(\xi_0^W/\xi_0^B) \frac{DT_0}{T-T_0}$, 
but different for 3D ($d=3$). 
This difference should affect the estimation of $\xi$. Thus, further careful studies are necessary on this point. 

\subsection{Free energy governing crystallization and vitrification}

Here we briefly mention the free energies, which have been proposed to describe 
the state of a supercooled liquid and/or a glass, and then we describe our view.  
On the details on each theory, please refer \cite{GotzeB,das2011statistical,leuzziB,BerthierR}

\subsubsection{Mode coupling theory}

Mode coupling theory is a microscopic theory, which is extended from a liquid side to a low temperature 
nonergodic state \cite{GotzeB,das2004mode,das2011statistical}. The ergodic-to-nonergodic transition is basically kinetic, and  
the free energy employed is that describing a liquid state.  
The kinetic equations describing the particle density $\rho(\mbox{\boldmath$r$},t)$ and 
the momentum density $\mbox{\boldmath$j$}(\mbox{\boldmath$r$},t)$ of a one component liquid 
(particle mass $m$, average density $\rho_0$, and viscosity $\nu_0$) are given by  
\begin{eqnarray}
m \frac{\partial}{\partial t} \rho(\mbox{\boldmath$r$},t)&=&-\mbox{\boldmath$\nabla$}\cdot \mbox{\boldmath$j$}(\mbox{\boldmath$r$},t), \\
\frac{\partial}{\partial t} \mbox{\boldmath$j$}(\mbox{\boldmath$r$},t)&=&\mbox{\boldmath$f$}-\frac{\nu_0}{m\rho_0} \mbox{\boldmath$j$}(\mbox{\boldmath$r$},t)+\mbox{\boldmath$\zeta$}^0(\mbox{\boldmath$r$},t),
\end{eqnarray} 
where $\mbox{\boldmath$\zeta$}^0(\mbox{\boldmath$r$},t)$ represents thermal force noises satisfying the following 
fluctuation dissipation theorem:  
\begin{eqnarray}
\langle \mbox{\boldmath$\zeta$}^0(\mbox{\boldmath$r$},t)\mbox{\boldmath$\zeta$}^0(\mbox{\boldmath$r$}',t') \rangle
=2k_{\rm B}T \nu_0 \mbox{\boldmath$I$} \delta (\mbox{\boldmath$r$}-\mbox{\boldmath$r$}') \delta (t-t'). 
\end{eqnarray}
$\mbox{\boldmath$f$}(\mbox{\boldmath$r$},t)$ is the force density acting on the liquid, which is induced by   
the inhomogeneous density field, and given by 
\begin{eqnarray}
\mbox{\boldmath$f$}(\mbox{\boldmath$r$})=-\rho(\mbox{\boldmath$r$})\mbox{\boldmath$\nabla$} \frac{\delta F\{\rho \}}
{\delta \rho(\mbox{\boldmath$r$})}. 
\end{eqnarray}
Here $F\{\rho \}$ is the free energy functional expressed as (see sec. \ref{sec:DFT}) 
\begin{eqnarray}
F \{\rho \}=\int f(\rho(\mbox{\boldmath$r$})) d\mbox{\boldmath$r$}=k_{\rm B}T 
\int d\mbox{\boldmath$r$}\ \rho(\mbox{\boldmath$r$})[\ln \frac{\rho(\mbox{\boldmath$r$})}{\rho_0}-1] \nonumber \\
+\int \int d\mbox{\boldmath$r$} d\mbox{\boldmath$r$}'(\rho(\mbox{\boldmath$r$})-\rho_0)
c(\mbox{\boldmath$r$}-\mbox{\boldmath$r$}') (\rho(\mbox{\boldmath$r$}')-\rho_0). \label{eq:F}
\end{eqnarray}
Here $c(\mbox{\boldmath$r$}-\mbox{\boldmath$r$}')$ is the Ornstein-Zernike direct correlation function 
of the liquid. 
The localization cost is the same as for a perfect gas, whereas the interaction term involves the direct correlation function 
of the liquid, a renormalized form of the bare interaction potential. 
The direct correlation function is determined by the condition that the functional gives small 
fluctuations in density reproducing the static liquid structure factor. 
For examples, if one expands the logarithm of the above equation to second order in the density fluctuations, one obtains 
the usual expression relating the direct correlation function to the static structural factor $S(\mbox{\boldmath$k$})$. 
For hard spheres, the standard Percus-Yevick form is often employed for $c(\mbox{\boldmath$r$})$. 
Higher order terms in the density can also be included. 
However, we note that information on bond orientational correlations, which we believe is important for the description of 
a supercooled liquid, is missed as a result of various assumptions commonly used in the standard liquid-state theory.   
Thus, we may say that the free energy used in MCT is effectively that for a liquid together with the assumptions made. 

It may be worth noting here that Dasgupta and Valls \cite{dasgupta1994two} studied the dynamic behaviour of a dense hard-sphere 
liquid by numerically integrating a set of Langevin equations that incorporate a free energy functional of the 
Ramakrishnan-Yussouff form and found that an fcc configuration, which is a crystal, as an inhomogeneous minimum of the free energy by using 
a set of bond orientational order parameters. 
However, this type of model cannot reproduce the development of bond orientational order fluctuations (neither crystal nor translational order) 
in a `liquid' state. We speculate that in a supercooled liquid many body correlations represented by bond orientational order, 
which may be the origin of cooperativity, plays a more important role than (local) two-point density correlator.

\subsubsection{Random first order transition theory}

The random first order transition theory is based on 
the density functional approach, which relies on the Lindemann criterion for vitrification 
(see, e.g., refs. \cite{Kirkpatrick,Xia,lubchenko2007}). 
The density functional considers the cost of forming any density wave by breaking
the free energy into an entropic localization penalty and an interaction term.  
It is given by the same form as the above free energy given by eq. (\ref{eq:F}) (see, e.g., ref. \cite{singh1985hard}, for hard spheres). 
In the frozen aperiodic state, the density wave is decomposed
into a sum of Gaussians centred around random lattice sites,
$\rho(\mbox{\boldmath$r$})=\Sigma_i (\pi/3)^{3/2} \exp(-\alpha(\mbox{\boldmath$r$}-\mbox{\boldmath$r$}_i)^2)$, 
where $\alpha$ represents the effective local spring constant that determines the rms displacement 
from the fiducial lattice site. The localization sites are given by 
$\{ \mbox{\boldmath$r$}_i \}$. So the reference state of the free energy is chosen to be a glassy state with `amorphous order' 
in this theory. The above $\alpha$ can be regarded as the Lindemann parameter measuring the scale of vibrational motions 
in an amorphous state. This relies on the clear separation of fast vibrational motions from the slow motions, which 
is confirmed by the mode coupling theory, to which this theory is expected to be connected at 
a high temperature. 

\subsubsection{Frustrated spin glass theory based on hypothetical icosahedral ordering}

A model of glass transition was proposed on the basis of a picture 
that a liquid tends to have structural order, which is energetically preferred locally 
over simple crystalline packing, but frustrated over large distance 
\cite{Frank,Steinhardt,NelsonB,sadoc1999,Sethna}. 
An icosahedral structure was considered to be such a candidate. 
Although locally favoured structures play a central role in 
both this type of theory and our model, there is an important difference 
between them: In the above model, locally favoured structures 
tend to attain long-range order as shown below, 
whereas in our model they compete with the ordering into a crystal. 

Steinhardt et al. \cite{Steinhardt} suggested that structural glass can be 
modelled by the following model Hamiltonian representing frustration 
in a lattice model of interacting icosahedra: 
\begin{eqnarray}
H=-J_Q \Sigma_{<i,j>} \Sigma_{m=-6}^6 Q_{6m}(\mbox{\boldmath$r$}_i) 
Q^\ast_{6m}(\mbox{\boldmath$r$}_j) \nonumber \\
+ \Sigma_{i \neq j} \Sigma_{j} K_Q(\mbox{\boldmath$r$}_{ij})_m 
Q_{6m}(\mbox{\boldmath$r$}_i) 
Q^\ast_{6m}(\mbox{\boldmath$r$}_j). \label{eq:Hami}
\end{eqnarray}
Here $\mbox{\boldmath$r$}_i$ denotes a site on a regular lattice, 
and the sum in the first term denotes a sum over nearest-neighbour 
pairs of sites. The positive interaction energy $J_Q$ tends to 
align neighbouring icosahedra. Frustration is modelled by the 
much weaker long-range interaction $K_Q(\mbox{\boldmath$r$}_{ij})$ 
in the second term, whose sign is randomly changed, in analogy 
with spin glasses. 
The coupling $J_Q$ favours a transition to an orientationally 
ordered state, whereas the random long-range part 
acts like a temperature-dependent random field of strength 
\begin{eqnarray}
h_{6,m}^{eff}(\mbox{\boldmath$r$}_{ij})= \Sigma_{j \neq i} 
K_Q(\mbox{\boldmath$r$}_{ij}) <Q_{6m}^\ast(\mbox{\boldmath$r$}_j)>. 
\end{eqnarray}
This field becomes stronger as $Q_6(T)$ increases. 
Thus, long-range ordering is prevented by the ``feedback'' of 
extended orientational order into a random field \cite{Steinhardt} 
and a glassy state is formed upon cooling. 

This scenario might somewhat look similar to ours. However, 
there is a crucial difference between them: In the above model, 
the long-range ordering of $Q_{6m}$ is prevented by internal frustration of 
the order parameter itself. The crystallization has to be avoided 
purely kinetically in the above model. Note that this model 
does not describe the crystallization itself. 
On the other hand, our model describes both crystallization 
and vitrification in the same framework. 
This difference is related to the fundamental question of 
whether a liquid tends to attain long-range icosahedral order 
or long-range crystalline order in a supercooled state.

\subsubsection{Frustration limited domain theory}

Kivelson and Tarjus {\it et al.} 
demonstrated an interesting possibility 
of a universal description of the temperature dependence of 
viscosity \cite{Tarjus}. 
They employed the Hamiltonian with frustration, which is 
basically similar to that of Steinhardt et al. \cite{Steinhardt} 
[see eq. (\ref{eq:Hami})], 
and used the concept of the avoided critical point. 
Their Hamiltonian has the following form: 
\begin{eqnarray}
H=-J_S\Sigma_{<ij>} \mbox{\boldmath$S$}_i \cdot \mbox{\boldmath$S$}_j 
+K_S \Sigma_{i \neq j} \frac{\mbox{\boldmath$S$}_i \cdot \mbox{\boldmath$S$}_j}
{|\mbox{\boldmath$R$}_i -\mbox{\boldmath$R$}_j|^x},  
\end{eqnarray} 
where $J_S$ and $K_S$ are both positive and $0<x<3$. 
The first (short-range, ferromagnetic) term favours long-range order 
of the locally preferred structure (in their terminology), whereas  
the second (long-range antiferromagnetic) term represents 
the frustration effects. 
Note that the ordering is prevented by internal frustration 
of the order parameter itself in their case, which leads to the concept of the avoided critical point. 
More explicitly, this model postulates an icosahedral ordering free from frustration in a curved space, 
which has a critical point. 
In an Euclidian space, however, the critical point is avoided due to frustration and a system 
has frustration-limited domains, which cause dynamic heterogeneity and lead to slow dynamics.   
We note that the underlying transition is of second order in their model.

\subsubsection{Our standpoint}
As described above, previous theories of glass transition put a focus on the 
vitrification branch, and neglect the fact that a liquid may crystallize or regard crystallization as a separate phenomenon. 
Thus kinetic freezing (kinetically constrained model or mode coupling theory) 
or exotic ordering (spin-glass-type, random first order transition, 
and frustration limited domain theory) have been considered to be key to glass transition.   

As we mentioned in the beginning of this section (see fig. \ref{fig:cry_glass}), 
it may be physically natural to consider that both crystallization and glass transition are governed by the same free energy which is 
a function of density order parameter $\rho$, bond orientational order $\mbox{\boldmath$Q$}_{\rm CRY}$ 
whose symmetry is consistent with the equilibrium crystal, and bond orientational order $\mbox{\boldmath$Q$}_{\rm LFS}$ 
which has a symmetry of locally favoured structures: $f(\rho, \mbox{\boldmath$Q$}_{\rm CRY}, \mbox{\boldmath$Q$}_{\rm LFS})$. 
$\mbox{\boldmath$Q$}_{\rm CRY}$ and $\mbox{\boldmath$Q$}_{\rm LFS}$ correspond to 
interparticle interactions compatible to the equilibrium crystal 
and those incompatible to it, respectively. 
Crystallization takes place under cooperation between orderings of $\rho$ and $\mbox{\boldmath$Q$}_{\rm CRY}$. 
The presence of $\mbox{\boldmath$Q$}_{\rm LFS}$ and its coupling to the other order parameters 
increase the nucleation barrier, as mentioned before, but do not alter crystallization itself once it takes place due to strong constraint 
(or, filtering effects) of translational order. 
On the other hand, glass transition takes place when long-range $\mbox{\boldmath$Q$}_{\rm CRY}$ and $\rho$ ordering is prohibited 
(kinetically) by a large nucleation barrier caused by its coupling to $\mbox{\boldmath$Q$}_{\rm LFS}$, 
which causes frustration effects. 
In the absence of $\rho$ ordering, which is the case of a metastable supercooled liquid before crystal nucleation, 
we need to consider only the couplings between $\mbox{\boldmath$Q$}_{\rm CRY}$
and $\mbox{\boldmath$Q$}_{\rm LFS}$. As speculated above, this may lead to critical-like bond orientational 
ordering towards the ideal glass transition point $T_0$. 
The fragility and the glass-forming ability of a liquid can be explained at least partly by the degree of frustration 
between the two types of the bond orientational order parameters.  

When there are two competing crystal polymorphs (CRY1 and CRY2), the free energy may be described as 
$f(\rho, \mbox{\boldmath$Q$}_{\rm CRY1}, \mbox{\boldmath$Q$}_{\rm CRY2})$. 
In relation to this, we should mention a series of beautiful experiments on a supercooled state of 2D magnetic colloids 
with frustration \cite{Maret2006,Maret2006b,Maret2008,Maret2011}. These studies show that a supercooled liquid has the low free energy 
configurations locally, which include crystal-like patches, and these configurations have a clear link to slow dynamics. 
This may be regarded as a clear manifestation of the fact that the free energy controlling the crystallization also governs glass transition.  

For systems with random disorder (e.g., polydisperse colloids and atactic polymers), we should also consider random disorder effects 
on $\mbox{\boldmath$Q$}_{\rm CRY}$ and $\rho$ orderings. However, the basic physics should 
remain the same as the above. We note that the disorder can be annealed by phase separation or fractionation 
for binary mixtures or polydisperse colloidal systems. 

As explained above, in other theories of glass transition, the peculiar free energy responsible for glass transition, 
which is often chosen to favour glassy disordered structures, is 
considered separately from the free energy describing crystallization. Thus, it is difficult for this type of theories 
to explain the glass-forming ability. 
We believe that it is more natural to consider the same free energy for both crystallization and 
vitrification (and quasicrystal formation). 

The above argument may be valid at least for one-component liquids. 
However, we should note that for a system like a binary mixture, in which 
crystallization takes place only when accompanying phase separation, there is complexity. 
The metastability, i.e., the separation between structural relaxation, phase separation, and crystallization, 
plays an important role in the selection of structures of low local free energy within the metastable state. 
For such a case, we need to include the contributions of the concentration field, mixing entropy,  
and interactions between the components. 
Glassy structural order in this case is not necessarily linked to 
crystal-like structures, but should still be linked to structures of low local free energy, as mentioned above. 
For 2DBC, such structures may be linked to distorted triangular tiling without voids, or long-lived 
stress-bearing solid-like structures. 
We note that the driving force for this structural order is common to that in 2DPC: 
Reduction of the free energy by gaining the correlational (or vibrational) entropy while an expense 
of the configurational entropy. We emphasize that low configurational entropy has a link to low fluidity, or high 
solidity, which is further connected to the coherence of particle motion.

\subsection{Glassy structural ordering, dynamic heterogeneity, and the Kauzmann paradox} \label{sec:Kauzmann}

\begin{figure}
\begin{center}
\includegraphics[width=7.0cm]{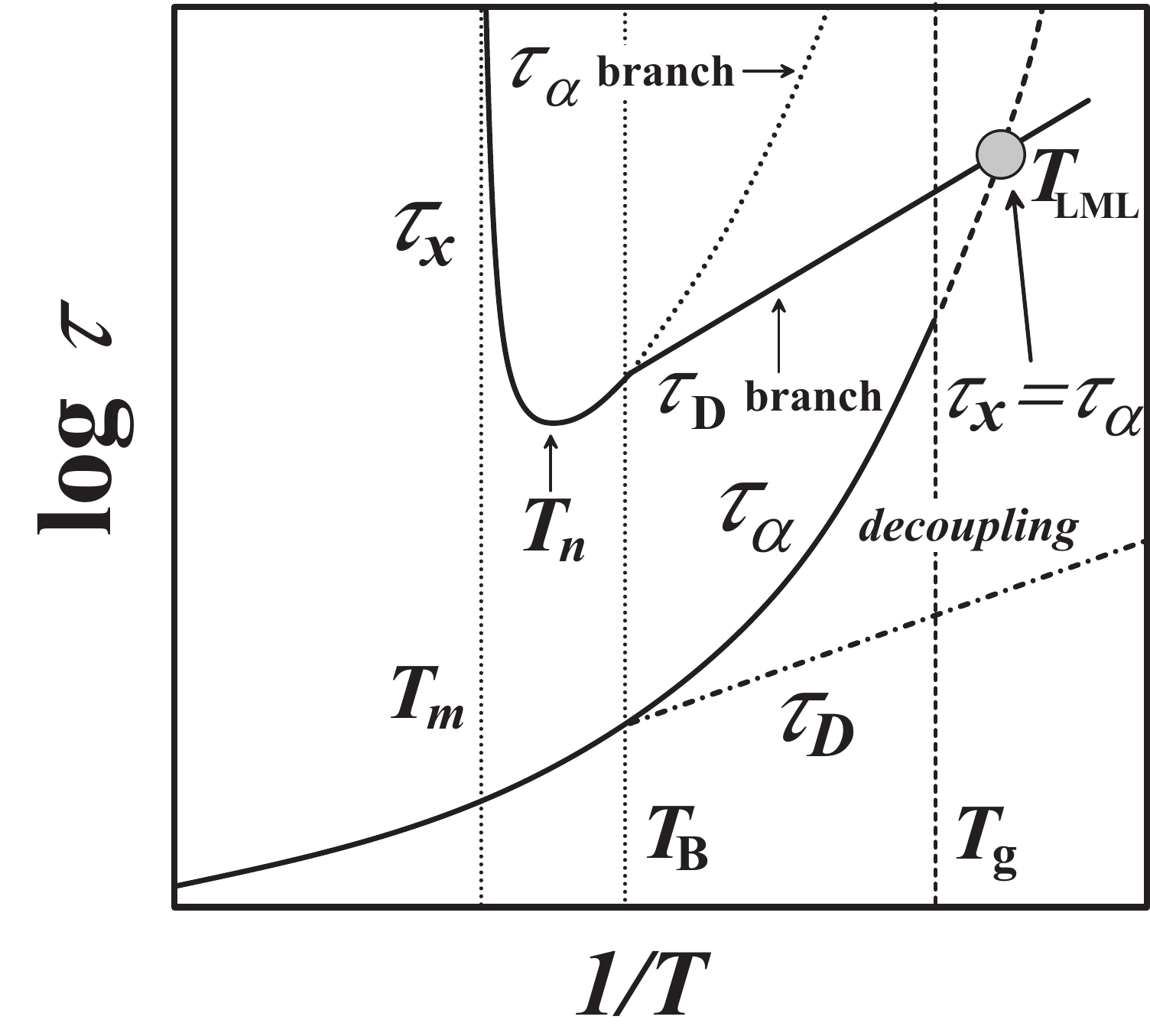}
\end{center}
\caption{
Schematic figure representing the temperature dependence of 
the characteristic times ($\tau_\alpha$, 
$\tau_D$, and $\tau_x$) of a glass-forming liquid. 
The structural relaxation time $\tau_\alpha$ 
obeys the Vogel-Fulcher-Tammann equation and diverges with approaching $T_0$. 
The translational diffusion mode $\tau_D$ is, on the other hand, 
decoupled from 
the structural relaxation mode $\tau_\alpha$ at $T_B$ upon cooling. 
Below a melting point $T_m$, a liquid tends to crystallize and thus 
the characteristic time of nucleation and crystallization, 
$\tau_x$, becomes finite below $T_m$. 
Reflecting the decoupling of $\tau_D$ from $\tau_\alpha$ at $T_B$, 
$\tau_x$ also changes its temperature dependence at $T_B$. 
Since the relevant transport process of crystallization is 
not the structural relaxation mode, but the translational 
diffusion one, the true $\tau_x$ ($\tau_D$ branch) is considerably 
shorter than 
$\tau_x$ estimated with the assumption that $\tau_t=\tau_\alpha$ 
($\tau_\alpha$ branch). 
This figure is reproduced from fig. 2 of ref. \cite{TanakaK}.
} 
\label{fig:kauzmann}
\end{figure}

So far we discuss the hypothetical ideal glass transition point $T_0$ as a hidden critical point 
of bond orientational ordering under frustration. 
However, this is based on the assumption that crystallization, or the long-range density ordering, 
is avoided at least practically. 
Here we consider the validity of this assumption. 
This problem is directly linked to the so-called Kauzmann paradox. 
It is known that the entropy of a supercooled liquid decreases more rapidly than that of the crystal 
and thus the extrapolated value of the former becomes equal to the latter at the Kauzmann temperature $T_{\rm K}$. 
Further extrapolation below $T_{\rm K}$ leads to an unphysical situation that the entropy of a disordered liquid is lower 
than the ordered crystal, which results in the violation of the third law of thermodynamics. This is known as the 
``Kauzmann paradox'' \cite{Kauzmann}, which has been one of the most fundamental problems of liquid-glass transition 
for more than 50 years \cite{DebenedettiB,DebenedettiN,donth,CavagnaR,Parisi2010,BerthierR}. 
Some time ago, we proposed its resolution by answering another fundamental question of how deeply we 
can supercool a liquid. We demonstrated that we can never supercool an ``equilibrium'' liquid below the lower metastable limit, 
$T_{\rm LML}$, since a liquid should crystallize before its equilibration. By proving that $T_{\rm LML}>T_{\rm K}$, thus, 
we resolved the Kauzmann paradox, or the entropy crisis. 
The scenario is schematically shown in fig. \ref{fig:kauzmann} 
(on its details, see ref. \cite{TanakaK}). 
We note that the presence of $T_{\rm LML}$ was recently suggested for a metallic glass experimentally \cite{mitrofanov2012relaxation}. 

The key to the above conclusion is that the kinetics of crystallization is controlled 
not by macroscopic viscosity, but rather by translational diffusion. 
In an ordinary liquid, the macroscopic viscosity is inversely proportional to 
the translational diffusion. However, this is not the case for a supercooled liquid. 
This breakdown of the Stokes-Einstein relation is considered to be a consequence of dynamic heterogeneity 
\cite{DebenedettiR,EdigerR,berthier2004,berthier2005,furukawa2009,furukawa2011,furukawa2012}. As schematically shown in fig. \ref{fig:kauzmann}, the crystal nucleation rate 
is empirically known to be controlled by translational diffusion rather than by viscosity \cite{TanakaK}. 
The fundamental origin of this decoupling may be related to the nonlocal nature of 
viscous transport and the intrinsic decoupling between the longitudinal and transverse dynamics, which were recently revealed 
by Furukawa and Tanaka \cite{furukawa2009,furukawa2011,furukawa2012}.  
The viscosity depends on the wavenumber $k$ even in a mesoscopic wavenumber regime \cite{kim2005,furukawa2009,puscasu2010,furukawa2011,furukawa2012}. 
It is low at a microscopic lengthscale, 
but increases towards the macroscopic value with a crossover length of $\xi$. 
This indicates that the viscous dissipation mainly comes from the length scale larger than $\xi$. 
Since crystallization proceeds by material transport by diffusion, it is natural to assume 
that its dynamics is governed by translational diffusion rather than by viscosity, although the very mechanism behind this 
remains elusive.  

Thus, our scenario indicates that it is ``dynamic heterogeneity'' that 
destabilizes a deeply supercooled liquid state as well as a glassy state against crystallization. 
Another effect of dynamic heterogeneity (or, more specifically, MRCO) on destablization of a metastable 
supercooled liquid state against crystallization will be discussed in sec. \ref{sec:crystallization}. 
This has a significant implication on the stability of a glassy state. 
The relevance of this scenario was also recently shown by numerical simulations \cite{saika2009crystal,ikeda2011,moore2011structural}. 
We note that a similar scenario for the resolution of the Kauzmann paradox was also proposed by Cavagna et al. \cite{Cavagna2003} 
on the basis of the numerical study of a lattice spin model.

In glass-forming systems, crystallization should take place before reaching the ideal glass transition point $T_0$, 
as far as we equilibrate a liquid for a time sufficiently longer than the structural relaxation time $\tau_\alpha$.  
More precisely, bond orientational ordering already starts in a supercooled liquid state and grows upon cooling, 
however, crystallization (cooperative bond orientational and translational ordering) 
takes place before reaching the hypothetical ideal glass state.  
Nevertheless, $T_0$ may have an important conceptual meaning 
in our understanding of glass transition as a hypothetical hidden ordering point of bond orientational 
order (under frustration). 

What we described above can be rephrased as follows: In a supercooled liquid state, before long-range crystalline bond orientational ordering takes place, 
crystal-like bond orientational orders compete with those incompatible to them, which leads to slow glassy dynamics. 
For systems suffering from random disorder effects bond orientational order still grows under frustration effects.  
When cooperative bond orientational and density ordering comes into play, however, a metastable supercooled state or a glass state 
transform into a crystal. This retardation of long-range density ordering, which is essential for vitrification, is controlled 
at least partly by the strength of frustration effects on crystal-like bond orientational ordering, 
which is prerequisite for translational ordering to take place (see sec. \ref{sec:crystallization}).  

For a system with quenched disorder, another process may be required for crystallization 
to take place, as mentioned before. For example, for colloidal systems with large polydispersity, phase separation and the resulting 
fractionation are prerequisite for crystallization \cite{sollich2010crystalline}. Although this complicates the situation, 
the basic behaviour should be the same. For binary mixtures, phase separation may lead to crystallization before 
reaching $T_0$ or $\phi_0$. In some cases where it is intrinsically difficult to remove 
quenched disorder, crystallization may never take place. An atactic polymer may be such a case. 
Then, the Kauzmann paradox may still be resolved by a thermodynamic transition 
to ideal glass or by slowing down of the rate of entropy decrease, which is realized by the presence of thermally activated 
processes, or the presence of defects until $T=0$ K.

\subsection{Relationship of the strength of frustration against crystallization 
to glass-forming ability and fragility and its link to the phase diagram}

Below we consider several model systems and material groups, focusing on the 
relation of the degree of frustration against crystallization to 
glass-forming ability and fragility. 
The systems we consider are 
(a) hard-sphere-like systems, (b) polymeric glasses, 
(c) 2D spin liquids, (d) water-type liquids, (e) eutectic mixtures, 
(f) chalcogenide glasses, and (g) metallic glass formers. 
Cases (a) and (b) suffer from both competing orderings 
and random disorder effects. In case (a), bond orientational 
order is a consequence of constraint due to dense packing of hard 
spheres. Cases (c)-(f), on the other hand, suffer from 
competing bond orderings. In case (c), for example, bond orientational ordering 
is a consequence of anisotropic interactions. In this sense, 
this model may be regarded as a model mimicking 
covalent bonding for oxides and chalcogenides or hydrogen bonding 
for molecular liquids and also the basis for cases (d)-(g). 
Case (g) is an example suffering from both random disorder effects 
(multicomponent effects) and competing bond orientational orderings (icosahedral ordering). 

Each material group (colloidal glasses, organic liquids, polymeric glasses, oxide glasses, chalcogenide glasses, 
or metallic glasses) has its own language and concept on its glass transition. 
By comparing these different cases, we aim at 
providing a general physical picture on the link between the thermodynamic 
phase behaviour and the nature of glass transition such as glass-forming 
ability and fragility, which may be applicable to any glass-forming systems in a universal manner. 
Note, however, that as mentioned below, the fragility in the following discussion 
does not include the energetic factor, which may cause inconsistency with experimental results.  
Furthermore, in reality, our argument might be too simplistic and miss other important factors 
controlling the glass-forming ability and fragility (see, e.g, ref. \cite{Paluch2011}). 

\subsubsection{The factors controlling the glass-forming ability}
Below we focus on the effects of the degree of frustration against crystallization on the glass-forming ability. 
However, we should mention that the ease of crystallization is determined by the kinetic factor and 
the thermodynamic factors controlling 
the barrier for crystal nucleation, which are the liquid-crystal interfacial tension and the difference in the 
chemical potential between the liquid and crystal. In the following discussion, we do not consider these factors, 
which are key to the classical nucleation theory, and just focus on the frustration effects. Thus, 
we definitely need to take into account these factors to make a better prediction 
(see refs. \cite{TanakaGJCP2,TanakaGJNCS} on this issue).

\subsubsection{The activation energy in an equilibrium liquid and the fragility of liquid} \label{sec:energy}
Before considering the glass transition behaviour of realistic glass formers, 
we need to mention an important factor controlling the fragility of liquid. 
So far we have emphasized that the strength of frustration against crystallization is a key factor 
controlling the fragility of a liquid. However, there is another important factor, which can practically 
be more important than the degree of frustration. 
It is the ratio between the energetic contribution and the contribution of the configurational entropy to 
the structural relaxation. This can be clearly seen in a few examples. One is the change of the fragility 
of colloidal systems as a function of the softness of the interaction \cite{mattsson2009soft}. The softer the interaction potential is, 
the stronger the liquid is. This can be viewed as the control of the ratio of the energetic (elastic) contribution to the 
configurational (centre of mass) entropy. Another example is our 2D spin liquid system \cite{ShintaniNP}, where we control 
the strength of the anisotropic part of the potential relative to the isotropic part. 
These examples clearly support the above-mentioned effect of the Arrhenius energy on the fragility. 

Here we explain why it is so. 
The glass transition point $T_{\rm g}$ is defined by a temperature (or pressure or density), at which the structural relaxation 
time $\tau_\alpha$ reaches the time scale of observation (in a typical experiment of molecular or atomic liquids it is the order of 100 s, 
whereas in colloidal experiments it is the order of 10$^3$-10$^4$ s \cite{Brambilla}). 
If a normal liquid has a high activation energy $E_{\rm a}$, the significant part of the slow dynamics comes from a simple activation process 
without cooperativity. 
As in the case of water dynamics (see sec. \ref{sec:waterdynamics}), structural relaxation must involve two processes: A simple energetic activation process 
and a process of cooperative motion linked to MRCO. Thus the total structural relaxation process should have an energy barrier, which is the sum of these processes. 
We proposed the following empirical function to incorporate the two factors \cite{TanakaGJPCM,TanakaII}: 
\begin{equation}
\tau_\alpha=\tau_0 \exp(\beta E_{\rm a}+f(T) D (\xi/\xi_0)^{d/2}),  
\end{equation}
Here $f(T)$ is introduced to ensure a smooth crossover from the activated regime to the critical-like regime, although there is no firm basis 
for it. For simplicity, we assume the following form for $f(T)$: 
$f(T)=1/(\exp (\kappa(T-T_{x})+1))$, where $\kappa$ is a positive constant that controls the sharpness of the crossover of $f(T)$ from 0 to 1 
upon cooling and $T_x$ is the crossover temperature, where the criticality disappears.  
Note that $D=\beta \Delta_{\rm a}$ and we expect that $\Delta_{\rm a} \cong E_{\rm a}$ (see the following section).  
In the above, we may also use $\xi=\xi_0 ((T-T_0)/T)^{-2/d}$ instead of $\xi=\xi_0 ((T-T_0)/T_0)^{-2/d}$ (see sec. \ref{sec:cross}). 
With this form of $\xi$, at a high temperature limit the liquid dynamics can smoothly change to a simple activation process.  

The $\tau_\alpha$ at $T_{\rm g}$ has a higher fraction of the simple activation process for a liquid with 
a higher activation energy $E_{\rm a}$. 
This leads to the more Arrhenius-like behaviour for a liquid having a larger ratio of the energetic contribution to the configurational entropy contribution 
in the Angell plot, where the temperature is scaled by $T_{\rm g}$. 

Thus, we may say that a strong liquid is a liquid which exhibits very slow dynamics due to its large activation energy before 
significant cooperativity associated with the growing length comes into play. In other words, for some cases this classification may be just a consequence 
of the definition of $T_{\rm g}$ and the resulting scaling used in the Angell plot. 
If we make a plot for $\tau_\alpha/(\tau_0 \exp(\beta E_{\rm a}))$, we can extract only the cooperative part.  
However, since the distance between $T_{\rm g}$ and $T_0$ should be large for a strong liquid, we cannot access the 
temperature range where a significant slowing down due to cooperativity takes place for a strong liquid. 
In this sense, the size of $\xi$ at $T_g$ may also be a good measure of the fragility of a liquid, although this quantity is not easy to access experimentally: 
The smaller $\xi(T_{\rm g})$ is, the stronger a liquid is. 

Although $E_{\rm a}$ can be a dominant factor determining the fragility of a real fluid, it is non-universal. 
Thus, hereafter, we do not consider this factor and concentrate only on the cooperativity associated with growing motional correlation. 
So in the following discussion on the fragility of realistic glass-forming liquids, please note that this energetic factor 
is neglected, although this might lead to incorrect predictions. If there is a relation of $E_a \cong \Delta_a =D k_BT$ 
(see sec. \ref{sec:activation} and below), it might be OK to neglect it. 

\subsubsection{A possible link between $E_{\rm a}$ and $\Delta_{\rm a}$}

The activation energy $E_{\rm a}$ at a high temperature region is related to the energy scale of interparticle interactions. 
On the other hand, $\Delta_{\rm a}$ is also the fundamental microscopic energy scale of a system, which controls 
the energy scale of frustration and thus the strength of the growing 
activation barrier, as discussed in sec. \ref{sec:activation}. 
This suggests a positive correlation between $E_{\rm a}$ and $\Delta_{\rm a}$. 
This is consistent with the correlation between the strength of the anisotropic potential $\Delta$ and the strong nature of 
a liquid, or the value of the fragility index $D$, for 2DSL \cite{ShintaniNP,STNM} as well as the widely known positive 
correlation between the high temperature activation energy $E_{\rm a}$, or the 
strong nature of the directional bonding, and $D$.

\subsubsection{A case of hard-sphere-like systems: Geometrical frustration and/or random disorder effects on 
crystal-like bond orientational ordering}

\begin{figure}
\begin{center}
\includegraphics[width=7cm]{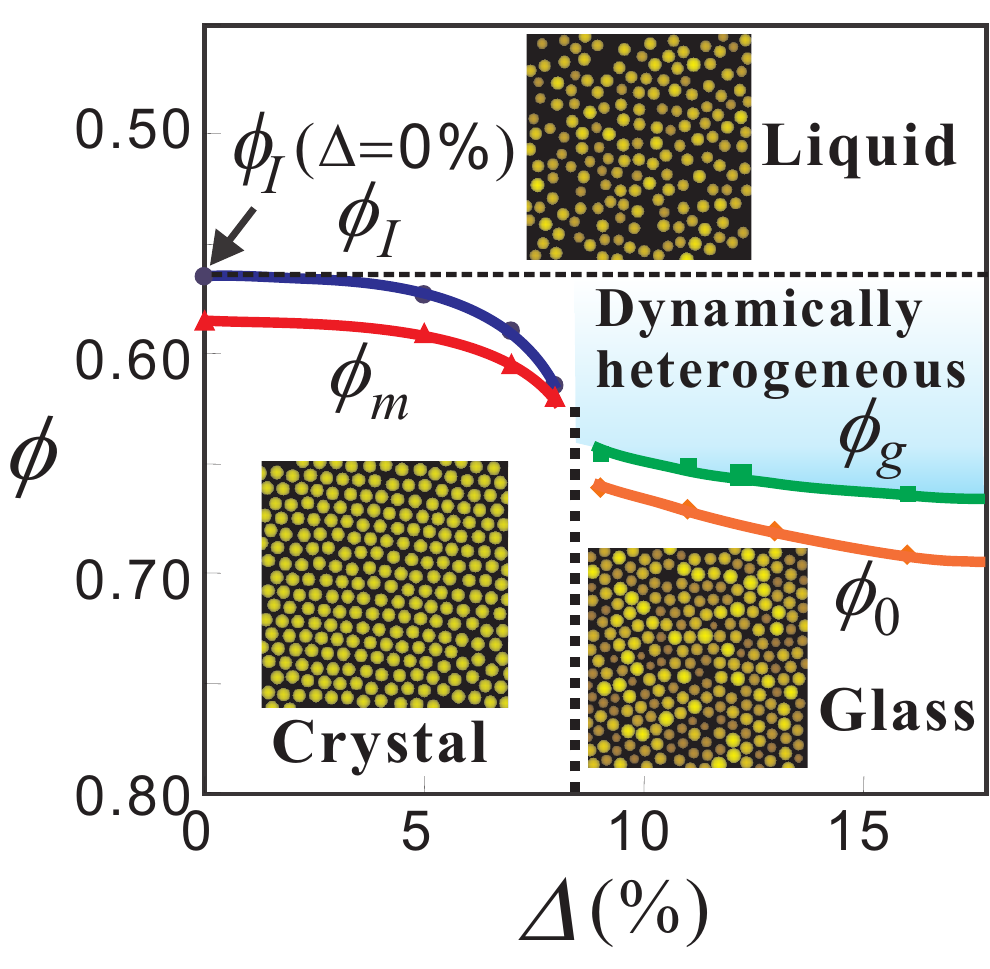} 
\end{center}
\caption{(Colour on-line) A state diagram for 2D polydisperse hard-sphere-like systems.  
Here $\phi$ is the volume fraction of colloidal particles and $\Delta$ 
is the degree of polydispersity, which can be regarded as the strength 
of frustration against crystallization. 
This figure is reproduced from fig. 1 of ref. \cite{KawasakiJPCM}. 
}
\label{fig:2DPC} 
\end{figure}

A state diagram for 2D polydisperse hard disks is shown in fig. \ref{fig:2DPC}. 
For a monodisperse case (the polydispersity $\Delta=0$\%), there are two sequential transitions: 
bond orientational ordering followed by translational ordering.  
Above $\Delta \geq 9$\% (the coloured region in fig. \ref{fig:2DPC}), 
a system starts to form glass without crystallization even for slow cooling. This shows the increase of glass-forming ability 
with an increase in $\Delta$. In the glass-forming region, the fragility 
monotonically decreases with an increase in $\Delta$. 

For a 2D system, the only source of frustration against 
crystallization is polydispersity $\Delta$. This is because hexatic order is the unique bond order parameter for a particle having 6 
nearest neighbours and this order does not suffer from any frustration upon its growth. 

\begin{figure}
\begin{center}
\includegraphics[width=8.5cm]{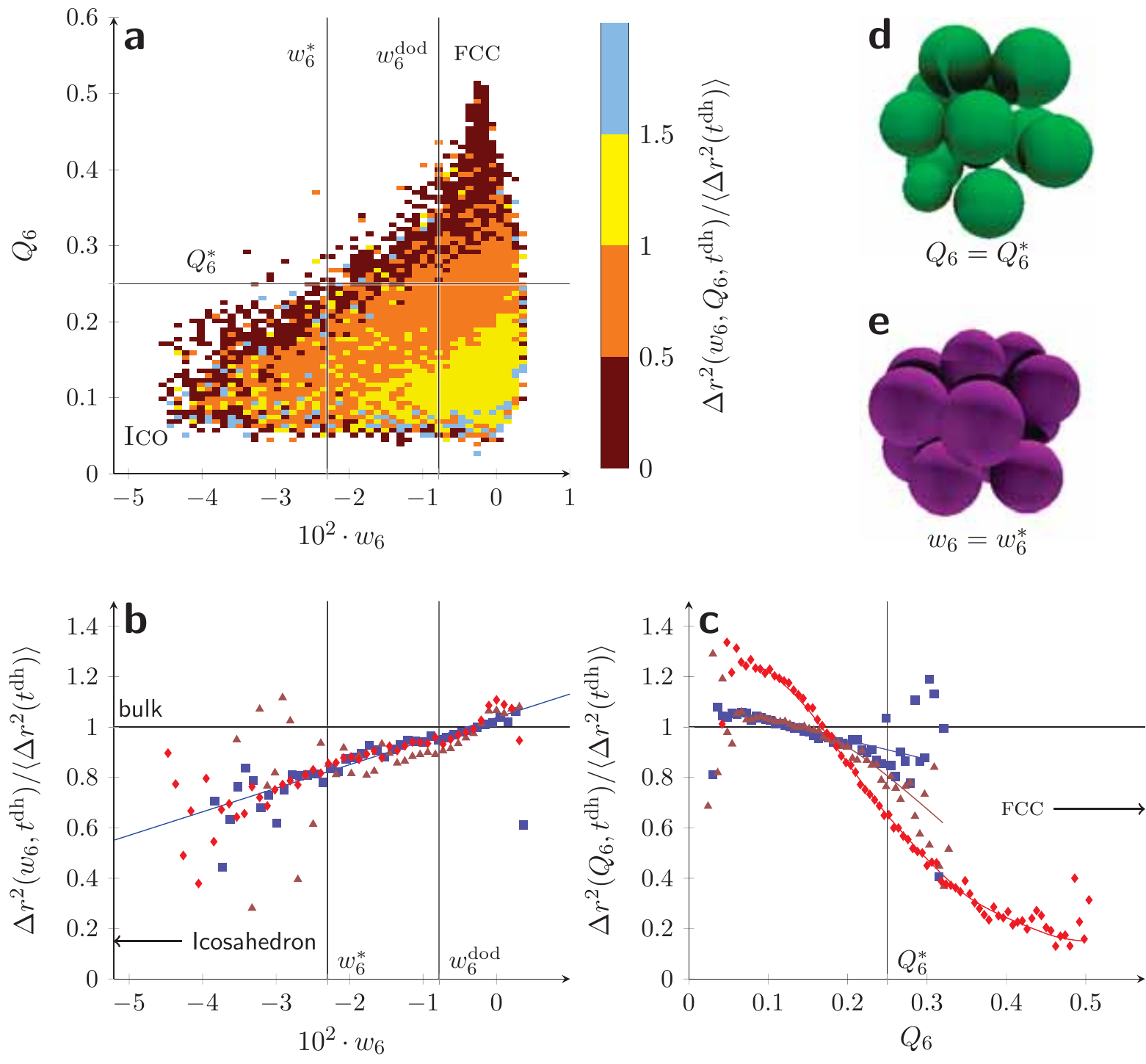}
\end{center}
\caption{(Colour on-line) Bond order mobility. (a) Normalised mobility in the $(w_6, Q_6)$-plane for a deeply supercooled sample of a colloidal suspension ($\phi = 0.575 \pm 0.03$). 
Here $Q_6$ is not $q_6$ and coarse-grained $Q_6$ (see sec. \ref{sec:defineBOO}). 
The colour scale is saturated at $1.5$ times the bulk mean square displacement. (b)-(c) Normalised mobility for icosahedral and crystalline order parameters respectively at volume fraction $0.535$ (squares), $0.555$ (triangles) and $0.575$ (diamonds), all $\pm 0.03$. Bulk mean square displacement is scaled to be at 1 (horizontal line). Perfect structures are on the edge of each plot. The lines are a guide for the eye, stressing the collapse of the $w_6$-mobility at all volume fractions in (b) and the absence of such collapse in (c). 
The collapse in (b) is a consequence of the non-extendable nature of icosahedral-like structures. 
The scattering at low volume fractions is due to poor averaging of rare structures. Straight lines in (a)-(c) corresponds to the important thresholds: $Q_6^\ast$, $w_6^\ast$ and $w_6^{\rm dod}$. 
For $Q_6 >Q_6^\ast$ and $w_6 < w_6^\ast$, we regard that particles have crystal-like bond orientational order and icosahedral-like order, respectively.  
$w_6^{\rm dod}$ is a measure for dodecahedral order. 
Examples of crystal-like cluster and distorted icosahedron at the respective threshold values are shown in (d) and (e), respectively. This figure is reproduced from fig.3 of ref. \cite{MathieuNM}.}
\label{fig:msd_Q6_w6}
\end{figure}

For a 3D polydisperse system, on the other hand, 
there are at least two origins of 
frustration against crystallization:  
One is local icosahedral ordering tendency and the other is 
random disorder effects originating from the polydispersity of particles. 
The importance of the former can be clearly seen in fig. \ref{fig:compare3D}(c). 
Note that for 3D hard spheres a particle having 12 nearest neighbours can have three types of bond orientational order 
(fcc, hcp, and ico) (see fig. \ref{fig:fcc}). 
Among them, local icosahedral ordering is not a major 
cause of slow dynamics due to its localized nature and the dominant one is crystal-like (fcc-like) 
bond orientational order. This can be seen in fig. \ref{fig:msd_Q6_w6} \cite{MathieuNM}. 
This tells us that only spatially extendable structural order is responsible for slow dynamics.  
So the scenario that icosahedral ordering is a major and unique 
underlying ordering 
behind vitrification may not be valid at least for a hard sphere system. 
Nevertheless, local icosahedral structures are formed, as shown in fig. \ref{fig:compare3D}(c), 
and their number density increases with an increase in $\phi$, which leads to stronger frustration effects on 
crystal-like ordering \cite{MathieuNM,russo2011}. In this sense, even a monodisperse hard sphere system 
is not free from frustration effects on crystallization and suffers from 
self-generated internal frustration controlled by entropy \cite{TanakaNM,russo2011}. 
This situation might be similar to metallic glass formers 
\cite{TanakaMJPCM,TanakaGJNCS}, 
although the tendency of icosahedral ordering may be 
more pronounced for these systems due to the chemical nature of bonding and the matching of atomic sizes. 

If the degree of the polydispersity becomes so large, 
even the bond orientational order parameter, which is valid for weakly polydisperse systems, 
cannot be applied any more for characterizing locally favoured structures in both 2D and 3D systems. 
The increase in the polydispersity $\Delta$ leads to an increase in 
fluctuations of the number of nearest neighbour particles, i.e., 
fluctuations of local bond orientational order parameters.  
For large $\Delta$, a one-to-one correspondence between high 
bond orientational order and slow mobility no longer holds \cite{KawasakiJPCM}. 
High order regions are always slow, but slow regions do not necessarily 
have high order. 

We confirmed that also in 3D systems the increase in the degree of polydispersity $\Delta$ leads to 
the increase in the glass-forming ability and the decrease in the fragility \cite{TanakaNM}. 
In both 2D and 3D systems, the growth of the correlation length of 
bond orientational order is suppressed by the increase in $\Delta$ for the same range of $\phi$ 
\cite{KAT,Kawasaki3D}. 

A binary hard sphere mixture with 
the size ratio around 1.4 corresponds to a situation suffering 
from strong random disorder effects \cite{Andersen,Hamanaka}. 
Even in such a case, geometrical packing of 
particles under volume constraint leads to some local structural features 
characterized by low configurational entropy 
to gain the correlational entropy \cite{TanakaNM}. 
This tendency that a system tends 
to have low local free-energy structures, which maximize the total entropy of a system,  
may be one of the most fundamental features of a `thermal' hard-sphere-like 
system whose free energy is controlled by the entropy alone, irrespective of the degree of polydispersity. 
For binary mixtures of soft spheres, the composition and size dependence of glass-forming ability was studied by Mountain and Thirumalai 
and the theoretical prediction of Egami and Waseda \cite{egami1984atomic} was confirmed. 
Recently, it was also shown by Hamanaka and Onuki \cite{Hamanaka} that the glass-forming ability can be controlled by the particle ratio 
and becomes maximum around 1.4.

\subsubsection{Random disorder effects in polymeric glass-forming systems}

In polymeric glass formers, energetic frustration on crystallization plays 
an important role as in any other systems. 
So bond orientational ordering may also play an important role in slow dynamics of glass-forming polymer systems. 
Such an example has recently been demonstrated by numerical simulations \cite{asai2011}. 
In addition to that, ends of a 
polymer chain causes disorder effects on crystallization. However, the most significant 
factor controlling the ease of vitrification is so-called stereoregularity of polymers, which is characterized by 
tacticity. Tacticity is the relative stereochemistry of adjacent chiral centres within a macromolecule. 
The regularity of the macromolecular structure influences the degree to which it has rigid crystalline 
long range order or flexible amorphous disorder. 
A tactic polymer is a macromolecule in which essentially all the configurational units are identical, 
whereas an atactic polymer is a macromolecule with a random sequence. 
Thus, atactic polymers suffer from strong random disorder effects, which prevent crystallization. 

\begin{figure}
\begin{center}
\includegraphics[width=7cm]{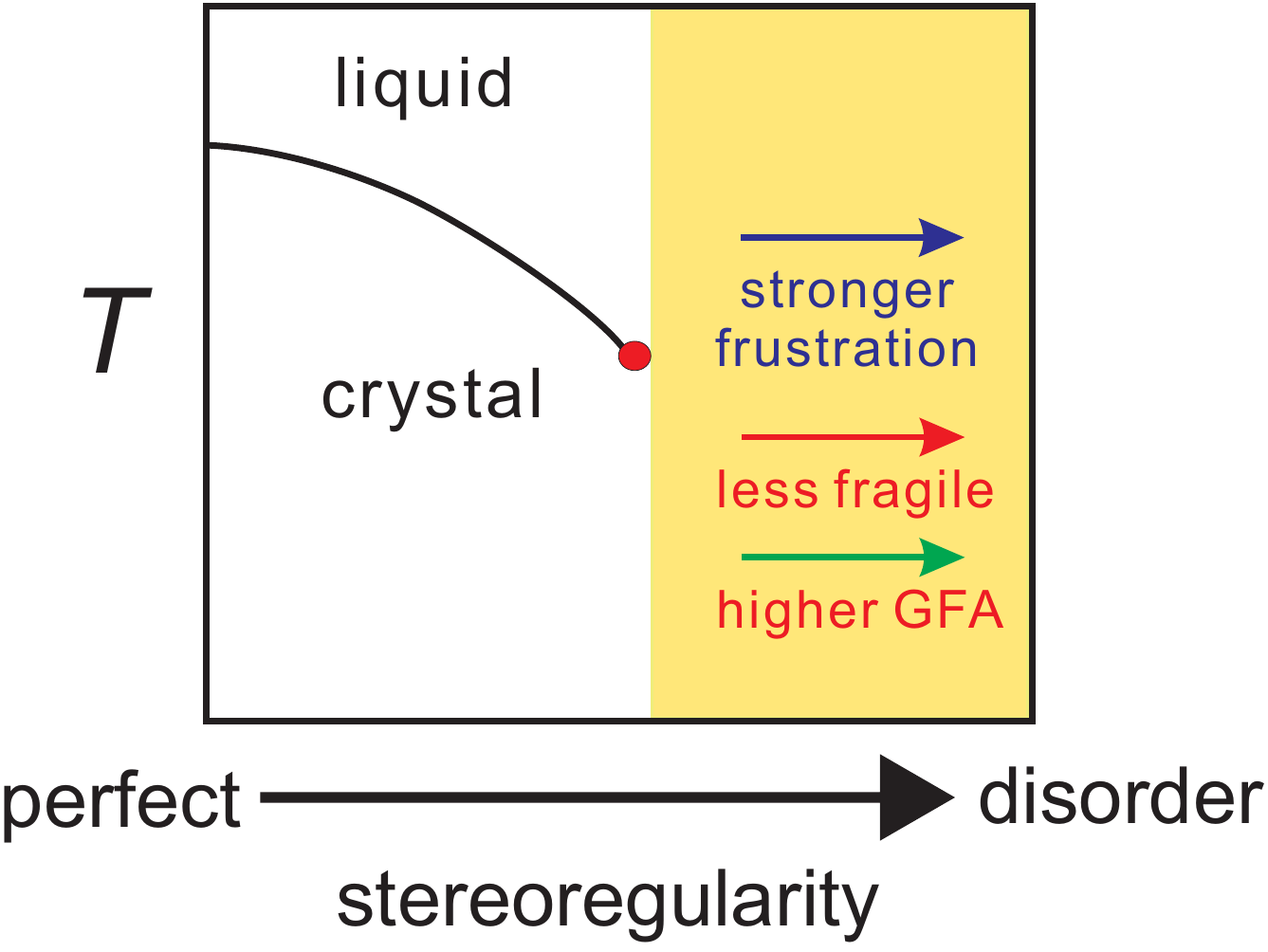} 
\end{center}
\caption{(Colour on-line) Schematic phase (or state) diagrams for polymer. 
Here the $x$-axis is the degree of stereoregularity, which increases 
towards the left-hand side. A polymer with perfect stereoregularity (isotactic or syndiotactic) 
should crystallizes into a crystal. However, with an increase in 
random disorder, or with a decrease in stereoregularity, a polymer 
becomes difficult to be crystallized, which is the case of so-called 
atactic polymers. So the basic behaviour is very similar to colloidal systems 
described above. 
The area painted in yellow is a glass-forming region. 
The solid curve represent a melting-point 
curves. The same behaviour should also be observed for random copolymerization. 
In this case, the $x$-axis should be replaced by the degree of random copolymerization in the above figure. 
}
\label{fig:polymer} 
\end{figure}

This situation is similar to that of polydisperse colloidal systems. 
However, since the disorder effects are quenched in the chemical structure, atactic polymers never crystallize unlike 
polydisperse colloids, which can crystallize after fractionation or phase separation (or, a sort of annealing). 
We speculate a state diagram of a polymer as a function of the degree of tacticity (see fig. \ref{fig:polymer}). 
It is highly desirable to make such a phase diagram experimentally or numerically. 
The basic structure of the phase diagram should be similar to that of polydisperse colloids. 
We predict that the increase in disorder increases the glass-forming ability and 
makes a polymeric liquid less fragile, which should be checked in the future. 
We note that for this case the Arrhenius-type activation energy affecting the fragility may be safely neglected 
since the basic physical interactions may not be strongly affected by the tacticity.   

It is empirically known that bigger side groups, chain complexity, branching, and random copolymerization 
reduce the ability of a polymer to crystallize. All these effects can be interpreted as random 
disorder effects on crystallization. So we also expect that an increase in these effects 
increases the glass-forming ability and makes a liquid stronger.

\subsubsection{A case of 2D spin liquids: competing bond orientational orderings}

\begin{figure}
\begin{center}
\includegraphics[width=7cm]{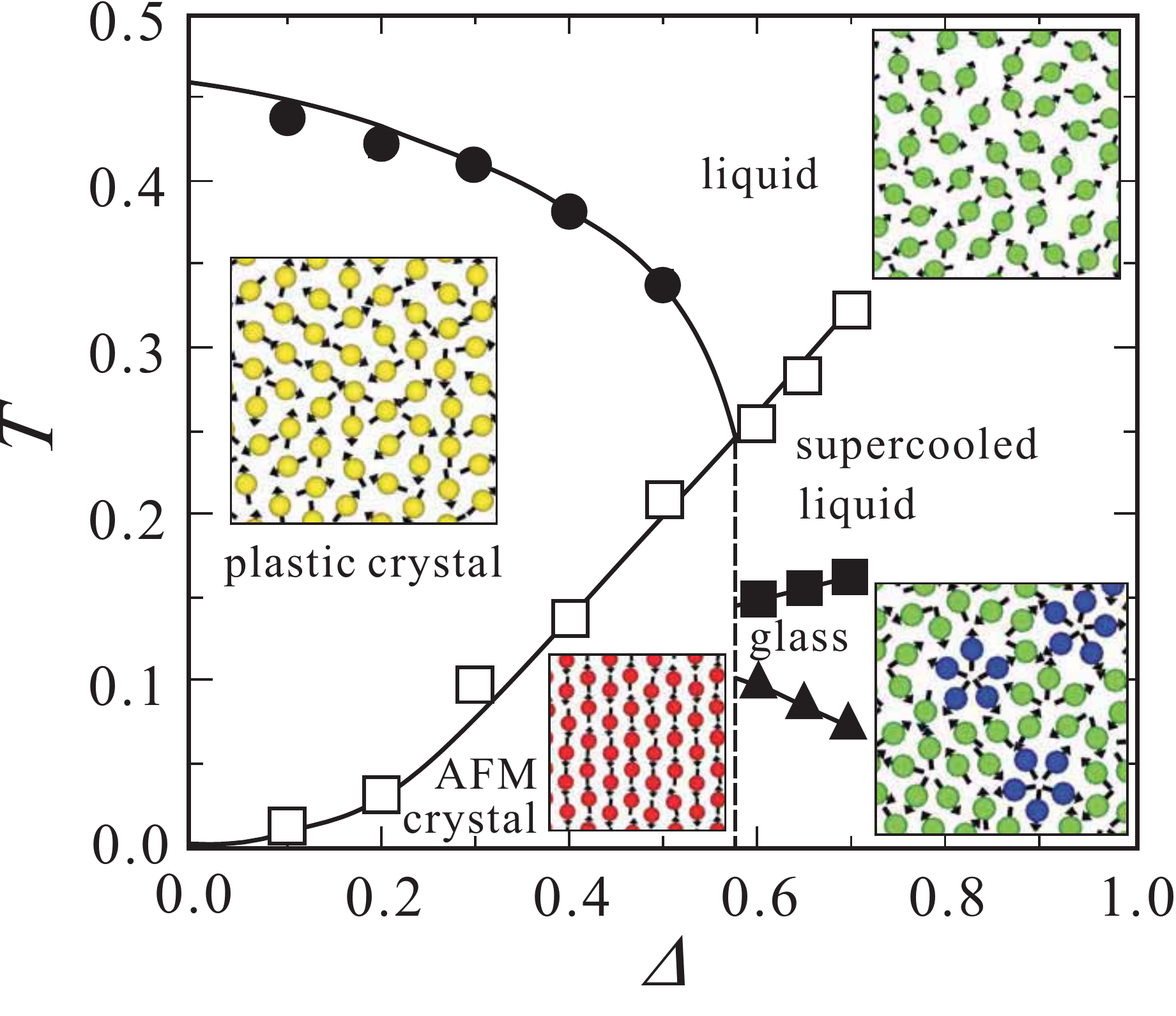}
\end{center}
\caption{(Colour on-line) Phase diagram of 2D spin liquid in the $T$-$\Delta$ plane. 
  Here $\Delta$ is a measure of the strength of frustration against 
  crystallization, or the strength favouring a locally favoured structure of 
  five fold symmetry. Energetic frustration 
  due to symmetric mismatch in the interacting potential is caused in this system. 
  The basic structure of the phase diagram is quite similar to that of water (see fig. \ref{fig:PD}). 
  For small $\Delta$, or weak frustration, the glass-forming ability is very low, 
  whereas with an increase in the frustration strength $\Delta$ the glass-forming 
  ability is increased and the fragility is decreased. This basic trend is also 
  very much consistent with the behaviour of water under pressure and 
  water/salt mixtures \cite{KobayashiPRL}.  This figure is reproduced from fig. 1 of 
  ref. \cite{ShintaniNP}.   
  }
  \label{fig:Two}
\end{figure}

Next we consider a case of 2D spin liquids (see fig. \ref{fig:Two}) \cite{ShintaniNP,STNM}. 
In this model, we put an anisotropic potential which forces  
particles having spins to favour the formation of pentagons (see fig. \ref{fig:SRO}). 
The statistical mechanics approaches to both thermodynamics and dynamics have been developed by Procaccia and his coworker for this model 
\cite{ilyin2007,lerner2009}. 
The ground state crystal has antiferromagnetic order 
on an uniaxially elongated hexagonal lattice. 
This crystal has a density higher than a liquid. 
In a supercooled liquid state of this model system, 
we found medium-range antiferromagnetic bond orientational 
order (see fig. 3), whose size $\xi$ grows almost as $\xi 
=\xi_0 ((T-T_0)/T_0)^{-1}$ when approaching $T_0$. 
We confirm that antiferromagnetic bond orientational ordering is almost 
completely decoupled with density ordering: Density change 
is not accompanied by the crystal-like bond orientational ordering and thus 
the density of a supercooled liquid is uniform 
in space after a certain level of coarse-graining, 
irrespective of the degree of antiferromagnetic order. 
Here it should be mentioned that pentagons have a larger specific volume. 
There are strong competing orderings between antiferromagnetic 
and five-fold pentagonal ordering (see fig. \ref{fig:Two}).

\begin{figure}
\begin{center}
\includegraphics[width=8.5cm]{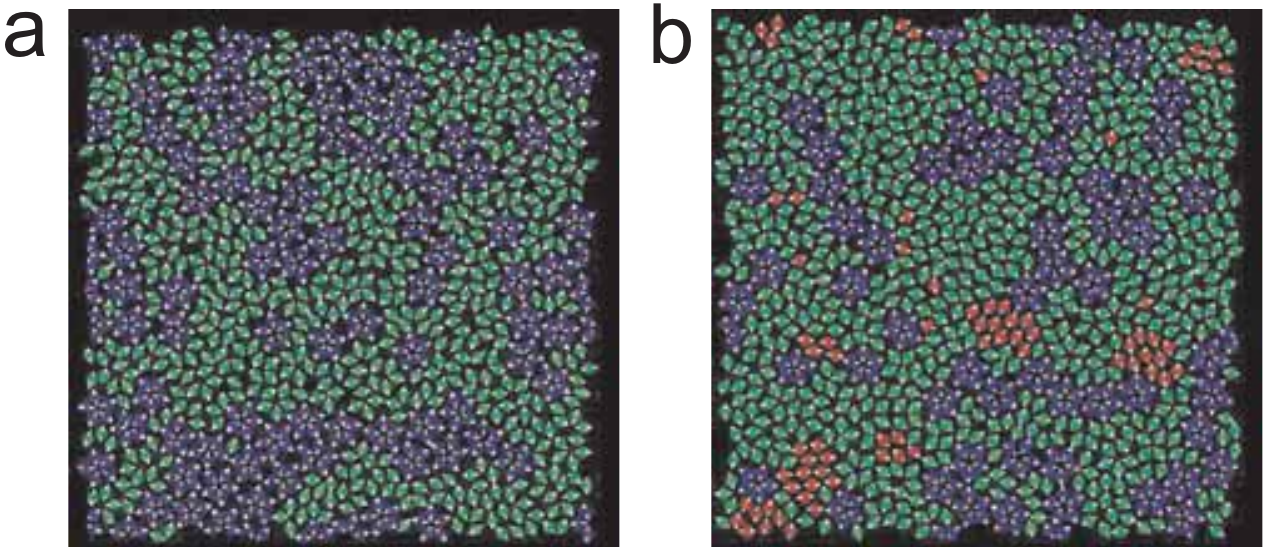}
\end{center}
\caption{(Colour on-line) Liquid states of 2DSL with $\Delta=0.8$ at $P=0.15$ and $T=0.27$ ($T_{\rm g}/T=0.78$) (a) and 
at $P=3.0$ and $T=0.29$ ($T_{\rm g}/T=0.82$) (b). With an increase in $P$, the number density of 
locally favoured structures (pentagon) decreases since $\Delta v>0$. This leads to the 
decrease in frustration, which causes enhancement of medium-range crystal-like bond 
orientational (antiferromagnetic) order. 
}
  \label{fig:2DSL_pictures}
\end{figure}

\begin{figure}
\begin{center}
\includegraphics[width=7cm]{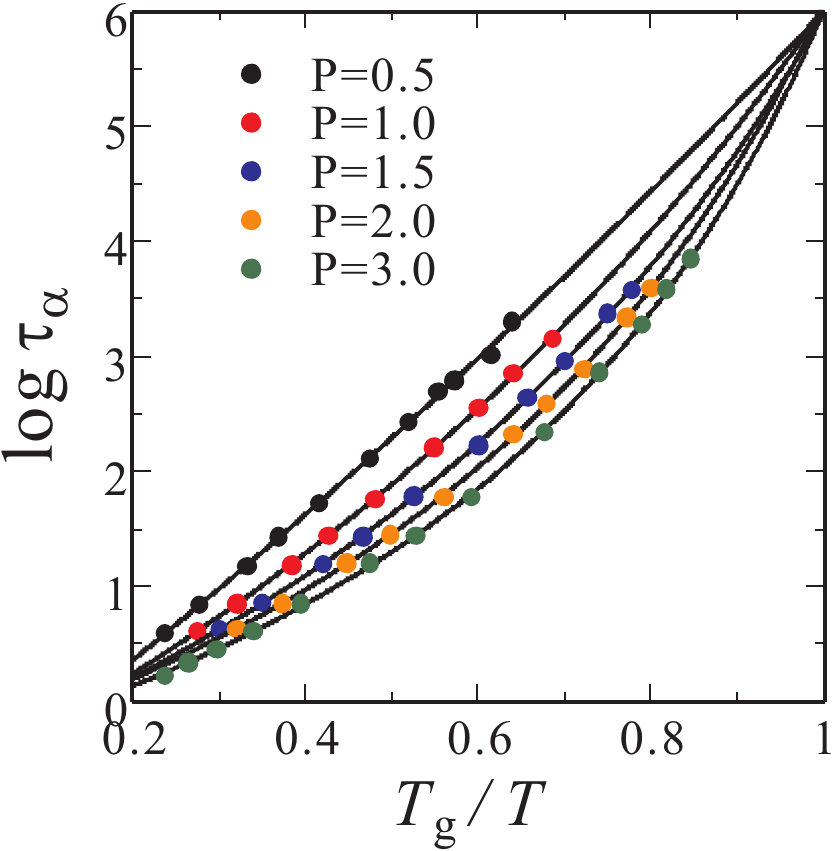}
\end{center}
\caption{(Colour on-line) The pressure dependence of the temperature dependence of the structural relaxation time 
$\tau_\alpha$. We can clearly see that the fragility decreases with an increase in $P$. 
This figure is reproduced from fig. 1 of ref. \cite{STNM}. 
}
  \label{fig:2DSL_fra}
\end{figure}

In this system, the strength of the frustration, 
which we express by $\Delta$,  
controls the glass forming ability, fragility, and criticality 
\cite{ShintaniNP,STNM}. 
The state diagram is shown in fig. \ref{fig:Two}. 
For small $\Delta$, a system easily crystallizes into the plastic crystal. 
For large $\Delta$, where the melting point of the antiferromagnetic crystal 
is higher than that of the plastic crystal, a system can be vitrified rather easily. 
Thus, the increase in $\Delta$ leads to the 
increase in the glass-forming ability. 
We also found that the increase in $\Delta$ decreases the fragility. 
This may be largely due to the increase in the activation energy $E_{\rm a}$ dominating the high temperature Arrhenius regime. 
Applying pressure leads to the decrease in pentagons (see eq. (\ref{eq:S}) and note 
that $\Delta v>0$ for a pentagon), as shown in fig. \ref{fig:2DSL_pictures}. 
This leads to the increase in the fragility \cite{STNM} (see fig. \ref{fig:2DSL_fra}). 
Since pressure does not alter the energy itself, 
this clearly indicates that the degree of frustration is a controlling factor of the fragility. 

It is interesting that the number density of pentagons has a distinct correlation with the growth of 
crystal-like bond orientational order and the fragility, indicating that pentagons disturb the growth of the 
correlation length of crystal-like order. 
These behaviours are essentially the same as that in the above case 
of polydisperse colloids and support our scenario.

Finally, we note that in this system crystal nucleation takes place 
preferentially in a region of high bond orientational order for $\Delta=0.6$, although it is a very rare event. 
Once a crystal nucleus is formed, the density in the nucleus increases and becomes higher than the surrounding liquid. 
We stress that before nucleation starts, there are few density fluctuations associated with antiferromagnetic bond orientational ordering. 
This suggests that translational ordering, or crystal nucleation, is initiated selectively inside regions of high crystalline 
bond orientational order (see also sec. \ref{sec:crystallization}), because of low interfacial energy.

\subsubsection{A case of water and water-type atomic liquids}

\begin{figure}
\begin{center}
\includegraphics[width=6cm]{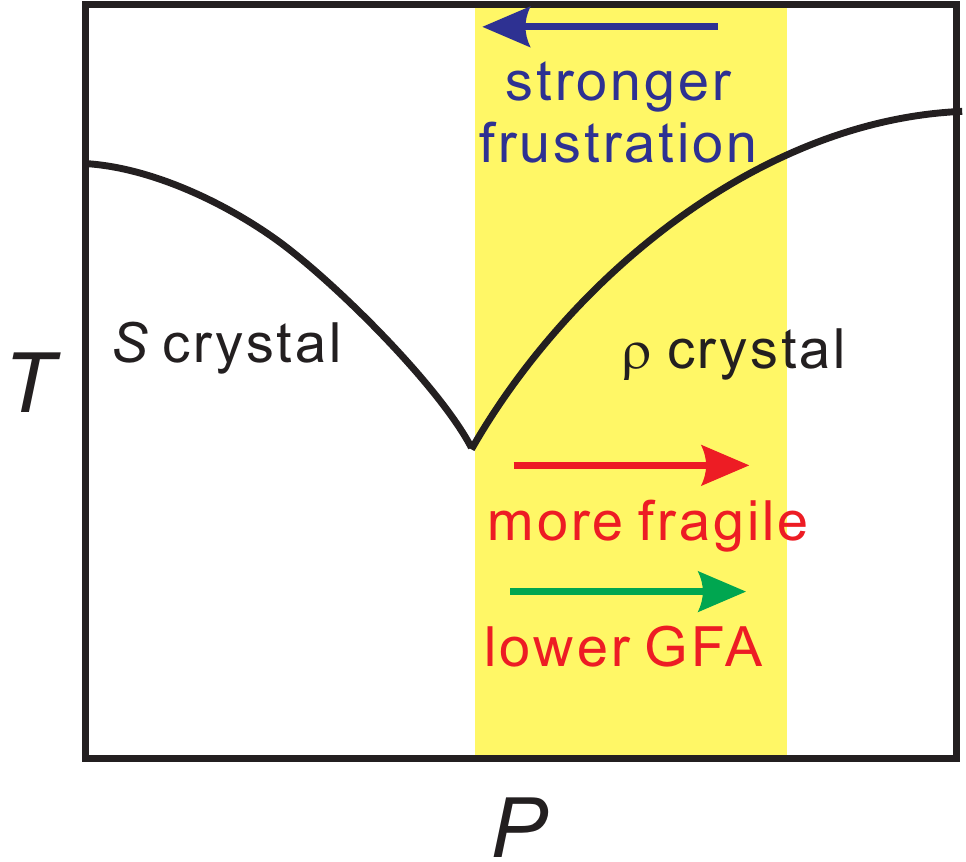} 
\end{center}
\caption{(Colour on-line) Schematic phase (or state) diagrams for water-type liquids. 
$S$ crystal is a crystal favoured by tetrahedral bond orientational 
ordering (note that tetrahedral symmetry is consistent with 
diamond-like order), whereas $\rho$ crystal is a crystal favoured 
by density ordering. The system volume expands upon crystallization 
of a liquid to $S$ crystal, 
whereas shrinks upon its crystallization to $\rho$ crystal. 
The area painted in yellow is a glass-forming region. 
Solid curves represent phase transition curves such as melting-point 
curves. 
}
\label{fig:WaterPD} 
\end{figure}

The V-shaped phase diagram schematically shown in fig. \ref{fig:WaterPD} (see also fig. \ref{fig:PD}) 
is common to water-type liquids such as water, Si, Ge, Sb, Bi, and Ga 
\cite{TanakaWPRB}. 
At low pressure, a system crystallizes into $S$ crystal, which 
is favoured by bond orientational (tetrahedral) ordering. 
It may be worth noting that below $P_{\rm x}$ there is almost no frustration since tetrahedral locally favoured structures having 
a lower energy than normal liquid structures are compatible to $S$-crystal. 
Reflecting the open structure of tetrahedral order, 
the system volume expands upon crystallization 
of a liquid to $S$ crystal, which leads to the negative slope of the 
melting point of $S$ crystal in the $T$-$P$ phase diagram. 
Under high pressures, a crystal into which a liquid crystallizes generally 
tends to have a more compact, denser structure. 
Thus, pressure destabilizes $S$-crystal and instead stabilizes 
$\rho$-crystal. Accordingly, the equilibrium crystal 
switches from $S$-crystal to $\rho$-crystal with increasing 
pressure at the crossover pressure $P_{\rm x}$, as shown in fig. \ref{fig:WaterPD}. 
In other words, the primary order parameter responsible for crystallization 
into the equilibrium crystal switches from the bond order parameter 
$S$ to the density order parameter $\rho$ there, or more precisely, bond order parameters linked to $\rho$-crystal. 

Above $P_{\rm x}$, thus, the melting point of $\rho$ crystal 
becomes higher than that of $S$ crystal. 
In this situation, we expect that locally favoured structures linked to tetrahedral ordering work as 
a source of frustration against crystallization to 
$\rho$ crystal because of the mismatch between the symmetries 
and thus helps vitrification. 
We note that the structure of the first shell is not enough to describe the locally favoured structure and that 
of the second shell may be necessary to specify it. 
Thus, water should tend to behave as an 
ordinary glass-forming liquid at very high pressures, which is consistent  
with the experimental indication \cite{Bett,Lang}. 
This tendency is difficult to explain in terms of the other existing theories of 
liquid-glass transition. 

We note that at a low pressure, 
local structural ordering simply helps crystallization to $S$ crystal since  
the symmetry of locally favoured structures is basically consistent with that of $S$ crystal. 
Since an open tetrahedral structure has a specific volume larger than 
a normal-liquid structure, the increase in the pressure 
decreases the number density of locally favoured structures, i.e., 
$\bar{S}$ (see eq. (\ref{eq:S})), which should lead to the decrease in the 
strength of frustration against crystallization to $\rho$ crystal. 
The situation is thus very similar to the case of 2D spin liquids under pressure discussed above. 
Our scenario tells us that the glass-forming ability and the fragility are 
positively correlated with an increase and a decrease 
in frustration, respectively. 
As shown in fig. \ref{fig:WaterPD}, we see high glass-forming ability 
and low fragility near the minimum of a melting curve 
\cite{ShintaniNP,STNM}. 
Thus our physical scenario predicts a high glass-forming ability 
around the minimum of a melting curve, a triple point 
\cite{TanakaWPRB,TanakaWJPCM}. 
Furthermore, in the glass-forming region, 
the glass-forming ability should decrease and the fragility should 
increase with an increase in $P$. 

\begin{figure}
\begin{center}
\includegraphics[width=8.5cm]{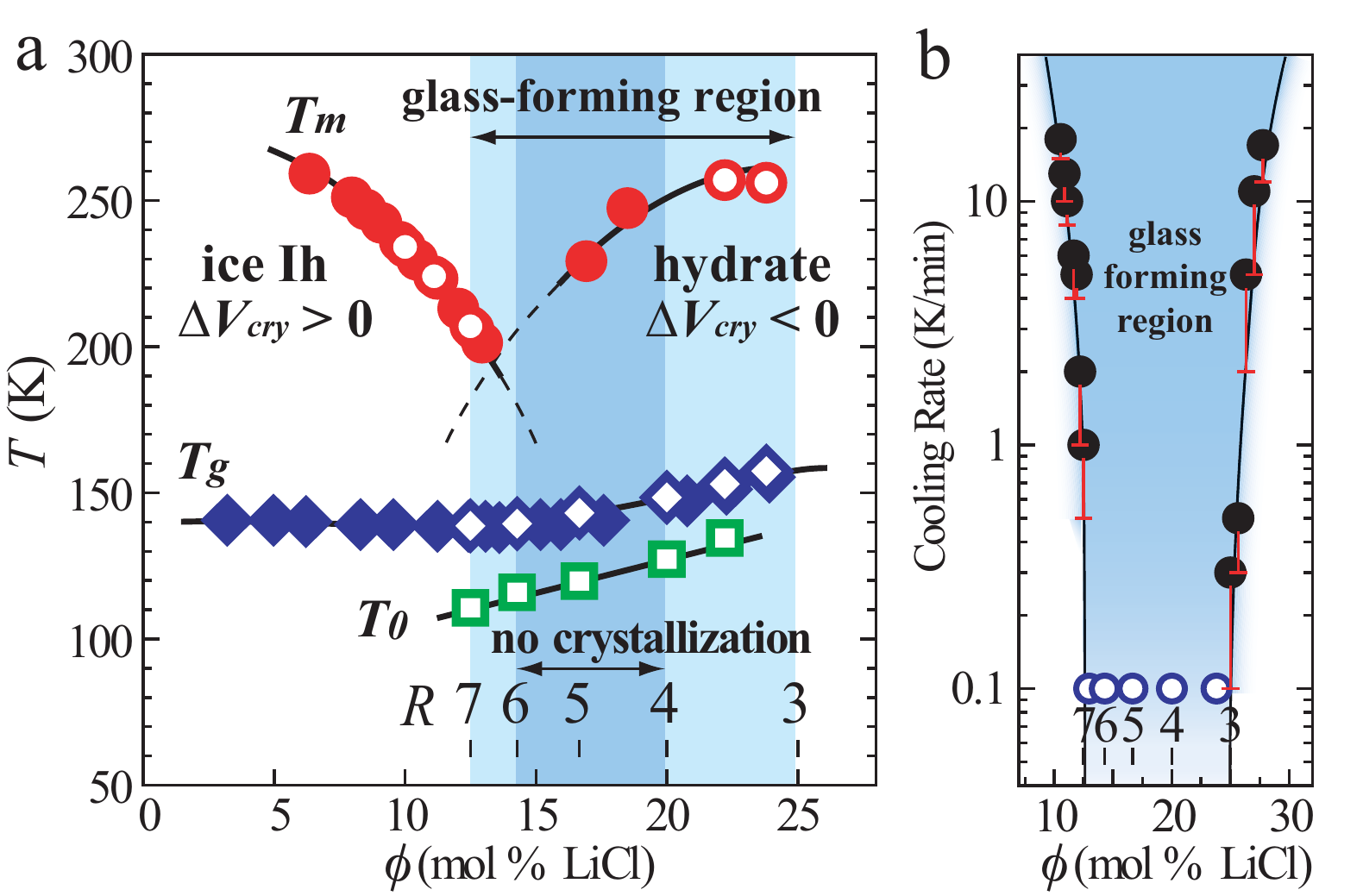} 
\end{center}
\caption{(Colour on-line) (a) Phase diagram of water/LiCl mixtures. Red circles: $T_{\rm m}$, 
blue diamonds: $T_{\rm g}$, green squares: $T_0$. 
Closed symbols are taken from \cite{Prevel}, which 
agree well with our data. 
(b) Critical cooling rate for vitrification, 
determined by DSC.  
Open circles means no sign of crystallization even for the slowest 
cooling rate (0.1 K/min). 
Solid curves are guides to eyes. 
The figure is reproduced from fig. 1 of ref. \cite{KobayashiPRL}. 
}
\label{fig:water_salt1} 
\end{figure}

\begin{figure}
\begin{center}
\includegraphics[width=8.5cm]{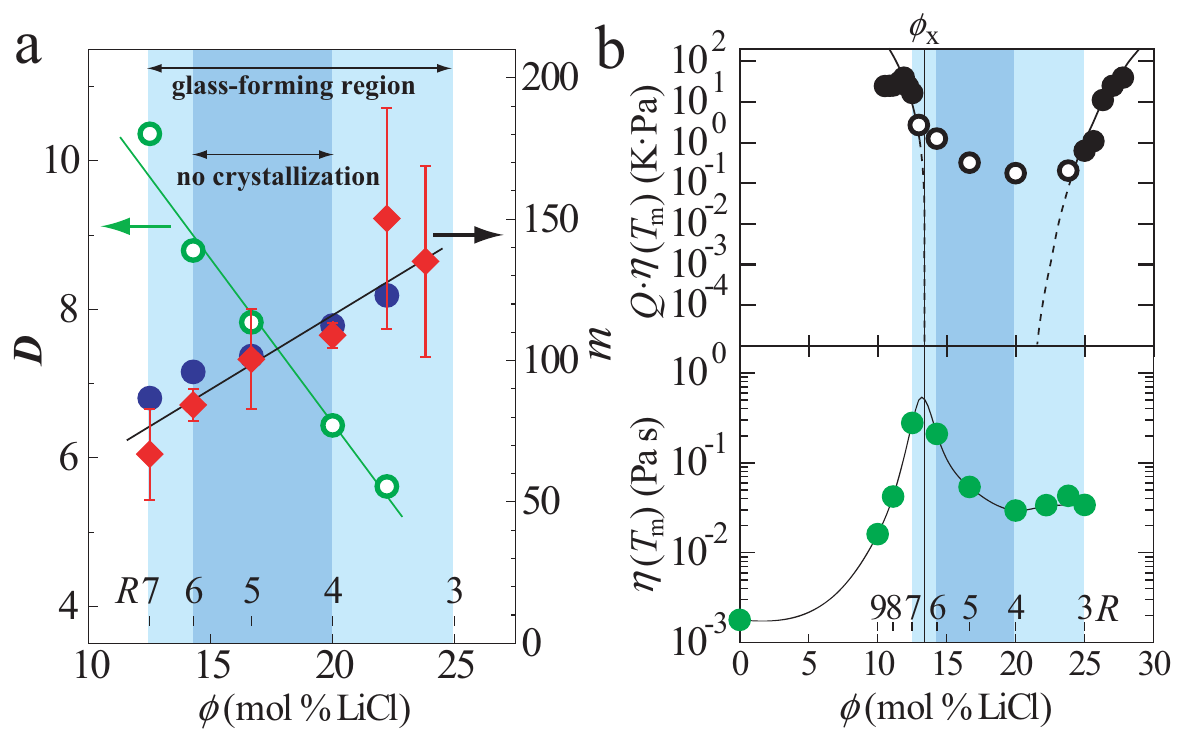} 
\end{center}
\caption{(Colour on-line) (a) $\phi$ (or $R$)-dependence of $D$ (open circles) and $m$. 
Diamonds are $m$ estimated from DSC, whereas 
filled circles are $m$ converted from $D$ (see text). 
Solid lines are guides to eyes. 
(b) Bottom: Dependence of $\eta$ at $T_{\rm m}$ 
as a function of $\phi$ [$\eta(T_{\rm m}(\phi))$]. 
Top: $\phi$-dependence of $Q \eta$. 
The figure is reproduced from fig. 4 of ref. \cite{KobayashiPRL}. 
}
\label{fig:water_salt2} 
\end{figure}

We confirm this scenario experimentally by using a water/LiCl mixture \cite{KobayashiPRL,kobayashi2011}. 
In this mixture, 
the addition of the salt leads to the decrease in local tetrahedral order of water and the increase in hydrated structures. Thus, 
the salt basically acts as the breaker of locally favoured tetrahedral structures as pressure does \cite{Leber364_1995}. 
Figure \ref{fig:water_salt1}(a) shows the phase diagram of this mixture as a function of 
the salt concentration $\phi$, which has 
a V-shape as the $T$-$P$ phase diagram of pure water. As can be seen in figs. \ref{fig:water_salt1}(a) and (b), 
the glass-forming ability becomes maximum slightly above $\phi_{\rm x}$, where 
the melting point has a minimum. Figure \ref{fig:water_salt2}(a) shows how the fragility 
index $D$ decreases with an increase in $\phi$ in the glass-forming region. 
Figure \ref{fig:water_salt2}(b) shows the $\phi$-dependences of the viscosity and the 
thermodynamic driving force of crystallization which does not include the kinetic factor. 
The results clearly indicate that the eutectic-like deep minimum of the melting 
point and the resulting slow dynamics upon crystallization there alone cannot explain 
the enhancement of the glass-forming ability, suggesting the importance of a thermodynamic 
factor (energetic frustration) in glass transition. 
Furthermore, this conclusion is also supported 
by the large discrepancy between 
$\phi_{\rm x}$ ($\sim 12$ mol\%) where the viscosity at $T_{\rm m}$ has a maximum and $\phi$ 
($\sim 20$ mol\%) for the maximum glass-forming ability. 
This discrepancy may be a consequence of the fact that local tetrahedral ($S$) ordering 
has random disorder effects only for $\rho$-crystal and not for $S$-crystal.

Consistent with our prediction, 
Molinero et al. \cite{Molinero} succeeded in vitrifying 
a monoatomic Si-like liquid by weakening the tetrahedrality 
in the Stillinger-Weber potential 
in their molecular dynamics simulations: 
The glass-forming ability increases 
around the triple point between diamond cubic (dc) crystal, 
body centred cubic (bcc) crystal, and liquid. 
Furthermore, Bhat et al. \cite{Bhat} succeeded in experimentally obtaining 
a monoatomic `metallic' glass of Ge at a pressure near the triple point (see also refs. \cite{Zhang1995,he1998electrical}). 

Our scenario provides a possibility to 
predict the glass-forming ability and fragility 
from the shape of the equilibrium phase diagram. 
The key is the relationship between 
global minimization of the free energy towards crystal 
and local minimization towards locally favoured structures. 
Depending upon the consistency of these two symmetries, 
locally favoured structures can be either a promoter  
of crystallization or its preventer. 
A physical factor making water so unusual among `molecular' liquids 
is the V-shaped $P$-$T$ phase diagram (see fig. \ref{fig:PD}): Water may be only such a molecule. 
Instead of changing pressure, we can add additives to 
a liquid to modify the number density of locally favoured structures, which 
opens up a new possibility to control the glass-forming ability  
and the fragility of a liquid in a systematic way, as shown next.

\subsubsection{A case of eutectic mixtures}

The same scenario may also be applied to various materials 
having a V-shaped phase diagram of a mixture as a function of the 
composition, e.g., metallic systems having 
a deep eutectic point (see fig. \ref{fig:eutectic}). 
Instead of changing pressure, for example, we can add additives to 
a liquid to modify the number density of locally favoured structures, which 
allows us to control the glass-forming ability 
and the fragility of a liquid in a systematic way. 
Typical examples are salt for water 
\cite{Angell_Sare,AngellCR,Prevel}, Na$_2$O for SiO$_2$ \cite{Angell_Sare}, and Au for Si \cite{Jakse}. 
The similar behaviour has also been reported for 
Ca(NO$_3$)$_2$-KNO$_3$ systems \cite{Senapati}.

It has been known empirically that the glass-forming ability of 
a multicomponent metallic glass former increases while approaching 
a deep eutectic point. This was explained by the fact that the viscosity of 
an equilibrium liquid just above the melting point is higher 
simply because the melting point is minimum at the eutectic point. 
As shown above, this kinetic argument may not be enough \cite{TanakaGJNCS}. 
We propose that there is another important factor: 
Near an eutectic point the two types of local structures linked to 
the two types of crystals compete, leading to 
stronger energetic frustration against crystallization, 
which helps glass formation \cite{TanakaMJPCM,KobayashiPRL,kobayashi2011}. 
Such behaviour was indeed observed by simulations \cite{Jakse_Si,Jakse}. 
In other words, the thermodynamic factors, more specifically, energetic 
and geometrical frustration, also play a crucial role 
in the enhancement of the glass-forming ability near an eutectic point 
in addition to the kinetic factors. 
We argue that a V-shaped phase diagram is in general a manifestation 
of underlying competing orderings. 

As discussed in the case of water-salt mixtures (see above and refs. \cite{KobayashiPRL,kobayashi2011}), 
the best glass-forming and lowest fragility region is not necessarily located at the eutectic point, 
but is a bit shifted from it. Such behaviour has recently been confirmed for several systems \cite{eutectic2010}. 

Here we note some cases which are not consistent with our prediction. Gong et al. \cite{eutectic2011} recently reported 
that in ZnCl$_2$-AlCl$_3$ and glycerol-water mixtures, 
there is no correlation between the fragility minimum and the eutectic point.  
The composition of the minimum fragility and the best glass-forming ability is located near 
pure ZnCl$_2$ and glycerol, respectively.  
We speculate that this is because these mixtures are a mixture of good and poor glass formers 
and thus mixing a poor glass former simply leads to the decrease of frustration effects 
on crystallization, which already exists in the pure substance before mixing the poor glass former. 
This may result in the monotonic decrease and increase in the glass-forming ability and the 
fragility, respectively. 

Our scenario may also apply to strongly correlated electronic systems 
(see, e.g., fig.~2 of \cite{Dagotto}), where glassy behaviour is also 
observed near the boundary between two competing phases. 
Our scenario may shed new light on a general mechanism of glassy behaviour 
itself on a wider perspective.

\subsubsection{A case of chalcogenide glasses} \label{sec:chalcogenide}

A chalcogenide glass is defined as a glass containing one or more chalcogenide elements, which 
belong to Group 16 in the periodic table, e.g., sulphur, selenium or tellurium. 
In these glasses, elements are covalently bonded and often regarded as network solids. 
Thus, chalcogenide liquids are typical network-forming liquids and categorized to strong glass formers. 
For example, sulphur based systems such as As-S and Ge-S easily form glasses in wide concentration regions. 
Semiconducting properties of chalcogenide glasses provide many applications. 
Unlike the classical chalcogenide glasses, modern ones such as GeSbTe, widely used in rewritable optical disks, 
are fragile glass-formers, which is useful for quick crystallization \cite{wuttig2006,wuttig2007,wuttig2008}. 
The relationship between vitrification and crystallization is thus a key to phase-change materials. 
A recent study suggested a similarity between amorphous and crystalline structures and pointed out 
its possible relevance to rapid crystallization \cite{matsunaga2011local}. 
We also point out that the correlation between the large fragility and the ability of rapid crystallization. 
It was also reported that near $T_{\rm g}$ there is evidence for decoupling of the crystal-growth
kinetics from viscous flow \cite{orava2012characterization}, matching the behaviour expected for a fragile liquid \cite{TanakaK}. 
These facts look consistent with our scenario (see also sec. \ref{sec:crystallization}). 
 
The glass-forming ability of chalcogenide systems such as Ge$_x$Se$_{(1-x)}$ and As$_x$Se$_{(1-x)}$ 
has been well explained by the so-called constraint 
theory, or the concept of mechanical rigidity \cite{Phillips1,Phillips2,Phillips3,AlexanderR,Thorpe,Boolchand}. 
The concept of mechanical rigidity is expressed by using the average coordination number 
$\langle r \rangle$ in covalent bonded inorganic glasses.  
This model can explain peculiar behaviours 
observed near $\langle r \rangle=2.4$, such as the maximum glass-forming ability, 
the minimum of fragility, the maximum of the Prigogine-Defey ratio, the maximum of 
the pronounced first sharp diffraction peak, and boson peak, in a coherent manner. 
Near $\langle r \rangle=2.4$, the number of constraints is almost the same as the number of degrees of freedom 
and a system is rigid but unstressed. This region is called ``a self-organized or intermediate state''. 
For $\langle r \rangle<2.4$, a system has floppy modes and is called ``floppy''. 
For $\langle r \rangle>2.4$, on the other hand, a system is over constrained and is called 
``rigid and stressed''.   

\begin{figure}
\begin{center}
\includegraphics[width=6cm]{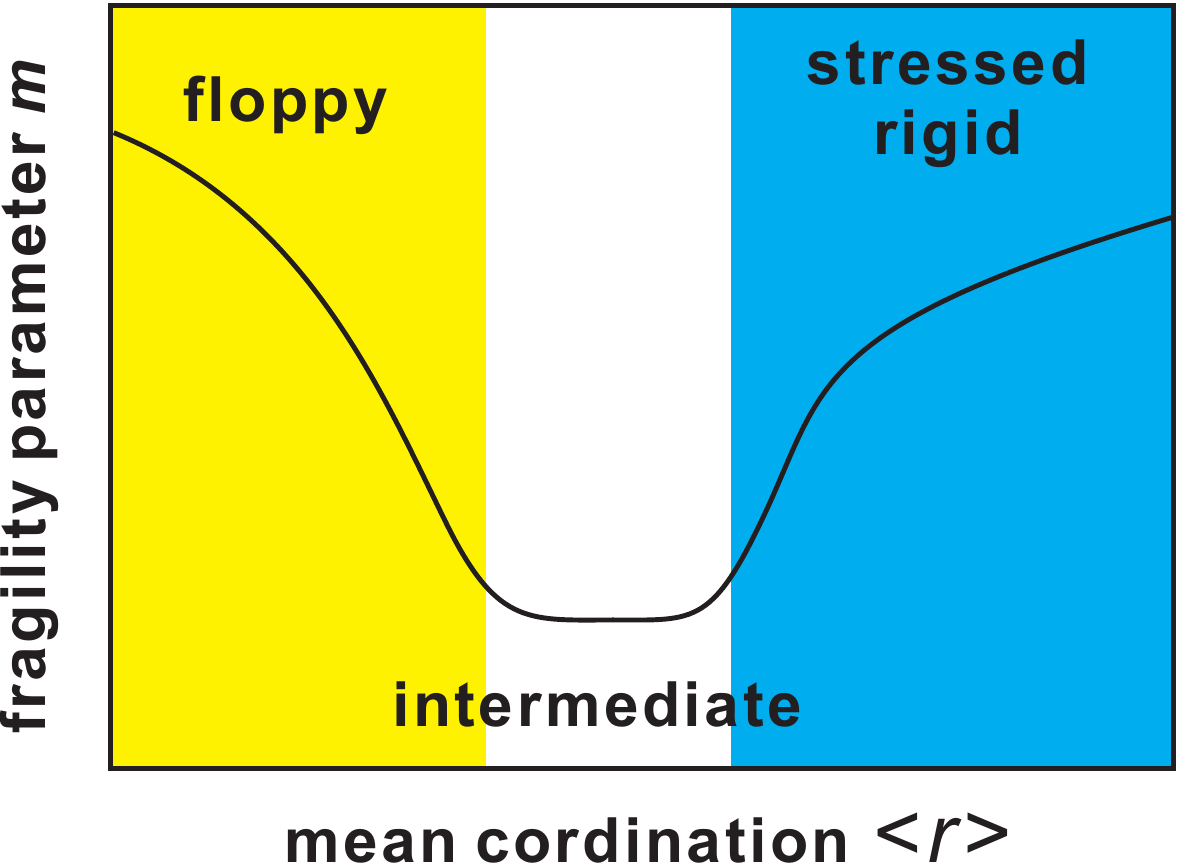} 
\end{center}
\caption{(Colour on-line) Schematic diagram showing the relationship between the fragility parameter $m$ 
and the mean coordination number $\langle r \rangle$. 
Near $\langle r \rangle=2.4$, there is a region called ``intermediate state'', 
where glass-forming liquids are strongest and have high glass-forming ability. 
The area painted in yellow is a floppy region, whereas 
the area painted blue is a rigid, stressed region. 
}
\label{fig:chalcogenide} 
\end{figure}

We propose that the condition of $\langle r \rangle \sim 2.4$, which 
corresponds to the intermediate phase, is a condition 
for maximizing the formation of locally favoured structures (short-range 
bond ordering), which is not consistent with the symmetry of the equilibrium crystal. 
According to our two-order-parameter model of liquids, 
it is this short-range bond order that produces frustration effects against 
crystallization (long-range density and bond orientational ordering), which leads to vitrification. 
A typical behaviour of chalcogenide glasses is schematically shown in fig. \ref{fig:chalcogenide}. 

Unlike the case of water, the dependence of the fragility and the glass-forming ability is rather 
symmetric about the mean coordination number $\langle r \rangle$.  
This may be a consequence of the fact that locally favoured structures are not consistent with 
crystals formed in both sides of $\langle r \rangle \sim 2.4$. 
In the case of water, on the other hand, locally favoured tetrahedral-type structures 
are consistent with $S$-crystal, but inconsistent with $\rho$-crystal. 
However, since the discussion here is highly speculative, this argument needs to be confirmed carefully.

\subsubsection{A case of metallic glass-forming systems} \label{sec:metallic}

The glass-forming ability of metallic liquids has been increased dramatically 
and good glass formers are called bulk metallic glasses \cite{WangR,Ma_review}: 
The critical cooling rate was reduced from 10$^6$ K/s to 1 K/s or even a slower rate. 
Here we consider the glass-forming ability of metallic liquids 
\cite{TanakaMJPCM,TanakaGJNCS}, which is related to the topic discussed here. 
Metallic glass formers usually possess two types of sources 
of frustration: multicomponent effects 
and competing orderings. 
We proposed that local icosahedral chemical 
ordering in a liquid state is directly linked to quasicrystal ordering \cite{TanakaMJPCM}. 
Crystal-like bond orientational ordering (e.g., fcc-like order) competes with 
local icosahedral ordering because of the mutual inconsistency of the symmetry, 
as in the case of hard spheres \cite{MathieuNM,TanakaJSP,TanakaNara}. 
The importance of structural features in metallic glass formers has also be emphasized from different viewpoints \cite{Miracle2004,Ma2006}. 

The importance of such competing bond orientational ordering has recently been confirmed both numerically 
and experimentally. 
Jakse and Pasturel \cite{Jakse2008,Jakse0} showed by ab initio molecular dynamics simulations of Cu-Zr alloy that 
a supercooled liquid is characterized by pronounced icosahedral short-range order, which 
increases the structural incompatibility of liquid and amorphous
states with competing crystalline phases. They suggested that good glass formers have
high degree of icosahedral short-range order already present in the liquid, consistent with our scenario. 
The similar conclusion was drawn by Wu et al. \cite{Wu2011}. 
L\"u and Entel also found by numerical simulations that 
a Ni-Si system has fcc-like MRCO, which is perturbed by the addition of Si atoms \cite{Entel2011}. 
Liu et al. found that glass transition can be seen as a process of increasing MRCO \cite{Liu2010}. 
Fujita et al. \cite{fujita2009} also reported that the atomic-scale heterogeneity caused by chemical short- and medium-range order plays a key role in 
stabilizing the liquid phase and in improving the glass forming ability of the multicomponent alloy. 
Hwang et al. showed that hybrid reverse Monte Carlo simulations of the structure of Zr$_{50}$Cu$_{45}$Al$_5$ bulk metallic glass, 
which incorporate medium-range structures from fluctuation electron microscopy data and short-range structure from
an embedded atom potential, produce structures with significant fractions of icosahedral- and crystal-like
atomic clusters \cite{hwang2012nanoscale}. 
If we focus on competing orderings in metallic glass formers, the situation looks very similar to those of 2D spin liquid \cite{ShintaniNP} 
and hard-sphere liquids in which `local' icosahedral order also develops \cite{MathieuNM}.  
In relation to this, it is worth mentioning a simulation study by Shimono and Onodera \cite{Shimono2012}. 
They reported that in a model binary mixture the stability of
the icosahedral cluster and the number density of the clusters in supercooled liquids
increases as the  atomic  size  difference between the constituent atoms increases. 
They also found that an increase in atomic size difference changes liquid
property from fragile into strong. This is quite consistent with our scenario. 

In addition to competing bond orderings, there are also random disorder effects induced by mixing multicomponents. 
The effects should be similar to those in a binary mixture \cite{KawasakiJPCM}. The coexistence of the two types of frustration effects 
and specific interactions between different species make the origin of frustration in metallic liquids a bit complicated.  

The above consideration provides a theoretical background to the Inoue's  three empirical rules for bulk metallic glass formation 
\cite{inoue_rule}, i.e., 1) multi-component 
consisting of more than three elements, 2) significant atomic size mismatches above 12 \% 
among the main three elements, and 3) negative heats of mixing among the main elements. 
Rules 1) and 2) can be regarded as random disorder effects, whereas rule 3) may be regarded as a condition for 
the effective formation of locally favoured structures, which are icosahedral structures in many metallic glass formers.  

\begin{figure}
\begin{center}
\includegraphics[width=6cm]{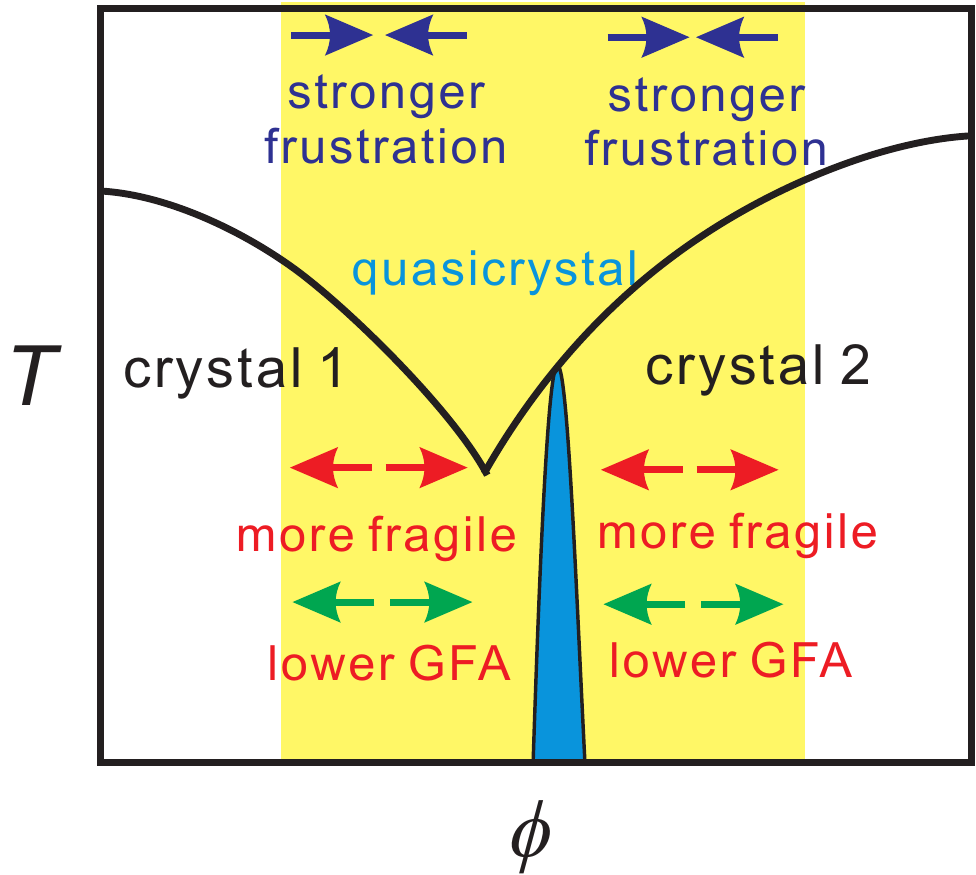} 
\end{center}
\caption{(Colour on-line) Schematic phase (or state) diagrams for eutectic mixtures with a quasicrystal formation region. 
Here $\phi$ is the concentration of one of the two components of a mixture. 
The area painted in yellow is a glass-forming region. 
Solid curves represent phase transition curves such as melting-point 
curves. 
}
\label{fig:eutectic} 
\end{figure}

Next we consider quasicrystal formation in bulk metallic glass formers. 
We now have much evidence that metastable quasicrystals are formed upon annealing of many  bulk metallic glass formers. 
Our scenario naturally explains why the composition region of bulk metallic glass formers is closely 
related to the region of quasicrystal formation: $\bar{S}$ is large in the quasicrystal-forming 
composition region, which leads to the high glass formability and the strong nature of liquids \cite{TanakaMJPCM,TanakaGJNCS}.  
As far as the primary crystallization of supercooled melt is the formation of intermetallic crystals, 
local icosahedral structures act as frustration and/or random impurity effects on crystallization. 
This situation is schematically shown in fig. \ref{fig:eutectic}. 
We expect that larger $\bar{S}$ (or, $\mbox{\boldmath$Q$}_{ico}$) increases the glass-forming ability and 
makes liquid stronger as far as it does not attain a long-range order, i.e., quasicrystal formation. 
Formation of local icosahedral structures reduces the Gibbs free-energy 
difference between the supercooled liquid and the crystal 
and also increases the interface tension, 
both of which make crystallization more difficult \cite{TanakaMJPCM,TanakaGJNCS}. 
This scenario was recently confirmed by careful experiments 
\cite{Kelton}. 
For a deeper supercooling, on the other hand, 
local icosahedral structures of the liquid increases, which is characterized by large $\bar{S}$: 
the liquid becomes more similar to the quasicrystal. 
Thus, the interface tension between liquid and quasicrystal becomes 
smaller, which makes quasicrystal formation easier there. 
This may explain quasicrystal formation upon heating of a glass \cite{TanakaMJPCM,TanakaGJNCS}. 

If a tendency of icosahedral 
chemical ordering is too strong, however, a stable quasicrystal can be directly 
formed instead of intermetallic crystals, or 
a liquid may become unstable against 
quasicrystal formation itself even upon normal cooling. 
This again leads to a poor glass formability. 
This corresponds to the composition region of direct quasicrystal formation 
upon cooling. 
This is because icosahedral ordering helps quasicrystal formation 
rather than disturbs it. This situation is similar to 
the case of water-type liquids, whose glass-forming ability 
is very poor at ambient pressure. 
This may lead to a rather complicated dependence of the glass-forming ability and fragility 
on $\phi$ for such a case, as schematically shown in fig. \ref{fig:eutectic}. 
The validity of this prediction needs to be confirmed.

\subsection{Origin of the strong nature of glass-forming liquids: Locally favoured structures or networks}

In literature, strong glass formers such as silica and germania, 
where covalent bonding plays a crucial role in the liquid structure,  
and molecular liquids with hydrogen bonding such as water are often 
called ``network-forming liquids''. 
This apparently looks natural, however, we argue that network-forming liquids 
are not appropriate to express the very nature of these liquids. 
We prefer to call them liquids with a strong tendency towards the formation of locally favoured structures. 

The important point is that the directional bonding is only transient 
and temporally fluctuating. The structural relaxation of these liquids 
is dominated by the bond lifetime. Even if there exists a percolated network 
in a system, the structural relaxation just takes place at the time scale of 
the bond lifetime. Unlike chemical or physical gel, percolation does not have any 
significant meaning in the structural or stress relaxation dynamics \cite{tanaka_lap}. 

As described above, we argue that the two state picture is more appropriate 
to describe the structure and dynamics of liquid. The symmetry of locally favoured structures 
is selected by the symmetry of directional interactions or the competition 
between attraction and repulsion. 
For example, a system where atoms interact with sp$_3$-type electronic functions 
favours the formation of tetrahedral structures. Hydrogen bonding in water 
also favours the same symmetry. For metallic systems, on the other hand, 
icosahedral structures are favoured. Complex molecules may form complex local structures 
to lower the free energy locally. 
Locally favoured structures have a longer relaxation time (lifetime)
than normal liquid structures, but they are also transient.  
So we argue that a basic spatio-temporal picture illustrated in fig. \ref{fig:SRO} is generic to any liquid. 

Here we emphasize that the network-forming ability is not a necessary condition to have 
a very strong liquid. We can see an example supporting this in our 2D spin liquid. 
As shown in fig. \ref{fig:2DSL_fra}, 2D spin liquid of $\Delta=0.8$ can be very strong 
at a low pressure: The structural relaxation time almost obeys the Arrhenius relation. 
In this case, the Arrhenius behaviour largely comes from strong energetic interactions, which also set the energy scale associated with 
frustration against crystallization.  
However, locally favoured structures do not form a network (see fig. \ref{fig:2DSL_pictures}(a)) since the pentagon structures (isolated structures)
cannot form it (please note that all spins in five particles forming a pentagon point outwards). We speculate that 
the situations in silica and water may be essentially the same as 
the case of 2D spin liquid: tetrahedral structures with a long lifetime are created and annihilated 
in the sea of normal liquid structures, which may also have rather a high degree of tetrahedrality.  
The validity of this picture needs to be confirmed.

\subsection{Relation between the location of the melting point minimum pressure and 
glass transition behaviour}

\begin{figure}
\begin{center}
\includegraphics[width=6cm]{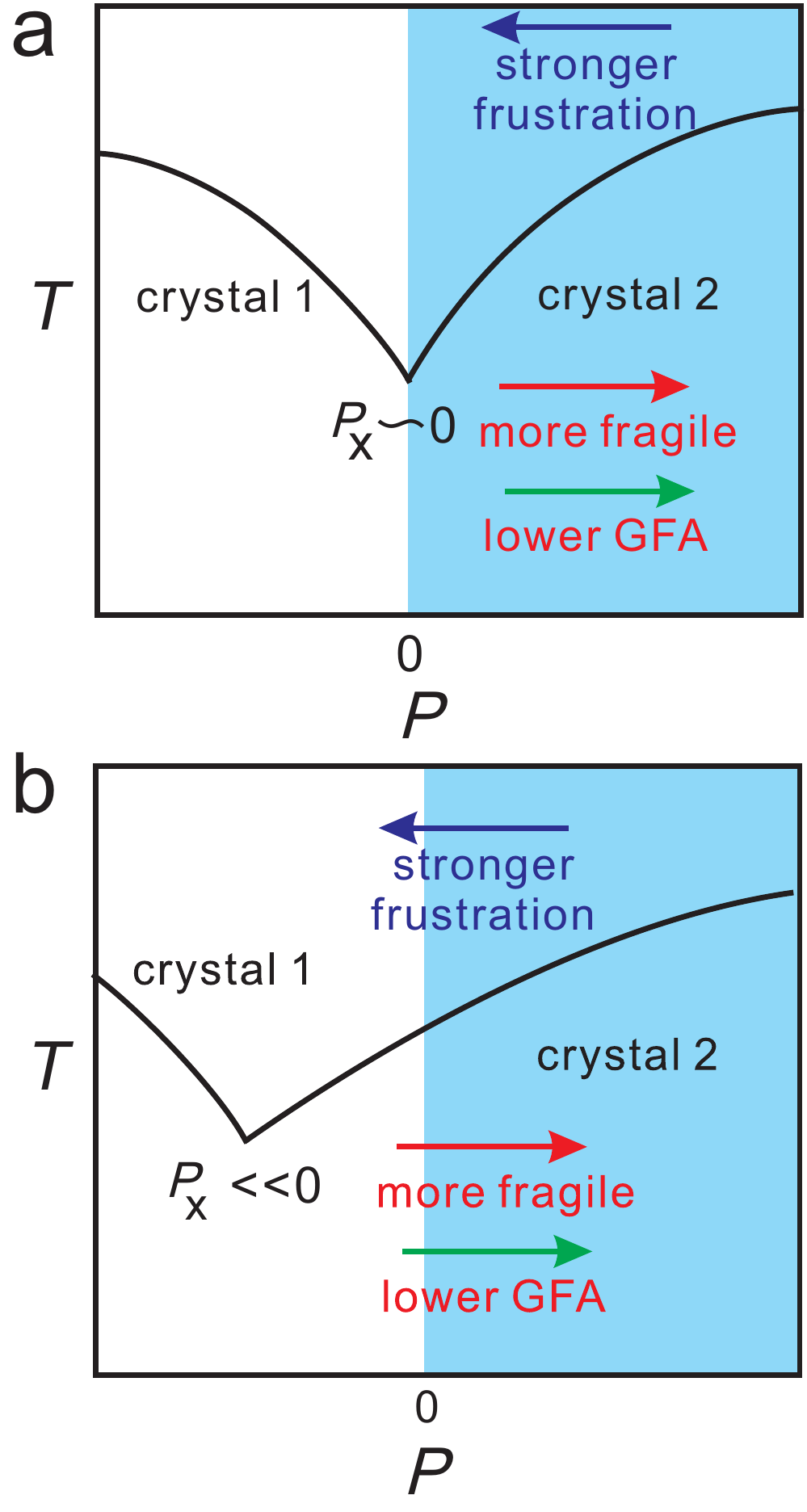}
\end{center}
\caption{(Colour on-line) Phase diagram and its relation to glass-forming ability and fragility. 
Here we consider the case of $\Delta v>0$. In this case, the number density of locally favoured structures, 
i.e., the strength of frustration, decreases with an increase in $P$.  
For $\Delta v<0$, the pressure dependence of the behaviour is opposite (see text). 
(a) Liquid like silica, whose the melting point minimum is located slightly at 
negative pressure ($P_{\rm x} \sim <0$). 
(b) Ordinary liquids, whose melting point minimum is located 
at a very negative pressure ($P_{\rm x} \ll 0$). 
Blue areas represent a positive pressure region. 
}     
  \label{fig:PT}
\end{figure}

Here we propose that there is a link between the location of the melting point minimum 
pressure $P_{\rm x}$ and glass transition behaviour such as glass-forming ability and fragility. 
We already showed that water-type liquids having a V-shaped phase diagram 
share common features, which are summarized in fig. \ref{fig:WaterPD}. 
Some liquids have even more complicated phase diagrams such as those shown in fig. \ref{fig:ph3}(a). 
In such cases, more than two types of bond order parameters compete and the glass-forming ability 
and fragility are expected to show oscillatory behaviour as a function of pressure, 
reflecting the oscillation of the strength of frustration. 
As schematically shown in fig. \ref{fig:PT}, many liquids have the $T$-$P$ phase diagram, where the melting point monotonically 
increases with pressure and only exhibits some kinks, which reflect the presence of polymorphs. 

We argue that liquids like silica and germania, which are categorized as very strong liquids at ambient pressure, 
may have a melting point minimum pressure $P_{\rm x}$ very near at ambient pressure (see 
fig. \ref{fig:PT}(a)). 
For example, the phase diagram of silica was studied by experiments \cite{jackson1976melting,ghiorso2004equation} 
and simulations \cite{saika2004phase,saika64supercooled}, 
which showed $P_{\rm x}$ is located near ambient pressure. 
The volume of each crystalline phase was also measured, including both the $T$ and $P$ dependences 
\cite{mao2001volumetric}. 
The phase diagram of GeO$_2$ also has a similar shape \cite{jackson1976melting}. 
These liquids have locally favoured structures of tetrahedral symmetry, 
which are stabilized by covalent bonding. 
Thus, the situation is basically similar to water-type liquids at a pressure near $P_{\rm x}$. 
The phase diagram of silica and its link to the glass transition behaviour looks remarkably similar to those of a water/LiCl mixture 
(see fig. \ref{fig:water_salt1}). 
However, the glass-forming ability and fragility of these liquids are, respectively, much 
higher and stronger than those of water-type liquids. Although a reason for this is not so clear 
at this moment, the trend itself is consistent with our scenario. The reason may be 
in the differences in the strength of bonds (covalent vs. hydrogen bonding) and the crystalline structures, 
which may result in the large difference in the strength of frustration. 
This problem needs to be clarified in the future.  

Similarly to the case of water, our model predicts the decrease of viscosity with an increase 
in pressure for liquids like silica and germania, reflecting the decrease of locally favoured tetrahedral structures with an increase in pressure (since $\Delta v>0$). Our model also predicts the decrease in the 
glass-forming ability and the increase in the fragility, which seem to be consistent with 
experimental results for silica \cite{AngellR,Kushiro} and germania \cite{KushiroGeO}. Further careful experimental 
studies are highly desirable. 

For ordinary liquids, we speculate that $P_{\rm x}$ is located at a very negative pressure 
or there is no minimum before a liquid is destabilized to the gas state by lowering pressure 
(see fig. \ref{fig:PT}(b)). 
In this case, the pressure dependence of the glass-forming ability and the fragility 
depends on the nature of locally favoured structures. 
If there is only a weak tendency to form locally favoured structures, there are very weak 
$P$-dependence of the glass-forming ability and fragility. 
This is the case for a liquid whose fragility is high at ambient pressure. 
In such a case, the volume and the temperature can be scaled, reflecting  
strong pressure-energy correlation (see below). 

A liquid which is rather strong at ambient pressure (e.g., B$_2$O$_3$ and glycerol) is expected 
to have a rather strong tendency to form locally favoured structures. 
For these liquids, whether the glass-forming ability and the fragility increase or decrease 
crucially depends upon the sign of $\Delta v$. 
For $\Delta v>0$, the increase in pressure decreases the number density of locally favoured structures, which should decrease the glass-forming ability and 
increase the fragility. Such behaviour is observed for silica. For $\Delta v<0$, on the other hand, the increase in pressure 
increases the number density of locally favoured structures, which should increase the glass-forming ability and decrease the fragility. Such behaviour 
was reported for triphenyl phosphite (TPP), where $\Delta v<0$, by Paluch and his coworkers \cite{Paluch}.  

The link between the phase diagram and glass transition behaviour discussed here may be useful 
for a qualitative level of understanding of the phenomena, but further studies are necessary for 
a more quantitative understanding.

\subsection{Effects of pressure on glass-forming ability and fragility for 
systems without the melting point minimum pressure in a positive pressure region}

We note that applying pressure induces the increase or decrease 
of the number density of locally favoured structures 
for $\Delta v<0$ or $\Delta v>0$, respectively. 
Provided that locally favoured structures are not consistent with the symmetry of the 
equilibrium crystal, this leads to the increase or decrease of frustration strength 
against crystallization for $\Delta v<0$ or $\Delta v>0$, respectively. 
For the former, the fragility decreases and the glass-forming ability increases 
with an increase in $P$. For the latter, on the other hand, the fragility increases and the glass-forming ability decreases 
with an increase in $P$. 

The relationship between these predictions and the recent findings 
on the density scaling expression for the relaxation time $\tau$ \cite{Roland1,Roland2,pedersen2008,Dyre_one} is interesting. 
For example, Dyre and his coworkers argued that a description with a single ‘order’ parameter applies to a good approximation whenever thermal equilibrium fluctuations of fundamental variables like energy and pressure are strongly
correlated. Results from computer simulations showing that this is the case for a number of
simple glass-forming liquids, as well as a few exceptions. 
They also conjectured that the relaxation time should follow the density scaling expression, 
$\tau=F(\rho^x/T)$, if and only if the liquid is ‘strongly correlating’, 
i.e., is described by a single ‘order’ parameter to a good approximation.

\subsection{Glass transition and jamming transition}

As already stated above, we argue that glass transition is governed by the same free energy  
as crystallization. This automatically means that glass transition, which is a nonequilibrium but 
thermal transition, 
is intrinsically different from jamming transition, which is athermal transition. 
The jamming state is defined as an isostatic state where mechanical forces are balanced. 
On the other hand, jamming transition in driven granular matter with continuous energy input is similar to thermal glass 
transition \cite{WT}, as far as there is no significant (correlated) energy dissipation \cite{head2010}.  
The effects of energy dissipation has recently been studied in detail \cite{kranz2010}. 

\begin{figure}
\begin{center}
\includegraphics[width=7cm]{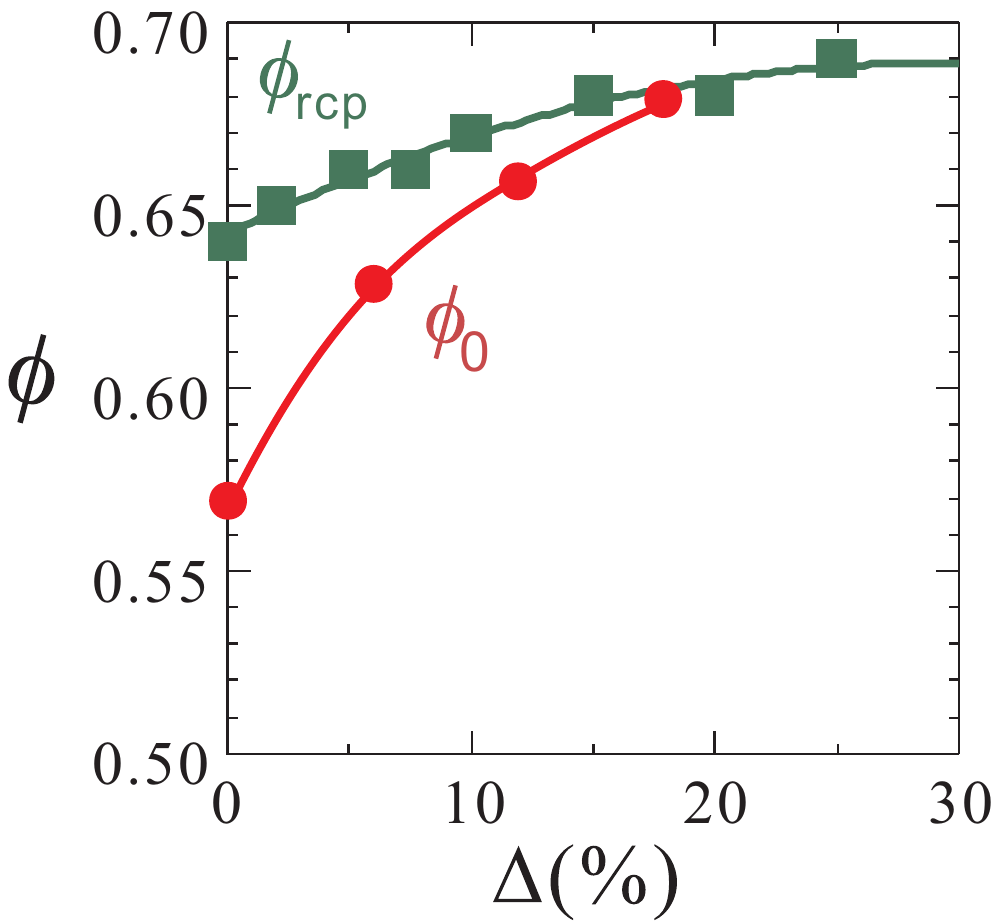} 
\end{center}
\caption{(Colour on-line) The dependence of $\phi_0$ \cite{Kawasaki3D} and $\phi_{\rm rcp}$ 
\cite{Sillescu2} on the polydispersity $\Delta$.  
This figure is reproduced from fig. 4 of ref. \cite{TanakaNara}. 
}
\label{fig:jamming} 
\end{figure}

In relation to this, we point out that the hypothetical ideal glass state is a state 
where the configurational 
entropy is vanished but considerable correlational entropy still remains. 
In other words particles do not necessarily freeze there unlike 
the state of random close packing, where 
particle positions are severely constrained geometrically (the nature 
of the isostatic jamming transition \cite{Nagel2010}). 
The importance of entropic contributions should make the glass transition 
of a thermal system distinct from the jamming transition of an (undriven) 
athermal system. 
For polydisperse hard-sphere-like systems, the ideal glass transition point $\phi_0$ deduced from the dynamics 
is quite different from the random close packing volume fraction $\phi_{\rm RCP}$ 
(see fig. \ref{fig:jamming}), 
although there might still remain a deep link between them \cite{Liu}. 
We should also note that the determination of $\phi_0$ involves a large extrapolation, and thus there remains some ambiguity on the 
location of $\phi_0$ or even its presence. 
This problem is also linked to the nature of ``amorphous'' 
packings of hard spheres \cite{Parisi2010,Jamming}. Our study suggests 
that packings in an amorphous state are not necessarily 
a perfectly random state (see ref. \cite{kurchan2011order} on exotic amorphous order), but possess 
bond orientational order for hard spheres as far as the polydispersity 
is not so large. 
We emphasize that bond orientational ordering in hard spheres is a consequence of dense packing under thermal fluctuations 
with the presence of excluded volume effects (entropic effects).  
On the nature of the hypothetical ideal glass state, see our discussion in sec. \ref{sec:Speculation}.  

\subsection{Glassy structural order and transient shear elasticity} \label{sec:transient}

When a liquid is cooled below its glass transition point, a nonzero shear modulus
emerges at least on the observation time scale $\tau_{\rm o}$. The origin of the emergence of this rigidity
at the glass transition point is one of the key unsolved problems. 
Shear modulus is directly linked to mechanical stability of materials. 
Recently this problem was studied theoretically \cite{yoshino2010,szamel2011}.  

\begin{figure}
\begin{center}
\includegraphics[width=8.5cm]{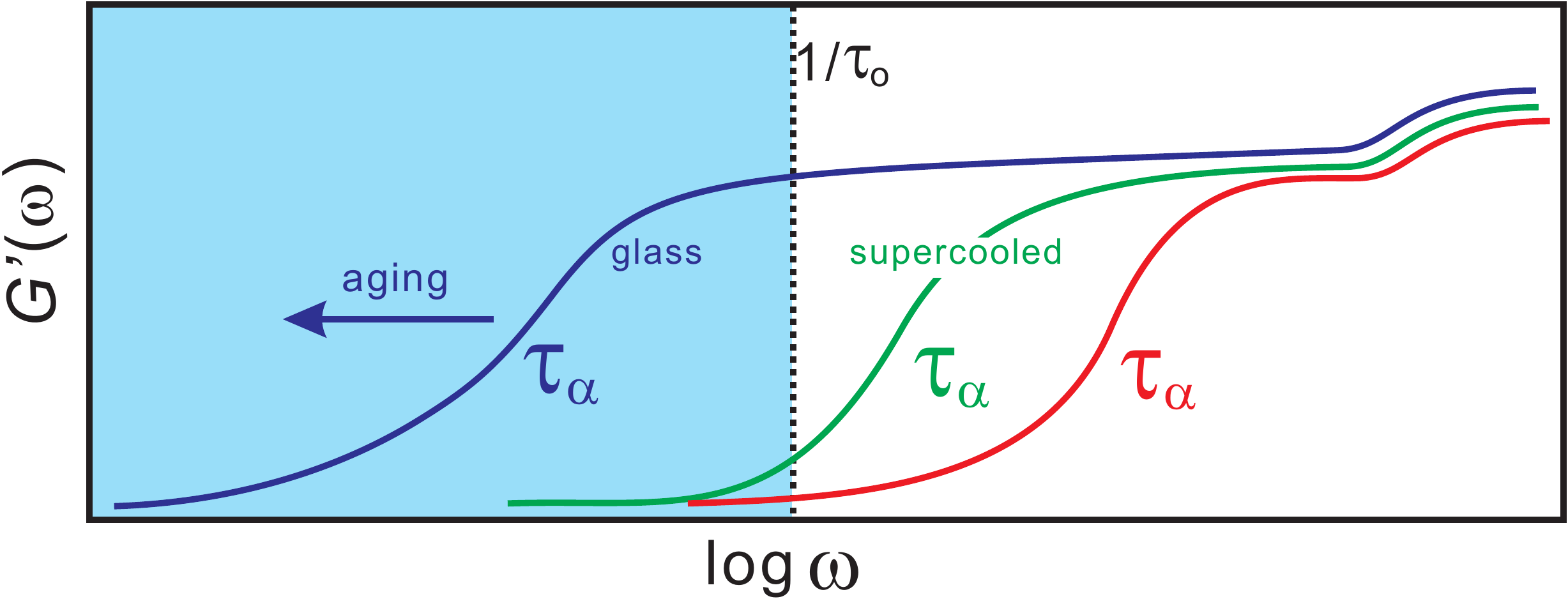} 
\end{center}
\caption{(Colour on-line) Schematic figure showing the dependence of the shear modulus $G'(\omega)$ on the angular frequency $\omega$. 
The three curves show $G'(\omega)$ for a high temperature supercooled liquid ($\tau_\alpha \ll \tau_{\rm o}$), 
a supercooled liquid near $T_{\rm g}$ ($\tau_\alpha < \tau_{\rm o}$), 
and a glass ($\tau_\alpha \gg \tau_{\rm o}$) from left to right.  Here $\tau_{\rm o}$ is the observation time scale. We note that the ratio 
$\tau_\alpha/\tau_{\rm o}$ is the so-called Deborah number. 
}
\label{fig:G(omega)} 
\end{figure}

Here we take a different standpoint that even a glass does not have nonzero static shear elasticity and the stress 
should eventually relaxes to zero if we can wait for a time longer than the structural relaxation time $\tau_\alpha$ 
(see fig. \ref{fig:G(omega)}). 
Note that $\tau_\alpha \gg \tau_{\rm o}$ for a glassy state. 
The static shear stress may emerge at the ideal glass transition point in this scenario, but this does not have a practical meaning 
since crystallization should take place before reaching this ideal glass state (see sec. \ref{sec:Kauzmann} ). 
This view is based on a physical picture that a glass is in a process of ageing towards the hypothetical equilibrium liquid 
(at least above $T_0$, or more strictly $T_{\rm LML}$). 
In the case of crystal, the static shear elasticity is a consequence of translational order: the displacement of one particle 
should accompany the displacement of an infinite number of particles. 
In our picture, the mechanical properties of a glass are characterized by the correlation length of glassy structural order as well as the amount of defects. 
Glassy structural order may be regarded as structures with low fluidity (or, high solidity) without voids or defects.  
Although glassy structural order whose characteristic lengthscale is finite cannot support `static' shear deformation in a direct manner, it can help in supporting it 
transiently. The state without defects (or, with an infinite correlation length) 
may exhibit `static' elasticity even without translational order. 
This implies that our glassy structural order parameter (including bond orientational order parameter) should be anticorrelated with fluidity. 

Here we should also consider glassy states prepared by strongly nonequilibrium processes, such as 
hyperquenched glasses and glasses made by densification of crystals. 
Hyperquenched glass may preserve the structure of a high temperature liquid and thus be very disordered. 
Thus, the correlation length of glassy structural order $\xi$ may be very short. 
This implies a rather small activation barrier $B \sim \Delta_{\rm a}(\xi/\xi_0)^\theta$. Thus, to realize 
$\tau_\alpha \gg \tau_0$, we need a low temperature, which is a condition to have a solid glassy state. 
This consideration also implies that the activation type relation between $\xi$ and $\tau_\alpha$ may be 
more appropriate than a power-law type relation between them (see sec. \ref{sec:Link}).

\subsection{Difference in the roles of bond orientational order between (i) glass transition 
and (ii) water-like anomaly and liquid-liquid transition} 

So far we consider the glass transition phenomena and the phenomena of water-like anomaly and liquid-liquid transition (LLT) rather 
separately. However, we also show that both phenomena are linked to bond orientational ordering in a liquid state. 
So we need to consider what is the relation between them. 
When we consider bond orientational ordering for water-like anomaly or LLT, we treat it as a scalar order parameter 
and thus it is not coupled to glassy slow dynamics. Critical phenomena associated with this ordering is similar to ordinary critical phenomena, 
where the dynamics of the order parameter itself slows down reflecting the growth of the characteristic length of the order parameter 
fluctuations. Modes which contributes to slow dynamics are restricted to low wavenumber ones satisfying $k<\xi^{-1}$. 
However, this type of critical slowing down does not accompany slowing down of motion at a particle scale: 
no link to solidity. Here we emphasize that the solidity here is not linked to the elastic strength, but the structural lifetime. 
The above situation is the case for liquid-liquid transition. 
On the other hand, in glass transition the medium-range bond orientational order with a tensorial nature, which is a manifestation of voidless packing, 
is regarded as long-lived stress-bearing order in the sense that it can bear shear stress transiently (see above). 
This directly leads to the slow dynamics not only at a particle scale, but also in a mesoscopic scale (see sec. \ref{sec:Critical}). 
In this sense, we speculate that the nature of the bond order parameter, whether the order parameter is coupled to solidity 
or not, plays a crucial role in whether it is coupled to glassy dynamics or not. 
This leads to a marked difference from ordinary critical phenomena, where the order parameter is not coupled to local mobility. 
We stress that the coupling is a consequence of frustration effects (see sec. \ref{sec:activation} and below). 
 
This problem may also be related to the origin of cooperativity in $S$ ordering and $\mbox{\boldmath$Q$}$ ordering, 
or the type of the order parameter coupling, reflecting the nature of the order parameter. In relation to this, we note that 
structural or stress relaxation can take place in a time scale shorter than the lifetime of MRCO, as discussed in sec. \ref{sec:decoupling}. 
Note that the lifetime is measured for a wavenumber corresponding to the inverse of the characteristic domain size. 
More importantly, structural configuration can relax with fluctuations of the order parameter over the lengthscale of a particle size and stress relaxation 
takes place via local reconfiguration of particles from anisotropic to isotropic one. 
This does not necessarily mean that such relaxation takes place microscopically. 
Coherent motion of mesoscopic solid-like structures over a particle size is enough for structural relaxation (see sec. \ref{sec:decoupling}). 
In relation to this, Widmer-Cooper et al. \cite{widmer2009localized} 
showed that more than half of particle movements that have contributed irreversibly 
to relaxation belong to strain-like, meaning that they involved the loss of no more than one of the initial neighbours \cite{harrowell2011}. 
This is markedly different from a simple cage-based picture, but rather consistent with what we discussed in sec. \ref{sec:rotation}.

In some sense, we may also say that the relation of bond orientational order parameter between LLT\&water anomaly and glass transition is very similar to 
the relation of density order parameter between gas-liquid transition and crystallization.   
The lower-temperature transitions are associated with the breakdown of translational and/or rotational symmetry for both cases, 
which is linked to the tensorial nature of the relevant order parameters. 

Unlike positional ordering, however, 
the nearly second-order nature of bond orientational ordering under `frustration' may lead to unconventional critical-like behaviour 
in a supercooled liquid (see sec. \ref{sec:activation}), although this needs to be confirmed. 
This leads to the drastic difference in dynamics between the two types of orderings. 
The ordering of $S$ is governed by a competition between energy and entropy as in ordinary phase transition and thus the relevant energy scale 
is that of entropy which is  $k_{\rm B}T$. On the other hand, glass transition is controlled by frustration between two competing orderings 
and the activation energy can become huge ($\sim \Delta_a(\xi/\xi_0)^\theta$) (see sec. \ref{sec:activation}).  
From this respect, polyamorphic transitions may be rather complicated since the order parameter governing liquid-liquid transition 
may be linked to fluidity as well. A recent study by Limmer and Chandler \cite{limmer2011} on water might be related 
to this issue. This remains a topic for future investigation.  

\section{Crystallization} \label{sec:crystallization}

Crystallization, more strictly, crystal nucleation in a supercooled liquid, is a process in which a new ordered phase emerges 
from a disordered state. It is important not only as 
a fundamental problem of nonequilibrium statistical physics, 
but also as that of materials science \cite{kelton2010,AuerR,SearR,GasserR,das2011statistical}. 
Crystallization has been basically described by the classical nucleation theory. 
However, nature provides intriguing ways to help crystallization beyond such a simplified picture.  
An important point is that the initial and final states are not necessarily 
the only players. This idea goes back to the step rule of 
Ostwald \cite{Ostwald}, which was formulated more than a century ago. 
He argued that the crystal phase nucleated from a liquid is not necessarily 
the thermodynamically most stable one, but the one whose free energy 
is closest to the liquid phase. Stranski and Totomanow \cite{Stranski}, 
on the other hand,  
argued that the phase that will be nucleated should be the one that has the 
lowest free energy barrier. 
Later Alexander and McTague \cite{Alexander} 
argued, on the basis of the Landau 
theory, that the cubic term of the Landau free energy favours nucleation 
of a body-centred cubic (bcc) phase in the early stage of 
a weak first order phase transition of a simple liquid (see sec. 2.3.1). 
Since then there have been a lot of simulation studies on this problem, 
but with controversy (see, e.g., \cite{tenWolde,tenWoldeJCP} 
and the references therein). Here we show a new scenario of crystal nucleation 
which focuses on structural ordering intrinsic to the supercooled state of liquid. 

\subsection{Density functional theory of crystallization}

First we mention a phenomenological approach based on the density functional theory 
beyond classical nucleation theory (see refs.~\cite{oxtoby1998nucleation,gunton1999homogeneous} for review), 
since it is related to our two-order-parameter model. 

Density functional theory treats the solid as an inhomogeneous fluid.
The starting point for a calculation of crystal nucleation rates is a Fourier
expansion of $\rho(\mbox{\boldmath$r$})$ in terms of the reciprocal lattice vectors $\mbox{\boldmath$G$}_i$ \cite{ramakrishnan1979first}:  
\begin{equation}
\rho(\mbox{\boldmath$r$})=\rho_\ell[1+\mu_s +\Sigma_i \mu_i \exp(i \mbox{\boldmath$G$}_i \cdot \mbox{\boldmath$r$})], \label{eq:DFT}
\end{equation}
where $\rho_\ell$ is the mean-field liquid density and $\mu_s$ is the
average density change on freezing. The parameters $\mu_i$ 
are the amplitudes that describe the periodic structure in the
crystal; they are zero in the liquid. The transition is thus
characterized by an infinite set of order parameters {$\mu_i$}
instead of the single parameter (the average density) 
characterizing the gas-liquid transition.
The saddle point is found as usual by minimizing the grand canonical potential functional with respect to $\rho(\mbox{\boldmath$r$})$ 
with an approximation that the density can be written as as a sum of Gaussians, centred at the lattice
sites of the crystal. 

Oxtoby and his coworkers \cite{oxtoby1994crystal,oxtoby1996nucleation} showed 
that the classical theory for the free energy of formation of the critical droplet
is found to exceed that obtained in the density functional calculation.
They introduced an order parameter that continuously distorts a crystal with fcc
symmetry into one with bcc symmetry, to allow for the possibility that
precritical bcc crystallites form which then transform to critical fcc droplets.
The latter had been found in an earlier simulation of a Lennard-Jones
system \cite{tenWolde,tenWoldeJCP}.
Their calculation of the free energy functional showed a
metastable bcc state close to the stable fcc phase. This metastable bcc phase
induces a saddle point which serves as the lowest free energy barrier
between the liquid and crystal, with the minimum free energy interface
passing close to this saddle point. This has significant consequences for
nucleation, in that a small critical droplet is largely of bcc structure at the
centre and evolves into the stable fcc structure as it grows. 
We note that the similar framework was also applied to crystallization of hard spheres \cite{granasy2003phase}. 

The above approach has some similarity to ours in the sense that both consider the presence of 
at least two order parameters (density and a structural order parameter which has a link to the crystal 
structure). At the same time, there is a crucial difference: their order parameter is linked to 
translational order, whereas ours is linked not to it but to bond orientational order. 
For example, a supercooled liquid locally has high bond orientational order, but no 
translational order ($\mu_i=0$), as we have shown in sec. \ref{sec:glass}. 
In our scenario, the liquid state prior to crystal nucleation is `not' homogeneous and quite heterogeneous.  
In the density functional theory, on the other hand, it is treated as completely homogeneous: 
There the liquid state is simply characterized by a constant density $\rho_{\ell}$ (see eq. (\ref{eq:DFT})).  
Thus, we emphasize that despite the apparent similarity, the physical picture 
is essentially different between the two scenarios.

\subsection{A supercooled liquid: a metastable state prepared for crystallization}

As we saw in the preceding section, a metastable supercooled liquid from which crystal nuclei emerge is not homogeneous, 
but already possesses significant crystal-like bond orientational order. This forces us to change the basic 
physical picture of crystallization. 
Recently experiments~\cite{schoppe,savage,iacopini} and simulations~\cite{KTPNAS,Kawasaki3D,snook2005,Schilling,bolhuis} 
have started to point out deviations from
the classical picture of crystallization, suggesting that this
process is more complex than the `one-step discontinuous' classical nucleation scenario. 

\begin{figure}
\begin{center}
\includegraphics[width=8cm]{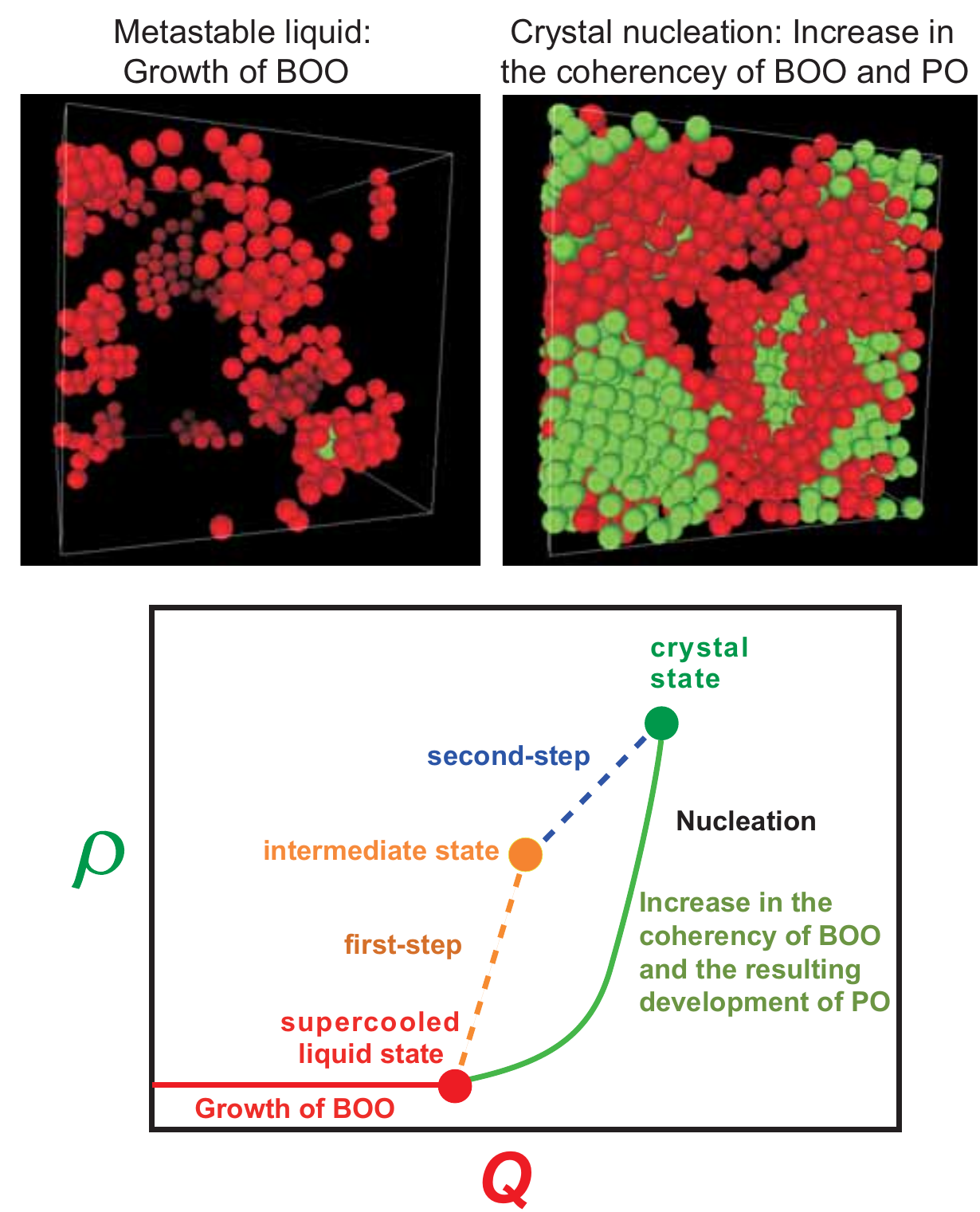} 
\end{center}
\caption{(Colour on-line) Top: A process of crystal nucleation in a supercooled liquid.  
Metastable liquid (top left): In a supercooled liquid state before crystal nucleation, 
a system attains only bond orientational ordering, which has a finite 
lifetime.  Crystal nucleation (top right): Crystal nucleation preferentially 
occurs in a region of high bond orientational order (BOO) by continuously increasing the spatial coherency of BOO 
and positional order (PO). 
Bottom: The microscopic kinetic pathway of crystal nucleation 
in a two-order-parameter plane. For simplicity, 
we consider only one type of bond orientational order $Q$. In reality, 
this process may occur in a multi-dimensional space. 
The two-step and the continuous scenarios of crystal nucleation are compared. 
According to the two-step crystallization scenario \cite{schoppe,savage,iacopini,snook2005,lutsko,Schilling,bolhuis}, the formation of precursors accompanies  
the density change from a liquid state and thus leads to a path along the $\rho$ axis (see orange dashed line). 
Such behaviour was not observed in our simulations 
at least in a mesoscopic scale \cite{TanakaNM,Kawasaki3D,KTPNAS,russo2011}.  
On a microscopic scale, on the other hand, there is continuous development 
of the coherency of crystal-like bond orientational order in high density regions, which accompanies a gradual 
increase in positional order (PO) and the resulting densification (see text) \cite{russo2011}. 
}
\label{fig:CRYTOP} 
\end{figure}

Recently it was suggested that crystal nuclei are not formed spontaneously
in one step from random fluctuations, but rather in a two step through preordered precursors 
of high density with structural order ~\cite{lutsko,Schilling}. 
This two-step crystal nucleation scenario now becomes very popular \cite{schoppe,savage,iacopini,snook2005,lutsko,Schilling,bolhuis}. 
However, Russo and Tanaka \cite{russoSM,russo2011} recently showed that the process of crystal nucleation does not consist of discrete steps but 
is rather continuous at the microscopic level: Crystal nucleation is a consequence of the continuous increase in the coherency of crystal-like bond orientational order 
in a high density region, which is already developed spontaneously in a supercooled liquid prior to crystal nucleation. 
The positional ordering follows the enhancement of the coherency of crystal-like bond orientational order. 
In other words, the latter is necessary for the development of the former. The difference between these two scenarios are schematically drawn in fig. 
\ref{fig:CRYTOP} (see also below).  
On the basis of our finding that a metastable liquid is structurally inhomogeneous, thus, we can say that one of the weakest points of
the classical nucleation theory and the density functional theory is the assumption that a supercooled liquid is in a homogeneous disordered state \cite{KTPNAS}.

For example, we found that a supercooled state of a hard-sphere-like liquid 
does not have a homogeneous random structure, contrary to 
the common belief, but has transient crystal-like bond orientational order together with icosahedral order \cite{TanakaNM,Kawasaki3D,KTPNAS,MathieuNM} (see fig. \ref{fig:compare3D}).  
This is also the case of 2D spin liquid (see fig. \ref{fig:KTTP}). 
The spatial correlation length $\xi_Q$ and the amplitude 
of fluctuations of the bond orientational order parameter 
`at least apparently' diverge towards the ideal glass transition volume fraction $\phi_0$, 
where the structural relaxation time $\tau_\alpha$ hypothetically 
diverges following the VFT relation: 
$\tau_\alpha \propto \exp(D\phi/(\phi_0-\phi))$, 
where $D$ is the fragility index. 
We showed \cite{Kawasaki3D,KTPNAS} that crystal-like bond orientational ordering accompanies  
little density change on average and should not be regarded as prenuclei or 
small crystallites \cite{Kawasaki3D,KTPNAS,MathieuNM,russoSM,russo2011}. 
Thus, we conclude that bond orientational ordering is an intrinsic structural 
feature of a supercooled state, which is also 
confirmed from the presence of such ordering even in 
a system of polydispersity $\Delta >7$\% for 3D \cite{TanakaNM,Kawasaki3D}, which 
never crystallizes in a simulation period. 

Here we explain our finding, which strongly 
suggests the importance of bond orientational 
ordering in crystallization \cite{Kawasaki3D,KTPNAS,russo2011,russoSM}. 
We stress that this particular crystal-like bond orientational order is linked to the rotational symmetry that will be broken upon crystallization. 
A crystal nucleus is formed by thermal fluctuations selectively inside 
regions of high crystal-like bond orientational order. The reason is as follows. 
Nucleation in a region of high crystal-like order leads to a small 
free-energy gain 
upon crystal ordering, but decreases the crystal-liquid 
interfacial energy drastically, which in total results in a substantial 
decrease in the nucleation barrier, i.e., the enhancement of the nucleation 
probability. This may be regarded as wetting-induced crystallization. 
This preferential crystal nucleation 
in regions of high crystal-like bond orientational order \cite{Kawasaki3D,KTPNAS}  
may be consistent with the above described view that bond orientational 
ordering plays a crucial role in crystallization.

Our physical scenario of crystallization can be summarized as follows \cite{KTPNAS,russo2011}. 
After a quench from an equilibrium liquid state to a supercooled state, 
medium-range bond orientational order 
whose symmetry has a connection to an equilibrium crystal structure 
(fcc or hcp in hard-sphere colloids with more weight in fcc \cite{MathieuNM,russo2011}) 
first develops as spontaneous thermal fluctuations. 
When high bond orientational regions accidentally have high local density as a consequence of thermal fluctuations,  
crystal nucleation is initiated with a high probability by accompanying the increase in the coherence of 
bond orientational order without a discontinuous density jump \cite{russo2011}. 
Here we note that although regions of high crystal-like bond orientational order does not have high density on average, 
some regions can have high density as a result of thermal fluctuations. 
Thermal density fluctuations allows a system to access a density of crystal nuclei. 
Thus, crystal nucleation always happens in a region of a supercooled liquid simultaneously having high crystal-like 
bond orientational order and high density, as shown in fig. \ref{fig:CRYTOP}. 
However, we stress that a factor triggering crystal nucleation is the former and not the latter: the latter is a necessary condition, 
but not a sufficient condition.  
The sequence of crystallization from melt induced by a temperature quench is thus described as follows: 
(i) an initial homogeneous equilibrium liquid at a high temperature $\rightarrow$ 
(ii) an `inhomogeneous' supercooled liquid with crystal-like bond orientational 
order fluctuations after the equilibration after the quench $\rightarrow$ 
(iii) a continuous increase in the phase coherency of crystal-like bond orientational order in a region of high density $\rightarrow$ 
(iv) the formation of a crystalline phase due to the development of translational order induced by the growth of crystal-like bond orientational order.
In the conventional scenario, step (ii) is replaced by 
`homogeneous disordered supercooled liquid' and step (iii) has not been considered. 
Furthermore, we emphasize that processes (i)-(iv) continuously take place at the microscopic level. 
The kinetic pathway in the $\rho$-$Q$ plane is schematically shown in fig \ref{fig:CRYTOP}. 

Since the Ostwald's seminal argument, intermediate states between 
the initial liquid and the final crystal state has been searched 
from the crystal side 
\cite{Ostwald,Stranski,Alexander,tenWolde,tenWoldeJCP}. 
However, our study demonstrates that it is crucial  
to consider hidden structural ordering in a supercooled liquid. 
We argue that the slowness of these structural fluctuations is also crucial for nucleation to efficiently take place. 

This hidden ordering in a supercooled liquid further suggests 
an intimate link between crystallization and glass transition. 
Namely, a supercooled liquid is intrinsically heterogeneous and, 
in this sense, homogeneous nucleation may necessarily be 
``heterogeneous''. 
The state of a supercooled liquid is prepared, or self-organized, for future crystallization. 
This feature can be seen in a glass-forming liquid: although crystal nuclei whose size exceed the critical nucleus size 
are usually not formed in a good glass former, small transient nuclei are spontaneously formed selectively in regions of high crystal-like bond orientational order, 
as can be observed in fig. \ref{fig:compare3D}(c) and (d). Such crystal nuclei should be regarded as a part of bond order parameter fluctuations. 

\begin{figure}[h]
\begin{center}
\includegraphics[width=8.5cm]{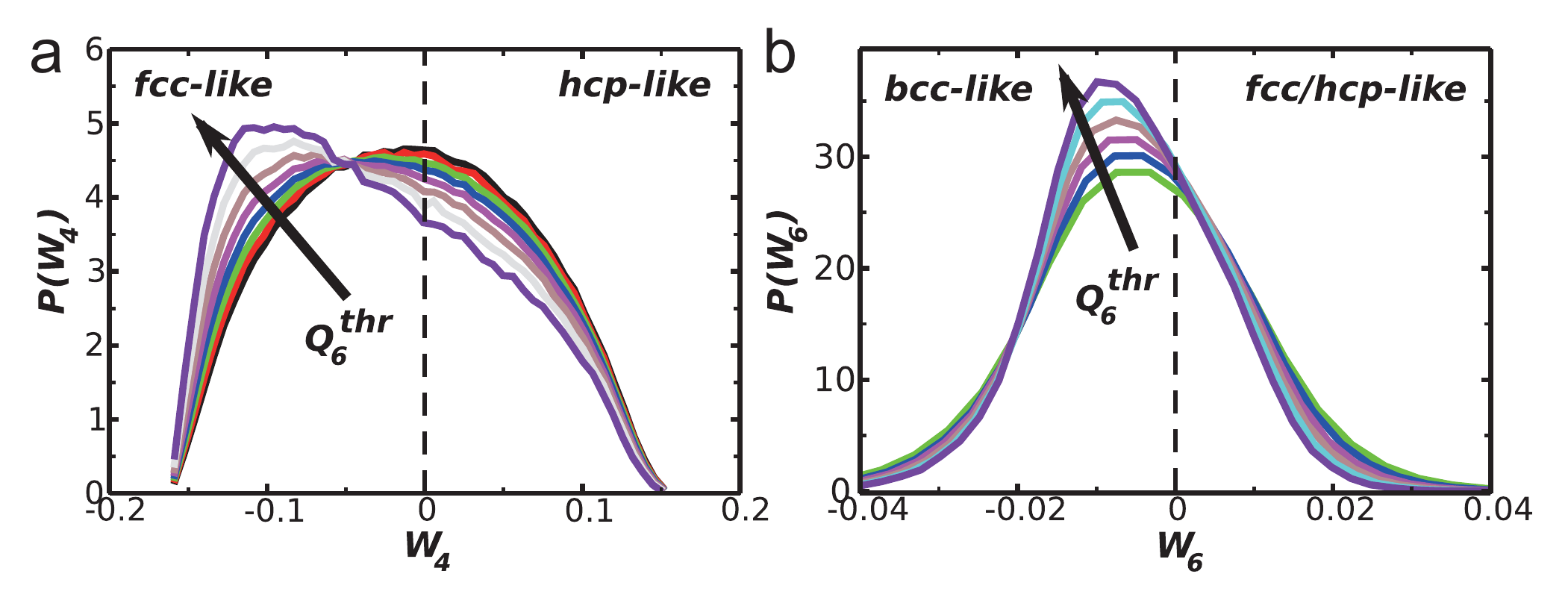}
\end{center}
\caption{(Colour on-line) Mechanism of polymorph selection.
(a) Order parameter $W_4$ for liquid particles
having $Q_6>0.25,0.26,0.27,0.28,0.29,0.30,0.31,0.32$ (the order is given by the arrow). 
As $Q_6$ increases, the regions
of high structural order in the liquid are characterized by a growing population of fcc-like clusters.
(b) Order parameter $W_6$ for liquid particles
having $Q_6>0.27,0.28,0.29,0.30,0.31,0.32$. As $Q_6$ increases, the distributions move to lower
and negative values of $W_6$, thus showing no preference for the bcc symmetry ($W_6>0$). 
Here $Q_l$ and $W_l$ are both coarse-grained ones. 
This figure is reproduced from fig. 3 of ref. \cite{russo2011}.}   
\label{fig:polymorphism}
\end{figure}

Recently Russo and Tanaka also revealed that the selection of polymorphs is made by crystal-like bond orientational 
order formed in a supercooled liquid \cite{russo2011,russoSM}. Thus, it is the state of a supercooled state that determines 
the fate of crystallization by imposing the initial condition on crystal nucleation (more specifically, 
symmetry selection).  Here we consider crystallization of hard spheres as an example to see the selection mechanism of 
crystal polymorphs. 
For hard spheres, it is known that fcc crystal has almost the same energy as hcp crystal in bulk. 
However, we found that fcc is 1.5 times more abundant than hcp. The similar result was also reported by Filion et al. \cite{Filion}. 
Crystals repeatedly appear, grow and melt as represented by the fluctuations in the bond orientational order parameter $Q_6$.
Since crystal nuclei appear from regions
of high bond orientational order (see fig. \ref{fig:compare3D}(c) and (d)), the study of such regions should provide important information on the forming nuclei.
It was found \cite{russo2011} that not only the precursor regions act as seed for crystal growth, but they also determine
which polymorph will be nucleated from them.
We used the order parameters $W_4$ and $W_6$, which are very useful in the detection of polymorphs. 
$W_6$ is a good order parameter
to distinguish between bcc crystals and close-packed crystals (hcp and/or fcc), since it is positive in the first case, and
negative for the latter. $W_4$ is instead good to distinguish between fcc crystals (for which it has negative values) and hcp
crystals (for which it has positive values).
Figure \ref{fig:polymorphism}(a) shows the probability distribution for the
order parameter $W_4$ in liquid regions having $Q_6$ higher than a fixed threshold, $Q_6^{\rm thr}$.
The $W_4$ distribution was obtained by considering
only liquid particles (crystal particles are not included in the histogram)
in the metastable state (before the critical nucleus is formed),
and the $Q_6^{\rm thr}$ threshold values are always within the liquid
distribution. While the metastable liquid has on average a symmetrical distribution around
$W_4=0$, fig. \ref{fig:polymorphism}(a) reveals that the high $Q_6$ regions have a predominant contribution from
negative values of $W_4$, which correspond to the fcc symmetry. 
Since we have shown that crystals form from particles of high $Q_6$, the following scenario emerges for
the nucleation of hard-sphere crystals: the supercooled melt develops regions of high orientational order,
whose symmetry favours the nucleation of the fcc phase (fig.~\ref{fig:polymorphism}(a)).
Figure \ref{fig:polymorphism}(b) plots the probability distribution for the
order parameter $W_6$, showing that indeed the regions of high $Q_6$ display no preference for the bcc symmetry (characterized by $W_6>0$).

\begin{figure}
\begin{center}
\includegraphics[width=8cm]{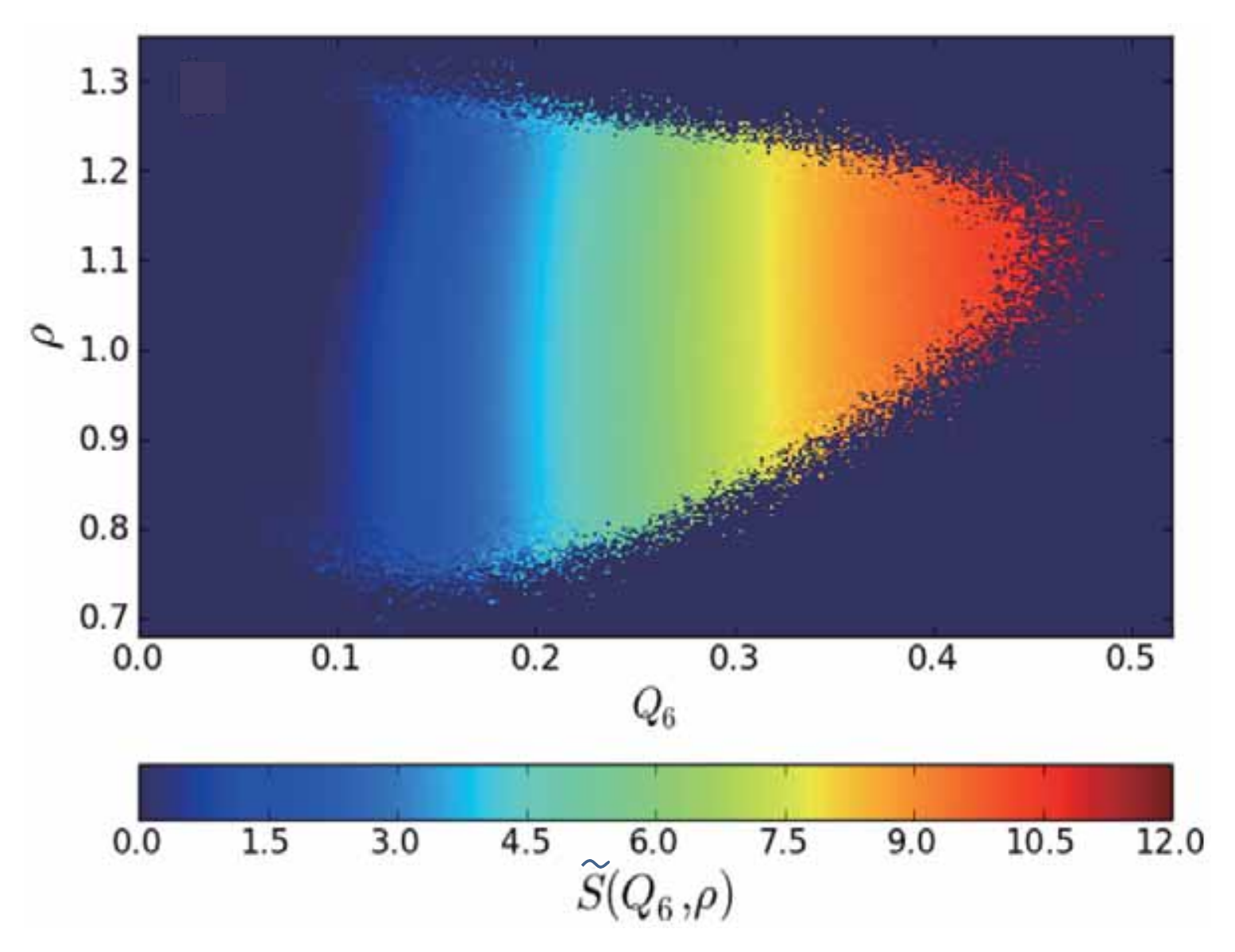}
\end{center}
\caption{(Colour on-line) Probability density for the structural order parameter $\tilde{S}$ in the $(Q_6,\rho)$ plane. 
The structural order parameter $\tilde{S}$ expresses the number of connected neighbours in a continuous way (for its definition, 
see ref. \cite{russo2011}). 
The number of connected neighbours
grows continuously from $0$ to $12$ from the fluid to the crystal phase. 
This figure is reproduced from fig. 2 of ref. \cite{russo2011}. 
}
\label{fig:rho_S_cry}
\end{figure}

To summarize, crystallization starts from $\mbox{\boldmath$Q$}_{\rm CRY}$ 
ordering and then density ordering (positional ordering) comes into play later \cite{russo2011}:  
Microscopically, crystallization starts from locally high density regions inside 
the regions of high bond orientational order, both of which 
are spontaneously formed by thermal fluctuations \cite{russo2011}. 
We note that density fluctuations whose amplitude is determined by isothermal compressibility $K_T$, 
can often allow a system to locally access the lower bound of crystal density. 
The importance of locally high density regions as precursors was also pointed out by ref. \cite{Schilling}. 
However, our study \cite{russo2011} shows that high local density is a necessary condition for crystal nucleation, but not 
a sufficient condition. On a microscopic scale it is bond order parameter 
and neither density nor translational order that triggers crystal nucleation. 
This can be clearly seen in fig. \ref{fig:rho_S_cry}: 
(1) Contour lines are almost parallel to the $\rho$ axis signalling
that crystallization is promoted mostly by bond orientational order. (2) Regions of high $\rho$ contain particles in a range
of environments from fluid-like to crystal-like, which means that density fluctuations alone are not sufficient to promote crystallization. 
Our finding is markedly different from the conventional view based on macroscopic observation where we can see 
a discontinuous change in the density upon crystal nucleation. This clearly indicates the crucial role of bond orientational ordering 
in crystallization. 

\begin{figure}
\begin{center}
\includegraphics[width=7cm]{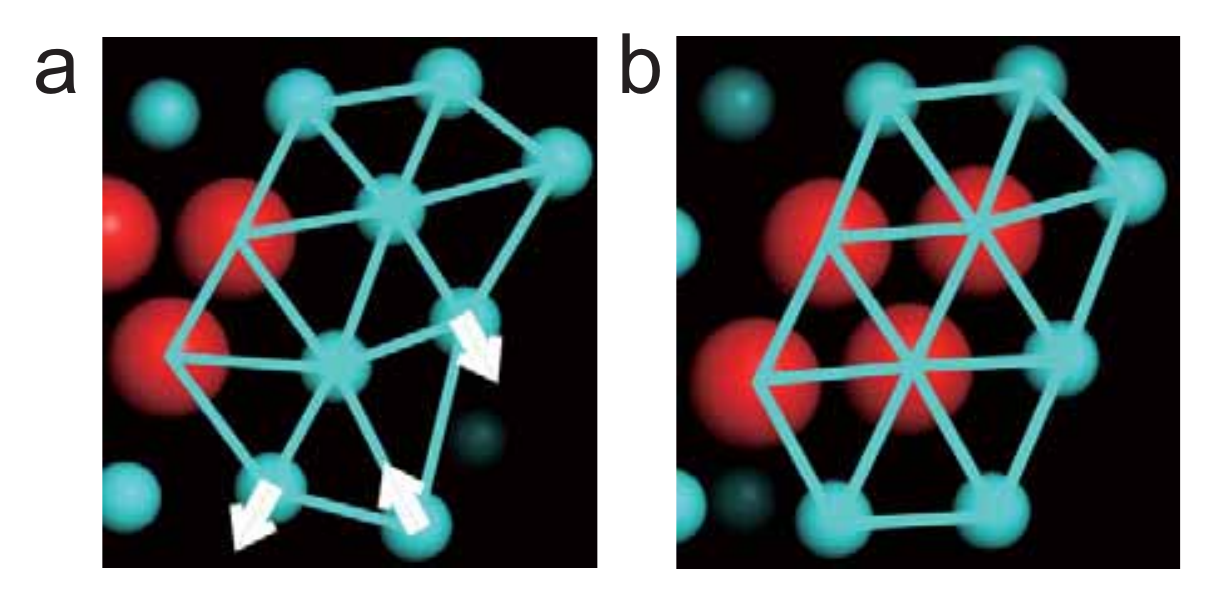}
\end{center}
\caption{(Colour on-line) An example of elementary particle rearrangements triggering crystal nucleation in hard spheres. 
The transition from fluid-like to crystal-like structures can happen at constant density, and can be rationalized
by the small cage rearrangements, which are sufficient to promote the transition with very little density change. 
This figure is reproduced from fig. 2 of ref. \cite{russo2011}. 
}
 \label{fig:elementary}
\end{figure}

Crystal nucleation is triggered by the enhancement of the phase coherence of bond orientational order 
in high density regions in a metastable liquid and then translational order follows afterwards \cite{russo2011}. 
An example of such local crystallization without accompanying density change is illustrated in fig. \ref{fig:elementary}: 
Here small displacement of particles (white arrow in fig. \ref{fig:elementary}(a)) can induce the increase in the coherency of crystal bond orientational order 
(see fig. \ref{fig:elementary}(b)).  
This looks natural, considering that crystal nucleation starts from a very small size: It is difficult to define 
translational order for such a small region, since it is characterized by periodicity over a long distance. 
Translational order can be attained in the growth process of nuclei, but not in the nucleation process. 
The theory of crystallization may need to be fundamentally modified to incorporate these findings probably along the line described in sec. \ref{sec:densitybond}. 
How universal this scenario is to more complex liquids 
remains for future investigation, but our preliminary studies on soft sphere and water suggests the universality \cite{russoSM,russo2011}. 

In relation to this, we mention that a recent simulation study \cite{desgranges2011} 
showed a pivotal role played by liquid
polymorphism, through the formation of the low-density-liquid droplet, during
the crystal nucleation process in liquid Si, which has a liquid-liquid transition.   

Finally, we mention a recent work which suggests critical concentration fluctuations may help crystallization of protein solutions \cite{Daan1,talanquer1998crystal}. 
In analogy to this, we may view the above phenomenon  
as enhanced crystal nucleation (positional ordering) 
by critical fluctuations associated 
with another phase ordering (bond orientational ordering), 
if we regard glass transition as a sort of critical phenomena associated 
with bond orientational ordering, which is supposed to occur 
at the ideal glass transition point $T_0$ \cite{ShintaniNP,KAT,WT,TanakaNM}. 
Unlike the case considered in \cite{Daan1}, the order parameter is tensorial, which is the key to the 
selection of crystal symmetry (polymorph).

\subsection{Other interesting topics on crystallization}

Usually it is believed that growth of crystal in a glassy state 
is extremely slow because of its solid-like nature, or extremely slow transport \cite{DebenedettiB} 
and thus we can practically assume that a glassy state 
is stable against crystallization. 
For example, this is a basis for storing materials 
in a glassy state for a long period of time while avoiding 
crystallization. 
Contrary to this common belief, 
Greet and Turnbull discovered a striking phenomena in o-terphenyl \cite{greet1967} 
and later it was systematically studied by Oguni and coworkers \cite{OguniE,OguniE2}. 
They observed the growth of a crystal from the surface of a crystal 
previously formed above $T_{\rm g}$. 
They found discontinuous enhancement of crystal growth behaviour just below $T_{\rm g}$ for some molecular liquids. 
The increase of the growth rate across $T_{\rm g}$ is far more than one order of magnitude (see fig. 61). 
This phenomenon has remained as a mysterious phenomenon. 

\begin{figure}
\begin{center}
\includegraphics[width=8.5cm]{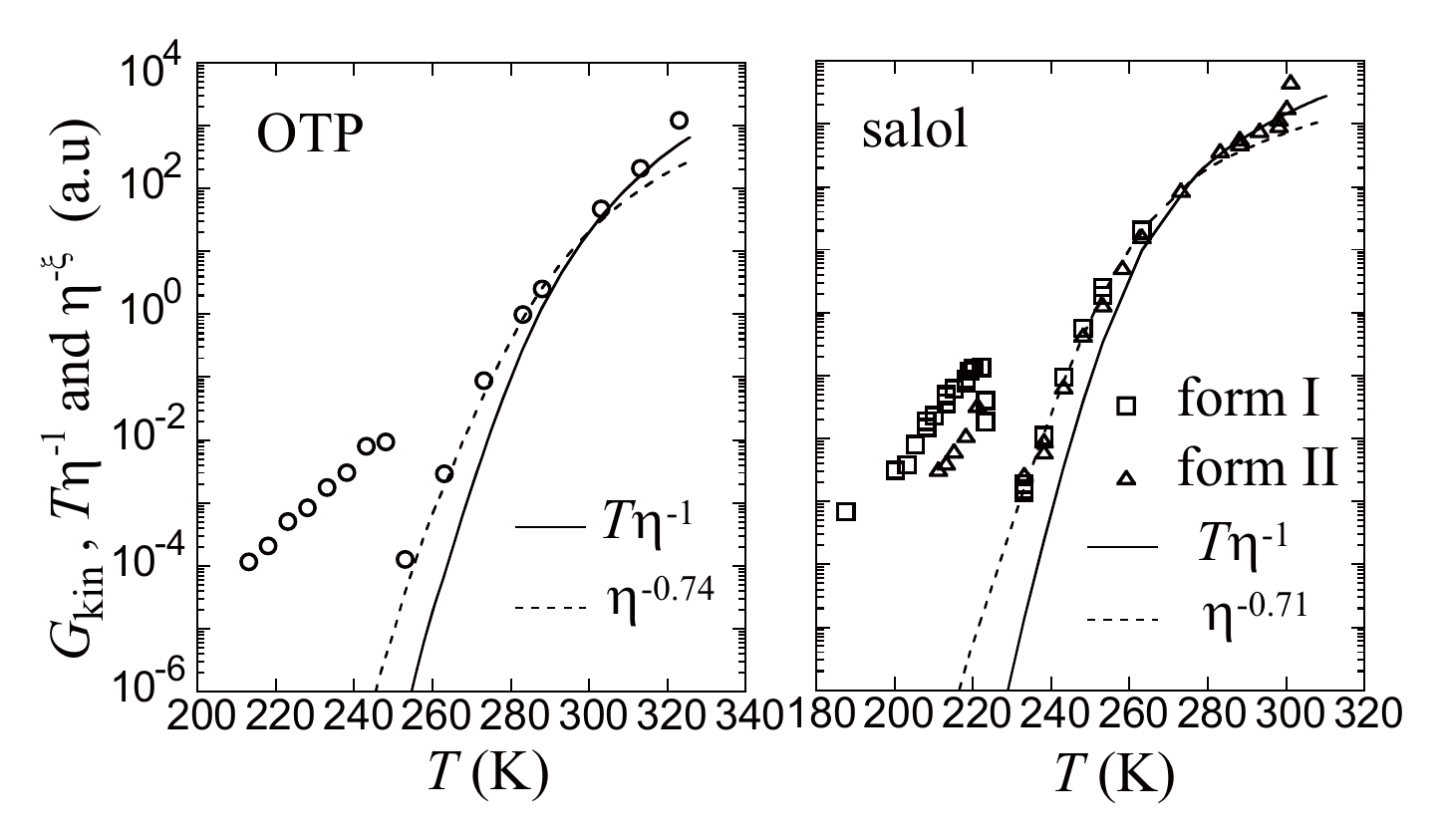}
\end{center}
\caption{$T$-dependence of the crystal growth rate divided by  the difference in the 
free energy between liquid and crystal, $G_{kin}$, which should be proportional to 
to the translational diffusion $D$ [left: OTP; right: salol]. 
For salol, squares are for form I crystal, and triangles for form II crystal. 
The solid curves represent $T/\eta$ ($\eta$: viscosity) calculated from 
the viscosity data, and the broken lines represent $\eta^{-\xi}$ 
with $\xi=0.74$ for OTP and $\xi=0.71$ for salol. 
The lines are vertically shifted to fit the data. 
This figure is reproduced from fig. 1 of ref. \cite{konishi}.
} 
\label{growth rate}
\end{figure}

Recently we proposed the following mechanism: the volume contraction upon crystallization and the resulting mechanical stress 
provides a region near the crystal-glass interface with large excess free volume, which results in the mobility 
increase at the growth front and leads to enhancement of the crystal growth \cite{TanakaK,konishi}. 
The role of high bond orientational order near the crystal surface \cite{russo2011} is also to be clarified. 
A different scenario was also proposed by Yu and his coworkers \cite{Yu2006,Yu2011}. 
This problem may also be related to crystallization of colloids in the glassy state \cite{CatesX}. 

Another interesting topic is wall-induced crystallization. 
Recently it was shown that heterogeneous nucleation of colloids near the wall is induced by a sort of wetting effects \cite{Schope}. 
This may be explained by combining our scenario with wall-induced bond orientational ordering \cite{watanabe2011}. Crystal nucleation may be 
helped by pre-existing crystal-like bond orientational order 
with wall-induced enhancement of the order (see sec. \ref{sec:wall}). 
Since translational order is also enhanced in the form of layering near a wall, it is interesting to study roles of these two types of orderings 
induced by a wall in wall-induced crystallization. 

Finally, we should mention the large discrepancy between theory and experiment on the crystal nucleation rate \cite{kelton2010,AuerN}. 
This has still remained as an unsolved mystery \cite{Filion,Filion2}. We note that on this point there was a mistake in the estimation of the volume fraction 
in our paper \cite{KTPNAS}, which led us to a wrong estimation of the volume fraction dependence of the nucleation rate \cite{kawasaki2011corr,Filion2}.

\subsection{Intimate link between crystallization and vitrification}

The above novel scenario of crystal nucleation strongly supports our physical picture (at least for the systems studied) that 
both liquid-glass transition and crystallization are governed 
by the same free energy, e.g., $f(\rho, \mbox{\boldmath$Q$}_{\rm CRY}, \mbox{\boldmath$Q$}_{\rm LFS})$, 
and the state of a supercooled liquid is prepared  
for future crystallization. 

The crystalline state can be described by translational order alone once it is formed. 
This is because translational ordering automatically accompanies long-range bond orientational ordering. 
However, this does not necessarily mean that bond orientational ordering is not important. 
We argue that bond orientational ordering is a key to the physical description of liquid and plays crucial roles in crystallization and glass transition. 
We emphasize that bond orientational order has local nature whereas translational order has global nature. 
Thus, it looks natural that crystal nucleation, which is a local event, is initiated by local bond orientational ordering 
and not by translational ordering. The increase in the coherency of the phase of bond orientational order may be prerequisite for 
the development of translational order. Glassy structural ordering may be viewed as ordering towards 
low free-energy local structures. Such structures have stress-bearing solid-like nature, 
which originates from its long structural lifetime.  
and thus has a deep link to low fluidity. 

\begin{figure}[h]
\begin{center}
\includegraphics[width=8.5cm]{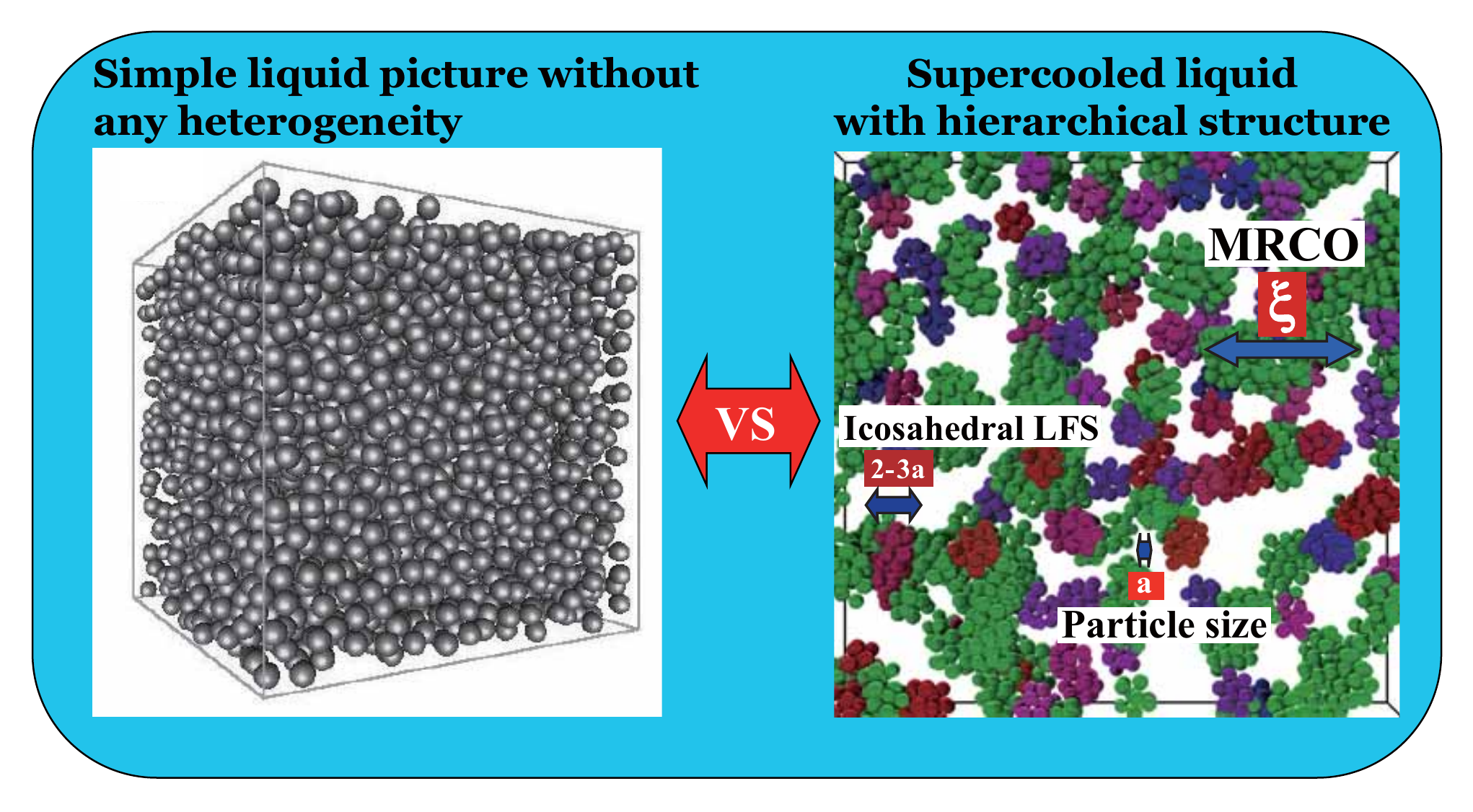}
\end{center}
\caption{ 
Schematic figure showing the difference between a classical picture of a supercooled liquid (the homogeneous liquid picture) and a picture based on our study 
(the spatio-temporally inhomogeneous liquid picture).  
For the latter we used a typical structure of a supercooled colloidal liquid \cite{MathieuNM}. }
\label{fig:softmatter}
\end{figure}

We can say that the rotational symmetry which is going to be broken upon crystallization 
is already broken `locally' in a supercooled liquid. Growth of its spatio-temporal fluctuations under frustration is an origin of glassy slow dynamics and 
dynamic heterogeneity. Furthermore, crystal-like bond orientational order triggers crystal nucleation with a high probability 
if regions of high order can reach a density required for crystallization by spontaneous thermal density fluctuations. 
Stronger frustration against crystallization leads to the lower probability of crystal nucleation since it 
reduces the degree of crystal-like bond orientational ordering. 
This not only suggests the intrinsic link between glass transition and crystallization, but also indicates 
that a supercooled liquid is `not' in a homogeneous state, but 
has mesoscopic spatio-temporal structures (see fig. \ref{fig:softmatter}). 
Thus, a supercooled liquid is not a simple liquid, but rather should be regarded as a soft-matter-like complex fluid. 

When the strength of frustration on crystallization is so strong that 
crystallization must involve phase separation, 
glassy order no longer has a direct link to the symmetry of the crystal, 
but may still be associated with low local free-energy configurations \cite{TanakaNara,TanakaJSP}. 
The validity of this physical picture is to be checked carefully for various glass-forming systems. 

We emphasize that (1) we cannot forget effects of bond orientational ordering associated with 
crystallization when considering glass transition at least for single-component liquids and 
(2) we cannot forget glassy structural ordering when considering crystallization.  
Namely, there is an intimate link between crystallization and vitrification 
at least in quasi-one-component systems (type I glass formers) (see fig. \ref{fig:cry_glass}). 
We argue that disturbance or frustration against crystal-like bond orientational order may be enough to prevent 
crystallization. 

\section{Summary and open questions}

In this review, we demonstrate that liquid is not in a random disordered, homogeneous state, 
but has local and mesoscopic structural orders, which can be characterized by bond orientational order parameters in many cases. 
Such orderings are a consequence of many-body correlations, particularly, bond angle correlations, which have not be properly considered for the physical description of the liquid state. 
There are two sources for such bond orientational correlations: energetic directional bonding and 
packing-induced constraint for bond orientation.   
On the basis of this picture, we have described our two (multi)-order-parameter model of liquid, 
which may describe water-like thermodynamic anomalies of liquids, liquid-liquid transition, liquid-glass transition, and crystallization 
in a coherent and unified manner. We argue that all these phenomena can be described by the same free energy, 
which implies that their unified description may be possible. 
We also discuss the relationship between these phenomena on the basis of our model. 

In particular, we put focus on spontaneous bond orientational ordering in liquid, 
which plays important roles in water-like anomalies, liquid-liquid transition, glass transition, 
crystallization, and quasicrystal formation. The number density of locally favoured structures 
plays a significant role in water-like anomaly and the cooperative excitation of 
locally favoured structures is the origin of liquid-liquid transition. 
In these cases, the order parameter can be treated as a scalar, the rotationally invariant form 
of the relevant bond orientational order parameter. 
On the other hand, the symmetry of bond orientational order may play a crucial role 
in crystallization, quasicrystal formation, and liquid-glass transition. 
In particular, the tensorial nature of the bond orientational order plays a crucial role in the symmetry selection upon crystallization 
and its initiation. The difference in the nature of the order parameter between the former two phenomena 
and the latter two phenomena may reflect that 
the former phenomena are liquid phenomena, which can be described by a scalar order parameter, 
whereas the latter phenomena are linked to crystallization accompanying symmetry breaking, which needs to be described by a tensorial order parameter and its spatial coherence. 
This difference together with whether there is frustration or not 
may explain why the former does not have a direct link to fluidity, whereas the order parameter responsible for glass transition 
has a direct link to fluidity, or solidity. However, the very origin of the link between glassy structural order and solidity is not perfectly clear yet, 
although a possible link has been discussed in this article. Thus we need further studies on this point. 

In glass transition, competition between bond orientational order parameter compatible 
to the crystal symmetry and that incompatible to it  
leads to strong frustration effects against crystallization. 
We show that crystal-like bond orientational order (more generally, glassy structural order) exhibits critical-like fluctuations apparently 
diverging towards the ideal glass transition point and the resulting growing activation energy may be responsible for slow glassy dynamics. Yet, the nature of glassy structural order for multi-component systems (type III systems, e.g., binary mixtures) remains elusive.  

Since our model involves many speculative arguments, 
further experimental, numerical, and theoretical studies are highly desirable 
to check the validity of this physical view.

\section*{Acknowledgements}
The author is very grateful to M. Kobayashi for her collaboration on glass transitions in water/salt mixtures, to 
R. Kurita, H. Mataki, K. Murata, and R. Shimizu  for their collaboration 
on experimental studies on liquid-liquid transitions, to 
T. Araki, A. Furukawa, T. Kawasaki, M. Leocmach, A. Mulins, C. P. Royall, H. Shintani, and K. Watanabe 
for their collaboration on glass transition, 
 and to T. Kawasaki, T. Konishi, and  J. Russo 
for their collaboration on crystallization. 
I also thank A. Furukawa for stimulating discussion on the nature of glass transition and S. F. Edwards 
for his continuous encouragements from the initial stage of this work. 
Finally, I would like to thank S. Sastry for providing a chance to write this review article  
and for his continuous encouragement and patience.  
This study was partly supported by a grant-in-aid from 
the Ministry of Education, Culture, Sports, Science and Technology, Japan (Kakenhi)
and by the Japan Society for the Promotion of
Science (JSPS) through its ``Funding Program for World-Leading
Innovative R\&D on Science and Technology (FIRST Program)''.

\end{document}